\documentclass[runningheads,a4paper]{llncs}

\usepackage{times}
\usepackage{amsmath,bm}
\usepackage{amsfonts}
\usepackage{amssymb}
\usepackage{graphicx}
\usepackage{color}
\usepackage{url}
\usepackage[algo2e,ruled,vlined,linesnumbered]{algorithm2e}
\usepackage{paralist}
\usepackage{verbatim}
\usepackage{mathtools}
\usepackage{subfigure}
\usepackage{multirow}
\usepackage{listings}
\newcommand{\om}{DC-Prophet\xspace}
\newcommand{\firstfailure}{Immediate-Reboot\xspace}
\newcommand{\secondfailure}{Slow-Reboot\xspace}
\newcommand{\thirdfailure}{Forcible-Decommission\xspace}

\newcommand{\fthree}{F\textsubscript{3}-score\xspace}
\newcommand{\add}{\lstinline{ADD}\xspace}
\newcommand{\remove}{\lstinline{REMOVE}\xspace}
\newcommand{\update}{\lstinline{UPDATE}\xspace}

\newcommand{\tf}{$t_{f}$\xspace}
\newcommand{\xrt}{$x_{r,t}$\xspace}
\newcommand{\mrt}{$m_{r,t}$\xspace}
\newcommand{\Xj}{\textbf{$x_{i}$}\xspace}
\newcommand{\yjone}{$y_{j,1}$\xspace}
\newcommand{\yjtwo}{$y_{j,2}$\xspace}
\newcommand{\ocsvm}{One-Class SVM\xspace}
\newcommand{\rf}{Random Forest\xspace}
\newcommand{\modified}[1]{{\textcolor{black}{#1}}}
\newcommand{\precision}{$Precision$\xspace}
\newcommand{\recall}{$Recall$\xspace}
\newcommand{\figcaption}{\def\@captype{figure}\caption} 
\newcommand{\tabcaption}{\def\@captype{table}\caption} 

\newcommand{\dataset}{\mathcal{D}}
\newcommand{\realnum}{\mathbb{R}}

\newcommand{\bx}{\bm{x}}

\newtheorem{observation}{Observation}

\newcommand{\hide}[1]{}
\newcommand{\reminder}[1]{{\textsf{\textcolor{red}{[#1]}}}}

\newcommand{\bit}{\begin{compactitem}}
\newcommand{\eit}{\end{compactitem}}
\newcommand{\ben}{\begin{compactenum}}
\newcommand{\een}{\end{compactenum}}

\newcommand{\ie}{\textit{i.e.}}
\newcommand{\eg}{\textit{e.g.}}

\newcommand{\tsc}{\textsuperscript}

\begin{document}

\title{DC-Prophet: Predicting Catastrophic Machine Failures in \underline{D}ata\underline{C}enters}

\author{You-Luen Lee\inst{1} \and Da-Cheng Juan\inst{2}
\and Xuan-An Tseng\inst{1} \and Yu-Ting Chen\inst{2}\and Shih-Chieh Chang\inst{1}}

\institute{Dept. of Computer Science, National Tsing Hua University, Hsinchu, Taiwan.\\    
\email{\{peggy199382, killerjack003\}@gmail.com, scchang@cs.nthu.edu.tw}
\and
Google Inc., California, USA.\\
 \email{\{dacheng, yutingchen\}@google.com}}

\authorrunning{You-Luen Lee \and Da-Cheng Juan \and Xuan-An Tseng \and Yu-Ting Chen \and Shih-Chieh Chang}

\maketitle

\begin{abstract}
When will a server fail catastrophically in an industrial datacenter? Is it possible to forecast these failures so preventive actions can be taken to increase the reliability of a datacenter? To answer these questions, we have studied what are probably the largest, publicly available datacenter traces, containing more than {\em 104 million} events from {\em 12,500} machines. Among these samples, we observe and categorize three types of machine failures, all of which are catastrophic and may lead to information loss, or even worse, reliability degradation of a datacenter. We further propose a two-stage framework\textemdash{\textbf \om}\footnote{\om stands for \textbf{D}ata\textbf{C}enter-\textbf{Prophet}}\textemdash based on One-Class Support Vector Machine and \rf. \om extracts surprising patterns and accurately predicts the next failure of a machine. Experimental results show that \om achieves an AUC of 0.93 in predicting the next machine failure, and a \fthree\footnote{The ideal value of \fthree is 1, indicating perfect predictions. Also, the intuition behind \fthree is to value ''Recall'' about three times more than ''Precision''.\cite{powers2011evaluation}} of 0.88 (out of 1). On average, \om outperforms other classical machine learning methods by 39.45\% in \fthree.
\end{abstract}

\section{Introduction} 
\label{sec:Introduction}
\textit{``When will a server fail catastrophically in an industrial datacenter?'' ``Is it possible to forecast these failures so preventive actions can be taken to increase the reliability of a datacenter?''} These two questions serve as the motivation for this work.

To meet the increasing demands for cloud computing, Internet companies such as Google, Facebook, and Amazon generally deploy a large fleet of servers in their datacenters. These servers bear heavy workloads and process various, diversified requests \cite{reiss2012heterogeneity}. For such a high-availability computing environment, when an unexpected machine failure happens upon a clustered partition, its workload is typically transferred to another machine in the same cluster, which increases the possibility of other failures as a chain effect \cite{miller2010terminating}. Also, this unexpected failure may cause (a) processed data loss, and (b) resource congestion due to machines being suddenly unavailable. In the worst case, these failures may paralyze a datacenter, causing an unplanned outage that requires a very high cost to recover \cite{2016costofdatadowntime}: on average \$9,000/minute, and up to \$17,000/minute. 
\begin{table}[tbp]
\centering
\small
\caption{Misprediction issues and the associated costs\label{tbl:fact_prediction}}
\begin{tabular}{|c||c|c|}
\hline 
\multicolumn{1}{|l||}{}                                        & \begin{tabular}[c]{@{}c@{}}Actual: failed \end{tabular}                                   & \begin{tabular}[c]{@{}c@{}}Actual: normal \end{tabular}                                                                                                     \\ 
\hline \hline
\begin{tabular}[c]{@{}c@{}}Predicted: \\ failed\end{tabular} & \begin{tabular}[c]{@{}c@{}}True Positive \\ (Correct inference)\end{tabular}              & \multicolumn{1}{l|}{\begin{tabular}[c]{@{}l@{}}False Positive: low cost \\ (e.g., extra rescheduling)\end{tabular}} \\ \hline
\begin{tabular}[c]{@{}c@{}}Predicted: \\ normal\end{tabular}  & \textbf{\begin{tabular}[c]{@{}c@{}}False Negative: high cost \\ (up to \$17,000/min)\end{tabular}} & \begin{tabular}[c]{@{}c@{}}True Negative \\ (Correct inference)\end{tabular}                                        \\ \hline
\end{tabular}
\vspace{-5mm}
\end{table}
To study machine failures in a modern datacenter, we analyze the traces from Google's datacenter \cite{clusterdata:Reiss2011}\cite{juan2014beyond}
; the traces contain more than 104 million events generated by 12,500 machines during 29 days. We observe that approximately 40\% of the machines have been removed (due to potential failures or maintenance) at least once during this period. This phenomenon suggests that potential machine failures happen quite frequently, and cannot be simply ignored. Therefore, we want to know: given the trace of a machine, can we accurately predict its next failure, ideally with low computing latency? If the answer is yes, the cloud scheduler (\eg, Borg  \cite{verma2015large} by Google) can take preventive actions to deal with incoming machine failures, such as by migrating tasks from the machine-to-fail to other machines. In this way, the cost of a machine failure is reduced to the very minimum: only the cost of task migration.

While predicting the next failure of a machine seems to be a feasible and promising solution for improving the reliability of a datacenter, it comes with two major challenges. The first challenge lies in high accuracy being required when making predictions, specifically for reducing false negatives. The false negatives (the machine actually failed but being predicted as normal) may incur a significant recovery cost \cite{2016costofdatadowntime} and should be avoided in Table~\ref{tbl:fact_prediction}. However, if the objective is set to minimize false negatives, the model will always predict a machine going to fail (so zero false negative), which introduces costs from false positives (the machine actually works but being predicted as failed). Therefore, one major challenge of designing a model is to better trade off between these two costs. The second challenge is the counts between normal events and failure events are highly imbalanced. Among 104 million events, only 8,957 events (less than 1\%) are associated with machine failures. In this case, most predictive models will trivially predict every event as normal to achieve a high accuracy (higher than 99\%). Consequently, this event-imbalance issue is the second roadblock that needs to be removed.

The contributions of this paper are as follows:
\bit
  \item We analyze probably the largest, publicly-available traces from an industrial datacenter, and categorize three types of machine failures: \firstfailure (IR), \secondfailure (SR), and \thirdfailure (FD). The frequency and duration of each type of failures categorized by our method further match experts' domain knowledge.
  \item We propose a two-stage framework: \textbf{\om} that accurately predicts the occurrence of next failure for a machine. \om first applies \ocsvm to filter out most normal cases to resolve the event-imbalance issue, and then deploys \rf to predict the type of failures that might occur for a machine. The experimental results show that \om accurately predicts machine failures and achieves an AUC of 0.93 and \fthree of 0.88, both on the test set. 
  \item To understand the effectiveness of \om, we also perform a comprehensive study on other widely-used machine learning methods, such as multi-class SVM, Logistic Regression, and Recurrent Neural Network. Experimental results show that, on average, \om outperforms other methods by 39.45\% in \fthree.
  \item Finally, we provide a practitioners' guide for using \om to predict the next failure of a machine. The latency of invoking \om to make one prediction is less 9 milliseconds (ms). Therefore, \om can be seamlessly integrated into a scheduling strategy of industrial datacenters to improve the reliability.
\eit

The remainder of this paper is organized as follows. Section \ref{sec:Problem Definition} provides the problem definition, and Section \ref{sec:Methodology} details the proposed \om framework. Section \ref{sec:Experimental Results} presents the implementation flow and experimental results, and Section \ref{sec:www at work} provides practitioners' guide.
Finally, Section \ref{sec:Conclusion} concludes this paper.

\section{Problem Definition} 
\label{sec:Problem Definition}
\subsection{Google Traces Overview}
\label{subsec:Google Traces Overview}
The Google traces \cite{clusterdata:Reiss2011} consist of the activity logs from 668,000 jobs during 29 days, and each job will spawn one or more tasks to be executed in a 12,500-machine cluster. For each machine, the traces record (a) computing resources consumed by all the tasks running on that machine, and (b) its machine state. Both resource consumption and machine states are recorded with associated time interval of one-microsecond (1 $\mu$s) resolution.

We focus on the usage measurements of six types of resources: (a) CPU usage, (b) disk I/O time, (c) disk space usage, (d) memory usage, (e) page cache, and (f) memory access per instruction. All these measurements are normalized by their respective maximum values and thus range from 0 to 1. In this work, the average and peak values during the time interval of 5 minutes are also calculated for each usage$-$the interval of 5 minutes is typically used to report the measured resource footprint of a task in Google's datacenter \cite{clusterdata:Reiss2011}. Furthermore, resource usages at minute-level provide a more macro view of a machine status \cite{guan2013adaptive}. We use $x_{r, t}$ to denote the average usage of resource type $r$ at time interval $t$; similarly, $m_{r, t}$ represents the peak usage. 
Both $x_{r, t}$ and $m_{r, t}$ are used to construct the training dataset, with further details provided in Section~\ref{subsec:Training/Test Datasets Construction}.


In addition, Google traces also contain three types of events to determine machine states: \add, \remove, and \update \cite{clusterdata:Reiss2011}. 
In this work, we treat each \remove event as an anomaly that could potentially be a machine failure. Detailed analyses are further provided in Section~\ref{subsec:Machine-Failure Analyses}.

\subsection{Problem Formulation}
\label{subsec:Problem Formulation}

The problem of predicting the next machine failure is formulated as follows:
{\problem[Categorize catastrophic failures] Given the traces of machine events, categorize the type of each machine failure at time interval $t$ (denoted as $y_t$).}
{\problem[Forecast catastrophic failures] Given the traces of resource usages\textemdash denoted as $x_{r,t}$ and $m_{r,t}$\textemdash up to time interval $\tau-1$, forecast the next failure and its type at time interval $\tau$ (denoted as $y_{\tau}$) for each machine. Mathematically, this problem can be expressed as:
\begin{equation}
	y_{\tau} = f(x_{r,t}, m_{r,t}), t = 1 \mbox{ to }\tau-1, r \in \mbox{resources}
\label{eq:original_problem_def}
\end{equation}
where $x_{r,t}$ and $m_{r,t}$ represent the respective average and peak usage of resource $r$ at time interval $t$.
}

\begin{figure}[htb]
\centering
\includegraphics[width=0.65\textwidth]{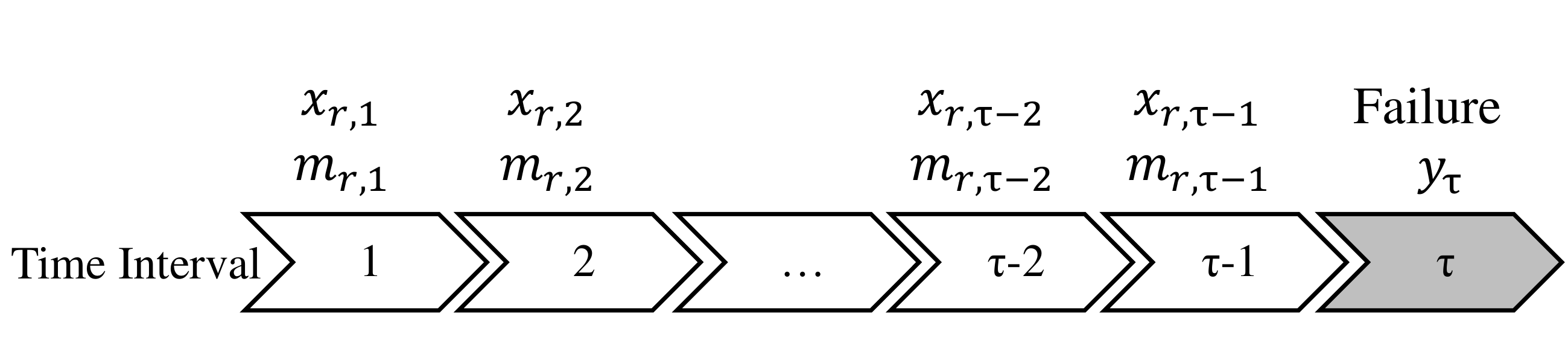}
\caption{\textit{Relationship among $y_{\tau}$, $x_{r,t}$ and $m_{r,t}$ for $t = 1$ to $\tau-1$.} 
\label{fig:eq1_description}}
\end{figure}

We use Figure~\ref{fig:eq1_description} to better illustrate the concept in Eq~\eqref{eq:original_problem_def}, specifically the temporal relationship among $y_{\tau}$, $x_{r,t}$ and $m_{r,t}$ for $t = 1$ to $\tau-1$. One goal here is to find a function $f$ that takes $x_{r,t}$ and $m_{r,t}$ as inputs to predict $y_\tau$. 

\vspace{-7mm}
\vspace{-7mm}

\subsection{Machine-Failure Analyses}
\label{subsec:Machine-Failure Analyses}
\begin{figure}[tb]
\centering
\subfigure[Rank-Frequency plot of machine failures (log-log scale).]{\includegraphics[width=0.49\textwidth]{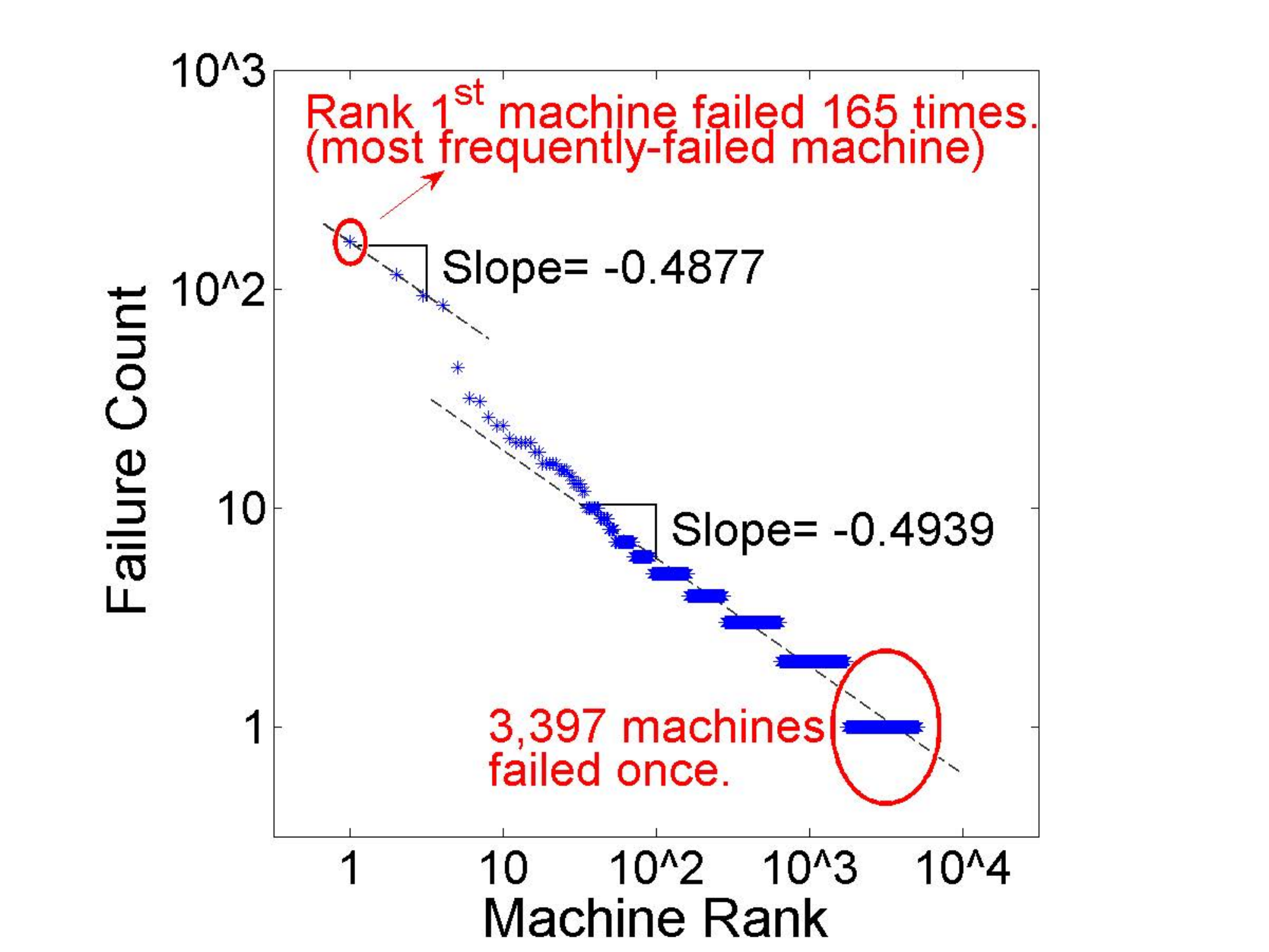}}
\subfigure[Three ``peaks'' in the distribution of failure durations (log-log scale).]{\includegraphics[width=0.49\textwidth]{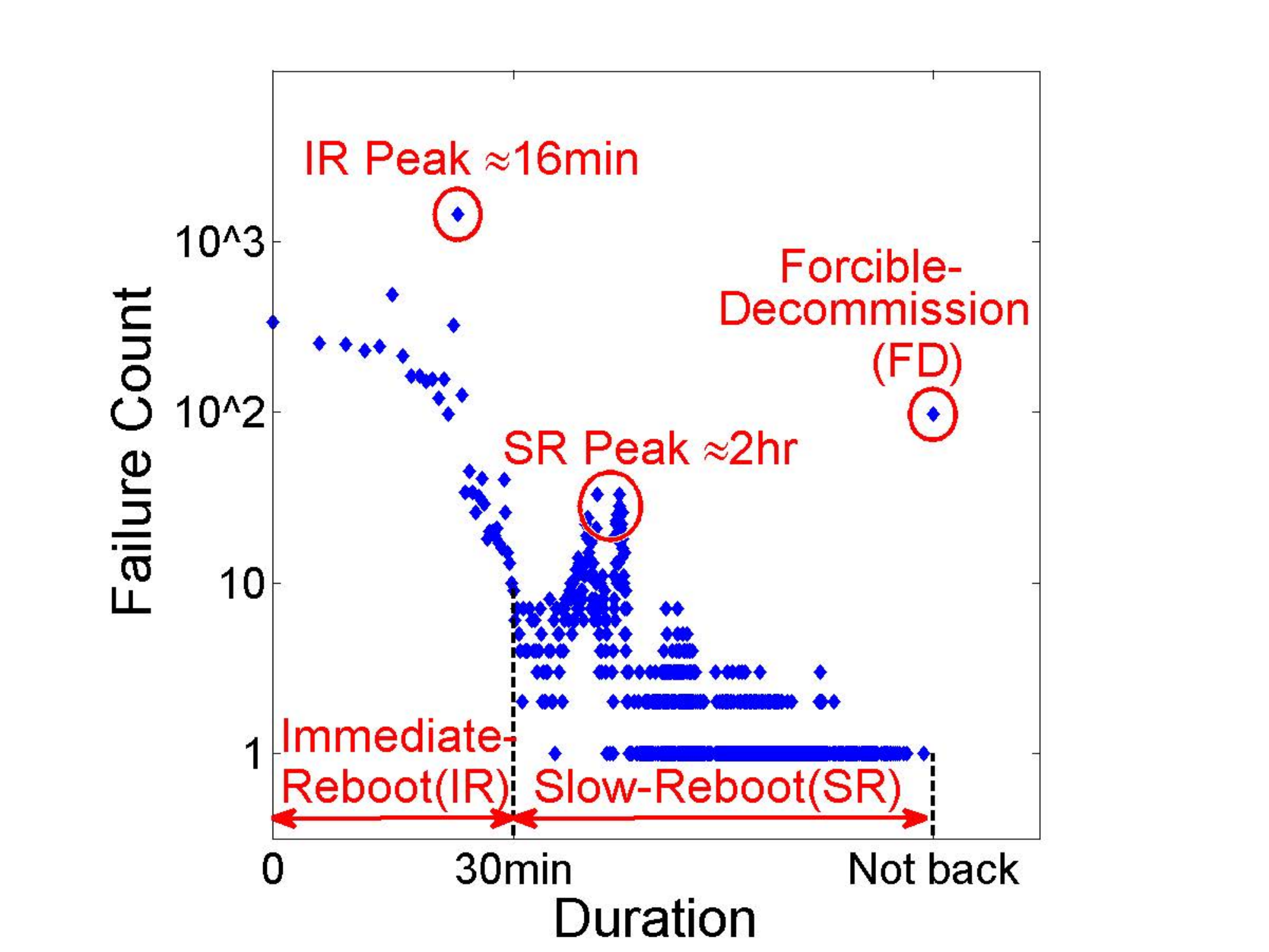}}
\vspace{-2mm}
\caption{\textit{}  (a) The x-axis represents the rank of each machine sorted based on the number of failures (high rank means more failures), whereas the y-axis is the number of failures. Both axises are in logarithmic scale. The distribution is power-law-like: three machines failed more than 100 times, whereas 3,397 machines failed only once. (b) Each dot represents the count of failures at a specific duration. The x-axis is duration and the y-axis represents the count. Both axises are in logarithmic scale. Notice the three peaks highlighted by the red circles: $\approx$ 16 minutes, $\approx$ 2 hours, and never back.
\label{fig:duration_of_failures}} 
\vspace{-3mm}
\end{figure}

Throughout the 29-day traces, we find a total of 8,957 potential machine failures from the \remove events, and Figure~\ref{fig:duration_of_failures}(a) illustrates the rank-frequency of these failures. 
The distribution is power-law-like and heavily skewed: the top-ranked machines failed more than 100 times, whereas the majority of machines (3,397 machines) failed only once. 
Overall, about 40\% (out of 12,500) machines have been removed at least once.
We further notice that the resource usages of these most frequently-failing machines are all zeros, indicating a clear abnormal behavior. 
These machines seem being marked as unavailable internally \cite{GoogleClusterData}, and hence are apparent anomalies.
They are excluded from the analysis later on.

\begin{observation}
Most frequently-failing machines have failed more than 100 times over 29 days, with usages of all resource types being zero. 
\end{observation}

To categorize the type of a failure, we further analyze its duration which is calculated by the time difference between the \remove and the following \add event. 
Figure~\ref{fig:duration_of_failures}(b) illustrates the distribution of durations for all machine failures. 
The failure duration can vary a lot, ranging from few minutes, to few hours, to never back\textemdash a machine is never added back to the cluster after its \remove event.
Furthermore, three ``peaks'' can be observed in failure durations: $\approx$16 minutes, $\approx$2 hours, and never back. 

\begin{observation}
Three ``peaks'' in the histogram of failure durations correspond to $\approx$16 minutes, $\approx$2 hours, and never back.
\end{observation}

This observation raises an intriguing question: why there are three peaks in failure durations? We correspond these three peaks ($\approx$ 16 minutes, $\approx$ 2 hours, and never back) to three types of machine failures:
\bit
  \item {\bf\firstfailure} (IR). This type of failures may occur with occasional machine errors and these machines can recover themselves in a short duration by rebooting. Here, failures of less than 30-minute downtime are categorized as IR failures \cite{barroso2013datacenter}.
  \item {\bf \secondfailure} (SR). This type of failures requires more than 30 minutes to recover. According to \cite{barroso2013datacenter}, the causes of slow reboots include file system integrity checks, machine hangs that require semiautomatic restart processes, and machine software reinstallation and testing. Also, a machine could be removed from a cluster due to system upgrades (\eg, automated kernel patching) or network down \cite{liu2012characterizing} \cite{chen2014failure}. 
  We categorize SR failures as the ones with longer than 30-minute downtime and will eventually be added back to the cluster.
  
  \item {\bf \thirdfailure} (FD). This type of failures may occur when either a machine (\eg, part of hardware) is broken and not repaired before the end of the traces, or a machine is taken out from the cluster for some reasons, such as a regular machine retirement (or called ``decommission'') \cite{GoogleClusterData}\cite{barroso2013datacenter}.
  We categorize this type of failures that a machine is removed permanently from the cluster, as FD failures.
\eit 

Among 8,771 failure events (186 obvious anomalies are removed beforehand as Observation 1 described), we summarize 5,894 to be IR failures, 2,783 SR failures, and 94 FD failures. On the other hand, there are 104,644,577 normal operations.

One important goal of this work is to predict the next failure for a machine. If a failure is mispredicted as a normal operation (a false negative), a high cost can incur. For example, the user jobs can be killed unexpectedly, leading to processed data loss. If these failures can be predicted accurately in advance, the cloud/cluster scheduler can perform preventive actions such as rescheduling jobs to another available machine to mitigate the negative impacts. Compared to the cost incurred from false  negatives, \ie, mispredicting a failure as a normal operation, the cost of ``misclassifying'' one failure type as another is relatively low. Still, if the right types of failures can be correctly predicted, the cloud/cluster scheduler can plan and arrange the computing resources accordingly. 

\subsection{Construct Training Dataset}
\label{subsec:Training/Test Datasets Construction}
\begin{figure}[tb]
\centering
\subfigure[Partial autocorrelation of CPU usage.]
{\includegraphics[width=0.40\textwidth]{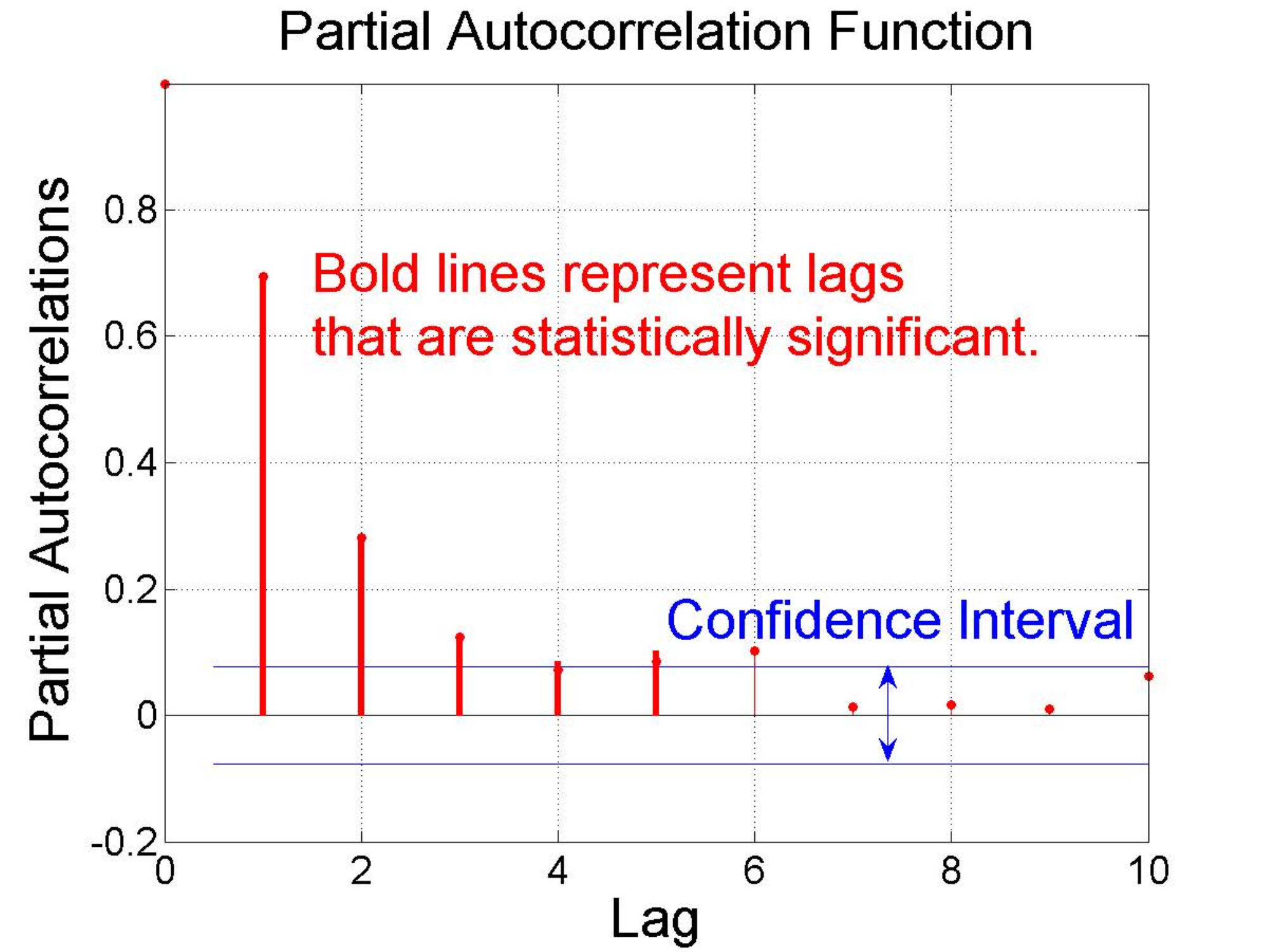}}
\subfigure[Histogram of statistically-significant partial autocorrelations on all machines.]
{\includegraphics[width=0.40\textwidth]{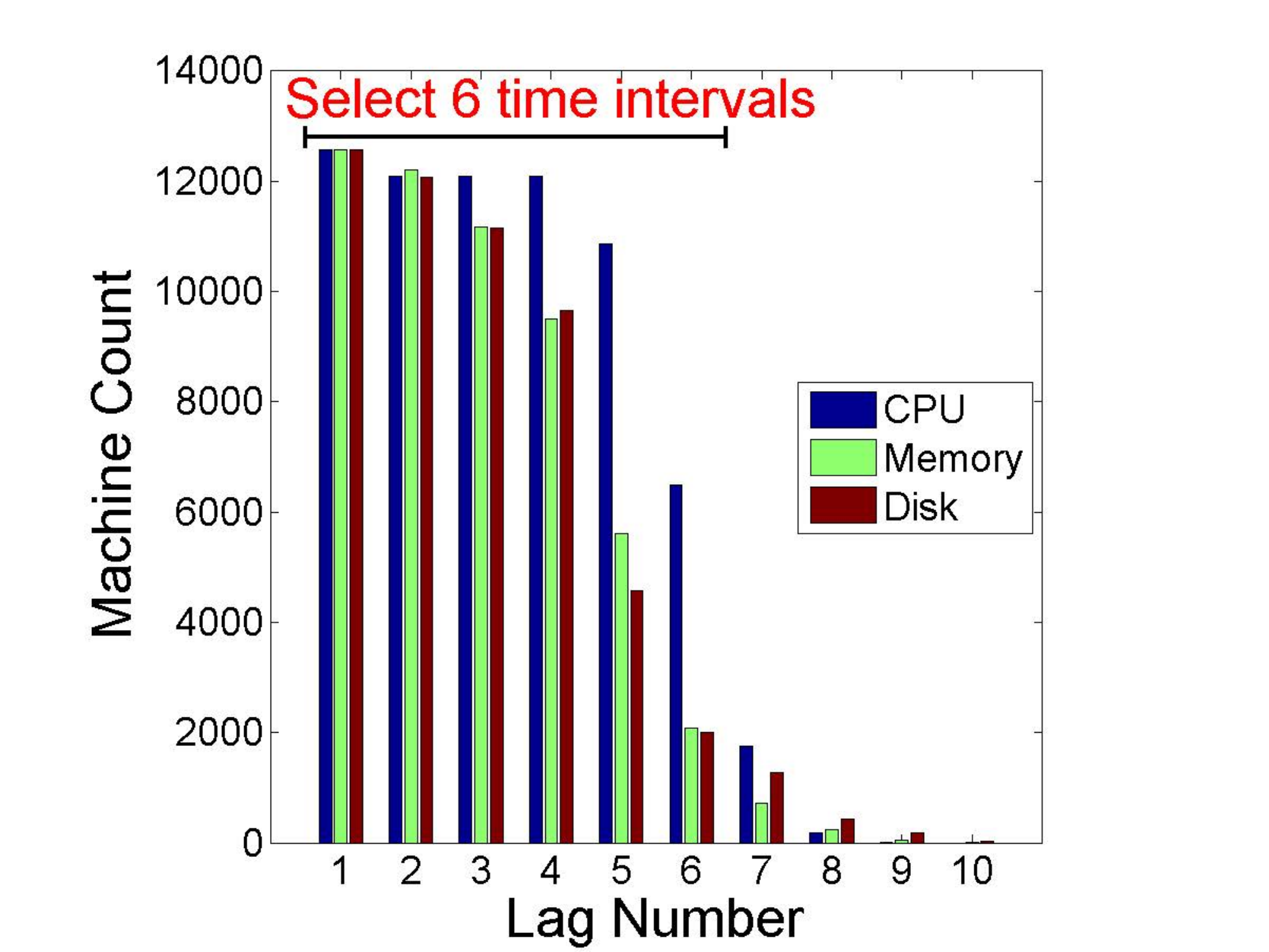}}
\vspace{-2mm}
\caption{\textit{} (a) Lags of 1, 2, 3, 5 and 6 correlate with lag 0, \ie, $x_{cpu, \tau}$, and these correlations are statistically significant. (b) For each machine, partial autocorrelations with up to 10 lags are calculated; only statistically-significant lags are reported for plotting this histogram. Notice in general, after 6 lags (or time intervals) the resource usages are less relevant\textemdash only few machines report partial autocorrelations with 6+ lags that are statistically significant.} 
\label{fig:pacf_single}
\vspace{-5mm}
\end{figure}
We model the prediction of the next machine failure from Eq~\eqref{eq:original_problem_def} as a multi-class classification and construct the training dataset accordingly. Each instance in the dataset consists of a label $y_\tau$ that represents the failure type at time interval $\tau$, and a set of predictive features $\bm{x}$ (or called a feature vector) extracted from the resource usages up to time interval $\tau-1$.

The type of a label $y_\tau$ is determined based on the failure duration described in Section~\ref{subsec:Machine-Failure Analyses}. If there is no machine failure at time interval $\tau$, label $y_\tau$ is marked as ``normal operation.'' Therefore, we defined $y_\tau \in \{0, 1, 2, 3\}$, which represents normal operation, IR, SR, and FD, respectively.   

For the predictive features $\bm{x}$, we leverage both the average $x_{r,t}$ and peak values $m_{r,t}$ of six resource types as mentioned in Section~\ref{subsec:Google Traces Overview}. Now the question is: how to select the number of time intervals needed to be included in the dataset for an accurate prediction? We propose to calculate the partial autocorrelation to determine the number of intervals, or called ``lags'' in time series, to be included in the predictive features $\bm{x}$. Assume target interval is $\tau$, the interval with ``one lag'' will be $\tau-1$ (and the interval with two lags will be $\tau-2$, etc.).  Partial autocorrelation is a type of conditional correlation between $x_{r,\tau}$ and $x_{r,t}$, with the linear dependency of $x_{r,t+1}$ to $x_{r,\tau-1}$ removed \cite{box2015time}. 
Since the partial autocorrelation can be treated as ``the correlation between $x_{r,\tau}$ and $x_{r,t}$, with other linear dependency removed,'' it suggests how many time intervals (or lags) should be included in the predictive features.

Figure~\ref{fig:pacf_single}(a) illustrates the partial autocorrelation of the CPU usage on one machine, and Figure~\ref{fig:pacf_single}(b) represents the histogram of partial autocorrelations with certain lags. Both the figures show statistical significance. Notice in general, after 6 lags (30 minutes), the resource usages are less relevant.


\begin{observation}
Resource usages from 30 minutes ago are less relevant to the current usage in terms of partial autocorrelation.
\end{observation}

Based on this observation, we include resource usages within 30 minutes as features to predict failure type $y_\tau$. In other words, 6 time intervals (lags) are selected for both $x_{r,t}$ and $m_{r,t}$ to construct the predictive features $\bm{x}_t$. Specifically, $\bm{x}_t = \{x_{r,t}, m_{r,t}\}$, $r \in \mbox{resources}$ and $t = \tau-j$ where $j = 1$ to $6$. Therefore, $\bm{x}$ has 2 (average and peak usages) $\times$ 6 (number of resources) $\times$ 6 (intervals) $=$ 72 predictive features.  

Now we have constructed the training dataset, and are ready to proceed to the proposed framework. For conciseness, in the rest of this paper each instance will be presented as $(y, \bm{x})$ instead of $(y_{\tau}, \bm{x}_t)$ with $t = \tau-1...\tau-6$.

\section{Methodology} 
\label{sec:Methodology}
\subsection{Overview: Two-Stage Framework}
\label{subsec:Two-stage Anomaly Detection}
\begin{figure}[htb]
\vspace{-7mm}
\centering
\includegraphics[width=0.65\textwidth]{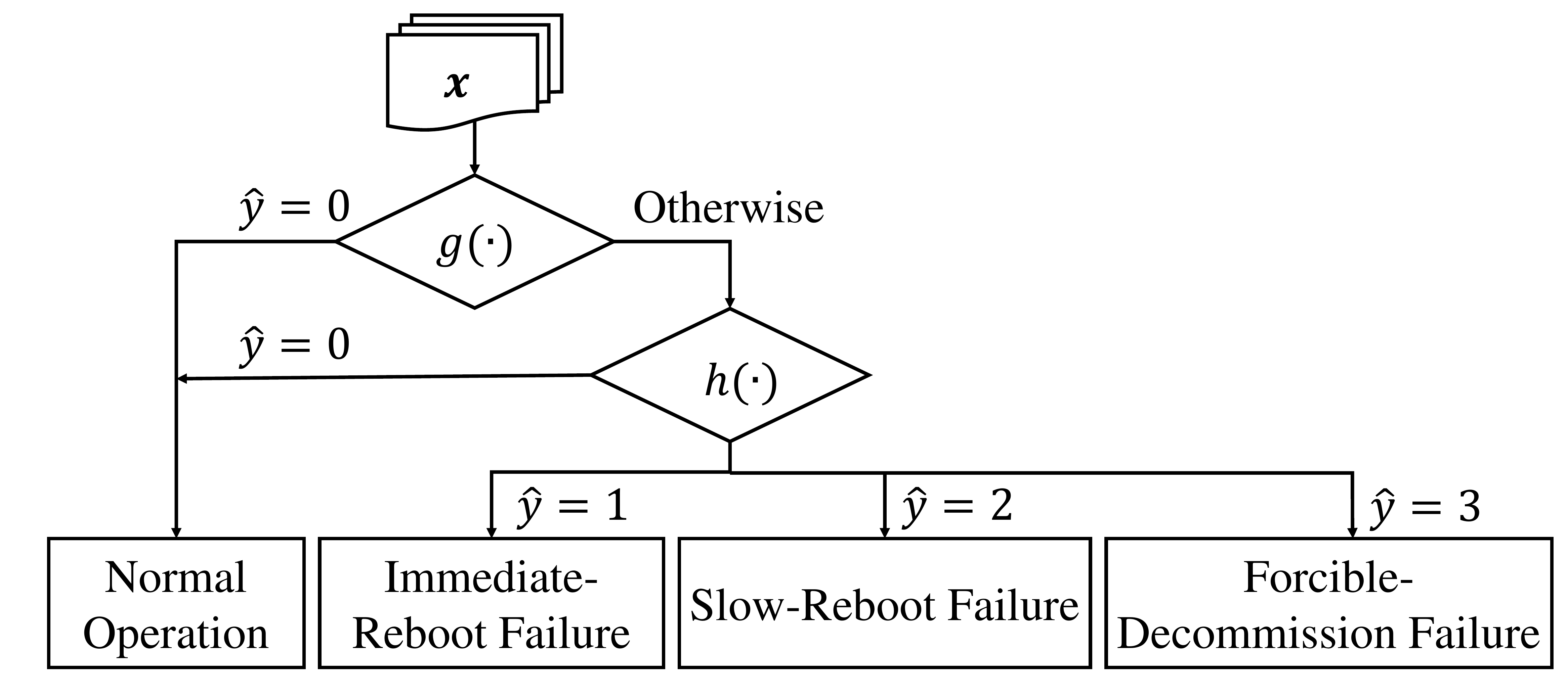}
\vspace{-2mm}
\caption{\textit{Flow Chart of \om: two-stage framework.} At the first stage, a sample $\bm{x}$ is sent to One-Class SVM $g(\cdot)$ for anomaly detection (\ie, potential machine failure or normal operation). If $\bm{x}$ is classified as a potential machine failure, then this sample will be further sent to \rf $h(\cdot)$ for multi-class (IR, SR, FD, or normal) classification.}
\label{fig:two_step_flow_chart}
\vspace{-3mm}
\end{figure}
Begin immediately, we illustrate the proposed two-stage framework with Figure~\ref{fig:two_step_flow_chart}. In the first stage, One-Class Support Vector Machine (OCSVM) is deployed for anomaly detection. All the detected anomalies are then sent to \rf for multi-class classification. Mathematically, \om can be expressed as a two-stage framework:
\begin{equation}
f(\bm{x}) =  g(\bm{x})\cdot h(\bm{x}) = \left\{\begin{array}{ll}
                 0, 	& \mbox{if $g(\bm{x}) = 0$} \\  
                 h(\bm{x}), 	& \mbox{if $g(\bm{x}) = 1$} \\  
                \end{array} \right.
 	\label{eq:fx}
\end{equation}
where $g(\cdot) \in \{0,1\}$ is OCSVM and $h(\cdot) \in \{0, 1, 2, 3\}$ is \rf. For an incoming instance $\bm{x}$, it will first be sent to $g(\cdot)$ for anomaly detection. If $\bm{x}$ is detected as an anomaly, \ie, a potential machine failure, it will be further sent to $h(\cdot)$ for multi-class classification.

In Google traces, the distribution of four label types is extremely unbalanced: 104 millions of normal cases versus 8,771 failures that are treated as anomalies (including all three types of failures). Therefore, OCSVM is applied to filter out most of normal operations and detect anomalies, \ie, potential machine failures. Without doing so, classifiers will be swamped by normal operations, learn only the ``normal behaviors,'' and choose to ignore all the failures. This will cause significant false negatives as mentioned in Table~\ref{tbl:fact_prediction} since most machine failures are mispredicted as normal operations.

\subsection{One-Class SVM}
\label{subsec:One-Class SVM}
One-class SVM (OCSVM) is often applied for novelty (or outlier) detection \cite{bishop2006pattern} and deployed as $g(\cdot)$ in \om. OCSVM is trained on instances that have only one class, which is the ``normal'' class; given a set of normal instances, OCSVM detects the soft boundary of the set, for classifying whether a new incoming instance belongs to that set (\ie, ``normal'') or not. 
Specifically, OCSVM computes a non-linear decision boundary, using appropriate kernel functions; in this work, radial basis function (RBF) kernel is used \cite{scholkopf1997comparing}. 
Eq~\eqref{eq:gx} below show how OCSVM makes an inference: 
\begin{equation}
 g(\bm{x}) =  \left\{\begin{array}{ll}
              1, 	& \mbox{$\widehat{ g}(\bm{x}) \geq 0$} \\  
              0, 	& \mbox{$\widehat{ g}(\bm{x}) < 1$} \\  
                \end{array} \right. \mbox{    where    } \hbox{  } \widehat{g}(\bm{x}) = \langle \bm{w},\phi(\bm{x}) \rangle + \rho
 	\label{eq:gx}
\end{equation}
\noindent where $\bm{w}$ and $\rho$ are learnable weights that determine the decision boundary, and the function $\phi(\cdot)$ maps the original feature(s) into a higher dimensional space, to determine the optimal decision boundary. By further modifying the hard-margin SVM to tolerate some misclassifications, we have:
\begin{align}
	\min_{\bm{w}, \bm{\rho}} \hspace{6pt}& \frac{1}{2}||\bm{w}||_2^2 + C\sum_{i}^n \xi_i -\rho \nonumber \\
	\mbox{s.t.} \hspace{6pt} & \widehat{g}(\bm{x_{i}}) = \langle \bm{w},\phi(\bm{x_{i}}) \rangle -\rho \leq \xi_i  \nonumber \\
	& \xi_i \ge 0
	\label{eq:primal_SVM_soft_margin}	
\end{align}
\noindent where $\xi_i$ represents the classification error of $i$\tsc{th} sample, and $C$ represents the weight that trades off between the maximum margin and the error-tolerance.

\subsection{Random Forest}
\label{subsec:Random Forests}
In the second stage of \om, \rf \cite{breiman2001random} is used for multi-class classification. \rf is a type of ensemble model that leverages the classification outcomes from several (say $B$) decision trees for making the final classification. In other words, \rf is an ensemble of $B$ trees \{$T_1(\bm{x})$, ..., $T_B(\bm{x})$\}, where $\bm{x}$ is the vector of predictive features described in Section~\ref{subsec:Training/Test Datasets Construction}. This ensemble of $B$ trees predicts $B$ outcomes \{$\hat y_1$ = $T_1(\bm{x})$, ..., $\hat y_B$ = $T_B(\bm{x})$\}. Then the outcomes of all trees are aggregated for majority voting, and the final prediction $\hat y$ is made based on the highest (\ie, most popular) vote. Empirically, \rf is robust to overfitting and achieves a very high accuracy.

Given a dataset of $n$ instances \{($\bm{x}_1$, $y_1$), ..., ($\bm{x}_n$, $y_n$)\}, the training procedure of \rf is as follows:
\begin{enumerate}
	\item Randomly sample the training data \{($\bm{x}_1$, $y_1$), ..., ($\bm{x}_n$, $y_n$)\}, and then draw $n$ samples to form a bootstrap batch.
    \item Grow a decision tree from the bootstrap batch using the Decision Tree Construction Algorithm \cite{bishop2006pattern}. 
    \item Repeat the above two steps until the whole ensemble of $B$ trees \{$T_1(\bm{x})$, ..., $T_B(\bm{x})$\} are grown.
\end{enumerate}

After \rf is grown, along with the OCSVM in the first stage, \om is ready for predicting the type of a machine failure.

\section{Experimental Results} 
\label{sec:Experimental Results}
\subsection{Experimental Setup}
\label{subsec:Experiment Settings}
To best compare the proposed \om with other machine learning models, we manage to search for the best hyperparameters by using 5-fold cross-validation for all the methods. Then the accuracy of each method is evaluated on the test set. All the experiments are conducted via MATLAB, running on Intel I5 processor (3.20GHz) with 16GB of RAM.

For the evaluation metrics, we report \precision, \recall, $F$-$score$, and $AUC$ (area under ROC curve) to provide a comprehensive study on the performance evaluation for different models. $F$-$score$ is defined as:
\begin{equation}
    F_{\beta} = (1 + \beta ^{2})  \frac{Precision*Recall}{(\beta ^{2} * Precision) + Recall}
    \label{eq:f_score}
\end{equation}
where $\beta$ is the parameter representing the relative importance between \recall and \precision \cite{Van-Rijsbergen79Information}. In this work, $\beta$ is selected to be 3, which means \recall is approximately three times more important than \precision. Since the false negative (machine failure mispredicted as normal event) is much more costly as mentioned in Table~\ref{tbl:fact_prediction}, \fthree is used as the main criterion to select the best framework for predicting failure types.

\subsection{Results Summary}
\label{subsec:Results Summary}

Table \ref{tbl:experiment_result} shows the experimental results from  different methods. We calculate and report \precision, \recall, \fthree and AUC for comprehensive comparisons. The results demonstrate that the two-stage algorithms have better performance on both \fthree and AUC. It also shows that using \ocsvm for anomaly detection as the first stage is necessary. Among 8,771 failures, \ocsvm only  mispredicts 11 failures as normal events, which serves as an excellent filter. Furthermore, our proposed framework, \om, which combines \ocsvm and \rf, has the best \fthree and AUC among all the two-stage methods. 

However, it seems that all the algorithms have very limited capability to recognize FD failures. One reason could be that several FD failures are found to share similar patterns with the other two failure types\textemdash IR and SR; also out of 18 FD failures in the test set, 4 failures are predicted and categorized as SR failures. We suspect that for these FD cases, the machines are eventually added back; therefore they should be categorized as SR instead of FD failures. However, the \add events occur after the end of traces. 

We also notice that by simply applying \rf algorithm, we can already achieve great results in \precision. However, our proposed \om still outperforms \rf in failure prediction, especially for the IR failures.

To evaluate the capability of \om in industrial datacenters during serving, we measure the amortized runtime of one single prediction. Table \ref{tbl:experiment_result} shows that \om only requires 8.7ms to make one prediction, which is almost negligible for most of the services in datacenters. This short latency allows the cloud scheduler to make preventive actions to deal with possible incoming machine failures. Furthermore, \om is memory efficient---only 72 features are stored for making a prediction. 


\begin{table*}[tb]
\centering
\caption{Experimental Result}
\label{tbl:experiment_result}
\begin{tabular}{|l|c|c|cccc|cccc|c|}
\hline
\multicolumn{1}{|c|}{\multirow{2}{*}{Algorithm}} & \multicolumn{1}{c|}{\multirow{2}{*}{F$_{3}$-score}} & \multicolumn{1}{c|}{\multirow{2}{*}{AUC}} & \multicolumn{4}{c|}{Precision}                                                                         & \multicolumn{4}{c|}{Recall}                                                                            & \multicolumn{1}{c|}{\multirow{2}{*}{\begin{tabular}[c]{@{}c@{}}Runtime\\ (ms)\end{tabular}}} \\ \cline{4-11}
\multicolumn{1}{|c|}{}                           & \multicolumn{1}{c|}{}                          & \multicolumn{1}{c|}{}                     & \multicolumn{1}{c}{Normal} & \multicolumn{1}{c}{IR} & \multicolumn{1}{c}{SR} & \multicolumn{1}{c|}{FD} & \multicolumn{1}{c}{Normal} & \multicolumn{1}{c}{IR} & \multicolumn{1}{c}{SR} & \multicolumn{1}{c|}{FD} & \multicolumn{1}{c|}{}                                                                        \\ \hline
\multicolumn{12}{|c|}{One-Stage Method}                                                                                                                                                                                                                                                                                                                                                                                                                                               \\ \hline
DT                              & 0.846   		  & 0.920           & 0.995           & 0.663                                      & 0.438                                      & \multicolumn{1}{c|}{0.222}                                      & 0.995           & 0.684                                      & 0.423                                      & 0.111                                            & 0.002\\
LR                              & 0.344   		  & 0.660           & 0.978           & 0.756                                      & 0.642                                      & \multicolumn{1}{c|}{0}                                          & 0.999           & 0.336                                      & 0.077                                      & 0                                                & 0.001 \\
SVM                             & 0.184   		  & 0.584           & 0.973           & 0.624                                      & 0.521                                      & \multicolumn{1}{c|}{0}                                          & 0.998           & 0.154                                      & 0.068                                      & 0                                                & 18.62 \\
RNN                             & 0.505   		  & 0.740           & 0.983           & 0.742                                      & 0.689                                      & \multicolumn{1}{c|}{0}                                          & 0.999           & 0.464                                      & 0.184                                      & 0                                                & 0.471\\ 
RF                              & 0.848   		  & 0.918           & 0.995           & 0.785                                      & 0.710                                      & \multicolumn{1}{c|}{0}                                          & 0.999           & 0.786                                      & 0.410                                      & 0                                                & 0.117\\ \hline
\multicolumn{12}{|c|}{Two-Stage Method}                                                                                                                                                                                                                                                                                                                                                                                                                                               \\ \hline
OCSVM+DT                        & 0.856   		  & 0.919           & 0.986           & 0.591                                      & 0.378                                      & \multicolumn{1}{c|}{0.046}                                      & 0.969           & 0.666                                      & 0.449                                      & 0.111                                            & 8.711\\
OCSVM+LR                        & 0.442   		  & 0.707           & 0.940           & 0.735                                      & 0.640                                      & \multicolumn{1}{c|}{0}                                          & 0.998           & 0.406                                      & 0.131                                      & 0                                                & 8.816\\
OCSVM+SVM                       & 0.202   		  & 0.591           & 0.919           & 0.654                                      & 0.519                                      & \multicolumn{1}{c|}{0}                                          & 0.996           & 0.173                                      & 0.074                                      & 0                                                & 17.46\\
OCSVM+RNN                       & 0.542   		  & 0.757           & 0.950           & 0.766                                      & 0.639                                      & \multicolumn{1}{c|}{0}                                          & 0.998           & 0.469                                      & 0.256                                      & 0                                                & 9.247\\
\textbf{OCSVM+RF}               & \textbf{0.878}  & \textbf{0.933}  & \textbf{0.986}  & \textbf{0.729}                             & \textbf{0.591}                             & \multicolumn{1}{c|}{\textbf{0.667}}                             & \textbf{0.991}  & \textbf{0.795}                             & \textbf{0.408}                             & \textbf{0.111}                                   & \textbf{8.714}\\ \hline
\end{tabular}
\vspace{-3mm}
\end{table*}
\subsection{Feature Analysis}
\label{subsec:Feature Analysis}
\begin{figure}[tb]
\centering
\subfigure[Histogram of The Number of $x_{r,t}$ Being Selected (Average-Value)]{
    \includegraphics[width=0.47\textwidth]{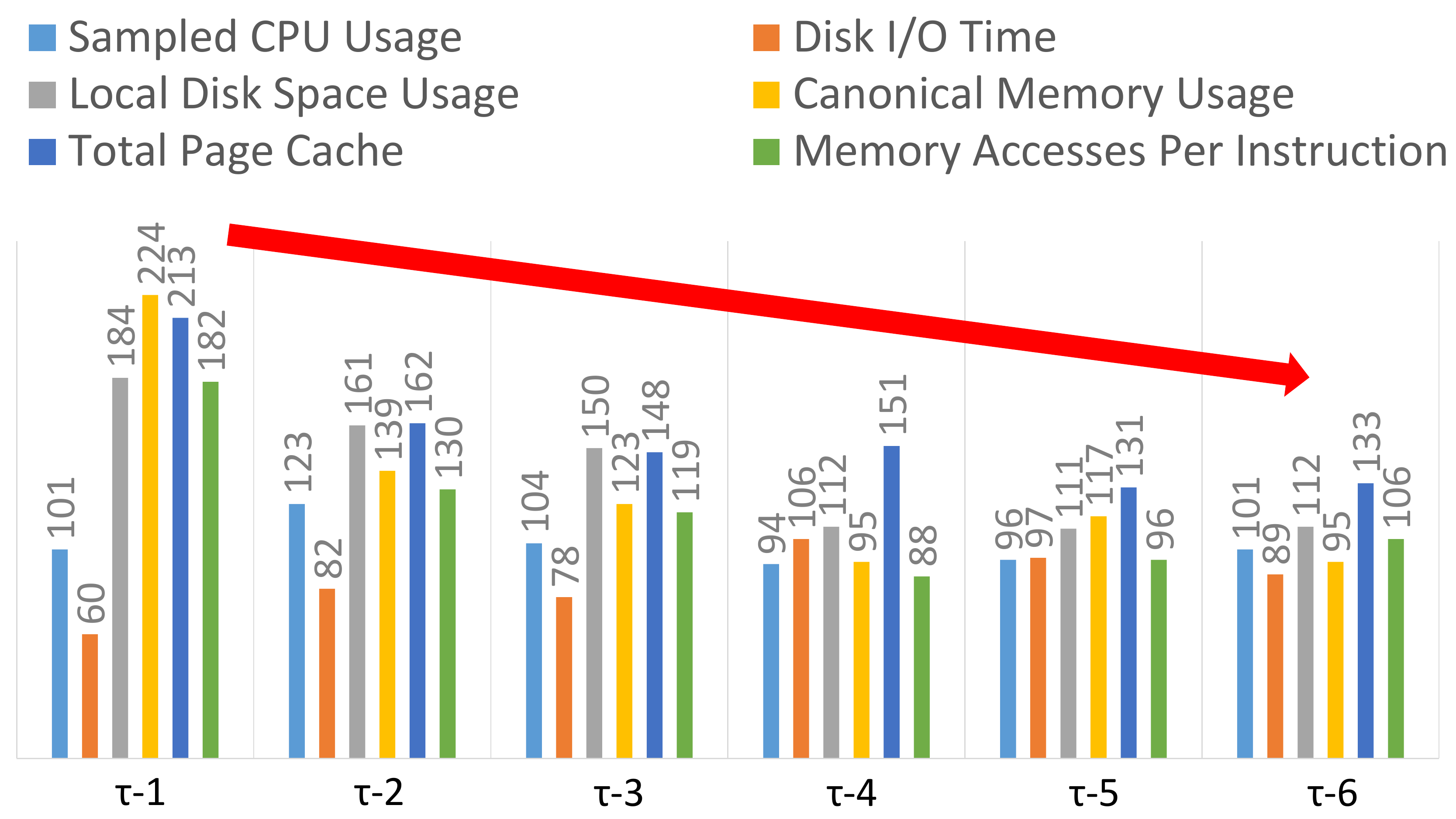}
}
\subfigure[Histogram of The Number of $m_{r,t}$ Being Selected (Peak-Value)]{
    \includegraphics[width=0.47\textwidth]{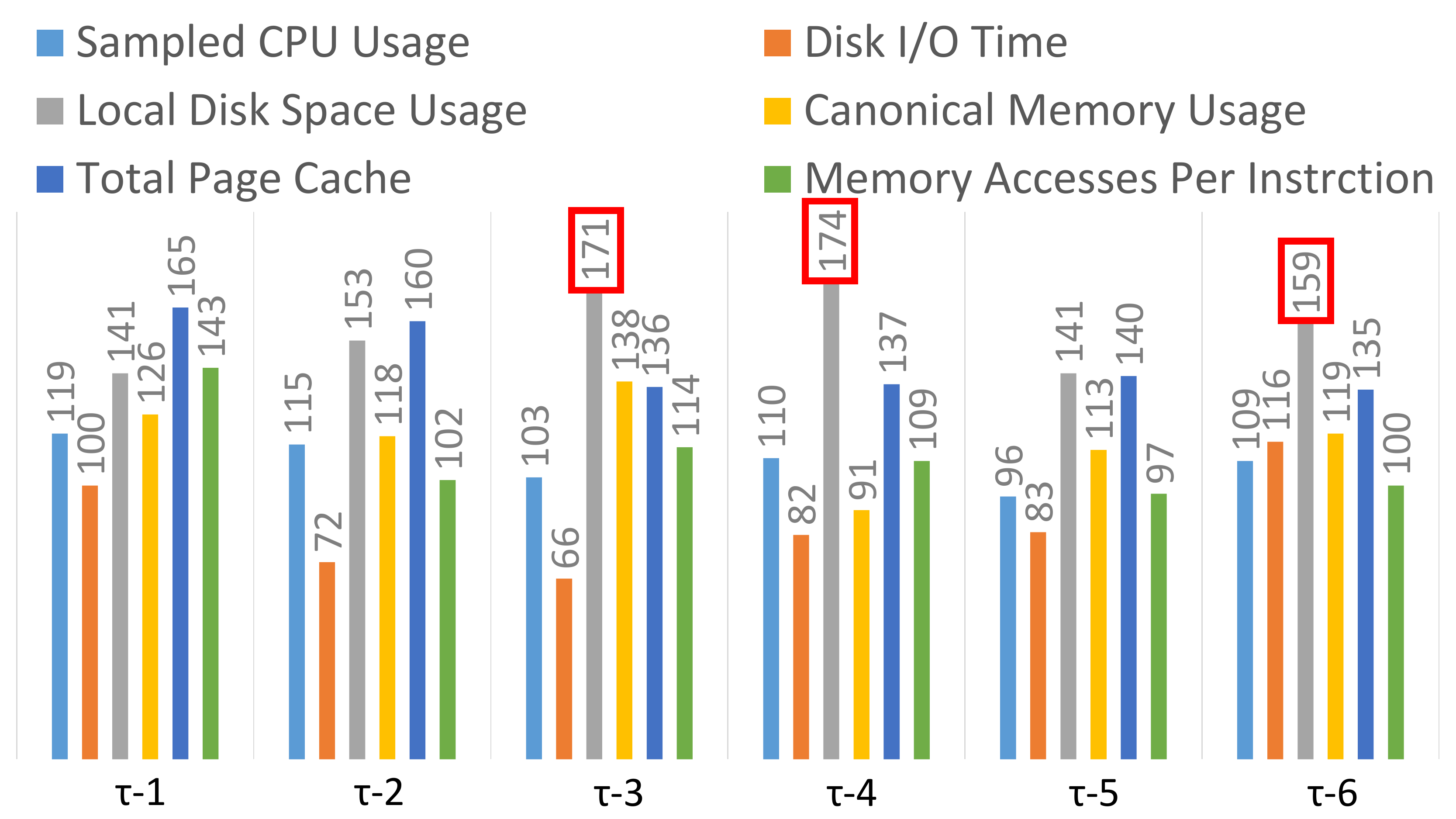}
}
\vspace{-2mm}
\caption{\textit{Counts of Features Selected by \rf}: (a) shows the number of average-value features $x_{r,t}$ being selected. We observe a trend that more recent features are more discriminative. (b) shows the number of peak-value features $m_{r,t}$ being selected.} 
\label{fig:feature_chosen_rf}
\end{figure}

Among all the predictive features, we observe several features to be more discriminative than others. Figure~\ref{fig:feature_chosen_rf} shows how many times a feature in $\bm{x}$ is selected to be split on in \rf. Figure \ref{fig:feature_chosen_rf}(a) shows the number of average-value features $x_{r,t}$ being selected in \rf while \ref{fig:feature_chosen_rf}(b) illustrates the number of peak-value features $m_{r,t}$ being selected. For average-value features, we observe a trend that recent features are more discriminative. In addition, the features related to memory usages are more discriminative than the others. 

We also discover that the number of peak-value features is more discriminative than the average-value ones in general. Furthermore, the peak-value features have similar predictive capabilities over six time intervals, as shown in Figure \ref{fig:feature_chosen_rf}(b). In addition, we observe that the peak usage of local disk is an important feature for predicting machine failures (see red circles in Figure~\ref{fig:feature_chosen_rf}(b)).

\vspace{-2mm}
\section{Practitioners' Guide}
\label{sec:www at work}
Here we provide the practitioners' guide to applying \om for forecasting machine failures in a datacenter: 
\bit
	\item \textbf{Construct training dataset:} Given the traces of machines in a datacenter, extract abnormal events representing potential machine failures, and determine their types based on the observations in Section~\ref{subsec:Machine-Failure Analyses} for obtaining label $y$. Then calculate the partial autocorrelation for each resource measurement (\eg, CPU usage, disk I/O time, etc.) to determine the number of time intervals (or lags) to be included as the predictive features $\bm{x}$. 
	\item \textbf{One-Class SVM:} After constructing the dataset of ($y$, $\bm{x}$), train OCSVM with the instances labeled as ``normal'' only, and find the best hyperparameters via grid-search and cross-validation.
	\item \textbf{\rf:} After OCSVM is trained, remove the instances detected as normal from the training dataset. Use the rest of dataset (treated as anomalies) to train \rf. Choose the number of trees in the ensemble and optimize it by cross-validation.
\eit

After both components of \om are trained, each new incoming instance will follow the flow in Figure~ \ref{fig:two_step_flow_chart} for failure prediction. Thanks to \om's low latency (8.71 milliseconds per invocation), it can be used for both (a) offline analysis in other similar datacenters, and (b) serving as a failure predictor integrated into a cloud/cluster scheduler, with training via historical data offline.


\section{Conclusion} 
\label{sec:Conclusion}
In this paper, we propose \om: a two-stage framework for forecasting machine failures. Thanks to \om, we now can answer the two motivational questions: ``When will a server fail catastrophically in an industrial datacenter?'' ``Is it possible to forecast these failures so preventive actions can be taken to increase the reliability of a datacenter?'' Experimental results show that \om accurately predicts machine failures and achieves an AUC of 0.93 and \fthree of 0.88. Finally, a practitioners' guide is provided for deploying \om to predict the next failure of a machine. The latency of invoking \om to make one prediction is less 9 milliseconds, and there can be seamlessly integrated into the scheduling strategy of industrial datacenters to improve the reliability.

%
%

@article{scholkopf1997comparing,
  title={Comparing support vector machines with Gaussian kernels to radial basis function classifiers},
  author={Scholkopf, Bernhard and Sung, Kah-Kay and Burges, Christopher JC and Girosi, Federico and Niyogi, Partha and Poggio, Tomaso and Vapnik, Vladimir},
  journal={IEEE transactions on Signal Processing},
  volume={45},
  number={11},
  pages={2758--2765},
  year={1997},
  publisher={IEEE}
}

@book{bishop2006pattern,
  title={Pattern Recognition and Machine Learning},
  author={Bishop, C.M.},
  isbn={9780387310732},
  lccn={2006922522},
  series={Information Science and Statistics},
  url={https://books.google.com.tw/books?id=kTNoQgAACAAJ},
  year={2006},
  publisher={Springer}
}

@book{breiman1984classification,
  title={Classification and regression trees},
  author={Breiman, Leo and Friedman, Jerome and Stone, Charles J and Olshen, Richard A},
  year={1984},
  publisher={CRC press}
}

@book{box2015time,
  title={Time series analysis: forecasting and control},
  author={Box, George EP and Jenkins, Gwilym M and Reinsel, Gregory C and Ljung, Greta M},
  year={2015},
  publisher={John Wiley \& Sons}
}

@misc{stat510,
 Title = {STAT510 - Applied Time Series Analysis},
 note={\url{https://onlinecourses.science.psu.edu/stat510/node/41}}
 }

@misc{GoogleClusterData,
 Title = {Google cluster data - discussions},
 year = {2011},
 note={\url{https://groups.google.com/forum/#!forum/googleclusterdata-discuss}}
 }

@article{barroso2013datacenter,
  title={The datacenter as a computer: An introduction to the design of warehouse-scale machines},
  author={Barroso, Luiz Andr{\'e} and Clidaras, Jimmy and H{\"o}lzle, Urs},
  journal={Synthesis lectures on computer architecture},
  volume={8},
  number={3},
  pages={1--154},
  year={2013},
  publisher={Morgan \& Claypool Publishers}
}

@inproceedings{guan2013adaptive,
  title={Adaptive anomaly identification by exploring metric subspace in cloud computing infrastructures},
  author={Guan, Qiang and Fu, Song},
  booktitle={Reliable Distributed Systems (SRDS), 2013 IEEE 32nd International Symposium on},
  pages={205--214},
  year={2013},
  organization={IEEE}
}

@misc{GoogleComputeEngine,
 Title = {Google Cloud Platform},
 note={\url{https://cloud.google.com}}
 }
 
@article{armbrust2010view,
  title={A view of cloud computing},
  author={Armbrust, Michael and Fox, Armando and Griffith, Rean and Joseph, Anthony D and Katz, Randy and Konwinski, Andy and Lee, Gunho and Patterson, David and Rabkin, Ariel and Stoica, Ion and others},
  journal={Communications of the ACM},
  volume={53},
  number={4},
  pages={50--58},
  year={2010},
  publisher={ACM}
}


@Misc{clusterdata:Wilkes2011,
  author = {John Wilkes},
  title = {More {Google} cluster data},
  howpublished = {Google research blog},
  month = Nov,
  year = 2011,
  note = {Posted at
	\url{http://googleresearch.blogspot.com/2011/11/more-google-cluster-data.html}.},
}

@TechReport{clusterdata:Reiss2011,
  author = {Charles Reiss and John Wilkes and Joseph L. Hellerstein},
  title = {{Google} cluster-usage traces: format + schema},
  institution = {Google Inc.},
  year = 2011,
  month = Nov,
  type = {Technical Report},
  address = {Mountain View, CA, USA},
  note = {Revised 2014-11-17 for version 2.1.  Posted at
	\url{https://github.com/google/cluster-data}},
}

@book{opac-b1127878,
   title = "The elements of statistical learning : data mining, inference, and prediction",
   author = "Hastie, Trevor J. and Tibshirani, Robert John and Friedman, Jerome H.",
   series = "Springer series in statistics",
   publisher = "Springer",
   address = "New York",
   url = "http://opac.inria.fr/record=b1127878",
   isbn = "978-0-387-84857-0",
   note = "Autres impressions : 2011 (corr.), 2013 (7e corr.)",
   year = 2009
}

@inproceedings{Du:2016:CMY:2939672.2939687,
 author = {Du, Bowen and Liu, Chuanren and Zhou, Wenjun and Hou, Zhenshan and Xiong, Hui},
 title = {Catch Me If You Can: Detecting Pickpocket Suspects from Large-Scale Transit Records},
 booktitle = {Proceedings of the 22Nd ACM SIGKDD International Conference on Knowledge Discovery and Data Mining},
 series = {KDD '16},
 year = {2016},
 isbn = {978-1-4503-4232-2},
 location = {San Francisco, California, USA},
 pages = {87--96},
 numpages = {10},
 url = {http://doi.acm.org/10.1145/2939672.2939687},
 doi = {10.1145/2939672.2939687},
 acmid = {2939687},
 publisher = {ACM},
 address = {New York, NY, USA},
 keywords = {anomaly detection, automated fare collection, mobility patterns, public safety, travel behaviors},
} 

@inproceedings{verma2015large,
  title={Large-scale cluster management at Google with Borg},
  author={Verma, Abhishek and Pedrosa, Luis and Korupolu, Madhukar and Oppenheimer, David and Tune, Eric and Wilkes, John},
  booktitle={Proceedings of the Tenth European Conference on Computer Systems},
  pages={18},
  year={2015},
  organization={ACM}
}

@article{breiman1996bagging,
  title={Bagging predictors},
  author={Breiman, Leo},
  journal={Machine learning},
  volume={24},
  number={2},
  pages={123--140},
  year={1996},
  publisher={Springer}
}

@book{breiman1984classification,
  title={Classification and regression trees},
  author={Breiman, Leo and Friedman, Jerome and Stone, Charles J and Olshen, Richard A},
  year={1984},
  publisher={CRC press}
}

@article{svetnik2003random,
  title={Random forest: a classification and regression tool for compound classification and QSAR modeling},
  author={Svetnik, Vladimir and Liaw, Andy and Tong, Christopher and Culberson, J Christopher and Sheridan, Robert P and Feuston, Bradley P},
  journal={Journal of chemical information and computer sciences},
  volume={43},
  number={6},
  pages={1947--1958},
  year={2003},
  publisher={ACS Publications}
}

@article{powers2011evaluation,
  title={Evaluation: from precision, recall and F-measure to ROC, informedness, markedness and correlation},
  author={Powers, David Martin},
  year={2011},
  publisher={Bioinfo Publications}
}

@inproceedings{tan2010predictability,
  title={On predictability of system anomalies in real world},
  author={Tan, Yongmin and Gu, Xiaohui},
  booktitle={2010 IEEE International Symposium on Modeling, Analysis and Simulation of Computer and Telecommunication Systems},
  pages={133--140},
  year={2010},
  organization={IEEE}
}
@article{murray2005machine,
  title={Machine learning methods for predicting failures in hard drives: A multiple-instance application},
  author={Murray, Joseph F and Hughes, Gordon F and Kreutz-Delgado, Kenneth},
  journal={Journal of Machine Learning Research},
  volume={6},
  number={May},
  pages={783--816},
  year={2005}
}

@inproceedings{di2012characterization,
  title={Characterization and comparison of cloud versus grid workloads},
  author={Di, Sheng and Kondo, Derrick and Cirne, Walfredo},
  booktitle={2012 IEEE International Conference on Cluster Computing},
  pages={230--238},
  year={2012},
  organization={IEEE}
}

@inproceedings{hamerly2001bayesian,
  title={Bayesian approaches to failure prediction for disk drives},
  author={Hamerly, Greg and Elkan, Charles and others},
  booktitle={ICML},
  pages={202--209},
  year={2001},
  organization={Citeseer}
}

@article{ducatch,
  title={Catch Me If You Can: Detecting Pickpocket Suspects from Large-Scale Transit Records},
  author={Du, Bowen and Liu, Chuanren and Zhou, Wenjun and Hou, Zhenshan and Xiong, Hui}
}

@article{zhang2011characterizing,
  title={Characterizing task usage shapes in google compute clusters},
  author={Zhang, Qi and Hellerstein, Joseph and Boutaba, Raouf},
  year={2011}
}

@inproceedings{khan2012workload,
  title={Workload characterization and prediction in the cloud: A multiple time series approach},
  author={Khan, Arijit and Yan, Xifeng and Tao, Shu and Anerousis, Nikos},
  booktitle={2012 IEEE Network Operations and Management Symposium},
  pages={1287--1294},
  year={2012},
  organization={IEEE}
}

@article{breiman2001random,
  title={Random forests},
  author={Breiman, Leo},
  journal={Machine learning},
  volume={45},
  number={1},
  pages={5--32},
  year={2001},
  publisher={Springer}
}

@article{hsu2003practical,
  title={A practical guide to support vector classification},
  author={Hsu, Chih-Wei and Chang, Chih-Chung and Lin, Chih-Jen and others},
  year={2003}
}

@inproceedings{kohavi1995study,
  title={A study of cross-validation and bootstrap for accuracy estimation and model selection},
  author={Kohavi, Ron and others},
  booktitle={Ijcai},
  volume={14},
  number={2},
  pages={1137--1145},
  year={1995}
}

@article{scholkopf2001estimating,
  title={Estimating the support of a high-dimensional distribution},
  author={Sch{\"o}lkopf, Bernhard and Platt, John C and Shawe-Taylor, John and Smola, Alex J and Williamson, Robert C},
  journal={Neural computation},
  volume={13},
  number={7},
  pages={1443--1471},
  year={2001},
  publisher={MIT Press}
}

@article {ASMB:ASMB537,
author = {Chen, Pai-Hsuen and Lin, Chih-Jen and Schölkopf, Bernhard},
title = {A tutorial on ν-support vector machines},
journal = {Applied Stochastic Models in Business and Industry},
volume = {21},
number = {2},
publisher = {John Wiley & Sons, Ltd.},
issn = {1526-4025},
url = {http://dx.doi.org/10.1002/asmb.537},
doi = {10.1002/asmb.537},
pages = {111--136},
keywords = {ν-support vector machines, support vector regression, support vector implementation, statistical learning theory, positive definite kernels},
year = {2005},
}

@book{Jiuyong2013Trends,
  title={Trends and Applications in Knowledge Discovery and Data Mining},
  author={Li, Jiuyong and Cao, Longbing and Wang, Can and Tan, Kay Chen and Lir, Bo and Pei, Jian and Tseng, Vincent S.},
  year={2013},
  publisher={Springer Berlin Heidelberg}
}

@article{CC01a,
 author = {Chang, Chih-Chung and Lin, Chih-Jen},
 title = {{LIBSVM}: A library for support vector machines},
 journal = {ACM Transactions on Intelligent Systems and Technology},
 volume = {2},
 issue = {3},
 year = {2011},
 pages = {27:1--27:27},
 note =	 {Software available at \url{http://www.csie.ntu.edu.tw/~cjlin/libsvm}}
}

@inproceedings{Kavulya:2010:ATP:1844765.1845224,
 author = {Kavulya, Soila and Tan, Jiaqi and Gandhi, Rajeev and Narasimhan, Priya},
 title = {An Analysis of Traces from a Production MapReduce Cluster},
 booktitle = {Proceedings of the 2010 10th IEEE/ACM International Conference on Cluster, Cloud and Grid Computing},
 series = {CCGRID '10},
 year = {2010},
 isbn = {978-0-7695-4039-9},
 pages = {94--103},
 numpages = {10},
 url = {http://dx.doi.org/10.1109/CCGRID.2010.112},
 doi = {10.1109/CCGRID.2010.112},
 acmid = {1845224},
 publisher = {IEEE Computer Society},
 address = {Washington, DC, USA},
 keywords = {MapReduce, Workload characterization, Distributed systems},
} 

 @misc{ wiki:DT,
   author = "Wikipedia",
   title = "Decision tree --- Wikipedia{,} The Free Encyclopedia",
   year = "2016",
   url = "https://en.wikipedia.org/w/index.php?title=Decision_tree&oldid=739653063",
   note = "[Online; accessed 16-September-2016]"
 }
 
 @misc{ wiki:LR,
   author = "Wikipedia",
   title = "Logistic regression --- Wikipedia{,} The Free Encyclopedia",
   year = "2016",
   url = "https://en.wikipedia.org/w/index.php?title=Logistic_regression&oldid=749013747",
   note = "[Online; accessed 11-November-2016]"
 }
 
@misc{ wiki:RT,
   author = "Wikipedia",
   title = "Random forest --- Wikipedia{,} The Free Encyclopedia",
   year = "2016",
   url = "https://en.wikipedia.org/w/index.php?title=Random_forest&oldid=747497674",
   note = "[Online; accessed 2-November-2016]"
 }
 


@inproceedings{ho1995random,
  title={Random decision forests},
  author={Ho, Tin Kam},
  booktitle={Document Analysis and Recognition, 1995., Proceedings of the Third International Conference on},
  volume={1},
  pages={278--282},
  year={1995},
  organization={IEEE}
}

@article{Ho:1998:RSM:284980.284986,
 author = {Ho, Tin Kam},
 title = {The Random Subspace Method for Constructing Decision Forests},
 journal = {IEEE Trans. Pattern Anal. Mach. Intell.},
 issue_date = {August 1998},
 volume = {20},
 number = {8},
 month = aug,
 year = {1998},
 issn = {0162-8828},
 pages = {832--844},
 numpages = {13},
 url = {http://dx.doi.org/10.1109/34.709601},
 doi = {10.1109/34.709601},
 acmid = {284986},
 publisher = {IEEE Computer Society},
 address = {Washington, DC, USA},
 keywords = {Pattern recognition, decision tree, decision forest, stochastic discrimination, decision combination, classifier combination, multiple-classifier system, bootstrapping.},
} 

@Article{Cortes1995,
author="Cortes, Corinna
and Vapnik, Vladimir",
title="Support-vector networks",
journal="Machine Learning",
year="1995",
volume="20",
number="3",
pages="273--297",
abstract="Thesupport-vector network is a new learning machine for two-group classification problems. The machine conceptually implements the following idea: input vectors are non-linearly mapped to a very high-dimension feature space. In this feature space a linear decision surface is constructed. Special properties of the decision surface ensures high generalization ability of the learning machine. The idea behind the support-vector network was previously implemented for the restricted case where the training data can be separated without errors. We here extend this result to non-separable training data.",
issn="1573-0565",
doi="10.1007/BF00994018",
url="http://dx.doi.org/10.1007/BF00994018"
}


@inproceedings{liu2012characterizing,
  title={Characterizing machines and workloads on a Google cluster},
  author={Liu, Zitao and Cho, Sangyeun},
  booktitle={2012 41st International Conference on Parallel Processing Workshops},
  pages={397--403},
  year={2012},
  organization={IEEE}
}

@misc{2016costofdatadowntime,
  title = {2016 Cost of Data Center Outages Report},
  howpublished = "\url{https://goo.gl/OeNM4U}",
}

@article{botezatu2016predicting,
  title={Predicting Disk Replacement towards Reliable Data Centers},
  author={Botezatu, Mirela and Giurgiu, Ioana and Bogojeska, Jasmina and Wiesmann, Dorothea},
  year={2016}
}
@inproceedings{chen2014failure,
  title={Failure analysis of jobs in compute clouds: A google cluster case study},
  author={Chen, Xin and Lu, Charng-Da and Pattabiraman, Karthik},
  booktitle={2014 IEEE 25th International Symposium on Software Reliability Engineering},
  pages={167--177},
  year={2014},
  organization={IEEE}
}

@inproceedings{schroeder2007disk,
  title={Disk failures in the real world: What does an mttf of 1, 000, 000 hours mean to you?},
  author={Schroeder, Bianca and Gibson, Garth A},
  booktitle={FAST},
  volume={7},
  pages={1--16},
  year={2007}
}

@misc{miller2010terminating,
  title={Terminating a non-clustered workload in response to a failure of a system with a clustered workload},
  author={Miller, Troy Don and Crawford Jr, Isom Lawrence},
  year={2010},
  month=jan # "~26",
  publisher={Google Patents},
  note={US Patent 7,653,833}
}

@article{gupta2014outlier,
  title={Outlier Detection for Temporal Data},
  author={Gupta, Manish and Gao, Jing and Aggarwal, Charu and Han, Jiawei},
  journal={Synthesis Lectures on Data Mining and Knowledge Discovery},
  volume={5},
  number={1},
  pages={1--129},
  year={2014},
  publisher={Morgan \& Claypool Publishers}
}

@article{shie2013mining,
  title={Mining interesting user behavior patterns in mobile commerce environments},
  author={Shie, Bai-En and Philip, S Yu and Tseng, Vincent S},
  journal={Applied intelligence},
  volume={38},
  number={3},
  pages={418--435},
  year={2013},
  publisher={Springer}
}

@inproceedings{das2014commerce,
  title={E-commerce product search: personalization, diversification, and beyond},
  author={Das Sarma, Atish and Parikh, Nish and Sundaresan, Neel},
  booktitle={Proceedings of the companion publication of the 23rd international conference on World wide web companion},
  pages={189--190},
  year={2014},
  organization={International World Wide Web Conferences Steering Committee}
}

@inproceedings{szpektor2013relevance,
  title={When relevance is not enough: promoting diversity and freshness in personalized question recommendation},
  author={Szpektor, Idan and Maarek, Yoelle and Pelleg, Dan},
  booktitle={Proceedings of the 22nd international conference on World Wide Web},
  pages={1249--1260},
  year={2013},
  organization={International World Wide Web Conferences Steering Committee}
}

@inproceedings{vaca2014time,
  title={A time-based collective factorization for topic discovery and monitoring in news},
  author={Vaca, Carmen K and Mantrach, Amin and Jaimes, Alejandro and Saerens, Marco},
  booktitle={Proceedings of the 23rd international conference on World wide web},
  pages={527--538},
  year={2014},
  organization={International World Wide Web Conferences Steering Committee}
}

@inproceedings{castillo2008query,
  title={Query-log mining for detecting spam},
  author={Castillo, Carlos and Corsi, Claudio and Donato, Debora and Ferragina, Paolo and Gionis, Aristides},
  booktitle={Proceedings of the 4th international workshop on Adversarial information retrieval on the web},
  pages={17--20},
  year={2008},
  organization={ACM}
}

@article{becchetti2008link,
  title={Link analysis for web spam detection},
  author={Becchetti, Luca and Castillo, Carlos and Donato, Debora and Baeza-Yates, Ricardo and Leonardi, Stefano},
  journal={ACM Transactions on the Web (TWEB)},
  volume={2},
  number={1},
  pages={2},
  year={2008},
  publisher={ACM}
}

@article{barabasi2005origin,
  title={The origin of bursts and heavy tails in human dynamics},
  author={Barabasi, Albert-Laszlo},
  journal={Nature},
  volume={435},
  number={7039},
  pages={207--211},
  year={2005},
  publisher={Nature Publishing Group}
}

@incollection{DBLP:reference/dmkdh/RatanamahatanaLGKVD10,
  author    = {Chotirat Ann Ratanamahatana and
               Jessica Lin and
               Dimitrios Gunopulos and
               Eamonn J. Keogh and
               Michail Vlachos and
               Gautam Das},
  title     = {Mining Time Series Data},
  booktitle = {Data Mining and Knowledge Discovery Handbook},
  year      = {2010},
  pages     = {1049-1077},
  ee        = {http://dx.doi.org/10.1007/978-0-387-09823-4_56},
  bibsource = {DBLP, http://dblp.uni-trier.de}
}

@inproceedings{DBLP:conf/sdm/XingPYW11,
  author    = {Zhengzheng Xing and
               Jian Pei and
               Philip S. Yu and
               Ke Wang},
  title     = {Extracting Interpretable Features for Early Classification
               on Time Series},
  booktitle = {SDM},
  year      = {2011},
  pages     = {247-258},
  ee        = {http://dx.doi.org/10.1137/1.9781611972818.22},
  bibsource = {DBLP, http://dblp.uni-trier.de}
}

@article{DBLP:journals/kais/CamerraSPRK14,
  author    = {Alessandro Camerra and
               Jin Shieh and
               Themis Palpanas and
               Thanawin Rakthanmanon and
               Eamonn J. Keogh},
  title     = {Beyond one billion time series: indexing and mining very
               large time series collections with i SAX2+},
  journal   = {Knowl. Inf. Syst.},
  volume    = {39},
  number    = {1},
  year      = {2014},
  pages     = {123-151},
  ee        = {http://dx.doi.org/10.1007/s10115-012-0606-6},
  bibsource = {DBLP, http://dblp.uni-trier.de}
}

@inproceedings{faloutsos1999power,
  title={On power-law relationships of the internet topology},
  author={Faloutsos, Michalis and Faloutsos, Petros and Faloutsos, Christos},
  booktitle={ACM SIGCOMM Computer Communication Review},
  volume={29},
  number={4},
  pages={251--262},
  year={1999},
  organization={ACM}
}

@incollection{koutra2013patterns,
  title={Patterns amongst Competing Task Frequencies: Super-Linearities, and the Almond-DG model},
  author={Koutra, Danai and Koutras, Vasileios and Prakash, B Aditya and Faloutsos, Christos},
  booktitle={Advances in Knowledge Discovery and Data Mining},
  pages={201--212},
  year={2013},
  publisher={Springer}
}

@incollection{juan2014beyond,
  title={Beyond Poisson: Modeling Inter-Arrival Time of Requests in a Datacenter},
  author={Juan, Da-Cheng and Li, Lei and Peng, Huan-Kai and Marculescu, Diana and Faloutsos, Christos},
  booktitle={Advances in Knowledge Discovery and Data Mining},
  pages={198--209},
  year={2014},
  publisher={Springer}
}

@book{schweizer2011probabilistic,
  title={Probabilistic metric spaces},
  author={Schweizer, Berthold and Sklar, Abe},
  year={2011},
  publisher={Courier Dover Publications}
}

@inproceedings{pass2006picture,
  title={A picture of search.},
  author={Pass, Greg and Chowdhury, Abdur and Torgeson, Cayley},
  booktitle={InfoScale},
  volume={152},
  pages={1},
  year={2006},
  organization={Citeseer}
}

@inproceedings{bar2007position,
  title={Position paper: Access to query logs—an academic researcher’s point of view},
  author={Bar-Ilan, Judit},
  booktitle={Query Log Analysis Workshop, WWW},
  year={2007}
}

@article{fischer1993markov,
  title={The Markov-modulated Poisson process (MMPP) cookbook},
  author={Fischer, Wolfgang and Meier-Hellstern, Kathleen},
  journal={Performance Evaluation},
  volume={18},
  number={2},
  pages={149--171},
  year={1993},
  publisher={Elsevier}
}

@article{demarta2005t,
  title={The t copula and related copulas},
  author={Demarta, Stefano and McNeil, Alexander J},
  journal={International statistical review},
  volume={73},
  number={1},
  pages={111--129},
  year={2005},
  publisher={Wiley Online Library}
}

@article{saveski2011web,
  title={Web Services for Stream Mining: A Stream-Based Active Learning Use Case},
  author={Saveski, Martin and Gr{\v{c}}ar, Miha},
  journal={ECML PKDD 2011},
  pages={36},
  year={2011}
}

@inproceedings{ihler2006adaptive,
  title={Adaptive event detection with time-varying poisson processes},
  author={Ihler, Alexander and Hutchins, Jon and Smyth, Padhraic},
  booktitle={KDD},
  pages={207--216},
  year={2006},
  organization={ACM}
}

@inproceedings{leland1993self,
  title={On the self-similar nature of Ethernet traffic},
  author={Leland, Will E and Taqqu, Murad S and Willinger, Walter and Wilson, Daniel V},
  booktitle={ACM SIGCOMM Computer Communication Review},
  volume={23},
  number={4},
  pages={183--193},
  year={1993},
  organization={ACM}
}

@article{kleinberg2003bursty,
  title={Bursty and hierarchical structure in streams},
  author={Kleinberg, Jon},
  journal={Data Mining and Knowledge Discovery},
  volume={7},
  number={4},
  pages={373--397},
  year={2003},
  publisher={Springer}
}

@inproceedings{wang2002data,
  title={Data mining meets performance evaluation: Fast algorithms for modeling bursty traffic},
  author={Wang, Mengzhi and Madhyastha, Tara and Chan, Ngai Hang and Papadimitriou, Spiros and Faloutsos, Christos},
  booktitle={ICDE},
  pages={507--516},
  year={2002},
  organization={IEEE}
}

@article{bass2004comments,
  title={Comments on ``A New Product Growth for Model Consumer Durables The Bass Model''},
  author={Bass, Frank M},
  journal={Management science},
  volume={50},
  number={12 supplement},
  pages={1833--1840},
  year={2004},
  publisher={INFORMS}
}

@inproceedings{gokhale1998log,
  title={Log-logistic software reliability growth model},
  author={Gokhale, Swapna S and Trivedi, Kishor S},
  booktitle={HASE},
  pages={34--41},
  year={1998},
  organization={IEEE}
}

@incollection{de2010surprising,
  title={Surprising patterns for the call duration distribution of mobile phone users},
  author={De Melo, Pedro OS Vaz and Akoglu, Leman and Faloutsos, Christos and Loureiro, Antonio AF},
  booktitle={ECML PKDD},
  pages={354--369},
  year={2010},
  publisher={Springer}
}

@book{lawless2011statistical,
  title={Statistical models and methods for lifetime data},
  author={Lawless, Jerald F},
  volume={362},
  year={2011},
  publisher={John Wiley \& Sons}
}

@article{bennett1983log,
  title={Log-logistic regression models for survival data},
  author={Bennett, Steve},
  journal={Applied Statistics},
  pages={165--171},
  year={1983},
  publisher={JSTOR}
}

@article{massey1951kolmogorov,
  title={The Kolmogorov-Smirnov test for goodness of fit},
  author={Massey Jr, Frank J},
  journal={JASA},
  volume={46},
  number={253},
  pages={68--78},
  year={1951},
  publisher={Taylor \& Francis Group}
}

@book{casella1990statistical,
  title={Statistical inference},
  author={Casella, George and Berger, Roger L},
  volume={70},
  year={1990},
  publisher={Duxbury Press Belmont, CA}
}

@inproceedings{vaz2013self,
  title={The self-feeding process: a unifying model for communication dynamics in the web},
  author={Vaz de Melo, Pedro Olmo S and Faloutsos, Christos and Assun{\c{c}}{\~a}o, Renato and Loureiro, Antonio},
  booktitle={WWW},
  pages={1319--1330},
  year={2013},
  organization={International World Wide Web Conferences Steering Committee}
}

@article{greenberg2008cost,
  title={The cost of a cloud: research problems in data center networks},
  author={Greenberg, Albert and Hamilton, James and Maltz, David A and Patel, Parveen},
  journal={ACM SIGCOMM Computer Communication Review},
  volume={39},
  number={1},
  pages={68--73},
  year={2008},
  publisher={ACM}
}

@inproceedings{le2010managing,
  title={Managing the cost, energy consumption, and carbon footprint of internet services},
  author={Le, Kien and Bilgir, Ozlem and Bianchini, Ricardo and Martonosi, Margaret and Nguyen, Thu D},
  booktitle={ACM SIGMETRICS Performance Evaluation Review},
  volume={38},
  number={1},
  pages={357--358},
  year={2010},
  organization={ACM}
}

@article{armbrust2010view,
  title={A view of cloud computing},
  author={Armbrust, Michael and Fox, Armando and Griffith, Rean and Joseph, Anthony D and Katz, Randy and Konwinski, Andy and Lee, Gunho and Patterson, David and Rabkin, Ariel and Stoica, Ion and others},
  journal={CACM},
  volume={53},
  number={4},
  pages={50--58},
  year={2010},
  publisher={ACM}
}

@article{barroso2009datacenter,
  title={The datacenter as a computer: An introduction to the design of warehouse-scale machines},
  author={Barroso, Luiz Andr{\'e} and H{\"o}lzle, Urs},
  journal={Synthesis Lectures on Computer Architecture},
  volume={4},
  number={1},
  pages={1--108},
  year={2009},
  publisher={Morgan \& Claypool Publishers}
}

@article{kliazovich2012greencloud,
  title={GreenCloud: a packet-level simulator of energy-aware cloud computing data centers},
  author={Kliazovich, Dzmitry and Bouvry, Pascal and Khan, Samee Ullah},
  journal={The Journal of Supercomputing},
  volume={62},
  number={3},
  pages={1263--1283},
  year={2012},
  publisher={Springer}
}

@inproceedings{reiss2012heterogeneity,
  title={Heterogeneity and dynamicity of clouds at scale: Google trace analysis},
  author={Reiss, Charles and Tumanov, Alexey and Ganger, Gregory R and Katz, Randy H and Kozuch, Michael A},
  booktitle={SOCC},
  pages={7},
  year={2012},
  organization={ACM}
}

@inproceedings{reiss2011traces,
  title={Google cluster-usage traces},
  author={Reiss, Charles and Wilkes, John and Hellerstein, Joseph L},
  booktitle={https://code.google.com/p/googleclusterdata/wiki/TraceVersion2},
  year={2012},
  organization={Google}
}

@inproceedings{zaharia2011datacenter,
  title={The datacenter needs an operating system},
  author={Zaharia, Matei and Hindman, Benjamin and Konwinski, Andy and Ghodsi, Ali and Joesph, Anthony D and Katz, Randy and Shenker, Scott and Stoica, Ion},
  booktitle={Hot-Cloud},
  pages={17--17},
  year={2011},
  organization={USENIX Association}
}

@inproceedings{kozuch2009tashi,
  title={Tashi: location-aware cluster management},
  author={Kozuch, Michael A and Ryan, Michael P and Gass, Richard and Schlosser, Steven W and O'Hallaron, David and Cipar, James and Krevat, Elie and L{\'o}pez, Julio and Stroucken, Michael and Ganger, Gregory R},
  booktitle={Proceedings of the 1st workshop on Automated control for datacenters and clouds},
  pages={43--48},
  year={2009},
  organization={ACM}
}


@article{Aho79Optimal,
   author = {A.V. Aho and J.D. Ullman},
   title = {Optimal Partial Match Retrieval When Fields are Independently Specified},
   journal = {ACM TODS},
   volume = {4},
   number = {2},
   pages = {168-179},
   month = jun,
   year = {1979},
   keywords = {seckey mah},
   contents = {(not listed)}
}

@article{Aho75Fast,
   author = {A.V. Aho and M.J. Corasick},
   title = {Fast Pattern Matching: An Aid to Bibliographic Search},
   journal = {CACM},
   volume = {18},
   number = {6},
   pages = {333-340},
   month = jun,
   year = {1975},
   keywords = {fulltext},
   contents = {(not listed)}
}

@inproceedings{Ahuja80Associative,
   author = {S.R. Ahuja and C.S. Roberts},
   title = {An Associative / Parallel Processor for Partial Match Retrieval Using Superimposed Codes},
   booktitle = {Annual Symposium on Computer Architecture},
   pages = {218-227},
   year = {1980},
   keywords = {superimposed coding hardware},
   contents = {(not listed)}
}

@article{Batcher68Sorting,
   author = {K.E. Batcher},
   title = {Sorting Networks and their Applications},
   journal = {AFIPS SJCC},
   pages = {307-314},
   year = {1968},
   keywords = {hardware},
   contents = {(not listed)}
}

@inproceedings{Bird77Associative,
   author = {R.M. Bird and J.C. Tu and R.M. Worthy},
   title = {Associative/Parallel Processors for Searching Very Large Textual Data Bases},
   booktitle = {Proc. of the 3rd ACM Workshop on Computer Architecture for Non-numeric Processing},
   pages = {8-16},
   address = {New York},
   month = may,
   year = {1977},
   keywords = {fulltext hardware},
   contents = {(not listed)}
}

@book{Bourne63Methods,
   author = {C.P. Bourne},
   title = {Methods of Information Handling},
   publisher = {Wiley},
   address = {New York},
   year = {1963},
   keywords = {text superimposed coding},
   contents = {(not listed)}
}

@article{Boyer77Fast,
   author = {R.S. Boyer and J.S. Moore},
   title = {A Fast String Searching Algorithm},
   journal = {CACM},
   volume = {20},
   number = {10},
   pages = {762-772},
   month = oct,
   year = {1977},
   keywords = {fulltext},
   contents = {(not listed)}
}

@article{Burkowski82Hardware,
   author = {F.J. Burkowski},
   title = {A Hardware Hashing Scheme in the Design of a Multiterm String Comparator},
   journal = {IEEE Trans. on Computers},
   volume = {c-31},
   number = {9},
   pages = {825-834},
   month = sep,
   year = {1982},
   keywords = {text hardware},
   contents = {(not listed)}
}

@article{Christodoulaki84Design,
   author = {S. Christodoulakis and C. Faloutsos},
   title = {Design Considerations for a Message File Server},
   journal = {IEEE Trans. on Software Engineering},
   volume = {SE-10},
   number = {2},
   pages = {201-210},
   month = mar,
   year = {1984},
   keywords = {text superimposed coding},
   contents = {(not listed)}
}

@article{Cooper70Deriving,
   author = {W.S. Cooper},
   title = {On Deriving Design Equations for Information Retrieval Systems},
   journal = {JASIS},
   pages = {385-395},
   month = {Nov.-Dec.},
   year = {1970},
   contents = {(not listed)}
}

@book{Duda73Pattern,
   author = {R.O. Duda and P.E. Hart},
   title = {Pattern Classification and Scene Analysis},
   publisher = {Wiley},
   address = {New York},
   year = {1973},
   abstract = {Classic book on pattern recognition.
Interesting points:
1) p. 66, and p. 114: Mentions the problems with dimensionality curse.
2) p. 243-246: Mentions Multidimensional scaling (MDS), Karhunen-Loeve
and dimensionality reduction.
Also, has the spiral data-set as a sample.
3) p. 333: mentions SVD/eigenvalues for linear fitting.},
   contents = {(not listed)}
}

@article{Fagin79Extendible,
   author = {R. Fagin and J. Nievergelt and N. Pippenger and H.R. Strong},
   title = {Extendible Hashing - {A} Fast Access Method for Dynamic Files},
   journal = {ACM TODS},
   volume = {4},
   number = {3},
   pages = {315-344},
   month = sep,
   year = {1979},
   keywords = {exhash primkey},
   contents = {(not listed)}
}

@techreport{Faloutsos82Extending,
   author = {C. Faloutsos},
   title = {Extending {A} {DBMS} to Handle Text},
   institution = {Univ. of Toronto},
   type = {M.Sc. thesis},
   month = oct,
   year = {1982},
   contents = {(not listed)}
}

@inproceedings{Files69Information,
   author = {J.R. Files and H.D. Huskey},
   title = {An Information Retrieval System Based on Superimposed Coding},
   booktitle = {Proc. AFIPS FJCC},
   volume = {35},
   pages = {423-432},
   year = {1969},
   keywords = {text},
   contents = {(not listed)}
}

@article{Gravina78National,
   author = {C.M. Gravina},
   title = {National Westminster Bank Mass Storage Archiving},
   journal = {IBM Systems J},
   volume = {17},
   number = {4},
   pages = {344-358},
   year = {1978},
   contents = {(not listed)}
}

@article{Hall80Approximate,
   author = {P.A.V. Hall and G.R. Dowling},
   title = {Approximate String Matching},
   journal = {ACM Computing Surveys},
   volume = {12},
   number = {4},
   pages = {381-402},
   month = dec,
   year = {1980},
   keywords = {text spelling errors},
   contents = {(not listed)}
}

@article{Harrison71Implementation,
   author = {M.C. Harrison},
   title = {Implementation of the Substring Test by Hashing},
   journal = {CACM},
   volume = {14},
   number = {12},
   pages = {777-779},
   month = dec,
   year = {1971},
   keywords = {text superimposed coding},
   contents = {(not listed)}
}

@inproceedings{Haskin82Extending,
   author = {R.L. Haskin and R.A. Lorie},
   title = {On Extending the Functions of a Relational Database System},
   booktitle = {Proc. ACM SIGMOD},
   pages = {207-212},
   address = {Orlando, Florida},
   year = {1982},
   contents = {(not listed)}
}

@article{Haskin81Special,
   author = {R.L. Haskin},
   title = {Special-Purpose Processors for Text Retrieval},
   journal = {Database Engineering},
   volume = {4},
   number = {1},
   pages = {16-29},
   month = sep,
   year = {1981},
   keywords = {hardware fulltext},
   contents = {(not listed)}
}

@article{Hollaar78Specialized,
   author = {L.A. Hollaar},
   title = {Specialized Merge Processor Networks for Combining Sorted Lists},
   journal = {ACM TODS},
   volume = {3},
   number = {3},
   pages = {272-284},
   month = sep,
   year = {1978},
   keywords = {hardware},
   contents = {(not listed)}
}

@article{Hollaar79Text,
   author = {L.A. Hollaar},
   title = {Text Retrieval Computers},
   journal = {IEEE Computer Magazine},
   volume = {12},
   number = {3},
   pages = {40-50},
   month = mar,
   year = {1979},
   keywords = {hardware},
   contents = {(not listed)}
}

@article{Haskin83Operational,
   author = {R.L. Haskin and L.A. Hollaar},
   title = {Operational Characteristics of a Hardware-Based Pattern Matcher},
   journal = {ACM TODS},
   volume = {8},
   number = {1},
   pages = {15-40},
   month = mar,
   year = {1983},
   keywords = {hardware fulltext},
   contents = {(not listed)}
}

@incollection{Hollaar83Architecture,
   author = {L.A. Hollaar and K.F. Smith and W.H. Chow and P.A. Emrath and R.L. Haskin},
   editor = {D.K. Hsiao},
   title = {Architecture and Operation of a Large, Full-Text Information-Retrieval System},
   booktitle = {Advanced Database Machine Architecture},
   pages = {256-299},
   publisher = {Prentice-Hall},
   address = {Englewood Cliffs, New Jersey},
   year = {1983},
   keywords = {hardware fulltext},
   contents = {(not listed)}
}

@book{Knuth73Art,
   author = {D.E. Knuth},
   title = {The Art of Computer Programming, Vol. 3: Sorting and Searching},
   publisher = {Addison-Wesley},
   address = {Reading, Mass},
   year = {1973},
   contents = {(not listed)}
}

@article{Larson78Dynamic,
   author = {P. Larson},
   title = {Dynamic Hashing},
   journal = {BIT},
   volume = {18},
   pages = {184-201},
   year = {1978},
   keywords = {exhash},
   contents = {(not listed)}
}

@techreport{Lesk78Some,
   author = {M.E. Lesk},
   title = {Some Applications of Inverted Indexes on the {UNIX} System},
   institution = {Bell Laboratories},
   type = {UNIX Programmer's Manual},
   volume = {2b},
     address = {Murray Hill, New Jersey},
   year = {1978},
   keywords = {inversion text},
   contents = {(not listed)}
}

@inproceedings{Litwin80Linear,
   author = {W. Litwin},
   title = {Linear Hashing: {A} new Tool for File and Table Addressing},
   booktitle = {Proc. 6th International Conference on VLDB},
   pages = {212-223},
   address = {Montreal},
   month = oct,
   year = {1980},
   keywords = {exhash},
   contents = {(not listed)}
}

@article{Luhn57Statistical,
   author = {H.P. Luhn},
   title = {A Statistical Approach to Mechanized Encoding and Searching of Literary Information},
   journal = {IBM Journal of Research and Development},
   volume = {1},
   number = {4},
   pages = {309-317},
   month = oct,
   year = {1957},
   keywords = {hashing},
   contents = {(not listed)}
}

@techreport{Martin79Spiral,
   author = {G.N.N. Martin},
   title = {Spiral Storage: Incrementally Augmentable Hash Addressed Storage},
   institution = {Univ. of Warwick, Coventry, England},
   type = {Theory of Computation, Report No.},
   number = {27},
   month = mar,
   year = {1979},
   keywords = {exhash},
   contents = {(not listed)}
}

@article{Peterson80Computer,
   author = {J.L. Peterson},
   title = {Computer Programs for Detecting and Correcting Spelling Errors},
   journal = {CACM},
   volume = {23},
   number = {12},
   pages = {676-687},
   month = dec,
   year = {1980},
   keywords = {text},
   contents = {(not listed)}
}

@article{Pfaltz80Partial,
   author = {J.L. Pfaltz and W.H. Berman and E.M. Cagley},
   title = {Partial Match Retrieval Using Indexed Descriptor Files},
   journal = {CACM},
   volume = {23},
   number = {9},
   pages = {522-528},
   month = sep,
   year = {1980},
   keywords = {mah superimposed coding},
   contents = {(not listed)}
}

@article{Van-Rijsbergen71Algorithm,
   author = {C.J. Van-Rijsbergen},
   title = {An Algorithm for Information Structuring and Retrieval},
   journal = {Computer Journal},
   volume = {14},
   number = {4},
   pages = {407-412},
   year = {1971},
   keywords = {clustering},
   contents = {(not listed)}
}

@book{Van-Rijsbergen79Information,
   author = {C.J. Van-Rijsbergen},
   title = {Information Retrieval},
   edition = {2nd},
   publisher = {Butterworths},
   address = {London, England},
   year = {1979},
   contents = {(not listed)}
}

@book{Van-Rijsbergen75Information,
   author = {C.J. Van-Rijsbergen},
   title = {Information Retrieval},
   edition = {1st},
   publisher = {Butterworths},
   address = {London, England},
   year = {1975},
   contents = {(not listed)}
}

@article{Rivest76Partial,
   author = {R.L. Rivest},
   title = {Partial Match Retrieval Algorithms},
   journal = {SIAM J. Comput},
   volume = {5},
   number = {1},
   pages = {19-50},
   month = mar,
   year = {1976},
   keywords = {mah},
   contents = {(not listed)}
}

@article{Rothnie74Attribute,
   author = {J.B. Rothnie and T. Lozano},
   title = {Attribute Based File Organization in a Paged Memory Environment},
   journal = {CACM},
   volume = {17},
   number = {2},
   pages = {63-69},
   month = feb,
   year = {1974},
   keywords = {mah},
   contents = {(not listed)}
}

@book{Salton83Introduction,
   author = {G. Salton and M.J. McGill},
   title = {Introduction to Modern Information Retrieval},
   publisher = {McGraw-Hill},
   year = {1983},
   contents = {(not listed)}
}

@article{Salton72Experiments,
   author = {G. Salton},
   title = {Experiments in Automatic Thesaurus Construction for Information Retrieval},
   journal = {Information Processing 71},
   pages = {115-123},
   publisher = {North Holland Publishing Company},
   address = {Amsterdam},
   year = {1972},
   contents = {(not listed)}
}

@article{Salton80Automatic,
   author = {G. Salton},
   title = {Automatic Information Retrieval},
   journal = {IEEE Computer Magazine},
   volume = {13},
   number = {9},
   pages = {41-56},
   month = sep,
   year = {1980},
   contents = {(not listed)}
}

@article{Severance83Practitioner,
   author = {D.G. Severance},
   title = {A Practitioner's Guide to Data Base Compression},
   journal = {Information Systems},
   volume = {8},
   number = {1},
   pages = {51-62},
   year = {1983},
   contents = {(not listed)}
}

@article{Sparck-Jones72Statistical,
   author = {K. Sparck-Jones},
   title = {A Statistical Interpretation of Term Specificity and its Application in Retrieval},
   journal = {J. of Documentation},
   volume = {28},
   number = {1},
   pages = {11-20},
   month = mar,
   year = {1972},
   keywords = {clustering},
   contents = {(not listed)}
}

@article{Stellhorn77Inverted,
   author = {W.H. Stellhorn},
   title = {An Inverted File Processor for Information Retrieval},
   journal = {IEEE Trans. on Computers},
   volume = {C-26},
   number = {12},
   pages = {1258-1267},
   month = dec,
   year = {1977},
   keywords = {text inversion hardware},
   contents = {(not listed)}
}

@inproceedings{Tsichritzis83Multimedia,
   author = {D. Tsichritzis and S. Christodoulakis and P. Economopoulos and C. Faloutsos and A. Lee and D. Lee and J. Vandenbroek and C. Woo},
   title = {A Multimedia Office Filing System},
   booktitle = {Proc. 9th International Conference on VLDB},
   address = {Florence, Italy},
   month = {Oct.-Nov.},
   year = {1983},
   keywords = {superimposed coding},
   contents = {(not listed)}
}

@article{Tsichritzis83Message,
   author = {D. Tsichritzis and S. Christodoulakis},
   title = {Message Files},
   journal = {ACM Trans. on Office Information Systems},
   volume = {1},
   number = {1},
   pages = {88-98},
   month = jan,
   year = {1983},
   keywords = {signature files},
   contents = {(not listed)}
}

@techreport{Wahl81Block,
   author = {R.M. Wahl and K.Y. Wong and R.G. Casey},
   title = {Block Segmentation and Text Extraction in Mixed Text/Image Documents},
   institution = {IBM Research Lab},
   type = {Research Report RJ3356},
   number = {(40312)},
   address = {San Jose, California},
   month = dec,
   year = {1981},
   keywords = {text},
   contents = {(not listed)}
}

@article{Zahn71Graph,
   author = {C.T. Zahn},
   title = {Graph-Theoretical Methods for Detecting and Describing Gestalt Clusters},
   journal = {IEEE Trans. on Computers},
   volume = {C-20},
   number = {1},
   pages = {68-86},
   month = jan,
   year = {1971},
   keywords = {clustering},
   contents = {(not listed)}
}

@article{Knuth77Fast,
   author = {D.E. Knuth and J.H. Morris and V.R. Pratt},
   title = {Fast Pattern Matching in Strings},
   journal = {SIAM J. Comput},
   volume = {6},
   number = {2},
   pages = {323-350},
   month = jun,
   year = {1977},
   keywords = {fulltext},
   contents = {(not listed)}
}

@article{Damerau64Technique,
   author = {F.J. Damerau},
   title = {A Technique for Computer Detection and Correction of Spelling Errors},
   journal = {CACM},
   volume = {7},
   number = {3},
   pages = {171-176},
   month = mar,
   year = {1964},
   contents = {(not listed)}
}

@article{Bourne77Frequency,
   author = {C.P Bourne},
   title = {Frequency and Impact of Spelling Errors in Bibliographic Databases},
   journal = {Information Processing and Management},
   volume = {13},
   number = {1},
   pages = {1-12},
   year = {1977},
   contents = {(not listed)}
}

@article{Lowerance75Extension,
   author = {R. Lowerance and R.A. Wagner},
   title = {An Extension of the String-to-String Correction Problem},
   journal = {JACM},
   volume = {22},
   number = {2},
   pages = {3-14},
   month = apr,
   year = {1975},
   keywords = {spelling errors detection},
   contents = {(not listed)}
}

@article{Schuegraph76Query,
   author = {E.J. Schuegraph and H.S. Heaps},
   title = {Query Processing in a  Retrospective Document  Retrieval System that Uses Word Fragments as Language Elements},
   journal = {Information Processing and Management},
   volume = {12},
   pages = {283-292},
   year = {1976},
   keywords = {text},
   contents = {(not listed)}
}

@article{Angell83Automatic,
   author = {R.C. Angell and G.E. Freund and P. Willet},
   title = {Automatic Spelling Correction Using a Trigram Similarity Measure},
   journal = {Information Processing and Management},
   volume = {19},
   number = {4},
   pages = {255-261},
   year = {1983},
   keywords = {spelling errors},
   contents = {(not listed)}
}

@article{Barton74Information,
   author = {I.J. Barton and S.E. Creasey and M.F. Lynch and M.J. Snell},
   title = {An Information-Theoretic Approach to Text Searching in Direct Access Systems},
   journal = {CACM},
   volume = {17},
   number = {6},
   pages = {345-350},
   month = jun,
   year = {1974},
   contents = {(not listed)}
}

@inproceedings{Roberts79Partial,
   author = {C.S. Roberts},
   title = {Partial-Match Retrieval via  the Method of Superimposed Codes},
   booktitle = {Proc. IEEE},
   pages = {1624-1642},
   month = dec,
   year = {1979},
   keywords = {coding},
   note = {Published as Proc. IEEE, volume 67, number 12},
   contents = {(not listed)}
}

@book{Salton71SMART,
   author = {G. Salton},
   title = {The {SMART} Retrieval System - Experiments in Automatic Document Processing},
   publisher = {Prentice-Hall Inc},
   address = {Englewood Cliffs, New Jersey},
   year = {1971},
   keywords = {book clustering text},
   contents = {(not listed)}
}

@incollection{Salton71Information,
   author = {G. Salton and M.E. Lesk},
   editor = {G. Salton},
   title = {Information Analysis and Dictionary Construction},
   booktitle = {The SMART Retrieval System - Experiments in Automatic Document Processing},
   publisher = {Prentice-Hall Inc},
   address = {Englewood Cliffs, New Jersey},
   year = {1971},
   keywords = {text},
   note = {Chapter 6},
   contents = {(not listed)}
}

@article{Severance76Practitioner,
   author = {D.G. Severance and R.A. Duhne},
   title = {A Practitioner's Guide to Addressing Algorithms},
   journal = {CACM},
   volume = {19},
   number = {6},
   pages = {314-326},
   year = {1976},
   keywords = {hashing},
   contents = {(not listed)}
}

@article{Bayer72Organization,
   author = {R. Bayer and E. McCreight},
   title = {Organization and Maintenance of Large Ordered Indexes},
   journal = {Acta Informatica},
   volume = {1},
   number = {3},
   pages = {173-189},
   year = {1972},
   keywords = {primkey trees},
   contents = {(not listed)}
}

@book{Teorey82Design,
   author = {T.J. Teorey and J.P. Fry},
   title = {Design of Database Structures},
   publisher = {Prentice-Hall Inc},
   address = {Englewood Cliffs, New Jersey},
   year = {1982},
   contents = {(not listed)}
}

@article{Lum70Multi,
   author = {V.Y. Lum},
   title = {Multi-attribute Retrieval with Combined Indexes},
   journal = {CACM},
   volume = {13},
   number = {11},
   pages = {660-665},
   month = nov,
   year = {1970},
   keywords = {seckey},
   contents = {(not listed)}
}

@inproceedings{Robinson81k,
   author = {J.T. Robinson},
   title = {The k-{D}-{B}-Tree: {A} Search Structure for Large Multidimensional Dynamic Indexes},
   booktitle = {Proc. ACM SIGMOD},
   pages = {10-18},
   year = {1981},
   keywords = {kdtrees},
   contents = {(not listed)}
}

@article{Vallarino76Use,
   author = {O. Vallarino},
   title = {On the Use of Bit Maps for  Multiple Key Retrieval},
   journal = {Conf. Proc. on Data Abstraction, Definition and Structure in ACM SIGPLAN Notices (Special issue)},
   volume = {11},
   pages = {108-114},
   month = mar,
   year = {1976},
   keywords = {superimposed coding},
   contents = {(not listed)}
}

@article{Bentley75Multidimensional,
   author = {J.L. Bentley},
   title = {Multidimensional Binary Search Trees Used for Associative Searching},
   journal = {CACM},
   volume = {18},
   number = {9},
   pages = {509-517},
   month = sep,
   year = {1975},
   keywords = {kdtrees},
   contents = {(not listed)}
}

@article{Salton83Extended,
   author = {G. Salton and E.A. Fox and H. Wu},
   title = {Extended Boolean Information Retrieval},
   journal = {CACM},
   volume = {26},
   number = {11},
   pages = {1022-1036},
   month = nov,
   year = {1983},
   keywords = {text},
   contents = {(not listed)}
}

@article{Fox84Extended,
   author = {E.A. Fox},
   title = {Extended Information Retrieval with Data and Text},
   journal = {PODS},
   year = {1984},
   note = {submitted for publication},
   contents = {(not listed)}
}

@article{Comer79Ubiquitous,
   author = {D. Comer},
   title = {The Ubiquitous {B}-Tree},
   journal = {Computing Surveys},
   volume = {11},
   number = {2},
   pages = {121-137},
   month = jun,
   year = {1979},
   keywords = {trees btree},
   contents = {(not listed)}
}

@article{Bayer77Prefix,
   author = {R. Bayer and K. Unterauer},
   title = {Prefix {B}-Trees},
   journal = {ACM TODS},
   volume = {2},
   number = {1},
   pages = {11-26},
   month = mar,
   year = {1977},
   keywords = {trees btree},
   contents = {(not listed)}
}

@article{Johnson84Approximate,
   author = {J.H. Johnson and F.W. Tompa},
   title = {Approximate String Matching in Query Languages},
   journal = {PODS},
   year = {1984},
   keywords = {spelling errors},
   note = {submitted for publication},
   contents = {(not listed)}
}

@techreport{Rabitti83Studio,
   author = {F. Rabitti and F. Romano},
   title = {Studio di parametri statistici per archivi elettronici di testi},
   institution = {IEI-CNR},
   type = {Nota tecnica},
   address = {Pisa, Italy},
   year = {1983},
   contents = {(not listed)}
}

@book{Dewey50Relative,
   author = {C. Dewey},
   title = {Relative Frequency of English Speech Sounds},
   publisher = {Harvard University Press},
   address = {Cambridge, Massachusetts},
   year = {1950},
   keywords = {text},
   contents = {(not listed)}
}

@article{Grignetti64Note,
   author = {M. Grignetti},
   title = {A Note on the Entropy of Words in Printed English},
   journal = {Information and Control},
   volume = {7},
   pages = {304-306},
   year = {1964},
   contents = {(not listed)}
}

@article{Shannon51Prediction,
   author = {C. Shannon},
   title = {Prediction and Entropy of Printed English},
   journal = {Bell System Technical Journal},
   volume = {30},
   number = {1},
   pages = {50-64},
   year = {1951},
   contents = {(not listed)}
}

@book{Zipf49Human,
   author = {G.K. Zipf},
   title = {Human Behavior and Principle of Least Effort: An Introduction to Human Ecology},
   publisher = {Addison Wesley},
   address = {Cambridge, Massachusetts},
   year = {1949},
   keywords = {compression},
   contents = {(not listed)}
}

@unpublished{Larson83method,
   author = {P.A. Larson},
   title = {A method for speeding up text retrieval},
   type = {Unpublished manuscript},
   year = {1983},
   keywords = {signature files},
   note = {Univ. of Waterloo},
   contents = {(not listed)}
}

@inproceedings{Rabitti84Evaluation,
   author = {F. Rabitti and J. Zizka},
   title = {Evaluation of Access Methods to Text Documents in Office Systems},
   booktitle = {Proc. 3rd Joint ACM-BCS Symposium on Research and Development in Information Retrieval},
   address = {Cambridge, England},
   year = {1984},
   keywords = {signature files},
   contents = {(not listed)}
}

@article{Lloyd82Partial,
   author = {J.W. Lloyd and K. Ramamohanarao},
   title = {Partial-Match Retrieval for Dynamic Files},
   journal = {BIT},
   volume = {22},
   pages = {150-168},
   year = {1982},
   keywords = {mah},
   contents = {(not listed)}
}

@article{Sebestyen66Algorithm,
   author = {Sebestyen and Edie},
   title = {An Algorithm for Non-Parametric Pattern Recognition},
   journal = {IEEE Trans. on Electronic Computers},
   volume = {EC-15},
   number = {6},
   pages = {908-914},
   month = dec,
   year = {1966},
   contents = {(not listed)}
}

@article{Specht67Polynomial,
   author = {D. Specht},
   title = {Polynomial Discriminant Functions for Pattern Recognition},
   journal = {IEEE Trans. on Electronic Computers},
   volume = {EC-16},
   pages = {308-319},
   month = jun,
   year = {1967},
   contents = {(not listed)}
}

@article{Salton78Generation,
   author = {G. Salton and A. Wong},
   title = {Generation and Search of Clustered Files},
   journal = {ACM TODS},
   volume = {3},
   number = {4},
   pages = {321-346},
   month = dec,
   year = {1978},
   keywords = {clustering},
   contents = {(not listed)}
}

@incollection{Brauen71Document,
   author = {T. Brauen},
   editor = {G. Salton},
   title = {Document Vector Modification},
   booktitle = {The SMART Retrieval System - Experiments in Automatic Document Processing},
   publisher = {Prentice-Hall Inc},
   address = {Englewood Cliffs, New Jersey},
   year = {1971},
   keywords = {clustering},
   note = {Chapter 24},
   contents = {(not listed)}
}

@article{Yu77Analysis,
   author = {C.T. Yu and W.S. Luk},
   title = {Analysis of Effectiveness of Retrieval in Clustered Files},
   journal = {JACM},
   volume = {24},
   number = {4},
   pages = {607-622},
   month = oct,
   year = {1977},
   keywords = {clustering},
   contents = {(not listed)}
}

@article{Croft83Experiments,
   author = {W.B. Croft},
   title = {Experiments with Representation in a Document Retrieval System},
   journal = {Information Technology: Research and Development},
   volume = {2},
   number = {1},
   pages = {1-21},
   month = jan,
   year = {1983},
   keywords = {clustering},
   contents = {(not listed)}
}

@article{Croft80Model,
   author = {W.B. Croft},
   title = {A Model of Cluster Searching Based on Classification},
   journal = {Information Systems},
   volume = {5},
   pages = {189-195},
   year = {1980},
   keywords = {clustering},
   contents = {(not listed)}
}

@article{Swets69Effectiveness,
   author = {J.A. Swets},
   title = {Effectiveness of Information Retrieval Methods},
   journal = {American Documentation},
   volume = {20},
   number = {1},
   pages = {41-54},
   month = jan,
   year = {1969},
   keywords = {clustering},
   contents = {(not listed)}
}

@article{Cooper68Expected,
   author = {W.S. Cooper},
   title = {Expected Search Length: {A} Single Measure of Retrieval Effectiveness Based on the Weak Ordering Action of Retrieval Systems},
   journal = {American Documentation},
   volume = {19},
   number = {1},
   pages = {30-41},
   month = jan,
   year = {1968},
   keywords = {clustering},
   contents = {(not listed)}
}

@article{Can83Clustering,
   author = {F. Can and E.A. Ozkarahan},
   title = {A Clustering Scheme},
   journal = {Sixth Annual Intern. ACM SIGIR Conf},
   pages = {115-121},
   address = {Bethesda, Maryland},
   month = jun,
   year = {1983},
   contents = {(not listed)}
}

@article{Salton73Recent,
   author = {G. Salton},
   title = {Recent Studies in Automatic Text Analysis and Document Retrieval},
   journal = {JACM},
   volume = {20},
   number = {2},
   pages = {258-278},
   month = apr,
   year = {1973},
   contents = {(not listed)}
}

@incollection{Salton71Computer,
   author = {G. Salton and M.E. Lesk},
   editor = {G. Salton},
   title = {Computer Evaluation of Indexing and Text Processing},
   booktitle = {The SMART Retrieval System - Experiments in Automatic Document Processing},
   publisher = {Prentice-Hall Inc},
   address = {Englewood Cliffs, New Jersey},
   year = {1971},
   keywords = {clustering},
   note = {Chapter 7},
   contents = {(not listed)}
}

@incollection{Salton71Comparison,
   author = {G. Salton},
   editor = {G. Salton},
   title = {A Comparison Between Manual and Automatic Indexing Methods},
   booktitle = {The SMART Retrieval System - Experiments in Automatic Document Processing},
   publisher = {Prentice-Hall Inc},
   address = {Englewood Cliffs, New Jersey},
   year = {1971},
   keywords = {clustering},
   note = {Chapter 7}
}

@incollection{Rocchio71Relevance,
   author = {J.J. Rocchio},
   editor = {G. Salton},
   title = {Relevance Feedback in Information Retrieval},
   booktitle = {The SMART Retrieval System - Experiments in Automatic Document Processing},
   publisher = {Prentice-Hall Inc},
   address = {Englewood Cliffs, New Jersey},
   year = {1971},
   note = {Chapter 14},
   contents = {(not listed)}
}

@incollection{Friedman71Relevance,
   author = {S.R. Friedman and J.A. Maceyak and S.F. Weiss},
   editor = {G. Salton},
   title = {A Relevance Feedback System Based on Document Transformation},
   booktitle = {The SMART Retrieval System - Experiments in Automatic Document Processing},
   publisher = {Prentice-Hall Inc},
   address = {Englewood Cliffs, New Jersey},
   year = {1971},
   keywords = {clustering},
   note = {Chapter 23},
   contents = {(not listed)}
}

@incollection{Rocchio71Performance,
   author = {J.J. Rocchio},
   editor = {G. Salton},
   title = {Performance Indices for Document Retrieval},
   booktitle = {The SMART Retrieval System - Experiments in Automatic Document Processing},
   publisher = {Prentice-Hall Inc},
   address = {Englewood Cliffs, New Jersey},
   year = {1971},
   keywords = {clustering},
   note = {Chapter 3},
   contents = {(not listed)}
}

@article{Lloyd80Optimal,
   author = {J.W. Lloyd},
   title = {Optimal Partial-Match Retrieval},
   journal = {BIT},
   volume = {20},
   pages = {406-413},
   year = {1980},
   keywords = {mah},
   contents = {(not listed)}
}

@article{Stiassny60Mathematical,
   author = {S. Stiassny},
   title = {Mathematical Analysis of Various Superimposed Coding Methods},
   journal = {American Documentation},
   volume = {11},
   number = {2},
   pages = {155-169},
   month = feb,
   year = {1960},
   keywords = {superimposed coding},
   contents = {(not listed)}
}

@article{Orosz56Some,
   author = {G. Orosz and L. Tackacs},
   title = {Some Probability Problems Concerning the Marking of Codes into the Superimposed Field},
   journal = {J. of Documentation},
   volume = {12},
   number = {4},
   pages = {231-234},
   month = dec,
   year = {1956},
   keywords = {superimposed coding},
   contents = {(not listed)}
}

@article{Kautz64Nonrandom,
   author = {W.H. Kautz and R.C. Singleton},
   title = {Nonrandom Binary Superimposed Codes},
   journal = {IEEE Trans. Inform. Theory},
   volume = {IT-10},
   pages = {363-377},
   month = oct,
   year = {1964},
   keywords = {superimposed coding},
   contents = {(not listed)}
}

@article{Severance76Differential,
   author = {D.G. Severance and G.M. Lohman},
   title = {Differential Files: Their Application to the Maintenance of Large Databases},
   journal = {ACM TODS},
   volume = {1},
   number = {3},
   pages = {256-267},
   month = sep,
   year = {1976},
   keywords = {superimposed coding},
   contents = {(not listed)}
}

@article{Zobrist77Detection,
   author = {A.L. Zobrist and F.R. Carlson},
   title = {Detection of Combined Occurrences},
   journal = {CACM},
   volume = {20},
   number = {1},
   pages = {31-35},
   month = jan,
   year = {1977},
   keywords = {superimposed coding},
   contents = {(not listed)}
}

@article{Yu82Term,
   author = {C.T. Yu and K. Lam and G. Salton},
   title = {Term Weighting in Information Retrieval Using the Term Precision Model},
   journal = {JACM},
   volume = {29},
   number = {1},
   pages = {152-170},
   month = jan,
   year = {1982},
   keywords = {clustering},
   contents = {(not listed)}
}

@article{McLeod81Data,
   author = {I.A. McLeod},
   title = {A Data Base Management System for Document Retrieval Applications},
   journal = {Information Systems},
   volume = {6},
   number = {2},
   pages = {131-137},
   year = {1981},
   contents = {(not listed)}
}

@article{Dattola79,
   author = {R. Dattola},
   title = {{FIRST}: Flexible Information Retrieval System for Text},
   journal = {JASIS},
   volume = {30},
   pages = {9-14},
   month = jan,
   year = {1979},
   contents = {(not listed)}
}

@inproceedings{Christodoulaki83Access,
   author = {S. Christodoulakis},
   title = {Access Files for Batching Queries in Large Information Systems},
   booktitle = {Proc. ICOD II},
   month = aug,
   year = {1983},
   keywords = {signature files},
   contents = {(not listed)}
}

@article{Fredkin60,
   author = {E. Fredkin},
   title = {{TRIE} Memory},
   journal = {CACM},
   volume = {3},
   number = {9},
   pages = {490-500},
   month = sep,
   year = {1960},
   keywords = {primkey},
   contents = {(not listed)}
}

@techreport{King82Design,
   author = {R.P. King and H.F. Korth and B.E. Willner},
   title = {Design of a Document Filing and Retrieval Service},
   institution = {IBM T.J. Watson Research Center},
   type = {Research Report RC9696 (\#42815)},
   number = {11/18/82},
   address = {Yorktown Heights, NY},
   year = {1982},
   keywords = {superimposed coding},
   contents = {(not listed)}
}

@article{Chang81Extended,
   author = {J.M. Chang and K.S. Fu},
   title = {Extended {K}-d Tree Database Organization: {A} Dynamic Multiattribute Clustering Method},
   journal = {IEEE Trans. on Software Engineering},
   volume = {SE-7},
   number = {3},
   pages = {284-290},
   month = may,
   year = {1981},
   keywords = {kdtrees},
   contents = {(not listed)}
}

@techreport{Salton82Automatic,
   author = {G. Salton and S. Buckley and E.A. Fox},
   title = {Automatic Query Formulations in Information Retrieval},
   institution = {Cornell Univ},
   type = {TR},
   number = {82-524},
   address = {Ithaca, NY},
   month = oct,
   year = {1982},
   keywords = {clustering},
   contents = {(not listed)}
}

@incollection{Salton71Relevance,
   author = {G. Salton},
   editor = {G. Salton},
   title = {Relevance Feedback and the Optimization of Retrieval Effectiveness},
   booktitle = {The SMART Retrieval System - Experiments in Automatic Document Processing},
   publisher = {Prentice-Hall Inc},
   address = {Englewood Cliffs, New Jersey},
   year = {1971},
   keywords = {clustering},
   note = {Chapter 15},
   contents = {(not listed)}
}

@book{Salton75Dynamic,
   author = {G. Salton},
   title = {Dynamic Information and Library Processing},
   publisher = {Prentice-Hall Inc},
   address = {Englewood Cliffs, N.J},
   year = {1975},
   keywords = {retrieval},
   contents = {(not listed)}
}

@book{Hopcroft79Introduction,
   author = {J.E. Hopcroft and J.D. Ullman},
   title = {Introduction to Automata Theory, Languages, and Computation},
   publisher = {Addison Wesley},
   address = {Reading, Mass.},
   year = {1979},
   contents = {(not listed)}
}

@techreport{Wing79Partial,
   author = {J.M. Wing},
   title = {Partial-Match Retrieval Using {TRIES}, Hashing and Superimposed Codes},
   institution = {MIT},
   type = {M.Sc. Thesis},
   month = jun,
   year = {1979},
   keywords = {superimposed coding text},
   contents = {(not listed)}
}

@article{Stonebraker83Document,
   author = {M. Stonebraker and H. Stettner and N. Lynn and J. Kalash and A. Guttman},
   title = {Document Processing in a Relational Database System},
   journal = {ACM Trans. on Office Information Systems},
   volume = {1},
   number = {2},
   pages = {143-158},
   month = apr,
   year = {1983},
   contents = {(not listed)}
}

@article{Severance74Identifier,
   author = {D.G. Severance},
   title = {Identifier Search Mechanisms: {A} Survey and Generalized Model},
   journal = {Computing Surveys},
   volume = {6},
   number = {3},
   pages = {175-194},
   month = sep,
   year = {1974},
   keywords = {primkey},
   contents = {(not listed)}
}

@inproceedings{Knott71Expandable,
   author = {G.D. Knott},
   title = {Expandable Open Addressing Hash Table Storage and Retrieval},
   booktitle = {Proc. SIGFIDET},
   pages = {187-206},
   address = {San Diego, Calif},
   year = {1971},
   keywords = {exhash},
   contents = {(not listed)}
}

@techreport{Gonnet82Unstructured,
   author = {G.H. Gonnet},
   title = {Unstructured Data Bases},
   institution = {Univ. of Waterloo},
   type = {Tech. Report},
   number = {CS-82-09},
   year = {1982},
   keywords = {signature files},
   contents = {(not listed)}
}

@inproceedings{Christodoulaki84Framework,
   author = {S. Christodoulakis},
   title = {A Framework for the Development of a Mixed-Mode Message System for an Office Environment},
   booktitle = {Proc. 3rd Joint ACM-BCS Symposium on Research and Development in Information Retrieval},
   address = {Cambridge, England},
   year = {1984},
   keywords = {mmm},
   contents = {(not listed)}
}

@article{Faloutsos84Signature,
   author = {C. Faloutsos and S. Christodoulakis},
   title = {Signature Files: An Access Method for Documents and its Analytical Performance Evaluation},
   journal = {ACM Trans. on Office Information Systems},
   volume = {2},
   number = {4},
   pages = {267-288},
   month = oct,
   year = {1984},
   contents = {(not listed)}
}

@inproceedings{Christodoulaki82Performance,
   author = {S. Christodoulakis and C. Faloutsos},
   title = {Performance Considerations in Message Files},
   booktitle = {Proc. CMG XIII International Conf},
   pages = {262-268},
   address = {San Diego, California},
   month = {Dec. 14-17},
   year = {1982},
   keywords = {superimposed coding},
   contents = {(not listed)}
}

@article{Sacks-Davis83Two,
   author = {R. Sacks-Davis and K. Ramamohanarao},
   title = {A Two Level Superimposed Coding Scheme for Partial Match Retrieval},
   journal = {Information Systems},
   volume = {8},
   number = {4},
   pages = {273-280},
   year = {1983},
   contents = {(not listed)}
}

@article{Yao77Approximating,
   author = {S.B. Yao},
   title = {Approximating Block Accesses in Database Organizations},
   journal = {CACM},
   volume = {20},
   number = {4},
   pages = {260-261},
   month = apr,
   year = {1977},
   keywords = {performance primkey},
   contents = {(not listed)}
}

@article{Cardenas75Analysis,
   author = {A.F. Cardenas},
   title = {Analysis and Performance of Inverted Data Base Structures},
   journal = {CACM},
   volume = {18},
   number = {5},
   pages = {253-263},
   month = may,
   year = {1975},
   keywords = {performance primkey},
   contents = {(not listed)}
}

@inproceedings{Christodoulaki83Estimating,
   author = {S. Christodoulakis},
   title = {Estimating Block Transfers and Join Sizes},
   booktitle = {Proc. ACM SIGMOD},
   pages = {40-54},
   address = {San Jose, California},
   year = {1983},
   keywords = {performance},
   contents = {(not listed)}
}

@book{Papoulis65Probability,
   author = {A. Papoulis},
   title = {Probability, Random Variables and Stochastic Processes},
   publisher = {McGraw-Hill Inc},
   year = {1965},
   contents = {(not listed)}
}

@article{Bloom70Space,
   author = {B. H. Bloom},
   title = {Space / Time Trade-offs in Hash Coding with Allowable Errors},
   journal = {CACM},
   volume = {13},
   number = {7},
   pages = {422-426},
   month = jul,
   year = {1970},
   keywords = {superimposed coding},
   contents = {(not listed)}
}

@inproceedings{Gustafson71Elements,
   author = {R. A. Gustafson},
   title = {Elements of the Randomized Combinatorial File Structure},
   booktitle = {ACM SIGIR, Proc. of the Symposium on Information Storage and Retrieval},
   pages = {163-174},
   address = {Univ. of Maryland},
   month = apr,
   year = {1971},
   keywords = {superimposed coding mah},
   contents = {(not listed)}
}

@book{Abramson63Information,
   author = {N. Abramson},
   title = {Information Theory and Coding},
   pages = {33-38},
   publisher = {McGraw-Hill},
   address = {New York},
   year = {1963},
   keywords = {infotheory},
   contents = {(not listed)}
}

@techreport{Goldberg61Multiple,
   author = {J. Goldberg and et al},
   title = {Multiple instantaneous response file},
   institution = {Stanford Res. Inst},
   type = {Final Rep. RADC-TR-61-233, ADD-266},
   number = {169},
   address = {Stanford, California},
   year = {1961},
   keywords = {superimposed coding},
   contents = {(not listed)}
}

@article{Kirstein80New,
   author = {P.T. Kirstein},
   title = {New Text and Message Services},
   journal = {in Proc. IFIP},
   pages = {521-535},
   address = {Tokyo, Melbourne},
   year = {1980},
   contents = {(not listed)}
}

@article{Shneiderman76Batched,
   author = {B. Shneiderman and V. Goodman},
   title = {Batched Searching of Sequential Files and Tree Structured Files},
   journal = {ACM TODS},
   volume = {1},
   pages = {268-275},
   month = sep,
   year = {1976},
   keywords = {performance primkey},
   contents = {(not listed)}
}

@article{Tsichritzis82Form,
   author = {D. Tsichritzis},
   title = {Form Management},
   journal = {CACM},
   volume = {25},
   pages = {453-478},
   month = jul,
   year = {1982},
   contents = {(not listed)}
}

@book{Taube57Superimposed,
   author = {M. Taube and A. Kreithan and L.B. Heilprin},
   title = {Superimposed Coding for Data Storage with an Appendix of Dropping Fraction Tables},
   volume = {4},
   publisher = {Washington, DC: Documentation, Inc.},
   year = {1957},
   note = {in Mechanization of Data Retrieval (Studies in Coordinate Indexing Series) ch. 5},
   contents = {(not listed)}
}

@book{Kleinrock75Queueing,
   author = {L. Kleinrock},
   title = {Queueing Systems, Vol. 1: Theory},
   publisher = {Wiley},
   address = {New York},
   year = {1975},
   contents = {(not listed)}
}

@article{McIlroy82Development,
   author = {M.D. McIlroy},
   title = {Development of a Spelling List},
   journal = {IEEE Trans. on Communications},
   volume = {COM-30},
   number = {1},
   pages = {91-99},
   month = jan,
   year = {1982},
   keywords = {superimposed coding},
   contents = {(not listed)}
}

@techreport{Mooers49Application,
   author = {C. Mooers},
   title = {Application of Random Codes to the Gathering of Statistical Information},
   institution = {Zator Co},
   type = {Bulletin},
   number = {31},
   address = {Cambridge, Mass},
   year = {1949},
   keywords = {superimposed coding},
   note = {based on M.S. thesis, MIT, January 1948},
   contents = {(not listed)}
}

@article{Gallager75Optimal,
   author = {R.G. Gallager and D.C. Van Voorhis},
   title = {Optimal Source Codes for Geometrically Distributed Integer Alphabets},
   journal = {IEEE Trans. on Information Theory},
   volume = {IT-21},
   pages = {228-230},
   month = mar,
   year = {1975},
   keywords = {infotheory},
   contents = {(not listed)}
}

@article{Golomb66Run,
   author = {S.W. Golomb},
   title = {Run Length Encodings},
   journal = {IEEE Trans. on Information Theory},
   volume = {IT-12},
   pages = {399-401},
   month = jul,
   year = {1966},
   keywords = {infotheory},
   contents = {(not listed)}
}

@article{Fujitani84Laser,
   author = {L. Fujitani},
   title = {Laser Optical Disk: The Coming Revolution in On-Line Storage},
   journal = {CACM},
   volume = {27},
   number = {6},
   pages = {546-554},
   month = jun,
   year = {1984},
   contents = {(not listed)}
}

@inproceedings{Faloutsos85Signature,
   author = {C. Faloutsos},
   title = {Signature Files: Design and Performance Comparison of some Signature Extraction Methods},
   booktitle = {Proc. ACM SIGMOD},
   pages = {63-82},
   address = {Austin, Texas},
   month = may,
   year = {1985},
   contents = {(not listed)}
}

@article{Faloutsos87Optimal,
   author = {C. Faloutsos and S. Christodoulakis},
   title = {Optimal Signature Extraction and Information Loss},
   journal = {ACM TODS},
   volume = {12},
   number = {3},
   pages = {395-428},
   month = sep,
   year = {1987},
   keywords = {signature files infotheory},
   contents = {(not listed)}
}

@article{Faloutsos85Access,
   author = {C. Faloutsos},
   title = {Access Methods for Text},
   journal = {ACM Computing Surveys},
   volume = {17},
   number = {1},
   pages = {49-74},
   month = mar,
   year = {1985},
   contents = {(not listed)}
}

@article{Standish84Essay,
   author = {T.A. Standish},
   title = {An Essay on Software Reuse},
   journal = {IEEE Trans. on Software Engineering},
   volume = {SE-10},
   number = {5},
   pages = {494-497},
   month = sep,
   year = {1984},
   keywords = {information retrieval for software components},
   contents = {(not listed)}
}

@book{Feller68Introduction,
   author = {W. Feller},
   title = {An Introduction to Probability Theory and Its Applications},
   edition = {3rd},
   publisher = {John Wiley and Sons},
   year = {1968},
   contents = {(not listed)}
}

@inproceedings{Faloutsos85Design,
   author = {C. Faloutsos and S. Christodoulakis},
   title = {Design of a Signature File Method that Accounts for Non-Uniform Occurrence and Query Frequencies},
   booktitle = {Proc. 11th International Conference on VLDB},
   pages = {165-170},
   address = {Stockholm, Sweden},
   month = aug,
   year = {1985},
   keywords = {files},
   contents = {(not listed)}
}

@article{Nievergelt84Grid,
   author = {J. Nievergelt and H. Hinterberger and K.C. Sevcik},
   title = {The Grid File: An Adaptable, Symmetric Multikey File Structure},
   journal = {ACM TODS},
   volume = {9},
   number = {1},
   pages = {38-71},
   month = mar,
   year = {1984},
   keywords = {mah},
   contents = {(not listed)}
}

@inproceedings{De-la-Briandai59File,
   author = {S.R. De-la-Briandais},
   title = {File Searching Using Variable Length Keys},
   booktitle = {Proc. WJCC},
   pages = {295-298},
   year = {1959},
   keywords = {primkey tries},
   contents = {(not listed)}
}

@article{Knott75Hashing,
   author = {G.D. Knott},
   title = {Hashing Functions},
   journal = {Computer Journal},
   volume = {18},
   number = {3},
   pages = {265-278},
   year = {1975},
   contents = {(not listed)}
}

@book{Marshall79Inequalities,
   author = {A.W. Marshall and I. Olkin},
   title = {Inequalities: Theory of Majorization and Its Applications},
   publisher = {Academic Press Inc},
   address = {New York},
   year = {1979},
   contents = {(not listed)}
}

@techreport{Johnson83Formal,
   author = {J.H. Johnson},
   title = {Formal Models for String Similarity},
   institution = {Univ. of Waterloo},
   type = {Research Report},
   number = {CS-83-32},
   address = {Waterloo, Ontario, Canada},
   month = nov,
   year = {1983},
   keywords = {errors spelling},
   note = {Ph.D. Dissertation},
   contents = {(not listed)}
}

@book{Beyer79Standard,
   author = {W.H. Beyer},
   title = {Standard Mathematical Tables},
   edition = {25th},
   publisher = {CRC Press},
   address = {Florida},
   year = {1979},
   keywords = {Chemical Rubber Company},
   contents = {(not listed)}
}

@article{Solomon82CSNET,
   author = {M. Solomon and L. Landweber and D. Neuhengen},
   title = {The {CSNET} Name Server},
   journal = {Computer Networks},
   volume = {6},
   number = {3},
   pages = {161-172},
   month = jul,
   year = {1982},
   contents = {(not listed)}
}

@techreport{Mockapetris84Domain,
   author = {P. Mockapetris},
   title = {The Domain Name Server},
   institution = {Univ. of Southern California / Information Science Institute},
   type = {ISI/RS-84-133},
   month = jun,
   year = {1984},
   contents = {(not listed)}
}

@article{Larson82Performance,
   author = {P.A. Larson},
   title = {Performance Analysis of Linear Hashing with Partial Expansions},
   journal = {ACM TODS},
   volume = {7},
   number = {4},
   pages = {566-587},
   month = dec,
   year = {1982},
   keywords = {exhash},
   contents = {(not listed)}
}

@book{MacKay69Information,
   author = {D. M. MacKay},
   title = {Information, Mechanism and Meaning},
   publisher = {The MIT Press},
   address = {Cambridge, Mass.},
   year = {1969},
   contents = {(not listed)}
}

@article{Welch84Technique,
   author = {T.A. Welch},
   title = {A Technique for High-Performance Data Compression},
   journal = {IEEE Computer Magazine},
   volume = {17},
   number = {6},
   pages = {8-19},
   month = jun,
   year = {1984},
   contents = {(not listed)}
}

@article{Ziv77Universal,
   author = {J. Ziv and A. Lempel},
   title = {A Universal Algorithm for Sequential Data Compression},
   journal = {IEEE Trans. Information Theory},
   volume = {IT-23},
   number = {3},
   pages = {337-343},
   month = may,
   year = {1977},
   contents = {(not listed)}
}

@book{Bogart83Introductory,
   author = {K.P. Bogart},
   title = {Introductory Combinatorics},
   publisher = {Pitman Publishing Inc},
   year = {1983},
   contents = {(not listed)}
}

@book{Reingold77Combinatorial,
   author = {E.M. Reingold and J. Nievergelt and N. Deo},
   title = {Combinatorial Algorithms: Theory and Practice},
   publisher = {Prentice-Hall Inc},
   address = {Englewood Cliffs, New Jersey},
   year = {1977},
   contents = {(not listed)}
}

@book{Hardy52Inequalities,
   author = {G.H. Hardy and J.E. Littlewood and G. Polya},
   title = {Inequalities},
   edition = {Second},
   publisher = {Cambridge University Press},
   year = {1952},
   contents = {(not listed)}
}

@book{Ballard82Computer,
   author = {D. Ballard and C. Brown},
   title = {Computer Vision},
   publisher = {Prentice Hall},
   year = {1982},
   contents = {(not listed)}
}

@inproceedings{Christodoulaki80,
   author = {S. Christodoulakis},
   title = {{IPRL}: An interactive Pattern Recognition Laboratory},
   booktitle = {Proc. ACM SIGCSE},
   year = {1980},
   contents = {(not listed)}
}

@article{Christodoulaki84Implication,
   author = {S. Christodoulakis},
   title = {Implication of Certain Assumptions in Data Base Performance Evaluation},
   journal = {ACM TODS},
   month = jun,
   year = {1984},
   keywords = {performance},
   contents = {(not listed)}
}

@misc{Christodoulaki85Similarity,
   author = {S. Christodoulakis and K. Elles},
   title = {Similarity Retrieval of Images in a Multimedia Server},
   year = {1985},
   keywords = {mmm},
   note = {in preparation},
   contents = {(not listed)}
}

@inproceedings{Christodoulaki84Development,
   author = {S. Christodoulakis and J. Vandenbroek and J. Li and T. Li and S. Wan and Y. Wang and M. Papa and E. Bertino},
   title = {Development of a Multimedia Information System for an Office Environment},
   booktitle = {Proc. 10th International Conference on VLDB},
   year = {1984},
   keywords = {mmm},
   contents = {(not listed)}
}

@techreport{Christodoulaki84Interactive,
   author = {S. Christodoulakis and M. Theodoridou and M. Papa},
   title = {Interactive Multimedia Document Formation},
   institution = {CSRI, Univ. of Toronto},
   type = {Internal Report},
   year = {1984},
   keywords = {mmm},
   contents = {(not listed)}
}

@article{Erman80Hearsy,
   author = {L.D. Erman and F. Hays-Roth and V.R. Lesser and D.R. Ready},
   title = {The Hearsy {II} Speech-Understanding System: Integrating Knowledge to Resolve Uncertainty},
   journal = {ACM Computing Surveys},
   volume = {12},
   number = {2},
   month = jun,
   year = {1980},
   contents = {(not listed)}
}

@book{Gonzalez77Digital,
   author = {R. Gonzalez and P. Wintz},
   title = {Digital Image Processing},
   publisher = {Addison-Wesley},
   year = {1977},
   contents = {(not listed)}
}

@inproceedings{Horowitz74Picture,
   author = {S. Horowitz and T. Pavlidis},
   title = {Picture Segmentation by a Directed Split-and-Merge Procedure},
   booktitle = {Proc. Second Intern. Joint Conf. on Pattern Recognition},
   pages = {424-433},
   month = aug,
   year = {1974},
   contents = {(not listed)}
}

@inproceedings{Kulick75Image,
   author = {J. Kulick and T. Challis and C. Brace and S. Christodoulakis and I. Merrit and P. Neelands},
   title = {An Image Processing Laboratory for Automated Screening of Chest {X}-rays},
   booktitle = {Proc. of IEEE Third Intern. Conf. on Pattern Recognition},
   month = nov,
   year = {1975},
   contents = {(not listed)}
}

@article{Lee72Proximity,
   author = {E.T. Lee},
   title = {Proximity Measures for the Classification of Geometric Figures},
   journal = {Journal of Cybernetics},
   volume = {2},
   number = {4},
   pages = {43-59},
   year = {1972},
   keywords = {spatial objects},
   contents = {(not listed)}
}

@book{McLeod83Management,
   author = {R. McLeod},
   title = {Management Information Systems},
   edition = {second},
   publisher = {SRA},
   year = {1983},
   contents = {(not listed)}
}

@book{Pavlidis77Structural,
   author = {T. Pavlidis},
   title = {Structural Pattern Recognition},
   publisher = {Springer-Verlag},
   year = {1977},
   contents = {(not listed)}
}

@book{Reddy75Speech,
   author = {D.R. Reddy},
   title = {Speech Recognition},
   publisher = {Academic Press},
   year = {1975},
   contents = {(not listed)}
}

@inproceedings{Reddy76Speech,
   author = {D.R. Reddy},
   title = {Speech Recognition by Machine: {A} review},
   booktitle = {Proc. IEEE},
   month = apr,
   year = {1976},
   note = {Published as Proc. IEEE, volume 64, number 4},
   contents = {(not listed)}
}

@book{Tou74Pattern,
   author = {J.T. Tou and R.C. Gonzalez},
   title = {Pattern Recognition Principles},
   publisher = {Addison Wesley},
   year = {1974},
   contents = {(not listed)}
}

@inproceedings{VLDB83Panel,
   author = {{VLDB}},
   title = {A Panel on Complex Data Objects: Text, voice, images: Can {DBMS} manage them?},
   booktitle = {Proc. 9th International Conference on VLDB},
   institution = {VLDB},
   year = {1983}
}

@inproceedings{VLDB83Panela,
   author = {{VLDB}},
   title = {A Panel on Office Information Systems: What is our Role?},
   booktitle = {Proc. 9th International Conference on VLDB},
   institution = {VLDB},
   year = {1983}
}

@inproceedings{VLDB84Panel,
   author = {{VLDB}},
   title = {Panel on Multimedia Management Systems},
   booktitle = {Proc. 10th International Conference on VLDB},
   institution = {VLDB},
   year = {1984}
}

@book{Wiederhold83Database,
   author = {G. Wiederhold},
   title = {Database Design},
   publisher = {McGraw Hill},
   year = {1983},
   contents = {(not listed)}
}

@article{Wong82Document,
   author = {K.Y. Wong and R.G. Casey and F.M. Wahl},
   title = {Document Analysis System},
   journal = {IBM Research and Development},
   volume = {26},
   number = {6},
   month = nov,
   year = {1982},
   keywords = {text},
   contents = {(not listed)}
}

@misc{Aho78Awk,
   author = {A.V. Aho and B.W. Kernighan and P.J. Weinberger},
   title = {Awk - {A} Pattern Searching and Processing Language},
   month = sep,
   year = {1978},
   contents = {(not listed)}
}

@inproceedings{Roussopoulos85Direct,
   author = {N. Roussopoulos and D. Leifker},
   title = {Direct Spatial Search on Pictorial Databases Using Packed {R}-Trees},
   booktitle = {Proc. ACM SIGMOD},
   address = {Austin, Texas},
   month = may,
   year = {1985},
   keywords = {spatial objects},
   contents = {(not listed)}
}

@inproceedings{Guttman84R,
   author = {A. Guttman},
   title = {R-Trees: {A} Dynamic Index Structure for Spatial Searching},
   booktitle = {Proc. ACM SIGMOD},
   pages = {47-57},
   address = {Boston, Mass},
   month = jun,
   year = {1984},
   keywords = {spatial objects},
   contents = {(not listed)}
}

@article{Christodoulaki85Design,
   author = {S. Christodoulakis and C. Faloutsos},
   title = {Design and Performance Considerations for a Multimedia Document Server},
   journal = {First Intern. Conf. on Supercomputing Systems},
   pages = {441-449},
   address = {Florida},
   month = dec,
   year = {1985},
   keywords = {mmm superimposed coding},
   contents = {(not listed)}
}

@article{Samet84Quadtree,
   author = {H. Samet},
   title = {The Quadtree and Related Hierarchical Data Structures},
   journal = {ACM Computer Surveys},
   volume = {16},
   number = {2},
   pages = {187-260},
   month = jun,
   year = {1984},
   keywords = {spatial objects kdtrees},
   contents = {(not listed)}
}

@inproceedings{Christodoulaki85Issues,
   author = {S. Christodoulakis},
   title = {Issues in the Architecture of a Document Archiver using Optical Disk Technology},
   booktitle = {Proc. ACM SIGMOD},
   pages = {34-50},
   address = {Austin, Texas},
   month = may,
   year = {1985},
   keywords = {mmm},
   contents = {(not listed)}
}

@techreport{Roussopoulos84,
   author = {N. Roussopoulos and C. Bader and J. O'Connor},
   title = {{ADMS}: An Advanced Database Management System},
   institution = {University of Maryland},
   type = {Department of Computer Science},
   month = jul,
   year = {1984},
   contents = {(not listed)}
}

@techreport{Roussopoulos85Multidimensional,
   author = {N. Roussopoulos and H. Kang},
   title = {Multidimensional Index-Trees},
   institution = {University of Maryland},
   type = {Department of Computer Science},
   month = oct,
   year = {1985},
   contents = {(not listed)}
}

@article{Finkel74Quadtrees,
   author = {R.A. Finkel and J.L. Bentley},
   title = {Quadtrees: {A} data structure for retrieval on composite keys},
   journal = {ACTA Informatica},
   volume = {4},
   number = {1},
   pages = {1-9},
   year = {1974},
   keywords = {kdtrees},
   contents = {(not listed)}
}

@book{Gray53Pulse,
   author = {F. Gray},
   title = {Pulse Code Communications},
   publisher = {U.S. Patent 2632058},
   month = {March 17},
   year = {1953},
   contents = {(not listed)}
}

@article{Gilbert58Gray,
   author = {E.N. Gilbert},
   title = {Gray Codes and Paths on the n-Cube},
   journal = {Bell System Technical Journal},
   volume = {37},
   number = {3},
   pages = {815-826},
   month = may,
   year = {1958},
   contents = {(not listed)}
}

@inproceedings{Orenstein84Class,
   author = {J.A. Orenstein and T.H. Merrett},
   title = {A Class of Data Structures for Associative Searching},
   booktitle = {Proc. of SIGACT-SIGMOD},
   pages = {181-190},
   address = {Waterloo, Ontario, Canada},
   month = {April 2-4},
   year = {1984},
   keywords = {seckey},
   contents = {(not listed)}
}

@article{Ramamohanarao83Partial,
   author = {K. Ramamohanarao and J.W. Lloyd and J.A. Thom},
   title = {Partial-Match Retrieval Using Hashing and Descriptors},
   journal = {ACM TODS},
   volume = {8},
   number = {4},
   pages = {552-576},
   month = dec,
   year = {1983},
   keywords = {mah},
   contents = {(not listed)}
}

@article{Moran83Complexity,
   author = {S. Moran},
   title = {On the Complexity of Designing Optimal Partial-Match Retrieval Systems},
   journal = {ACM TODS},
   volume = {8},
   number = {4},
   pages = {543-551},
   month = dec,
   year = {1983},
   keywords = {mah},
   contents = {(not listed)}
}

@article{Faloutsos88Gray,
   author = {C. Faloutsos},
   title = {Gray Codes for Partial Match and Range Queries},
   journal = {IEEE Trans. on Software Engineering},
   volume = {14},
   number = {10},
   pages = {1381-1393},
   month = oct,
   year = {1988},
   keywords = {mah},
   note = {early version available as UMIACS-TR-87-4, also CS-TR-1796},
   contents = {(not listed)}
}

@incollection{Minker78Experimental,
   author = {J. Minker},
   editor = {H. Gallaire and J. Minker},
   title = {An Experimental Relational Data Base System Based on Logic},
   booktitle = {Logic and Data Bases},
   publisher = {Plenum Press},
   year = {1978},
   contents = {(not listed)}
}

@inproceedings{Boral85Database,
   author = {H. Boral and S. Redfield},
   title = {Database Machine Morphology},
   booktitle = {Proc. 11th International Conference on VLDB},
   pages = {59-71},
   address = {Stockholm, Sweden},
   month = aug,
   year = {1985},
   keywords = {dbmachines},
   contents = {(not listed)}
}

@incollection{Kowalski78Logic,
   author = {R. Kowalski},
   editor = {H. Gallaire and J. Minker},
   title = {Logic for Data Description},
   booktitle = {Logic and Data Bases},
   pages = {77-103},
   publisher = {Plenum Press},
   year = {1978},
   contents = {(not listed)}
}

@article{Du80Symbolic,
   author = {H.C. Du and R.C.T. Lee},
   title = {Symbolic Gray Codes as a Multikey Hashing Function},
   journal = {IEEE Trans. on PAMI},
   volume = {PAMI-2},
   number = {1},
   pages = {83-90},
   month = jan,
   year = {1980},
   keywords = {mah},
   contents = {(not listed)}
}

@techreport{Roussopoulos85,
   author = {N. Roussopoulos and J. O'Connor},
   title = {{ADMS}: Advanced Database Management System Query Language and Programmatic Interface},
   institution = {Department of Computer Science},
   type = {Tech. Report},
   number = {\#1579},
   address = {University of Maryland, College Park},
   month = apr,
   year = {1985},
   contents = {(not listed)}
}

@inproceedings{Faloutsos86Multiattribute,
   author = {C. Faloutsos},
   title = {Multiattribute Hashing Using Gray Codes},
   booktitle = {Proc. ACM SIGMOD},
   pages = {227-238},
   address = {Washington D.C},
   month = may,
   year = {1986},
   keywords = {mah},
   note = {also available as UMIACS-TR-87-4, CS-TR-1796},
   contents = {(not listed)}
}

@article{Faloutsos87Description,
   author = {C. Faloutsos and S. Christodoulakis},
   title = {Description and Performance Analysis of Signature File Methods},
   journal = {ACM TOOIS},
   volume = {5},
   number = {3},
   pages = {237-257},
   year = {1987},
   contents = {(not listed)}
}

@incollection{Faloutsos85Accessa,
   author = {C. Faloutsos and S. Christodoulakis},
   editor = {D. Tsichritzis},
   title = {Access Methods for Documents},
   booktitle = {Office Automation},
   publisher = {Springer Verlag},
   year = {1985},
   contents = {(not listed)}
}

@article{Hinrichs85Implementation,
   author = {K. Hinrichs},
   title = {Implementation of the Grid File: Design Concepts and Experience},
   journal = {BIT},
   volume = {25},
   number = {4},
   pages = {569-592},
   year = {1985},
   keywords = {seckey},
   contents = {(not listed)}
}

@article{Jakobsson85Compression,
   author = {M. Jakobsson},
   title = {Compression of Character Strings by an Adaptive Dictionary},
   journal = {BIT},
   volume = {25},
   number = {4},
   pages = {593-604},
   year = {1985},
   contents = {(not listed)}
}

@article{Chock84Database,
   author = {M. Chock and A.F. Cardenas and A. Klinger},
   title = {Database Structure and Manipulation Capabilities of a Picture Database Management System ({PICDMS})},
   journal = {IEEE Trans. on Pattern Analysis and Machine Intelligence},
   volume = {PAMI-6},
   number = {4},
   pages = {484-492},
   month = jul,
   year = {1984},
   contents = {(not listed)}
}

@article{Faloutsos88Signature,
   author = {C. Faloutsos},
   title = {Signature Files: An Integrated Access Method for Text and Attributes Suitable for Optical Disk Storage.},
   journal = {BIT},
   volume = {28},
   number = {4},
   pages = {736-754},
   year = {1988},
   keywords = {signature files},
   note = {early version available as UMIACS-TR-87-23, CS-TR-1867.}
}

@article{Furuta82Document,
   author = {R. K. Furuta and J. Scofield and A. C. Shaw},
   title = {Document Formatting Systems: Survey, Concepts, and Issues},
   journal = {ACM Computing Surveys},
   volume = {14},
   number = {3},
   pages = {417-472},
   month = sep,
   year = {1982},
   contents = {(not listed)}
}

@incollection{Furuta86Integrated,
   author = {R. K. Furuta},
   editor = {J. C. van Vliet},
   title = {An Integrated, but not Exact-Representation, Editor/Formatter},
   booktitle = {Text Processing and Document Manipulation},
   pages = {246-259},
   publisher = {Cambridge University Press},
   month = apr,
   year = {1986},
   keywords = {Furuta EP86},
   contents = {(not listed)}
}

@phdthesis{Furuta86Integrateda,
   author = {R. K. Furuta},
   title = {An Integrated, but not Exact-Representation, Editor/Formatter},
   school = {University of Washington, Department of Computer Science},
   type = {Ph.{D}. Dissertation},
   address = {Seattle, WA},
   year = {1986},
   keywords = {Furuta PhD},
   contents = {(not listed)}
}

@inproceedings{Faloutsos86Integrated,
   author = {C. Faloutsos},
   title = {Integrated access methods for messages using signature files},
   booktitle = {IFIP WG 8.4 Working Conference on Methods and Tools for Office systems},
   pages = {135-157},
   address = {Pisa, Italy},
   month = oct,
   year = {1986},
   contents = {(not listed)}
}

@article{Christodoulaki86Design,
   author = {S. Christodoulakis and C. Faloutsos},
   title = {Design and Performance Considerations for an Optical Disk Based Multimedia Object Server},
   journal = {IEEE Computer Magazine},
   volume = {19},
   number = {12},
   pages = {45-56},
   month = dec,
   year = {1986},
   keywords = {mmm},
   contents = {(not listed)}
}

@techreport{Naish83MU,
   author = {L. Naish and J.A. Thom},
   title = {The {MU}-Prolog Deductive Database},
   institution = {Dept. of Computer Science, The Univ. of Melbourne},
   type = {Tech. Report},
   number = {83/10},
   year = {1983},
   keywords = {mah},
   contents = {(not listed)}
}

@article{Ammon85High,
   author = {G. Ammon and G. Calabria and D. Thomas},
   title = {A High-Speed, Large Capacity, `Jukebox', Optical Disk System},
   journal = {Computer},
   volume = {18},
   number = {7},
   month = jul,
   year = {1985},
   contents = {(not listed)}
}

@inproceedings{Bell83Critical,
   author = {A. Bell},
   title = {Critical Issues in High Density Magnetic and Optical Storage},
   booktitle = {Proceedings of the SPIE, vol. 32, Optical Data Storage},
   month = jan,
   year = {1983},
   contents = {(not listed)}
}

@inproceedings{Christodoulaki86Multimedia,
   author = {S. Christodoulakis and F. Ho and M. Theodoridou},
   title = {The Multimedia Object Presentation Manager in {MINOS}: {A} Symmetric Approach},
   booktitle = {Proc. ACM SIGMOD},
   month = may,
   year = {1986},
   keywords = {mmm},
   contents = {(not listed)}
}

@inproceedings{Christodoulaki85Multimedia,
   author = {S. Christodoulakis},
   title = {Multimedia Database Management: Applications and Problems.  {A} Position Paper},
   booktitle = {Proc. ACM SIGMOD},
   month = may,
   year = {1985},
   contents = {(not listed)}
}

@article{Christodoulaki84Estimating,
   author = {S. Christodoulakis},
   title = {Estimating Block Selectivities},
   journal = {Informations Systems},
   volume = {9},
   number = {1},
   month = mar,
   year = {1984},
   keywords = {performance},
   contents = {(not listed)}
}

@misc{Christodoulaki80a,
   author = {S. Christodoulakis},
   title = {{IPRL}: An Interactive Pattern Recognition Laboratory Proceedings of Eleventh {ACM}-{SIGCSE} Technical Symposium},
   address = {Kansas},
   month = mar,
   year = {1980},
   contents = {(not listed)}
}

@article{Doddington81Speech,
   author = {G. Doddington and T. Schalk},
   title = {Speech Recognition: Turning Theory Into Practice},
   journal = {IEEE Spectrum},
   pages = {26-32},
   month = sep,
   year = {1981},
   contents = {(not listed)}
}

@article{Forsdick84Initial,
   author = {Forsdick and Thomas and Robertson and Trowers},
   title = {Initial Experience with Multimedia Documents in Diamond},
   journal = {IEEE Database Engineering},
   volume = {7},
   number = {3},
   year = {1984},
   keywords = {mmm},
   contents = {(not listed)}
}

@article{IEEE84Special,
   author = {{IEEE}},
   title = {Special Issue on Multimedia Databases},
   institution = {IEEE},
   journal = {IEEE Database Engineering},
   volume = {7},
   number = {3},
   year = {1984},
   keywords = {mmm},
   contents = {(not listed)}
}

@misc{Kamel??Recognition,
   author = {M. Kamel},
   title = {Recognition of Handwritten Numerals},
   note = {patent accepted by the NCR Pattent Office, filing of US Patent in Progress},
   contents = {(not listed)}
}

@misc{Kamel??Information,
   author = {M. Kamel},
   title = {Information Contents in Bar Codes},
   note = {patent accepted by the NCR Corporate Patent Office, application to US Patent Office submitted},
   contents = {(not listed)}
}

@misc{Kamel82Image,
   author = {M. Kamel},
   title = {An Image Based Archival and Retrieval System},
   year = {1982},
   note = {NCR Product Opportunity Proposal},
   contents = {(not listed)}
}

@inproceedings{Kamel85Computationally,
   author = {M. Kamel and M. Ismail},
   title = {Computationally Efficient Techniques for Multidimensional Data Clustering},
   booktitle = {Proceedings IEEE International Conference on Systems, Man and Cybernetics},
   year = {1985},
   keywords = {seckey},
   contents = {(not listed)}
}

@inproceedings{Kamel85Knowledge,
   author = {M. Kamel and M. Mansour and A. Wong},
   title = {Knowledge Based Office Modelling: {A} New Approach},
   booktitle = {Proceedings of the International Conference on Applied Simulation and Modelling},
   year = {1985},
   contents = {(not listed)}
}

@inproceedings{Kamel84Knowledge,
   author = {M. Kamel},
   title = {Knowledge Aquisition using Pattern Recognition Techniques},
   booktitle = {Proceedings of the Conference on Intelligent Systems and Machines},
   address = {Rochester},
   year = {1984},
   contents = {(not listed)}
}

@inproceedings{Kamel84Computing,
   author = {M. Kamel},
   title = {Computing the Singular Value Decomposition in Image Processing},
   booktitle = {Proceedings of Conference on Information Systems},
   address = {Princeton},
   year = {1984},
   contents = {(not listed)}
}

@inproceedings{Kulick75Imagea,
   author = {J. Kulick and T. Challis and C. Brau and S. Christodoulakis and I. Merrit and P. Neelands},
   title = {An Image Processing Laboratory for Automated Screening of {X}-rays},
   booktitle = {Proceedings of IEEE Third International Conference on Pattern Recognition},
   month = nov,
   year = {1975},
   contents = {(not listed)}
}

@article{Kenney79Optical,
   author = {C. Kenney and D. Lou and R. McFarlane and A. Chou and J. Nadah and T. Kohler and J. Wanger and F. Zernike},
   title = {An Optical Disk Replaces 25 mag tapes},
   journal = {IEEE Spectrum},
   pages = {33-38},
   month = feb,
   year = {1979},
   contents = {(not listed)}
}

@inproceedings{Lehman86Query,
   author = {T. Lehman and M. Carey},
   title = {Query Processing in Main Memory Database Management Systems},
   booktitle = {Proc. ACM SIGMOD},
   month = may,
   year = {1986},
   keywords = {primkey trees},
   contents = {(not listed)}
}

@inproceedings{Maier82Using,
   author = {D. Maier},
   title = {Using Write-Once Memory for Database Storage},
   booktitle = {Proc. ACM PODS},
   pages = {239-264},
   year = {1982},
   keywords = {optical disk},
   contents = {(not listed)}
}

@article{BNR85Meridian,
   author = {{BNR}},
   title = {Meridian System Information},
   institution = {BNR},
   journal = {in Telesis (BNR quarterly magazine)},
   volume = {12},
   number = {2},
   year = {1985},
   contents = {(not listed)}
}

@article{Miller85Future,
   author = {W. Miller},
   title = {The Future of Mass Storage Systems},
   journal = {IEEE Computer Magazine},
   volume = {18},
   number = {7},
   month = jul,
   year = {1985},
   contents = {(not listed)}
}

@article{Reddy80Words,
   author = {R. Reddy},
   title = {Words into Action {II}: {A} Task Oriented System},
   journal = {IEEE Spectrum},
   pages = {26-29},
   month = jun,
   year = {1980},
   contents = {(not listed)}
}

@article{Rossenberg85All,
   author = {R. Rossenberg},
   title = {The All-digital {PBX}'s Stake Their Claim on Office Automation},
   journal = {Electronics Week},
   month = {May 6},
   year = {1985},
   contents = {(not listed)}
}

@article{Smith81Optimization,
   author = {A. Smith},
   title = {Optimization of {I}/{O} Systems by Cache Disks and File Migration.  {A} Summary},
   journal = {Performance Evaluation},
   volume = {1},
   pages = {249-262},
   year = {1981},
   contents = {(not listed)}
}

@article{Smith81Long,
   author = {A. Smith},
   title = {Long Term File Migration: Development and Evaluation of Algorithms},
   journal = {CACM},
   volume = {24},
   number = {8},
   year = {1981},
   contents = {(not listed)}
}

@article{IEEE85Special,
   author = {{IEEE}},
   title = {Special Issue on Multimedia Communications},
   institution = {IEEE},
   journal = {IEEE Computer},
   month = oct,
   year = {1985},
   keywords = {mmm}
}

@inproceedings{Stonebraker86Design,
   author = {M. Stonebraker and L. Row},
   title = {The Design of {POSTGRESS}},
   booktitle = {Proc. ACM SIGMOD},
   month = may,
   year = {1986},
   contents = {(not listed)}
}

@book{Tsichritzis85Office,
   author = {D. Tsichritzis},
   title = {Office Automation},
   publisher = {Springer-Verlag},
   year = {1985},
   keywords = {book},
   contents = {(not listed)}
}

@book{Wiederhold77Database,
   author = {G. Wiederhold},
   title = {Database Design},
   publisher = {McGraw-Hill},
   year = {1977},
   contents = {(not listed)}
}

@inproceedings{Woelk86Object,
   author = {D. Woelk and W. Kim and W. Luther},
   title = {An Object Oriented Approach to Multimedia Databases},
   booktitle = {Proc. ACM SIGMOD},
   month = may,
   year = {1986},
   keywords = {mmm},
   contents = {(not listed)}
}

@article{Nicholson85Usage,
   author = {R. Nicholson},
   title = {Usage Patterns in an Integrated Voice and Data Communications System},
   journal = {ACM TOOIS},
   volume = {3},
   number = {3},
   month = jul,
   year = {1985},
   contents = {(not listed)}
}

@article{DeWitt79,
   author = {D. DeWitt},
   title = {{DIRECT}- {A} Multiprocessor Organization for Supporting a Relational Database Management System},
   journal = {IEEE Transactions on Computers},
   volume = {28},
   number = {6},
   year = {1979},
   keywords = {dbmachines},
   contents = {(not listed)}
}

@techreport{Loo86Intelligent,
   author = {W. Loo},
   title = {An Intelligent User Interface to Databases},
   institution = {Department of Systems Design, University of Waterloo},
   type = {MSc. Thesis (supervised by M. Kamel)},
   year = {1986},
   contents = {(not listed)}
}

@techreport{Wilkes85Data,
   author = {D. Wilkes},
   title = {Data Compression for Randomly Addressed Files},
   institution = {Department of Computer Science, University of Toronto},
   type = {MSc. Thesis},
   year = {1985},
   contents = {(not listed)}
}

@phdthesis{Weddel86Physical,
   author = {G. Weddel},
   title = {Physical Data Base Design for Main Memory Resident Data Bases},
   school = {Department of Computer Science, University of Toronto},
   type = {Ph.{D}. Thesis},
   year = {1986},
   contents = {(not listed)}
}

@inproceedings{Banerjee86Supporting,
   author = {J. Banerjee and W. Kim},
   title = {Supporting {VLSI} Geometry Operations in a Database System},
   booktitle = {Proceedings of International Conference on Data Engineering},
   year = {1986},
   keywords = {spatial objects},
   contents = {(not listed)}
}

@inproceedings{Kamel84Language,
   author = {M. Kamel and C. Chau and A. Wong},
   title = {A Language for Knowledge Based Office Modelling},
   booktitle = {Proceedings IEEE workshop on Languages for Automation},
   month = nov,
   year = {1984},
   contents = {(not listed)}
}

@misc{Kamel83Guidelines,
   author = {M. Kamel},
   title = {Guidelines for Designing an Image Database System},
   year = {1983},
   note = {NCR Advanced Development Technical Note},
   contents = {(not listed)}
}

@inproceedings{Black85Architecture,
   author = {J. Black and L. Marshall and B. Randell},
   title = {The Architecture of {UNIX} United},
   booktitle = {IEEE proceedings},
   year = {1985},
   note = {Special Issue on Distributed Data Base Systems},
   contents = {(not listed)}
}

@article{Taylor??Redundancy,
   author = {D. Taylor and D. Morgan and J. Black},
   title = {Redundancy in Data Structures: Improving Software Fault Tolerence},
   journal = {IEEE Transactions on Software Engineering},
   volume = {6},
   number = {6},
   year = {Nov. 80},
   contents = {(not listed)}
}

@article{Chamberlin81Janus,
   author = {D. Chamberlin and J. King and D. Slutz and S. Todd and B. Wade},
   title = {Janus: An Interactive System for Document Composition},
   journal = {ACM},
   year = {1981},
   contents = {(not listed)}
}

@article{Christodoulaki84Experimental,
   author = {S. Christodoulakis},
   title = {An Experimental Multimedia System for an Office Environment},
   journal = {IEEE Database Engineering},
   volume = {7},
   number = {3},
   pages = {43-48},
   month = sep,
   year = {1984},
   note = {Special Issue on Multimedia},
   contents = {(not listed)}
}

@article{Feiner82Experimental,
   author = {S. Feiner and S. Nagy and A. Van Dam},
   title = {An Experimental System for Creating and Presenting Interactive Graphical Documents},
   journal = {ACM Trans. Graphics},
   volume = {1},
   number = {1},
   pages = {59-77},
   month = jan,
   year = {1982},
   keywords = {mmm},
   contents = {(not listed)}
}

@article{Gregory83Xanadu,
   author = {R. Gregory},
   title = {Xanadu: Hypertext from the Future},
   journal = {Dr. Dobb's J. 75},
   pages = {28-35},
   month = jan,
   year = {1983},
   contents = {(not listed)}
}

@article{Herot80Spatial,
   author = {C. Herot},
   title = {Spatial Management of Data},
   journal = {ACM TODS},
   volume = {5},
   number = {4},
   pages = {493-514},
   month = dec,
   year = {1980},
   keywords = {spatial objects},
   contents = {(not listed)}
}

@article{Izawa84Document,
   author = {K. Izawa},
   title = {Document Image Filing System Utilizing Optical Disc Memories},
   journal = {IEEE Database Engineering},
   volume = {7},
   number = {3},
   pages = {3-10},
   month = sep,
   year = {1984},
   note = {Special Issue on Multimedia},
   contents = {(not listed)}
}

@article{Knuth79,
   author = {D. Knuth},
   title = {{TEX}: {A} System for Technical Text},
   journal = {American Mathematical Society, Providence, RI},
   year = {1979},
   contents = {(not listed)}
}

@book{McLeod83Managementa,
   author = {R. McLeod},
   title = {Management Information Systems},
   publisher = {SRA, Second edition},
   year = {1983},
   contents = {(not listed)}
}

@article{Poggio85,
   author = {A. Poggio and Garcia Luna Aceves and E. Craghill and D. Moran and L. Aguilar and D. Worthington and J. Hight},
   title = {{CCWS}: {A} Computer Based Multimedia Information System},
   journal = {IEEE Computer},
   pages = {92-103},
   month = oct,
   year = {1985},
   keywords = {mmm},
   contents = {(not listed)}
}

@inproceedings{Reid80High,
   author = {B. Reid},
   title = {A High-Level Approach to Computer Document Formatting},
   booktitle = {Conference Record of the Seventh Annual ACM Symposium on Principles of Programming Languages},
   pages = {24-31},
   address = {Las Vegas, NV},
   year = {1980},
   contents = {(not listed)}
}

@article{Reynolds85ARPA,
   author = {J. Reynolds and J. Postel and A. Katz and G. Finn and A. DeShon},
   title = {The {ARPA} Experimental Multimedia Mail System},
   journal = {IEEE Computer},
   pages = {82-89},
   month = oct,
   year = {1985},
   contents = {(not listed)}
}

@article{Sakata85Distributed,
   author = {S. Sakata and T. Ueda},
   title = {A Distributed Interoffice Mail System},
   journal = {IEEE Computer},
   pages = {106-116},
   month = oct,
   year = {1985},
   contents = {(not listed)}
}

@article{Sarin85Computer,
   author = {S. Sarin and I. Grief},
   title = {Computer Based Real-Time Conferencing Systems},
   journal = {IEEE Computer},
   pages = {33-45},
   month = oct,
   year = {1985},
   contents = {(not listed)}
}

@techreport{Thacker79Alto,
   author = {C. Thacker and E. McGreight and B. Lampson and R. Sproull and D. Boggs},
   title = {Alto: {A} Personal Computer},
   institution = {Xerox Palo Alto Research Center},
   type = {Technical Report},
   number = {CSL-79-11},
   year = {1979},
   contents = {(not listed)}
}

@article{Thomas??Diamond,
   author = {R. Thomas and H. Forsdick and T. Crowley and R. Schaat and R. Tomlinson and V. Travers and G. Robertson:},
   title = {Diamond: {A} Multimedia Message System Built on a Distributed Architecture},
   journal = {IEEE Computer},
   pages = {65-78},
   year = {Dec. 85},
   keywords = {mmm},
   contents = {(not listed)}
}

@techreport{Theodoridou85Multimedia,
   author = {M. Theodoridou},
   title = {Multimedia Document Presentation and Formation in {MINOS}},
   institution = {Department of Electrical Engineering, University of Toronto},
   type = {M.Sc. Thesis},
   year = {1985},
   keywords = {mmm},
   contents = {(not listed)}
}

@article{Weyner85Prototype,
   author = {S. Weyner and A. Borning},
   title = {A Prototype Electronic Encyclopedia},
   journal = {ACM TOOIS},
   volume = {3},
   number = {1},
   pages = {63-88},
   month = jan,
   year = {1985},
   keywords = {mmm},
   contents = {(not listed)}
}

@article{Yankelovich85Reading,
   author = {N. Yankelovich and N. Meyrowitz and N. Van Dam},
   title = {Reading and Writing the Electronic Book},
   journal = {IEEE Computer},
   pages = {15-30},
   month = oct,
   year = {1985},
   keywords = {mmm},
   contents = {(not listed)}
}

@article{Koved86Embedded,
   author = {L. Koved and B. Shneiderman},
   title = {Embedded Menus: Selecting Items in Context},
   journal = {Comm. of the ACM},
   volume = {29},
   number = {4},
   pages = {312-318},
   month = apr,
   year = {1986},
   keywords = {mmm cacm ties},
   contents = {(not listed)}
}

@article{Naffah86Agora,
   author = {N. Naffah and A. Karmouch},
   title = {Agora - An Experiment in Multimedia Message Systems},
   journal = {IEEE Computer Magazine},
   volume = {19},
   number = {5},
   pages = {56-66},
   month = may,
   year = {1986},
   keywords = {mmm},
   contents = {(not listed)}
}

@techreport{Rieger79,
   author = {C. Rieger},
   title = {{ZMOB}: {A} Mob of 256 Cooperative Z80A-Based Microcomputers},
   institution = {Computer Science Department, Univ. of Maryland},
   type = {TR-825},
   address = {College Park},
   month = nov,
   year = {1979},
   keywords = {parallel},
   contents = {(not listed)}
}

@article{Nofel8640,
   author = {P.J. Nofel},
   title = {40 Million Hits on  Optical Disk},
   journal = {Modern Office Technology},
   pages = {84-88},
   month = mar,
   year = {1986},
   keywords = {library congress},
   contents = {(not listed)}
}

@article{Price84Optical,
   author = {Joseph Price},
   title = {The Optical Disk Pilot Project At the Library of Congress},
   journal = {Videodisc and Optical Disk},
   volume = {4},
   number = {6},
   pages = {424-432},
   month = {Nov.-Dec.},
   year = {1984},
   contents = {(not listed)}
}

@inproceedings{Larson85Hash,
   author = {P.A. Larson},
   title = {Hash Files: Some Recent Developments},
   booktitle = {Proc. of the First Intern. Conference on Supercomputing Systems},
   pages = {671-679},
   address = {St. Petersburg, Florida},
   month = dec,
   year = {1985},
   keywords = {hashing},
   contents = {(not listed)}
}

@inproceedings{Kawagoe85Modified,
   author = {K. Kawagoe},
   title = {Modified Dynamic Hashing},
   booktitle = {Proc. ACM SIGMOD},
   pages = {201-213},
   address = {Austin, Texas},
   month = may,
   year = {1985},
   keywords = {exhash},
   contents = {(not listed)}
}

@article{Lomet83Bounded,
   author = {D.B. Lomet},
   title = {Bounded Index Exponential Hashing},
   journal = {ACM TODS},
   volume = {8},
   number = {1},
   pages = {136-165},
   year = {1983},
   keywords = {exhash},
   contents = {(not listed)}
}

@article{Knott82Fixed,
   author = {G.D. Knott},
   title = {Fixed-Bucket Binary Storage Trees},
   journal = {Journal of Algorithms},
   volume = {3},
   pages = {276-287},
   year = {1982},
   contents = {(not listed)}
}

@article{Knott81Procedures,
   author = {G.D. Knott},
   title = {Procedures for Managing Extendible Array Files},
   journal = {Software - Practice and Experience},
   volume = {11},
   pages = {63-84},
   year = {1981},
   contents = {(not listed)}
}

@article{Litwin86Bounded,
   author = {W. Litwin and D.B. Lomet},
   title = {The Bounded Disorder Access Method},
   journal = {2nd Intern. Conf. on Data Engineering},
   pages = {38-48},
   address = {Los Angeles},
   month = feb,
   year = {1986},
   keywords = {exhash},
   contents = {(not listed)}
}

@article{Ramamohanarao84Recursive,
   author = {K. Ramamohanarao and R. Sacks-Davis},
   title = {Recursive Linear Hashing},
   journal = {ACM TODS},
   volume = {9},
   number = {3},
   pages = {369-391},
   month = sep,
   year = {1984},
   keywords = {exhash},
   contents = {(not listed)}
}

@inproceedings{Burkhard83Interpolation,
   author = {W.A. Burkhard},
   title = {Interpolation-based Index Maintenance},
   booktitle = {Proc. ACM Conf. on PODS},
   pages = {76-89},
   year = {1983},
   keywords = {exhash},
   contents = {(not listed)}
}

@inproceedings{Mullin84Unified,
   author = {J.K. Mullin},
   title = {Unified Dynamic Hashing},
   booktitle = {Proc. 10th International Conference on VLDB},
   pages = {473-480},
   month = aug,
   year = {1984},
   keywords = {exhash},
   contents = {(not listed)}
}

@article{Mullin81Tightly,
   author = {J.K. Mullin},
   title = {Tightly Controlled Linear Hashing without Separate Overflow Storage},
   journal = {BIT},
   volume = {21},
   number = {4},
   pages = {389-400},
   year = {1981},
   keywords = {exhash},
   contents = {(not listed)}
}

@article{Ramamohanarao81Dynamic,
   author = {K. Ramamohanarao and J.W. Lloyd},
   title = {Dynamic Hashing Schemes},
   journal = {the Computer Journal},
   volume = {25},
   number = {4},
   pages = {478-485},
   year = {1981},
   keywords = {exhash},
   contents = {(not listed)}
}

@inproceedings{Larson85External,
   author = {P.A. Larson and M.V. Ramakrishna},
   title = {External Perfect Hashing},
   booktitle = {Proc. ACM SIGMOD},
   pages = {190-199},
   address = {Austin, Texas},
   month = may,
   year = {1985},
   keywords = {exhash},
   contents = {(not listed)}
}

@techreport{Gonnet82External,
   author = {G.H. Gonnet and P.A. Larson},
   title = {External Hashing with Limited Internal Storage.},
   institution = {Univ. of Waterloo},
   type = {Technical report},
   number = {CS-82-38},
   year = {1982},
   contents = {(not listed)}
}

@article{Ewing86Experimental,
   author = {J. Ewing and S. Mehrabanzad and S. Sheck and D. Ostroff and B. Shneiderman},
   title = {An Experimental Comparison of a Mouse and Arrow-jump Keys for an Interactive Encyclopedia},
   journal = {Int. Journal of Man-Machine Studies},
   volume = {24},
   number = {1},
   pages = {29-45},
   month = jan,
   year = {1986},
   keywords = {TIES},
   contents = {(not listed)}
}

@book{Reza61Introduction,
   author = {F.M. Reza},
   title = {An Introduction to Information Theory},
   publisher = {McGraw-Hill},
   year = {1961},
   keywords = {infotheory},
   contents = {(not listed)}
}

@book{Peterson61Error,
   author = {W.W. Peterson},
   title = {Error Correcting Codes},
   publisher = {MIT Press and John Wiley \& Sons In.c},
   year = {1961},
   contents = {(not listed)}
}

@book{Peterson72Error,
   author = {W.W. Peterson and E.J. Weldon},
   title = {Error Correcting Codes},
   publisher = {MIT Press},
   year = {1972}
}

@book{Berlekamp68Algebraic,
   author = {E.R. Berlekamp},
   title = {Algebraic Coding Theory},
   publisher = {McGraw-Hill},
   year = {1968},
   contents = {(not listed)}
}

@article{Christodoulaki86Issues,
   author = {S. Christodoulakis and T. Velissaropoulos},
   title = {Issues in the Design of a Distributed Testbed for {MINOS}},
   journal = {Transactions on Management Information Systems},
   year = {1986},
   keywords = {mmm},
   note = {(to appear)},
   contents = {(not listed)}
}

@article{Christodoulaki86Multimediaa,
   author = {S. Christodoulakis and M. Theodoridou and F. Ho and M. Papa and A. Pathria},
   title = {Multimedia Document Presentation, Information Extraction and Document Formation in {MINOS}: {A} Model and a System},
   journal = {ACM TOOIS},
   volume = {4},
   number = {4},
   month = oct,
   year = {1986},
   keywords = {mmm},
   contents = {(not listed)}
}

@article{Noreault83Performance,
   author = {T. Noreault and M. McGill and M.B. Koll},
   editor = {R.N. Oddy and S.E. Robertson and C.J. van-Rijsbergen and P.W. Williams},
   title = {A Performance Evaluation of Similarity Measures, Document Term Weighting Schemes and Representation in a Boolean Environment},
   journal = {Information Retrieval Research},
   pages = {57-76},
   publisher = {Butterworths},
   year = {1983},
   contents = {(not listed)}
}

@article{Davies86Algorithms,
   author = {G. Davies and S. Bowsher},
   title = {Algorithms for Pattern Matching},
   journal = {Software-Practice and Experience},
   volume = {16},
   number = {6},
   pages = {575-601},
   month = jun,
   year = {1986},
   keywords = {fulltext},
   contents = {(not listed)}
}

@article{Yuen86Dynamic,
   author = {T.-S. Yuen and D. H.-C. Du},
   title = {Dynamic File Structure for Partial Match Retrieval Based on Overflow Bucket Sharing},
   journal = {IEEE Trans. on Software Engineering},
   volume = {SE-12},
   number = {8},
   pages = {801-810},
   month = aug,
   year = {1986},
   keywords = {mah},
   contents = {(not listed)}
}

@article{Cichelli80Minimal,
   author = {R.J. Cichelli},
   title = {Minimal Perfect Hashing Functions Made Simple},
   journal = {CACM},
   volume = {23},
   number = {1},
   pages = {17-19},
   year = {1980},
   contents = {(not listed)}
}

@article{Carter79Universal,
   author = {L.J. Carter and M.L. Wegman},
   title = {Universal Classes Of Hash Functions},
   journal = {Journal of Computer and System Sciences},
   volume = {18},
   pages = {143-154},
   year = {1979},
   contents = {(not listed)}
}

@article{Sprugnoli77Perfect,
   author = {R.J. Sprugnoli},
   title = {Perfect Hashing Functions: {A} Single Probe Retrieving Method for Static Sets},
   journal = {CACM},
   volume = {20},
   number = {11},
   pages = {841-850},
   year = {1977},
   contents = {(not listed)}
}

@article{Chang84Study,
   author = {C.C. Chang},
   title = {The Study of an Ordered Minimal Perfect Hashing Scheme},
   journal = {CACM},
   volume = {27},
   number = {4},
   pages = {384-387},
   month = apr,
   year = {1984},
   contents = {(not listed)}
}

@article{Jaeschke81Reciprocal,
   author = {G. Jaeschke},
   title = {Reciprocal Hashing: {A} Method for Generating Minimal Perfect Hashing Functions},
   journal = {CACM},
   volume = {24},
   number = {12},
   pages = {829-833},
   month = dec,
   year = {1981},
   contents = {(not listed)}
}

@inproceedings{Fredman82Storing,
   author = {M.L. Fredman and J. Komlos and E. Szemeredi},
   title = {Storing a Sparse Table with o(1) Worst Case Access Time},
   booktitle = {Proc. of 23rd Sumposium on Foundations of Computer Science (FOCS)},
   pages = {165-168},
   year = {1982},
   contents = {(not listed)}
}

@article{Larson84File,
   author = {P.A. Larson and A. Kajla},
   title = {File Organization - Implementation of a Method Guaranteeing Retrieval in One Access},
   journal = {CACM},
   volume = {27},
   number = {7},
   pages = {670-677},
   year = {1984},
   keywords = {hashing},
   contents = {(not listed)}
}

@book{Gallager68Information,
   author = {R.G. Gallager},
   title = {Information Theory and Reliable Communication},
   publisher = {John Wiley and Sons},
   year = {1968},
   keywords = {infotheory},
   contents = {(not listed)}
}

@inproceedings{Huffman62Method,
   author = {D.A. Huffman},
   title = {A Method for the Construction of Minimum Redundancy Codes},
   booktitle = {Proc. IRE},
   volume = {40},
   pages = {1098-1101},
   year = {1962},
   keywords = {compression},
   contents = {(not listed)}
}

@techreport{Faloutsos87Parallelism,
   author = {C. Faloutsos},
   title = {Parallelism and Multiattribute Hashing.},
   institution = {Univ. of Maryland},
   type = {working paper},
   year = {1987},
   keywords = {dbmachines},
   contents = {(not listed)}
}

@inproceedings{Sellis85Optimization,
   author = {T. Sellis and L. Shapiro},
   title = {Optimization of Extended Database Languages},
   booktitle = {Proc. ACM SIGMOD},
   address = {Austin, Texas},
   month = may,
   year = {1985},
   contents = {(not listed)}
}

@inproceedings{Sellis86Global,
   author = {T. Sellis},
   title = {Global Query Optimization},
   booktitle = {Proc. ACM SIGMOD},
   address = {Washington, DC},
   month = may,
   year = {1986},
   contents = {(not listed)}
}

@article{Sellis86Query,
   author = {T. Sellis},
   title = {Query Optimization in Extended Relational Database Systems},
   journal = {ACM TODS},
   year = {1986},
   note = {(submitted for publication)},
   contents = {(not listed)}
}

@phdthesis{Sellis86Optimization,
   author = {T. Sellis},
   title = {Optimization of Extended Relational Database Systems},
   school = {University of California},
   type = {Ph.{D}. Thesis},
   address = {Berkeley},
   month = jul,
   year = {1986},
   contents = {(not listed)}
}

@misc{Sellis86Analysis,
   author = {T. Sellis},
   title = {An Analysis of {R}+ Trees},
   year = {1986},
   note = {in preparation},
   contents = {(not listed)}
}

@inproceedings{Stonebraker86Rule,
   author = {M. Stonebraker and T. Sellis and E. Hanson},
   title = {Rule Indexing Implementations in Database Systems},
   booktitle = {Proceedings of the First International Conference on Expert Database Systems},
   address = {Charleston, SC},
   month = apr,
   year = {1986},
   contents = {(not listed)}
}

@article{Roussopoulos82View,
   author = {N. Roussopoulos},
   title = {View Indexing in Relational Databases},
   journal = {ACM TODS},
   volume = {7},
   number = {2},
   month = jun,
   year = {1982},
   contents = {(not listed)}
}

@article{Roussopoulos82Logical,
   author = {N. Roussopoulos},
   title = {The Logical Access Path Schema of a Database},
   journal = {IEEE Trans. on Software Engineering},
   volume = {SE-8},
   number = {6},
   month = nov,
   year = {1982},
   keywords = {lap},
   contents = {(not listed)}
}

@techreport{Mark85Self,
   author = {L. Mark},
   title = {Self-Describing Database Systems - Formalization and Realization},
   institution = {Computer Science Department, University of Maryland},
   type = {TR-1484},
   month = apr,
   year = {1985},
   contents = {(not listed)}
}

@inproceedings{Mark85Update,
   author = {L. Mark and N. Roussopoulos and B. Chu},
   title = {Update Dependencies},
   booktitle = {IFIP TC2 WG 2.6 Working Conference on Database Semantics},
   address = {Hasselt, Belgium},
   month = jan,
   year = {1985},
   contents = {(not listed)}
}

@inproceedings{Harhalakis86Integration,
   author = {G. Harhalakis and L. Mark and M. Bohse},
   title = {Integration of Manufacturing Resource Planning ({MRP} {II}) Systems with Computer Aided Design ({CAD})},
   booktitle = {SRC Conference proceedings},
   address = {College Park, Maryland},
   month = sep,
   year = {1986},
   keywords = {interoperability cad cam},
   contents = {(not listed)}
}

@techreport{Mark86TAME,
   author = {L. Mark},
   title = {The {TAME} Base},
   institution = {Department of Computer Science, University of Maryland},
   type = {Working note,},
   month = sep,
   year = {1986},
   keywords = {software engineering databases},
   contents = {(not listed)}
}

@article{Mark86Meta,
   author = {L. Mark and N. Roussopoulos},
   title = {Meta-Data Management},
   journal = {IEEE Computer Magazine Special Issue on "New Directions in Database Systems},
   month = dec,
   year = {1986},
   keywords = {active data dictionaries},
   contents = {(not listed)}
}

@inproceedings{Roussopoulos86Preliminary,
   author = {N. Roussopoulos and H. Kang},
   title = {Preliminary Design of {ADMS}+/-: {A} Workstation-Mainframe Integrated Architecture for Database management Systems},
   booktitle = {Proc. 12th International Conference on VLDB},
   address = {Kyoto, Japan},
   month = aug,
   year = {1986},
   note = {(also appeared as UMIACS-TR-86-13 CS-TR-1674)},
   contents = {(not listed)}
}

@inproceedings{Roussopoulos86Engineering,
   author = {N. Roussopoulos},
   title = {Engineering Information Systems},
   booktitle = {Proc. ACM SIGMOD},
   address = {Washington, DC},
   month = may,
   year = {1986},
   contents = {(not listed)}
}

@inproceedings{Roussopoulos87Using,
   author = {N. Roussopoulos and H. Kang},
   title = {Using 2-way Semijoins in Distributed Query Processing},
   booktitle = {Proc. of Third International Conference on Data Engineering, Los Angeles},
   month = feb,
   year = {1987},
   contents = {(not listed)}
}

@inproceedings{Burkhard84Index,
   author = {W.A. Burkhard},
   title = {Index Maintenance for Non-Uniform Record Distributions},
   booktitle = {Proc. ACM SIGACT-SIGMOD},
   pages = {173-180},
   address = {Waterloo, Ontario},
   month = {Apr. 2-4},
   year = {1984},
   keywords = {exhash},
   contents = {(not listed)}
}

@article{Garg86Order,
   author = {A.K. Garg and C.C. Gotlieb},
   title = {Order-Preserving Key Transformations},
   journal = {ACM Trans. on Database Systems (TODS)},
   volume = {11},
   number = {2},
   pages = {213-234},
   month = jun,
   year = {1986},
   keywords = {hashing},
   contents = {(not listed)}
}

@article{Christodoulaki86Multimediab,
   author = {S. Christodoulakis and C. Faloutsos},
   title = {A Multimedia Document Server},
   journal = {IEEE Aerospace and Electronic Systems Magazine},
   volume = {1},
   number = {11},
   pages = {2-9},
   month = nov,
   year = {1986},
   keywords = {mmm},
   contents = {(not listed)}
}

@phdthesis{Lee85Design,
   author = {D.L. Lee},
   title = {The Design and Evaluation of a Text-Retrieval Machine for Large Databases},
   school = {Univ. of Toronto},
   type = {Ph.{D}. Dissertation},
   year = {1985},
   keywords = {hardware superimposed coding},
   contents = {(not listed)}
}

@article{Stanfill86Parallel,
   author = {C. Stanfill and B. Kahle},
   title = {Parallel Free-Text Search on the Connection Machine System},
   journal = {CACM},
   volume = {29},
   number = {12},
   pages = {1229-1239},
   month = dec,
   year = {1986},
   keywords = {hardware parallel superimposed coding},
   contents = {(not listed)}
}

@book{Hillis85Connection,
   author = {D. Hillis},
   title = {The Connection Machine},
   publisher = {MIT Press},
   address = {Cambridge, Mass.},
   year = {1985},
   contents = {(not listed)}
}

@techreport{Gunther86Cell,
   author = {O. Gunther},
   title = {The Cell Tree: An Index for Geometric Data},
   institution = {Univ. of California, Berkeley},
   type = {Memorandum No. UCB/ERL},
   number = {M86/89},
   month = dec,
   year = {1986},
   keywords = {spatial objects},
   contents = {(not listed)}
}

@article{Terry87Caching,
   author = {D. Terry},
   title = {Caching Hints in Distributed Systems},
   journal = {IEEE Trans. on Software Engineering},
   volume = {SE-13},
   number = {1},
   pages = {48-54},
   month = jan,
   year = {1987},
   contents = {(not listed)}
}

@techreport{Chu86Analysis,
   author = {J.H. Chu and G.D. Knott},
   title = {An Analysis of {B}-trees and their Variants},
   institution = {Univ. of Maryland},
   type = {CAR-TR-238,},
   number = {CS-TR-1737},
   month = nov,
   year = {1986},
   keywords = {performance btrees},
   contents = {(not listed)}
}

@inproceedings{Faloutsos87Analysis,
   author = {C. Faloutsos and T. Sellis and N. Roussopoulos},
   title = {Analysis of Object Oriented Spatial Access Methods},
   booktitle = {Proc. ACM SIGMOD},
   pages = {426-439 426-439},
   address = {San Francisco, CA},
   month = {May 27-29},
   year = {1987},
   keywords = {spatial objects},
   note = {also available as SRC-TR-87-30, UMIACS-TR-86-27, CS-TR-1781.},
   contents = {(not listed)}
}

@inproceedings{Sellis87Efficiently,
   author = {T. Sellis},
   title = {Efficiently Supporting Procedures in Relational Database Systems},
   booktitle = {Proc. ACM SIGMOD},
   address = {San Francisco, CA},
   month = {May 27-29},
   year = {1987},
   keywords = {procedures},
   contents = {(not listed)}
}

@article{Tamminen82EXCELL,
   author = {M. Tamminen},
   title = {The {EXCELL} Method for Efficient Geometric Access to Data},
   journal = {Acta Polytech. Scand. Mathematics and Computer Science Series},
   number = {34},
   address = {Helsinki},
   year = {1982},
   keywords = {seckey hashing},
   contents = {(not listed)}
}

@article{Chang81Picture,
   author = {J.M. Chang and K.S. Fu},
   title = {Picture Query Languages for Pictorial Data-base Systems},
   journal = {IEEE Computer},
   volume = {14},
   number = {11},
   month = nov,
   year = {1981},
   contents = {(not listed)}
}

@techreport{Kedem81Quad,
   author = {G. Kedem},
   title = {The Quad-{CIF} tree: {A} Data Structure for Hierarchical on-line Algorithms},
   institution = {Computer Science Department, Univ. of Rochester},
   type = {Tech. Report},
   number = {91},
   address = {Rochester, NY},
   month = sep,
   year = {1981},
   keywords = {spatial objects},
   contents = {(not listed)}
}

@inproceedings{Hinrichs83Grid,
   author = {K. Hinrichs and J. Nievergelt},
   editor = {M. Nagl and J. Perl},
   title = {The Grid File: {A} Data Structure to Support Proximity Queries on Spatial Objects},
   booktitle = {Proc. of the WG'83 (Intern. Workshop on Graph Theoretic Concepts in Computer Science)},
   pages = {100-113},
   publisher = {Trauner Verlag},
   address = {Linz, Austria},
   year = {1983},
   contents = {(not listed)}
}

@article{Lauther784,
   author = {U. Lauther},
   title = {4-Dimensional Binary Search Trees as a Means to Speed Up Associative Searches in Design Rule Verification of Integrated Circuits},
   journal = {Journal of Desing Automation and Fault-Tolerant Computing},
   volume = {2},
   number = {3},
   pages = {241-247},
   month = jul,
   year = {1978},
   keywords = {spatial objects},
   contents = {(not listed)}
}

@article{Rosenberg85Geographical,
   author = {J.B. Rosenberg},
   title = {Geographical Data Structures Compared: {A} Study of Data Structures Supporting Region Queries},
   journal = {IEEE Trans. on Computer-Aided Design},
   volume = {4},
   number = {1},
   pages = {53-67},
   month = jan,
   year = {1985},
   keywords = {spatial objects},
   contents = {(not listed)}
}

@article{Peano90Sur,
   author = {G. Peano},
   title = {Sur une courbe qui remplit toute une aire plaine},
   journal = {Mathematische Annalen},
   volume = {36},
   pages = {157-160},
   year = {1890},
   contents = {(not listed)}
}

@inproceedings{Ousterhout84Magic,
   author = {J. K. Ousterhout and G. T. Hamachi and R. N. Mayo and W. S. Scott and G. S. Taylor},
   title = {Magic: {A} {VLSI} Layout System},
   booktitle = {21st Design Automation Conference},
   pages = {152 - 159},
   address = {Alburquerque, NM},
   month = jun,
   year = {1984},
   contents = {(not listed)}
}

@misc{Samet86Quadtrees,
   author = {H. Samet},
   title = {Quadtrees and Related Hierarchical Data Structures for Computer Graphics and Image Processing},
   year = {1986},
   keywords = {book},
   note = {under preparation},
   contents = {(not listed)}
}

@phdthesis{Guttman84New,
   author = {A. Guttman},
   title = {New Features for Relational Database Systems to Support {CAD} Applications},
   school = {University of California, Berkeley},
   type = {Ph.{D}. Thesis},
   month = jun,
   year = {1984},
   contents = {(not listed)}
}

@techreport{Stonebraker83Application,
   author = {M. Stonebraker and B. Rubenstein and A. Guttman},
   title = {Application of Abstract Data Types and Abstract Indices to {CAD} Data Bases},
   institution = {University of California, Berkeley},
   type = {Tech. Report UCB/ERL M83/3, Electronics Research Laboratory},
   month = jan,
   year = {1983},
   contents = {(not listed)}
}

@inproceedings{Orenstein86Spatial,
   author = {J. Orenstein},
   title = {Spatial Query Processing in an Object-Oriented Database System},
   booktitle = {Proc. ACM SIGMOD},
   pages = {326-336},
   address = {Washington D.C.},
   month = may,
   year = {1986},
   keywords = {spatial objects},
   contents = {(not listed)}
}

@book{White81N,
   author = {M. White},
   title = {N-trees: Large ordered Indexes for Multi-dimensional Space},
   publisher = {Application Mathematics Research Staff},
   address = {Statistical Research Division, U.S. Bureau of the Census},
   month = dec,
   year = {1981},
   keywords = {spatial objects},
   contents = {(not listed)}
}

@article{Tropf81Multidimensional,
   author = {H. Tropf and H. Herzog},
   title = {Multidimensional Range Search in Dynamically Balanced Trees},
   journal = {Angewandte Informatik},
   year = {1981},
   keywords = {spatial objects},
   contents = {(not listed)}
}

@inproceedings{Fang86Idea,
   author = {M.F. Fang and R.C.T. Lee and C.C. Chang},
   title = {The Idea of De-clustering and its Applications},
   booktitle = {Proc. 12th International Conference on VLDB},
   pages = {181-188},
   address = {Kyoto, Japan},
   month = aug,
   year = {1986},
   keywords = {parallel hashing mah},
   contents = {(not listed)}
}

@inproceedings{Thom86Superjoin,
   author = {J.A. Thom and K. Ramamohanarao and L. Naish},
   title = {A Superjoin Algorithm for Deductive Databases},
   booktitle = {Proc. 12th International Conference on VLDB},
   pages = {189-196},
   address = {Kyoto, Japan},
   month = aug,
   year = {1986},
   keywords = {mah},
   contents = {(not listed)}
}

@inproceedings{DeWitt86,
   author = {D. DeWitt and R.H. Gerber and G. Graefe and M.L. Heytens and K.B. Kumar and M. Muralikrishna},
   title = {{GAMMA} - {A} High Performance Dataflow Database Machine},
   booktitle = {Proc. 12th International Conference on VLDB},
   pages = {228-237},
   address = {Kyoto, Japan},
   month = aug,
   year = {1986},
   keywords = {dbmachines mah},
   contents = {(not listed)}
}

@inproceedings{Cheiney86Reliable,
   author = {J.P. Cheiney and P. Faudemay and Rodolphe Michel and J.M. Thevenin},
   title = {A Reliable Parallel Backend Using Multiattribute Clustering and Select-Join Operations},
   booktitle = {Proc. 12th International Conference on VLDB},
   pages = {220-227},
   address = {Kyoto, Japan},
   month = aug,
   year = {1986},
   keywords = {dbmachines mah},
   contents = {(not listed)}
}

@techreport{Faloutsos87Object,
   author = {C. Faloutsos and T. Sellis and N. Roussopoulos and D. Metaxas},
   title = {Object Oriented Access Methods for Spatial Objects: Analysis for Multi-dimensional Spaces},
   institution = {Dept. of Computer Science, Univ. of Maryland},
   type = {UMIACS-TR-87-54 (also available as},
   number = {CS-TR-1940)},
   year = {1987},
   keywords = {spatial objects},
   contents = {(not listed)}
}

@misc{Faloutsos87Objecta,
   author = {C. Faloutsos and T. Sellis and N. Roussopoulos and D. Metaxas},
   title = {Object Oriented Access Methods for Spatial Objects: Algorithms and Analysis},
   year = {1987},
   keywords = {spatial objects},
   note = {submitted to TODS. Early version available as UMIACS-TR-87-54, CS-TR-1940.},
   contents = {(not listed)}
}

@book{Preparata85Computational,
   author = {F.P. Preparata and M.I. Shamos},
   title = {Computational Geometry},
   publisher = {Springer-Verlag},
   address = {New York},
   year = {1985},
   keywords = {spatial objects},
   contents = {(not listed)}
}

@article{Bentley86Locally,
   author = {J.L. Bentley and D.D. Sleator and R.E. Tarjan and V.K. Wei},
   title = {A Locally Adaptive Data Compression Scheme},
   journal = {CACM},
   volume = {29},
   number = {4},
   pages = {320-330},
   month = apr,
   year = {1986},
   abstract = {Uses a self-organizing linked list, to assign codes to words},
   contents = {(not listed)}
}

@article{Malvestuto86Statistical,
   author = {F.M. Malvestuto},
   title = {Statistical Treatment of the Information Content of a Database},
   journal = {Information Systems},
   volume = {11},
   number = {3},
   pages = {211-223},
   year = {1986},
   keywords = {infotheory entropy},
   abstract = {Uses mutual information to detect functional dependencies},
   contents = {(not listed)}
}

@inproceedings{Matsuo86Efficient,
   author = {F. Matsuo and S. Futamura and T. Shinohara},
   title = {Efficient Storage and Retrieval of Very Large Document Databases},
   booktitle = {Proc. of IEEE Data Engineering Conference},
   pages = {456-463},
   address = {Los Angeles},
   month = feb,
   year = {1986},
   keywords = {text compression},
   abstract = {Achieves 4-to-1 compression!?},
   contents = {(not listed)}
}

@inproceedings{Sellis87R,
   author = {T. Sellis and N. Roussopoulos and C. Faloutsos},
   title = {The {R}+ Tree: {A} Dynamic Index for Multi-Dimensional Objects},
   booktitle = {Proc. 13th International Conference on VLDB},
   pages = {507-518},
   address = {England,},
   month = sep,
   year = {1987},
   note = {also available as SRC-TR-87-32, UMIACS-TR-87-3, CS-TR-1795},
   contents = {(not listed)}
}

@article{Bassiouni86Double,
   author = {M.A. Bassiouni and B. Ok},
   title = {Double Encoding - {A} Technique for Reducing Storage Requirements for Text},
   journal = {Information Systems},
   volume = {11},
   number = {2},
   pages = {177-184},
   year = {1986},
   keywords = {compression text},
   contents = {(not listed)}
}

@article{Batory85Modeling,
   author = {D.S. Batory},
   title = {Modeling the Storage Architectures of Commercial Database Systems},
   journal = {ACM TODS},
   volume = {10},
   number = {4},
   pages = {463-528},
   month = dec,
   year = {1985},
   contents = {(not listed)}
}

@article{Coffman70File,
   author = {E.G. Coffman and J. Eve},
   title = {File Structures Using Hashing Functions},
   journal = {CACM},
   volume = {13},
   number = {7},
   pages = {427-432, 436},
   month = jul,
   year = {1970},
   keywords = {primkey hashing},
   contents = {(not listed)}
}

@article{Adelson-Velski62Algorithm,
   author = {G.M. Adel'son-Vel'skij and Y.M. Landis},
   title = {An Algorithm   for   the   Organization  of  Information},
   journal = {Doklady Akademia Nauk  USSR},
   volume = {146},
   number = {2},
   pages = {263-266},
   year = {1962},
   keywords = {primkey trees},
   contents = {(not listed)}
}

@article{Knuth71Optimum,
   author = {D.E. Knuth},
   title = {Optimum binary search trees},
   journal = {Acta Informatica},
   volume = {1},
   number = {1},
   pages = {14-25},
   month = jan,
   year = {1971},
   keywords = {primkey trees},
   contents = {(not listed)}
}

@article{Hu71Optimal,
   author = {T.C. Hu and A.C. Tucker},
   title = {Optimal  Computer Search  Trees  and Variable-Length Alphabetical Codes},
   journal = {SIAM J.  Appl.  Math.},
   volume = {21},
   number = {4},
   pages = {514-532},
   month = dec,
   year = {1971},
   keywords = {trees primkey},
   contents = {(not listed)}
}

@article{Nievergelt74Binary,
   author = {J. Nievergelt},
   title = {Binary Search Trees and File Organization},
   journal = {ACM Computing Surveys},
   volume = {6},
   number = {3},
   pages = {195-207},
   month = sep,
   year = {1974},
   keywords = {primkey},
   contents = {(not listed)}
}

@article{Shneiderman78Jump,
   author = {B. Shneiderman},
   title = {Jump Searching:  {A} Fast Sequential Search Technique},
   journal = {CACM},
   volume = {21},
   number = {10},
   pages = {831-834},
   month = oct,
   year = {1978},
   keywords = {primkey},
   contents = {(not listed)}
}

@article{Perl78Interpolation,
   author = {Y. Perl and A. Itai and H. Avni},
   title = {Interpolation Search-{A} Log Log {N} Search},
   journal = {CACM},
   volume = {21},
   number = {7},
   pages = {550-553},
   month = jul,
   year = {1978},
   keywords = {primkey},
   contents = {(not listed)}
}

@article{Rivest76Self,
   author = {R.L. Rivest},
   title = {On Self-Organizing Sequential Search Heuristics},
   journal = {CACM},
   volume = {19},
   number = {2},
   pages = {63-67},
   month = feb,
   year = {1976},
   keywords = {primkey}
}

@inproceedings{Yao76Complexity,
   author = {A.C. Yao and F.F. Yao},
   title = {The Complexity of Searching an Ordered Random Table},
   booktitle = {Proc. Symp. on Foundations of Computer Science,},
   pages = {173-177},
   address = {Houston},
   month = oct,
   year = {1976},
   keywords = {interpolation primkey},
   contents = {(not listed)}
}

@inproceedings{Lindsay86Snapshot,
   author = {B. Lindsay and L. Haas and C. Mohan and H. Pirahesh and P. Wilms},
   title = {A Snapshot Differential Refresh Algorithm},
   booktitle = {Proc. ACM SIGMOD},
   pages = {53-60},
   address = {Washington, D.C.},
   month = may,
   year = {1986},
   contents = {(not listed)}
}

@inproceedings{Blakeley86Efficiently,
   author = {J.A. Blakeley and P.A. Larson and F.W. Tompa},
   title = {Efficiently Updating Materialized Views},
   booktitle = {Proc. ACM SIGMOD},
   pages = {61-71 53-60},
   address = {Washington, D.C.},
   month = may,
   year = {1986},
   contents = {(not listed)}
}

@article{Yu85Adaptive,
   author = {C.T. Yu and C.M. Suen and K. Lam and M.K. Siu},
   title = {Adaptive Record Clustering},
   journal = {ACM TODS},
   volume = {10},
   number = {2},
   pages = {180-204},
   month = jun,
   year = {1985},
   keywords = {seckey},
   contents = {(not listed)}
}

@incollection{Futo78Application,
   author = {I. Futo and al et},
   editor = {H. Gallaire and J. Minker},
   title = {The Application of Prolog to the Development of {QA} and {DBM} Systems},
   booktitle = {Logic and Data Bases},
   publisher = {Plenum Press},
   year = {1978},
   contents = {(not listed)}
}

@article{Ramamohanarao86Superimposed,
   author = {K. Ramamohanarao and J. Shepherd},
   title = {A Superimposed Codeword Indexing Scheme for Very Large Prolog Databases},
   journal = {Third Intern. Conf. on Logic Programming},
   publisher = {Springer Verlag},
   address = {London},
   year = {1986},
   contents = {(not listed)}
}

@article{Faloutsos89Tri,
   author = {C. Faloutsos and W. Rego},
   title = {Tri-cell: {A} Data Structure for Spatial Objects},
   journal = {Information Systems},
   volume = {14},
   number = {2},
   pages = {131-139},
   year = {1989},
   note = {early version available as UMIACS-TR-87-15, CS-TR-1829.}
}

@techreport{Samet86Comparison,
   author = {H. Samet and R.E. Webber},
   title = {A Comparison of the Space Requirements of Multi-dimensional Quadtree-Based File Structures},
   institution = {Computer Science Department, Univ. of Maryland},
   type = {Tech. Report CAR-TR-225, also CS-TR-1711,},
   number = {DCR-83-02118},
   address = {College Park},
   month = sep,
   year = {1986},
   contents = {(not listed)}
}

@techreport{Nelson86Population,
   author = {R. Nelson and H. Samet},
   title = {A Population Analysis of Quadtrees with Variable Node Size},
   institution = {Computer Science Department, Univ. of Maryland},
   type = {Tech. Report CAR-TR-241, also CS-TR-1740,},
   number = {DCR-86-05557},
   address = {College Park},
   month = dec,
   year = {1986},
   contents = {(not listed)}
}

@inproceedings{Deppisch86S,
   author = {U. Deppisch},
   title = {S-tree: {A} Dynamic Balanced Signature Index for Office Retrieval},
   booktitle = {Proc. of ACM "Research and Development in Information Retrieval"},
   pages = {77-87},
   address = {Pisa, Italy},
   month = {Sept. 8-10},
   year = {1986},
   contents = {(not listed)}
}

@techreport{Faloutsos87Grid,
   author = {C. Faloutsos and W. Rego},
   title = {A Grid File Structure for Spatial Objects},
   institution = {Univ. of Maryland},
   type = {Tech. Report UMIACS-TR-87-15,},
   number = {CS-TR-1829},
   address = {College Park},
   month = apr,
   year = {1987},
   contents = {(not listed)}
}

@article{Sellis87High,
   author = {T. Sellis and N. Roussopoulos and L. Mark and C. Faloutsos},
   title = {High Performance Expert Database Systems: Efficient Support for Engineering Environments},
   journal = {Conf. on Data and Knowledge Systems for Manufacturing and Engineering},
   pages = {134-143},
   address = {Hartford, Connecticut},
   month = {Oct. 18-20},
   year = {1987},
   note = {also available as SRC-TR-87-177, UMIACS-TR-87-14, CS-TR-1828.},
   contents = {(not listed)}
}

@techreport{Roussopoulos87High,
   author = {N. Roussopoulos and L. Mark and T. Sellis and C. Faloutsos},
   title = {High Performance Engineering Information Systems},
   institution = {University of Maryland},
   type = {Technical Report SRC-TR-87-178, Systems Research Center,},
   month = oct,
   year = {1987}
}

@inproceedings{Wong85Bit,
   author = {H.K.T. Wong and H.F Liu and F. Olken and D. Rotem and L. Wong},
   title = {Bit Transposed Files},
   booktitle = {Proc. 11th International Conference on VLDB},
   pages = {448-457},
   address = {Stockholm, Sweden},
   month = aug,
   year = {1985},
   keywords = {seckeys},
   contents = {(not listed)}
}

@article{Hirschberg87Improved,
   author = {D.S. Hirschberg and D.J. Volper},
   title = {Improved Update/Query Algorithms for the Interval Valuation Problem},
   journal = {Information Processing Letters},
   volume = {24},
   number = {5},
   pages = {307-310},
   month = mar,
   year = {1987},
   keywords = {ipl},
   contents = {(not listed)}
}

@techreport{McKeown83,
   author = {D.M. McKeown},
   title = {{MAPS}: The Organization of a Spatial Database System Using Imagery, Terrain, and Map Data},
   institution = {Dept. of Computer Science, Carnegie-Mellon University},
   type = {Technical Report},
   number = {CMU-CS-83-136},
   month = {July 17},
   year = {1983},
   contents = {(not listed)}
}

@inproceedings{Powell83Database,
   author = {M. Powell and M. Linton},
   title = {Database Support for Programming Environments},
   booktitle = {Proc. Engineering Design Applications of ACM-IEEE Database Week,},
   address = {San Jose, California},
   month = may,
   year = {1983},
   contents = {(not listed)}
}

@book{Roussopoulos84Intensional,
   author = {N. Roussopoulos},
   title = {Intensional Semantics of the Relational Model},
   publisher = {Dept. of Computer Science},
   address = {University of Maryland},
   month = jan,
   year = {1984},
   contents = {(not listed)}
}

@inproceedings{Stonebraker84,
   author = {M. Stonebraker and E. Anderson and E. Hanson},
   title = {{QUEL} as a Data Type},
   booktitle = {Proc. ACM SIGMOD},
   month = jun,
   year = {1984},
   contents = {(not listed)}
}

@inproceedings{Tang80Logical,
   author = {G.Y. Tang},
   title = {A Logical Data Organization for the {I} database of Pictures an Alphanumerical Data},
   booktitle = {IEEE Proc. of the Workshop on Picture Data Description and Management},
   address = {Asilomar, California},
   month = aug,
   year = {1980},
   contents = {(not listed)}
}

@inproceedings{Zaniolo83Database,
   author = {C. Zaniolo},
   title = {The Database Language {GEM}},
   booktitle = {Proc. ACM SIGMOD},
   address = {San Jose, California},
   month = may,
   year = {1983},
   contents = {(not listed)}
}

@article{Stonebraker83Applicationa,
   author = {M. Stonebraker and J. Rubenstein and A. Guttman},
   title = {Application of Abstract Data Types and Abstract Indices},
   journal = {Engineering Design \& Applications, Database Week, ACM SIGMOD},
   address = {San Jose},
   month = {May 23-26},
   year = {1983},
   contents = {(not listed)}
}

@inproceedings{Roussopoulos85Introduction,
   author = {N. Roussopoulos and D. Leifker},
   title = {An Introduction to {PSQL}: {A} Pictorial Structured Query Language},
   booktitle = {IEEE Workshop on Visual Languages},
   address = {Hiroshima, Japan},
   month = {December 6-8},
   year = {1985},
   contents = {(not listed)}
}

@article{Chamberlin76,
   author = {D.D. Chamberlin},
   title = {{SEQUEL} 2:  {A} Unified Approach to Data Definition, Manipulation, and Control},
   journal = {IBM J. Research and Development},
   volume = {20},
   number = {6},
   pages = {560-575},
   year = {1976},
   contents = {(not listed)}
}

@techreport{Roussopoulos84a,
   author = {N. Roussopoulos and al et},
   title = {{ADMS}: {A} Self-Describing and Self-Documenting Relational Database System},
   institution = {Dept. of Computer Science, Univ. of Maryland},
   type = {Technical Report},
   month = aug,
   year = {1984},
   contents = {(not listed)}
}

@inproceedings{Schmidt78Type,
   author = {J. Schmidt},
   title = {Type Concepts for Database Definition},
   booktitle = {Proc. of the International Conference on Data Bases},
   address = {Haifa, Israel},
   month = aug,
   year = {1978},
   contents = {(not listed)}
}

@article{Chang81Pictorial,
   author = {S.K. Chang},
   title = {Pictorial Information Systems: Guest Editor's Introduction},
   journal = {IEEE Computer},
   volume = {14},
   number = {11},
   month = nov,
   year = {1981},
   contents = {(not listed)}
}

@article{Chang81Pictoriala,
   author = {S.K. Chang and L.K. Kunii},
   title = {Pictorial Database Systems},
   journal = {IEEE Computer},
   volume = {14},
   number = {11},
   month = nov,
   year = {1981},
   contents = {(not listed)}
}

@article{Kent79Limitations,
   author = {W. Kent},
   title = {Limitations of Record-Based Information Models},
   journal = {ACM TODS},
   volume = {4},
   number = {1},
   pages = {107-131},
   year = {1979},
   contents = {(not listed)}
}

@inproceedings{Lin80,
   author = {B.S. Lin and S.K. Chang},
   title = {{GRAIN} - {A} Pictorial Database Interface},
   booktitle = {IEEE Proc. of the Workshop on Picture Data Description and Management},
   address = {Asilomar, California},
   month = aug,
   year = {1980},
   contents = {(not listed)}
}

@techreport{McKeown83Concept,
   author = {D.M. McKeown},
   title = {Concept Maps},
   institution = {Dept. of Computer Science, Carnegie-Mellon University},
   type = {Technical Report},
   number = {CMU-CS-83-117},
   month = {April 27},
   year = {1983},
   contents = {(not listed)}
}

@techreport{Ramesh87Optimal,
   author = {R. Ramesh and I.V. Ramakrishnan},
   title = {Optimal Speedups for Parallel Pattern Matching in Trees},
   institution = {Dept. of Computer Science, State University of New York at Stony Brook},
   type = {TR},
   number = {\#87/09},
   address = {Stony Brook},
   month = mar,
   year = {1987},
   contents = {(not listed)}
}

@article{Litwin87New,
   author = {W. Litwin and D.B. Lomet},
   title = {A New Method for Fast Data Searches with Keys},
   journal = {IEEE Software},
   pages = {16-24},
   month = mar,
   year = {1987},
   contents = {(not listed)}
}

@article{Lomet87Partial,
   author = {D.B. Lomet},
   title = {Partial Expansions for File Organizations with an Index},
   journal = {ACM TODS},
   volume = {12},
   number = {1},
   pages = {65-84},
   month = mar,
   year = {1987},
   keywords = {Trans. on Database Systems},
   contents = {(not listed)}
}

@article{Roussopoulos88Efficient,
   author = {N. Roussopoulos and C. Faloutsos and T. Sellis},
   title = {An Efficient Pictorial Database System for {PSQL}},
   journal = {IEEE Transactions on Software Engineering},
   volume = {14},
   number = {5},
   pages = {639-650},
   month = may,
   year = {1988},
   note = {also available as UMIACS-TR-87-47, CS-TR-1928},
   contents = {(not listed)}
}

@techreport{Roussopoulos87,
   author = {N. Roussopoulos and C. Faloutsos and T. Sellis},
   title = {{PSQL}: An Efficient Pictorial Database System},
   institution = {University of Maryland},
   type = {Technical Report SRC-TR-87-176, Systems Research Center,},
   month = oct,
   year = {1987}
}

@article{Edelsbrunner87Zooming,
   author = {H. Edelsbrunner and M.H. Overmars},
   title = {Zooming by Repeated Range Detection},
   journal = {Information Processing Letters},
   volume = {24},
   pages = {413-417},
   month = apr,
   year = {1987},
   keywords = {ipl},
   contents = {(not listed)}
}

@article{Berra87Computer,
   author = {P.B. Berra and S.M. Chung and N.I. Hachem},
   title = {Computer Architecture for a Surrogate File to a Very Large Data/Knowledge Base},
   journal = {IEEE Computer Magazine},
   volume = {20},
   number = {3},
   pages = {25-32},
   month = mar,
   year = {1987},
   contents = {(not listed)}
}

@article{Thoma85Prototype,
   author = {G.R. Thoma and S. Suthasinekul and F.A. Walker and J. Cookson and M. Rashidian},
   title = {A Prototype System for the Electronic Storage and Retrieval of Document Images},
   journal = {ACM TOOIS},
   volume = {3},
   number = {3},
   month = jul,
   year = {1985},
   contents = {(not listed)}
}

@article{Christodoulaki87Analysis,
   author = {S. Christodoulakis},
   title = {Analysis of Retrieval Performance for Records and Objects Using Optical Disk Technology},
   journal = {ACM TODS},
   volume = {12},
   number = {2},
   pages = {137-169},
   month = jun,
   year = {1987},
   contents = {(not listed)}
}

@techreport{IBM79,
   author = {{IBM}},
   title = {{IBM} System/370 ({OS}/{VS}), Storage and Information Retrieval System / Vertical Storage ({STAIRS}/{VS})},
   institution = {IBM},
   type = {Reference manual},
   publisher = {IBM World Trade Corporation},
   year = {1979}
}

@article{Karp87Efficient,
   author = {R.M. Karp and M.O. Rabin},
   title = {Efficient Randomized Pattern-Matching Algorithms},
   journal = {IBM Journal of Research and Development},
   volume = {31},
   number = {2},
   pages = {249-260},
   month = mar,
   year = {1987},
   keywords = {text signatures},
   contents = {(not listed)}
}

@article{Du86Disk,
   author = {H.C. Du},
   title = {Disk Allocation Methods for Binary Cartesian Product Files},
   journal = {BIT},
   volume = {26},
   pages = {138-147},
   year = {1986},
   contents = {(not listed)}
}

@book{Ozkarahan86Database,
   author = {E. Ozkarahan},
   title = {Database Machines and Database Management},
   publisher = {Prentice Hall},
   year = {1986},
   contents = {(not listed)}
}

@article{Wong87Modeling,
   author = {S.K.M. Wong and W. Ziarko and V.V. Raghavan and P.C.N. Wong},
   title = {On Modeling of Information Retrieval Concepts in Vector Spaces},
   journal = {ACM TODS},
   volume = {12},
   number = {2},
   pages = {299-321},
   month = jun,
   year = {1987},
   contents = {(not listed)}
}

@article{Alexandridis87Encoding,
   author = {N.A. Alexandridis and P.D. Tsanakas},
   title = {An Encoding Scheme for the Efficient Representation of Hierarchical Image Structures},
   journal = {Information Processing Letters},
   volume = {25},
   number = {3},
   pages = {199-206},
   month = may,
   year = {1987},
   keywords = {IPL},
   contents = {(not listed)}
}

@article{Dandamudi86Algorithms,
   author = {S.P. Dandamudi and P.G. Sorenson},
   title = {Algorithms for {BD} trees,},
   journal = {Software - Practice and Experience},
   volume = {16},
   number = {12},
   pages = {1077-1096},
   month = dec,
   year = {1986},
   contents = {(not listed)}
}

@article{Cooper83Usage,
   author = {M.D. Cooper},
   title = {Usage Patterns of an Online Search System},
   journal = {Journal of the American Society for Information Science (JASIS)},
   volume = {5},
   number = {34},
   pages = {343-349},
   month = sep,
   year = {1983},
   contents = {(not listed)}
}

@article{Bird77Two,
   author = {R.S. Bird},
   title = {Two Dimensional Pattern Matching},
   journal = {Information Processing Letters},
   volume = {6},
   number = {5},
   pages = {168-170},
   month = oct,
   year = {1977},
   keywords = {image},
   abstract = {Uses the Knuth-Morris-Pratt algorithm
on two dimensions}
}

@incollection{Faloutsos87Integrated,
   author = {C. Faloutsos},
   editor = {G. Bracchi and D. Tsichritzis},
   title = {Integrated Access Methods for Messages Using Signature Files},
   booktitle = {Office Systems: Methods and Tools},
   pages = {137-157},
   publisher = {North Holland},
   year = {1987},
   note = {also available as UMIACS-TR-87-23, CS-TR-1867.},
   contents = {(not listed)}
}

@article{Wu87Associative,
   author = {C.T. Wu and W.A. Burkhard},
   title = {Associative Searching in Multiple Storage Units},
   journal = {ACM TODS},
   volume = {12},
   number = {1},
   pages = {38-64},
   month = mar,
   year = {1987},
   keywords = {declustering},
   abstract = {Uses a multi-dimensional extension of interpolation hashing,
and suggests disk allocation methods.}
}

@techreport{Faloutsos87Gray,
   author = {C. Faloutsos},
   title = {Gray Codes for Partial Match and Range Queries},
   institution = {Univ. of Maryland, Dept. of Computer Science},
   type = {UMIACS-TR-87-4, also},
   number = {CS-TR-1796},
   address = {College Park, MD},
   month = feb,
   year = {1987},
   keywords = {mah techreport}
}

@inproceedings{Stubbs84Waterloo,
   author = {J. Stubbs and F.W. Tompa},
   title = {Waterloo and the New Oxford English Dictionary Project},
   booktitle = {Proc. of the Twentieth Annual Conference on Editorial Problems},
   address = {Toronto, Ontario},
   month = {Nov. 2-3},
   year = {1984},
   note = {in press},
   contents = {(not listed)}
}

@inproceedings{Gonnet87Mind,
   author = {G.H. Gonnet and F.W. Tompa},
   title = {Mind your grammar: a new approach to modelling text},
   booktitle = {Proc. of the Thirteenth Int. Conf. on Very Large Data Bases},
   pages = {339-346},
   address = {Brighton, England},
   month = {Sept. 1-4},
   year = {1987},
   keywords = {vldb},
   contents = {(not listed)}
}

@book{Weiner85Computerizing,
   author = {E. Weiner},
   title = {Computerizing the Oxford English Dictionary},
   pages = {240-253},
   publisher = {Scholarly Publishing},
   month = apr,
   year = {1985},
   contents = {(not listed)}
}

@article{Willard87Multidimensional,
   author = {D.E. Willard},
   title = {Multidimensional Search Trees That Provide New Types of Memory Reductions},
   journal = {JACM},
   volume = {34},
   number = {4},
   pages = {846-858},
   month = oct,
   year = {1987},
   keywords = {geometric},
   abstract = {O( N ( logN / loglogN ) sup k-1 ) space to achieve O(log sup k N )
retrieval on orthogonal queries}
}

@article{Perry87Hypermedia,
   author = {T.S Perry},
   title = {Hypermedia: finally here},
   journal = {IEEE Spectrum},
   volume = {24},
   number = {11},
   pages = {38-39},
   month = nov,
   year = {1987},
   contents = {(not listed)}
}

@article{Bruno87Making,
   author = {R. Bruno},
   title = {Making compact disks interactive},
   journal = {IEEE Spectrum},
   volume = {24},
   number = {11},
   pages = {39-45},
   month = nov,
   year = {1987},
   contents = {(not listed)}
}

@article{Stone87Parallel,
   author = {H.S. Stone},
   title = {Parallel Querying of Large Databases: {A} Case Study},
   journal = {IEEE Computer},
   volume = {20},
   number = {10},
   pages = {11-21},
   month = oct,
   year = {1987},
   keywords = {signature files inverted indices text},
   contents = {(not listed)}
}

@article{Artlip87Optical,
   author = {P.M. Artlip},
   title = {Optical Disk Formats Allow Option Range for End User Applications},
   journal = {Computer Technology Review},
   pages = {43-48},
   month = {Fall},
   year = {1987},
   abstract = {Interesting data on WORM and erasable O.D.},
   contents = {(not listed)}
}

@article{Gait88Optical,
   author = {J. Gait},
   title = {The Optical File Cabinet: {A} Random Access File System for Write Once Optical Disk},
   journal = {IEEE Computer Magazine},
   volume = {21},
   number = {6},
   pages = {11-22},
   month = jun,
   year = {1988},
   contents = {(not listed)}
}

@book{Horowitz76Fundamentals,
   author = {E. Horowitz and S. Sahni},
   title = {Fundamentals of Data Structures},
   publisher = {Computer Science Press, Inc.},
   year = {1976},
   contents = {(not listed)}
}

@article{Izawa84Documenta,
   author = {K. Izawa},
   title = {Document Image Filing System Utilizing Optical Disk Memories},
   journal = {Database Engineering},
   volume = {3},
   pages = {137-141},
   year = {1984},
   contents = {(not listed)}
}

@article{Levine87Optical,
   author = {R. Levine},
   title = {Optical Storage},
   journal = {DEC Professional},
   pages = {30-38},
   month = feb,
   year = {1987},
   contents = {(not listed)}
}

@article{Langeorthy86Mass,
   author = {G. Langeorthy},
   title = {Mass Storage},
   journal = {Digital Review},
   pages = {88-91},
   month = jun,
   year = {1986},
   contents = {(not listed)}
}

@article{Metzger87Erasable,
   author = {N. Metzger and N.J. Freundlich},
   title = {Erasable},
   journal = {Popular Science},
   pages = {56-59},
   month = may,
   year = {1987},
   contents = {(not listed)}
}

@article{OLear85Optical,
   author = {B.T. O'Lear and D.L. Kitts},
   title = {Optical Device Interfacing for a Mass Storage System},
   journal = {Computer},
   volume = {18},
   number = {7},
   pages = {24-32},
   month = jul,
   year = {1985},
   contents = {(not listed)}
}

@article{Ooi87Design,
   author = {B.C. Ooi and A.N. Narasimkalu and K.Y. Wang and I.F. Chang},
   title = {Design of a Mutimedia File Server Using Optical Disks for Office Applications},
   journal = {IEEE},
   pages = {157-163},
   month = jan,
   year = {1987},
   contents = {(not listed)}
}

@book{Standish80Data,
   author = {T.A. Standish},
   title = {Data Structure Techniques},
   publisher = {Addison Wesley},
   year = {1980},
   contents = {(not listed)}
}

@article{Walter85Optical,
   author = {G. Walter},
   title = {Optical Digital Data Disk Technology for the Management of Engineering Documents},
   journal = {Journal of Information \& Image Management},
   pages = {21-26},
   month = jan,
   year = {1985},
   contents = {(not listed)}
}

@book{Wiederhold77Databasea,
   author = {G. Wiederhold},
   title = {Database Design},
   publisher = {McGraw-Hill, Inc.},
   year = {1977},
   contents = {(not listed)}
}

@article{Chen87Processor,
   author = {M.S. Chen and K.G. Shin},
   title = {Processor Allocation in an {N}-Cube Multiprocessor Using Gray Codes},
   journal = {IEEE Trans. on Computers},
   volume = {C-36},
   number = {12},
   pages = {1396-1407},
   month = dec,
   year = {1987},
   contents = {(not listed)}
}

@article{Chen86Coompact,
   author = {P. Chen},
   title = {The Coompact Disk {ROM}: How it works},
   journal = {IEEE Spectrum},
   volume = {23},
   number = {4},
   pages = {44-49},
   month = apr,
   year = {1986},
   contents = {(not listed)}
}

@article{Berra87Optical,
   author = {P. Berra and N. Troullinos},
   title = {Optical Techniques and Data/Knowledge Base Machine},
   journal = {IEEE Computer Magazine},
   volume = {20},
   number = {10},
   pages = {59-70},
   month = oct,
   year = {1987},
   contents = {(not listed)}
}

@article{Langworthy86Inside,
   author = {G. Langworthy},
   title = {Inside a {WORM} Optical Disk Drive},
   journal = {Digital Review},
   volume = {3},
   number = {9},
   pages = {86-91},
   month = jun,
   year = {1986},
   contents = {(not listed)}
}

@incollection{Laub86What,
   author = {L. Laub},
   editor = {S. Lambert and S. Ropiequet},
   title = {What is {CD} {ROM}?},
   booktitle = {CD ROM},
   pages = {47-72},
   publisher = {Microsoft Press},
   year = {1986},
   contents = {(not listed)}
}

@article{Brandel87Erasable,
   author = {W.B. Brandel},
   title = {Erasable Optical Disk to Spin Out At Comdex/Fall},
   journal = {Digital Review},
   pages = {1,8},
   month = {Oct. 26},
   year = {1987},
   contents = {(not listed)}
}

@phdthesis{Hinterberger87Data,
   author = {H. Hinterberger},
   title = {Data Density: {A} Powerful Abstraction to Manage and Analyze Multivariate Data},
   school = {ETH},
   type = {Ph.{D}. Dissertation},
   address = {Zurich, Switzerland},
   year = {1987},
   keywords = {grid file}
}

@article{Murakami85Research,
   author = {K. Murakami and T. Kakuta and R. Onai and N. Ito},
   title = {Research on Parallel Machine Architecture for Fifth-Generation Computer Systems},
   journal = {IEEE Computer Magazine},
   volume = {18},
   number = {6},
   pages = {76-92},
   month = jun,
   year = {1985},
   keywords = {delta binary model dbmachine},
   abstract = {Uses binary relations for physical storage of knowledge bases}
}

@book{Kowalski79Logic,
   author = {R. Kowalski},
   title = {Logic for Problem Solving},
   publisher = {North Holland},
   year = {1979}
}

@article{Freese88Optical,
   author = {R.P. Freese},
   title = {Optical Disks Become Erasable},
   journal = {IEEE Spectrum},
   volume = {25},
   number = {2},
   pages = {41-45},
   month = feb,
   year = {1988},
   keywords = {worm cd-rom},
   abstract = {Laser beam heats the material, and helps
change the magnetic characteristics
of a "bit" on the disk.
Capacities of \~{}600Mb for 5 1/4'' disks,
price of medium \~{}\$200 (to drop down to \$50),
price of drive \~{}\$12K (to drop down to \$1K,
in a few years).
First models will appear in late 1988.}
}

@inproceedings{Copeland85Decomposition,
   author = {G.P. Copeland and S.N. Khoshafian},
   title = {A Decomposition Storage Model},
   booktitle = {Proc. ACM SIGMOD},
   pages = {268-279},
   address = {Austin, Texas},
   month = {May 28-31},
   year = {1985},
   keywords = {binary relationship model}
}

@article{Batory79Searching,
   author = {D.S. Batory},
   title = {On Searching Transposed Files},
   journal = {ACM TODS},
   volume = {4},
   number = {4},
   month = dec,
   year = {1979}
}

@article{Ramakrishna89File,
   author = {M.V. Ramakrishna and P.A. Larson},
   title = {File Organization Using Composite Perfect Hashing},
   journal = {ACM TODS},
   volume = {14},
   number = {2},
   pages = {231-263},
   month = jun,
   year = {1989},
   keywords = {external}
}

@article{Griffiths86Algorithm,
   author = {J.G. Griffiths},
   title = {An Algorithm for Displaying a Class of Space-filling Curves},
   journal = {Software-Practice and Experience},
   volume = {16},
   number = {5},
   pages = {403-411},
   month = may,
   year = {1986},
   keywords = {distance preserving}
}

@inproceedings{Katevenis83,
   author = {M.G.H. Katevenis and R.W. Sherburne and D.A. Patterson and C.H. Sequin},
   title = {\_ The {RISC} {II} Micro-Architecture},
   booktitle = {Proc. VLSI 83 Conference},
   address = {Trodheim, Norway},
   month = aug,
   year = {1983}
}

@inproceedings{Karp72Rapid,
   author = {R.M. Karp and R.E. Miller and A.L. Rosenberg},
   title = {Rapid Identification of Repeated Patterns in Strings, Trees and Arrays},
   booktitle = {Proc. of the 4th ACM Symposium on Theory of Computing},
   pages = {125-136},
   address = {Denver, Colorado},
   month = {May 1-3},
   year = {1972},
   keywords = {2-d pattern matching text}
}

@inproceedings{Katz86Version,
   author = {R.H. Katz and E. Chang and R. Bhateja},
   title = {Version Modeling Concepts for Computer-Aided Design Databases},
   booktitle = {Proc. of ACM SIGMOD},
   pages = {379-386},
   address = {Washington, D.C.},
   month = {May 28-30},
   year = {1986},
   keywords = {engineering cad},
   abstract = {They propose a model that supports: a) Version histories
(organized in tree-like networks, with currency indicator
for the prevailing design)
b) Time varying configurations (because a specific version of an object
consists of versions of its components)
c) Equivalence among objects of different type (e.g.,
source code, object code and manuals for a software module)}
}

@inproceedings{Hardwick87Why,
   author = {M. Hardwick},
   title = {Why {ROSE} is Fast: Five Optimizations in the Design of an Experimental Database System for {CAD}/{CAM} Applications},
   booktitle = {Proc. ACM SIGMOD Conf.},
   pages = {292-298},
   address = {San Francisco, California},
   month = {May 27-29},
   year = {1987},
   keywords = {version engineering},
   abstract = {Main ideas: a) each object is represented as a nested structure
(AND/OR tree) and is stored in a separate UNIX file
b) for each object, we have an INTERFACE (= specifications)
and one or more VERSIONS (implementations)}
}

@article{Batory85Modelinga,
   author = {D.S. Batory and W. Kim},
   title = {Modeling Concepts for {VLSI} Objects},
   journal = {ACM TODS},
   volume = {10},
   number = {3},
   pages = {289-321},
   month = sep,
   year = {1985},
   keywords = {cad engineering versions},
   abstract = {Important observation:
Design objects consist of the interface
(\~{} object type) and the implementation
(\~{} object instance).
.br
Failure of previous modeling efforts: due to absense
of a coherent model.
Present paper considers only VLSI objects.
.br
Modeling concepts proposed:
(i) molecular objects: contain heterogeneous records
(ii) version generalization (the relationship
between an object type and an object version)
(iii) instantiation (a copy of an object type or version
in a specific diagram (=larger object))
(iv) parameterized versions (a larger object,
that contains one or more object types (and not versions).
This objects serves as a template, with "sockets"
or "plugs")
.br
Future work: (a) include ADT. (b) function-valued
attributes (\~{} QUEL as data type).}
}

@article{Wang87Multiple,
   author = {J.-H. Wang and T.-S. Yuen and D.H.-C. Du},
   title = {On Multiple Random Accesses and Physical Data Placement in Dynamic Files},
   journal = {IEEE Trans. on Software Engineering},
   volume = {SE-13},
   number = {8},
   pages = {977-987},
   month = aug,
   year = {1987},
   keywords = {declustering parhash}
}

@misc{Lin88Performance,
   author = {C.-C. Lin and L. Mark and T. Sellis and C. Faloutsos},
   title = {Performance Issues in the Binary Relationship Model},
   year = {1988},
   keywords = {mah},
   note = {Available as TR: CS-2001, and UMIACS-88-21}
}

@article{Chang87Performance,
   author = {C.C. Chang and C.Y. Chen},
   title = {Performance of Two-Disk Partition Data Allocations},
   journal = {BIT},
   volume = {27},
   number = {3},
   pages = {306-314},
   year = {1987},
   keywords = {parhash declustering}
}

@inproceedings{Kim88Optimal,
   author = {M.H. Kim and S. Pramanik},
   title = {Optimal File Distribution for Partial Match Retrieval},
   booktitle = {Proc. ACM SIGMOD Conf.},
   pages = {173-182},
   address = {Chicago, Illinois},
   month = {June 1-3},
   year = {1988},
   keywords = {declustering parhash},
   abstract = {Create attribute signatures and EX-OR them,
to achieve declustering.
Elaborate heuristics, if the domain of an attribute
is too small.}
}

@article{McIlroy76Algorithm,
   author = {M.D. McIlroy and J.W. Hunt},
   title = {An Algorithm for Differential File Comparison},
   journal = {Bell Labs Computing Science TR 41},
   month = jun,
   year = {1976}
}

@inproceedings{Faloutsos88Fast,
   author = {C. Faloutsos and R. Chan},
   title = {Fast Text Access Methods for Optical and Large Magnetic Disks: Designs and Performance Comparison},
   booktitle = {Proc. 14th International Conf. on VLDB},
   pages = {280-293},
   address = {Long Beach, California},
   month = aug,
   year = {1988},
   note = {also available as UMIACS-TR-87-66, CS-TR-1958.}
}

@article{Barnsley88Better,
   author = {M.F. Barnsley and A.D. Sloan},
   title = {A Better Way to Compress Images},
   journal = {Byte},
   pages = {215-223},
   month = jan,
   year = {1988},
   keywords = {fractals space filling curves},
   abstract = {main idea: break down an image into homogeneous areas,
and approximate each area with a fractal.
A fractal needs very few parameters to be described.
They include a probabilistic algorithm,
that creates an image, given the parameters
of the fractal.}
}

@article{Li87Adjacency,
   author = {S-X Li and M.H. Loew},
   title = {Adjacency Detection Using Quadcodes},
   journal = {CACM},
   volume = {30},
   number = {7},
   pages = {627-631},
   month = jul,
   year = {1987},
   keywords = {quadtrees}
}

@article{Li87Quadcode,
   author = {S-X Li and M.H. Loew},
   title = {The Quadcode and Its Arithmetic},
   journal = {CACM},
   volume = {30},
   number = {7},
   pages = {621-626},
   month = jul,
   year = {1987},
   keywords = {quadtrees}
}

@inproceedings{Chou86Unifying,
   author = {H-T Chou and W. Kim},
   title = {A Unifying Framework for Version Control in a {CAD} Environment},
   booktitle = {Proc. VLDB},
   pages = {336-344},
   address = {Kyoto, Japan},
   month = aug,
   year = {1986},
   abstract = {- 3 levels (released, working \& transient versions)
- support for distributed environment
- idea of version change notification.
Also, they defined operations on versions
and proposed a language to support them.}
}

@article{Brown84System,
   author = {M.H. Brown and R. Sedgewick},
   title = {A System for Algorithm Animation},
   journal = {Computer Graphics},
   volume = {18},
   number = {3},
   pages = {177-186},
   month = jul,
   year = {1984}
}

@article{Lea88Digital,
   author = {D. Lea},
   title = {Digital and Hilbert k-d Trees},
   journal = {IPL},
   volume = {27},
   number = {1},
   pages = {35-41},
   month = feb,
   year = {1988},
   keywords = {space filling curves fractals}
}

@article{Garcia-Molina88Impact,
   author = {H. Garcia-Molina and K. Salem},
   title = {The Impact of Disk Stripping on Reliability},
   journal = {IEEE Database Engineering},
   volume = {11},
   number = {1},
   pages = {26-39},
   month = mar,
   year = {1988},
   keywords = {error correcting codes},
   abstract = {Main ideas: Distribute a block of a file over several
disks, to improve response time.
Use a clever scheme of error correction, to tolerate
the failure of one disk.}
}

@techreport{Faloutsos87Signature,
   author = {C. Faloutsos},
   title = {Signature Files: An Integrated Access Method for Text and Attributes Suitable for Optical Disk Storage.},
   institution = {Dept. of Computer Science, Univ. of Maryland},
   type = {Tech. Rep. UMIACS-TR-87-23,},
   number = {CS-TR-1867},
   address = {College Park},
   month = jun,
   year = {1987},
   keywords = {signature files},
   contents = {(not listed)}
}

@article{Manolopoulos88Estimating,
   author = {Y.P. Manolopoulos and J.G. Kollias},
   title = {Estimating disk head movement in batched searching},
   journal = {BIT},
   volume = {28},
   number = {1},
   pages = {27-36},
   year = {1988}
}

@article{Lasseter87Principles,
   author = {J. Lasseter},
   title = {Principles of Traditional Animation Applied to 3d Computer Animation},
   journal = {Computer Graphics},
   volume = {21},
   number = {4},
   pages = {35-44},
   month = jul,
   year = {1987},
   abstract = {Discusses 11 basic principles (squash \& stretch,
timing, anticipation, staging etc.)
to create realistic (cartoon) animation.}
}

@article{Lin92Frame,
   author = {Z. Lin and C. Faloutsos},
   title = {Frame Sliced Signature Files},
   journal = {IEEE Trans. on Knowledge and Data Engineering (TKDE)},
   volume = {4},
   number = {3},
   month = jun,
   year = {1992},
   note = {to appear. Also available as CS-TR-2146 and UMIACS-TR-88-88}
}

@techreport{Faloutsos87Fast,
   author = {C. Faloutsos and R. Chan},
   title = {Fast Text Access Methods for Optical and Large Magnetic Disks: Designs and Performance Comparison},
   institution = {Univ. of Maryland, College Park},
   type = {UMIACS-TR-87-66},
   number = {CS-TR-1958},
   month = dec,
   year = {1987}
}

@article{Kimbrell88Searching,
   author = {R.E. Kimbrell},
   title = {Searching for Text? Send an {N}-gram!},
   journal = {Byte},
   volume = {13},
   number = {5},
   pages = {297-312},
   month = may,
   year = {1988},
   keywords = {superimposed coding signature files},
   abstract = {Preprocess the collection of documents,
pick up \~{}12,000 n-grams
and represent each document as a 12,000-bit vector.}
}

@inproceedings{Papadimitriou76Some,
   author = {C. Papadimitriou and K. Steiglitz},
   title = {Some complexity results for the travelling salesman problem},
   booktitle = {Proc. Eighth Annual ACM Symposium on Theory of Computing},
   pages = {1-9},
   month = may,
   year = {1976}
}

@article{Du82Disk,
   author = {H.C. Du and J.S. Sobolewski},
   title = {Disk allocation for Cartesian product files on multiple disk systems},
   journal = {ACM Trans. Database Systems (TODS)},
   volume = {7},
   number = {1},
   pages = {82-101},
   month = mar,
   year = {1982}
}

@article{Sung87Performance,
   author = {Y.Y. Sung},
   title = {Performance Analysis of Disk Modulo Allocation Method for Cartesian Product Files},
   journal = {IEEE Transactions on Software Engineering},
   volume = {SE 13},
   number = {9},
   pages = {1018-1026},
   month = sep,
   year = {1987}
}

@inproceedings{Sung85Parallel,
   author = {Y.Y. Sung},
   title = {Parallel searching for Binary Cartesian Product Files},
   booktitle = {Proc. ACM CSC 1985 Conf.},
   pages = {163-172},
   address = {New Orleans},
   month = mar,
   year = {1985}
}

@article{Bially69Space,
   author = {T. Bially},
   title = {Space-Filling Curves: Their Generation and Their Application to Bandwidth Reduction},
   journal = {IEEE Trans. on Information Theory},
   volume = {IT-15},
   number = {6},
   pages = {658-664},
   month = nov,
   year = {1969},
   keywords = {hilbert}
}

@article{Quinqueton81Locally,
   author = {J. Quinqueton and M. Berthod},
   title = {A Locally Adaptive Peano Scanning Algorithm},
   journal = {IEEE Trans. on Pattern Analysis and Machine Intelligence},
   volume = {PAMI-3},
   number = {4},
   pages = {403-412},
   month = jul,
   year = {1981},
   keywords = {hilbert space filling curve}
}

@article{Croft88Implementing,
   author = {W.B. Croft and P. Savino},
   title = {Implementing Ranking Strategies Using Text Signatures},
   journal = {ACM Trans. on Office Informations Systems (TOOIS)},
   volume = {6},
   number = {1},
   pages = {42-62},
   month = jan,
   year = {1988}
}

@article{Sacks-Davis87Multikey,
   author = {R. Sacks-Davis and A. Kent and K. Ramamohanarao},
   title = {Multikey Access Methods Based on Superimposed Coding Techniques},
   journal = {ACM Trans. on Database Systems (TODS)},
   volume = {12},
   number = {4},
   pages = {655-696},
   month = dec,
   year = {1987},
   keywords = {text signature files}
}

@article{Chang88Application,
   author = {C.C. Chang},
   title = {Application of Principal Component Analysis to Multi-disk Concurrent Accessing},
   journal = {BIT},
   volume = {28},
   pages = {205-214},
   year = {1988},
   keywords = {declustering parhash}
}

@inproceedings{Gonnet77Analysis,
   author = {G.H. Gonnet and I. Munro},
   title = {The Analysis of an Improved Hashing Technique},
   booktitle = {Proc. 9th Ann. ACM Symposium on the Theory of Computing (STOC)},
   pages = {113-121},
   month = may,
   year = {1977},
   keywords = {primkey binary tree},
   abstract = {On collision, displace the record that will cause the fewest propagated collisions}
}

@article{Amble74Ordered,
   author = {O. Amble and D.E. Knuth},
   title = {Ordered Hash Tables},
   journal = {Computer Journal},
   volume = {17},
   number = {2},
   pages = {135-142},
   month = may,
   year = {1974},
   keywords = {primkey},
   abstract = {On collision, alphabetically smaller keys displace the larger ones.}
}

@article{Sleator85Self,
   author = {D.D. Sleator and R.E. Tarjan},
   title = {Self-adjusting binary search trees},
   journal = {JACM},
   volume = {32},
   number = {3},
   pages = {652-686},
   month = jul,
   year = {1985},
   keywords = {primkey},
   abstract = {The splay tree is introduced; it is the basis for a version
of persistent search trees.}
}

@inproceedings{VLDB88Panel,
   author = {{VLDB}},
   title = {Panel on the Lagouna Beach report},
   booktitle = {14th International Conference on Very Large Data Bases (VLDB)},
   institution = {VLDB},
   address = {Long Beach, CA},
   month = aug,
   year = {1988}
}

@article{Sellis88Expert,
   author = {T. Sellis and N. Roussopoulos and L. Mark and C. Faloutsos},
   title = {Expert Database Systems: Efficient Support for Engineering Environments},
   journal = {Data and Knowledge Engineering},
   volume = {3},
   number = {2},
   pages = {71-85},
   month = aug,
   year = {1988},
   note = {early version available as UMIACS-TR-87-14, CS-TR-1828.}
}

@article{Duff84Design,
   author = {I.S. Duff},
   title = {Design Features of a Frontal Code for Solving sparse Unsymmetric Linear Systems Out-of-core},
   journal = {SIAM J. Sci. Stat. Computing},
   volume = {5},
   number = {2},
   pages = {270-280},
   month = jun,
   year = {1984},
   keywords = {tricell fractals numerical analysis matrices}
}

@misc{Bartholdi86Heuristics,
   author = {J.J. Bartholdi and L.K. Platzman},
   title = {Heuristics Based on Spacefilling Curves for Combinatorial Problems in the Plane},
   year = {1986},
   keywords = {hilbert peano fractals},
   abstract = {Use a space filling curve, order the points in a 1-d space
and then solve the problem (eg., TSP) in 1-d,
which is trivial.},
   note = {unpublished manuscript}
}

@article{Bartholdi83Fast,
   author = {J.J. Bartholdi and L.K. Platzman},
   title = {A Fast Heuristic Based on Spacefilling Curves for Minimum-Weight Matching in the Plane},
   journal = {IPL},
   volume = {17},
   pages = {177-180},
   month = nov,
   year = {1983}
}

@article{Bartholdi82O,
   author = {J.J. Bartholdi and L.K. Platzman},
   title = {An {O}(NlogN) Travelling Salesman Heuristic Based on Spacefilling Curves},
   journal = {Operation Research Letters},
   volume = {1},
   number = {4},
   pages = {121-125},
   month = sep,
   year = {1982}
}

@article{Platzman89Spacefilling,
   author = {L.K. Platzman and J.J. Bartholdi},
   title = {Spacefilling Curves and the Planar Travelling Salesman Problem},
   journal = {Journal of ACM (JACM)},
   volume = {36},
   number = {4},
   pages = {719-737},
   month = oct,
   year = {1989}
}

@article{McCluskey59Error,
   author = {E.J. McCluskey},
   title = {Error-Correcting Codes - {A} Linear Programming Approach},
   journal = {The Bell System Technical Journal},
   volume = {38},
   pages = {1485-1512},
   month = nov,
   year = {1959},
   keywords = {ecc parhash declustering}
}

@book{Wirth86Algorithms,
   author = {N. Wirth},
   title = {Algorithms and Data Structures},
   publisher = {Prentice-Hall Inc},
   address = {Englewood Cliff, NJ},
   year = {1986}
}

@article{Larson88Linear,
   author = {P.A. Larson},
   title = {Linear Hashing with Separators - {A} Dynamic Hashing Scheme Achieving One-Access Retrieval},
   journal = {ACM Trans. on Database Systems (TODS)},
   volume = {13},
   number = {3},
   pages = {366-388},
   month = sep,
   year = {1988},
   keywords = {extendible},
   abstract = {Expand the idea of separators ("signatures") for each bucket:
For a static hash table, with double hashing,
keep a short signature for each bucket,
containing the largest signature in this bucket for the moment.
Examining the array of signatures in main memory, leads
to one only disk access in succ. and unsucc. search.
Drawbacks: Insertion is costly - the load factor must be kept
by 80-85\%.}
}

@article{Tucker88Architecture,
   author = {L.W. Tucker and G.G. Robertson},
   title = {Architecture and Applications of the Connection Machine},
   journal = {IEEE Computer Magazine},
   volume = {21},
   number = {8},
   pages = {26-38},
   month = aug,
   year = {1988},
   keywords = {datavault}
}

@article{Batory82Unifying,
   author = {D.S. Batory and C.C. Gotlieb},
   title = {A Unifying Model of Physical Databases},
   journal = {ACM Trans. on Database Systems (TODS)},
   volume = {7},
   number = {4},
   pages = {509-539},
   month = dec,
   year = {1982}
}

@book{Trivedi82Probability,
   author = {K.S. Trivedi},
   title = {Probability and Statistics with Reliability, Queuing and Computer Science Applications},
   publisher = {Prentice Hall, Inc.},
   address = {Englewood Cliffs, NJ},
   year = {1982},
   abstract = {contains formula for order statistics}
}

@article{Lelewer87Data,
   author = {D.A. Lelewer and D.S. Hirschberg},
   title = {Data Compression},
   journal = {ACM Computing Surveys},
   volume = {19},
   number = {3},
   pages = {261-296},
   month = sep,
   year = {1987},
   keywords = {infotheory}
}

@techreport{UCB84,
   author = {{UCB}},
   title = {{UNIX} User's Manual},
   institution = {UCB},
   type = {Version 4.2., Berkeley Software Distribution, Virtual VAX-11 Version},
   publisher = {Univ. of California},
   address = {Berkeley, CA},
   year = {1984},
   keywords = {compression},
   abstract = {information about "compress"}
}

@article{Gallager78Variations,
   author = {R.G. Gallager},
   title = {Variations on a Theme by Huffman},
   journal = {IEEE Trans. on Information Theory},
   volume = {24},
   number = {6},
   pages = {668-674},
   month = nov,
   year = {1978},
   abstract = {proposes adaptive huffman coding}
}

@article{Knuth85Dynamic,
   author = {D.E. Knuth},
   title = {Dynamic Huffman Coding},
   journal = {J. Algorithms},
   volume = {6},
   number = {2},
   pages = {163-180},
   month = jun,
   year = {1985},
   keywords = {compression},
   abstract = {proposes adaptive huffman coding}
}

@article{Faller73Adaptive,
   author = {N. Faller},
   title = {An Adaptive System for Data Compression},
   journal = {Record of the 7th Osilomar Conf. on Circuits, Systems and Computers},
   pages = {593-597},
   address = {Pacific Grove, CA},
   month = nov,
   year = {1973},
   keywords = {huffman},
   abstract = {proposes adaptive huffman coding}
}

@article{Blumer87Complete,
   author = {A. Blumer and J. Blumer and D. Haussler and R. McConnell and A. Ehrenfeucht},
   title = {Complete Inverted Files for Efficient Text Retrieval and Analysis},
   journal = {JACM},
   volume = {34},
   number = {3},
   pages = {578-595},
   month = jul,
   year = {1987},
   keywords = {dawg},
   abstract = {(Not read yet) - proposes a new data structure,
"compact directed acyclic word graph".
this allows searching for arbitrary strings
(I guess, like PATRICIA??)}
}

@inproceedings{Faloutsos89Fractals,
   author = {C. Faloutsos and S. Roseman},
   title = {Fractals for Secondary Key Retrieval},
   booktitle = {Eighth ACM SIGACT-SIGMOD-SIGART Symposium on Principles of Database Systems (PODS)},
   pages = {247-252},
   address = {Philadelphia, PA},
   month = {March 29-31},
   year = {1989},
   keywords = {hilbert space filling curves},
   note = {also available as UMIACS-TR-89-47 and CS-TR-2242}
}

@article{Frisse88From,
   author = {M. Frisse},
   title = {From Text to Hypertext},
   journal = {BYTE},
   volume = {13},
   number = {10},
   pages = {247-253},
   month = oct,
   year = {1988},
   abstract = {suggests indexing on hypertext databases;
proposes the "clustering" approach.}
}

@article{Orenstein88,
   author = {J.A. Orenstein and F.A. Manola},
   title = {{PROBE} Spatial Data Modeling and Query Processing in an Image Database Application},
   journal = {IEEE Trans. on Software Engineering},
   volume = {14},
   number = {5},
   pages = {611-629},
   month = may,
   year = {1988},
   keywords = {z-order fractals space filling curves}
}

@inproceedings{Faloutsos91,
   author = {C. Faloutsos and Y. Rong},
   title = {{DOT}: {A} Spatial Access Method Using Fractals},
   booktitle = {IEEE Data Engineering Conference},
   pages = {152-159},
   address = {Kobe, Japan},
   month = apr,
   year = {1991},
   keywords = {hilbert space filling curves},
   note = {early version available as UMIACS-TR-89-31, CS-TR-2214}
}

@techreport{Faloutsos89Spatial,
   author = {C. Faloutsos and Y. Rong},
   title = {Spatial Access Methods Using Fractals: Algorithms and Performance Evaluation.},
   institution = {Univ. of Maryland, Dept. of Computer Science},
   type = {UMIACS-TR-89-31,},
   number = {CS-TR-2214},
   address = {College Park},
   month = feb,
   year = {1989},
   keywords = {hilbert space filling curves}
}

@book{Feller68Introductiona,
   author = {W. Feller},
   title = {An Introduction to Probability Theory and its Applications},
   volume = {Vol. I},
   edition = {3rd},
   publisher = {John Wiley \& Sons Inc.},
   year = {1968},
   abstract = {capter IV discusses the inclusion-exclusion principle,
and the occupancy problem, ie.,
it calculates the probability that m slots will be empty,
after putting r balls among n slots.
Moreover, it approximates this probability with the poisson distribution.}
}

@book{Lass71Probability,
   author = {H. Lass and P. Gottlieb},
   title = {Probability and Statistics},
   publisher = {Addison Wesley},
   year = {1971},
   keywords = {order statistics},
   abstract = {chapter 8 discusses order statistics. p. 313 says that "For large samples
the probability of large deviations of Yk from its most probable value will
be small" (where Yk is the k-th smallest element of the sample)}
}

@article{Chan88Embedding,
   author = {M.Y. Chan and F.Y.L. Chin},
   title = {On Embedding Rectangular Grids in Hypercubes},
   journal = {IEEE Trans. on Computers},
   volume = {37},
   number = {10},
   pages = {1285-1288},
   month = oct,
   year = {1988},
   keywords = {gray codes fractals},
   abstract = {Use gray codes to assign grid points on the processors
of a hypercube, such that grid neighbors are kept within
a distance of two apart.}
}

@techreport{Hoppenstand88Technique,
   author = {G.S. Hoppenstand and D.K. Hsiao},
   title = {A Technique to Improve the Precision of Full-text Database Search},
   institution = {Naval Postgraduate School},
   type = {NPS52-88-029},
   address = {Monterey, California},
   month = sep,
   year = {1988},
   abstract = {For bibliographical records,
use multiattribute hashing for attributes,
full-text-scanning for text.
No new ideas.}
}

@article{Easton86Key,
   author = {M. Easton},
   title = {Key-sequence data sets on indelible storage},
   journal = {IBM J. of Research and Development},
   volume = {30},
   number = {3},
   pages = {230},
   month = may,
   year = {1986},
   keywords = {optical disks worm b-trees},
   abstract = {The "write-once B-tree" is introduced.}
}

@article{Tarjan85Amortized,
   author = {R. Tarjan},
   title = {Amortized Computational Complexity},
   journal = {SIAM J. of Algorithms and Disc. Meth.},
   volume = {6},
   number = {2},
   month = apr,
   year = {1985}
}

@article{Sarnak86Planar,
   author = {N. Sarnak and R. Tarjan},
   title = {Planar Point Location Using Persistent Search Trees},
   journal = {CACM},
   volume = {9},
   number = {7},
   pages = {669},
   month = jul,
   year = {1986}
}

@book{Mandelbrot77Fractal,
   author = {B. Mandelbrot},
   title = {Fractal Geometry of Nature},
   publisher = {W.H. Freeman},
   address = {New York},
   year = {1977}
}

@incollection{Lempel84Compression,
   author = {A. Lempel and J. Ziv},
   editor = {A. Apostolico and Z. Galil},
   title = {Compression of Two-dimensional Images},
   booktitle = {Combinatorial Algorithms on Words},
   pages = {141-154},
   publisher = {Springer-Verlag},
   address = {Malatea, Italy},
   month = {June 18-22},
   year = {1984},
   keywords = {hilbert curve fractals},
   abstract = {they suggest using the hilbert curve to scan the pixels,
and then compressing the resulting bit string
using a traditional compression method.
They also define the entropy of an image.},
   note = {Published as NATO ASI Series, volume F12}
}

@article{Faloutsos90Incorporating,
   author = {C. Faloutsos and R. Lee and C. Plaisant and B. Shneiderman},
   title = {Incorporating String Search in a Hypertext System: User Interface and Physical Design Issues},
   journal = {Hypermedia},
   volume = {2},
   number = {3},
   pages = {183-200},
   year = {1990}
}

@inproceedings{Faloutsos89Declustering,
   author = {C. Faloutsos and D. Metaxas},
   title = {Declustering Using Error Correcting Codes},
   booktitle = {Eighth ACM SIGACT-SIGMOD-SIGART Symposium on Principles of Database Systems (PODS)},
   pages = {253-258},
   address = {Philadelphia, Pennsylvania},
   month = {March 29-31},
   year = {1989},
   keywords = {parhash},
   note = {Also available as UMIACS-TR-88-91 and CS-TR-2157}
}

@inproceedings{Patterson88Case,
   author = {D.A. Patterson and G. Gibson and R.H. Katz},
   title = {A Case for Redundant Arrays of Inexpensive Disks ({RAID})},
   booktitle = {Proc. of ACM SIGMOD},
   pages = {109-116},
   address = {Chicago, Illinois},
   month = {June 1-3},
   year = {1988},
   keywords = {declustering parhash}
}

@inproceedings{Copeland88Data,
   author = {G. Copeland and W. Alexander and E. Boughter and T. Keller},
   title = {Data Placement in Bubba},
   booktitle = {Proc. of ACM SIGMOD},
   pages = {99-109},
   address = {Chicago, Illinois},
   month = {June 1-3},
   year = {1988},
   keywords = {parhash declustering}
}

@article{Samet88Hierarchical,
   author = {H. Samet},
   title = {Hierarchical Representations of Collections of Small Rectangles},
   journal = {ACM Computing Surveys},
   volume = {20},
   number = {4},
   pages = {271-309},
   month = dec,
   year = {1988},
   keywords = {spatial access methods}
}

@article{Voelcker89Personal,
   author = {J. Voelcker},
   title = {Personal Computers},
   journal = {IEEE Spectrum},
   volume = {26},
   number = {1},
   pages = {31-34},
   month = jan,
   year = {1989},
   keywords = {NeXT erasable optical disks},
   abstract = {information about the NeXT machine.}
}

@techreport{Johnson84Yacc,
   author = {S.C. Johnson},
   title = {Yacc: Yet Another Compiler-Compiler},
   institution = {4.2 Berkeley Software Distribution, VAX-11 Version},
   type = {Unix Programmer's Manual - Supplementary Documents},
   month = mar,
   year = {1984}
}

@inproceedings{Christodoulaki88Performance,
   author = {S. Christodoulakis and D.A. Ford},
   title = {Performance Analysis and Fundamental Performance Tradeoffs for {CVL} Optical Disks},
   booktitle = {Proc. ACM SIGMOD},
   pages = {286-294},
   address = {Chicago, Illinois},
   month = {June 1-3},
   year = {1988}
}

@book{McFadyen88Gray,
   author = {R. McFadyen},
   title = {Gray Code Linear Hashing},
   publisher = {Univ. of Manitoba, Dept. of CS},
   month = oct,
   year = {1988},
   abstract = {contains some proofs of the limit of the behavior,
as well as a cost function that takes into account the sequential
disk accesses as well.},
   note = {submitted to SIGMOD 1989}
}

@article{Jones89,
   author = {L.P. Jones and E.W. Gassie and S. Radhakrishnan},
   title = {{PORTREP}: {A} Portable Repeated String Finder},
   journal = {Software-Practice and Experience},
   volume = {19},
   number = {1},
   pages = {63-77},
   month = jan,
   year = {1989},
   keywords = {string searching trie automatic document indexing}
}

@article{Gumm85New,
   author = {H.P. Gumm},
   title = {A New Class of Check-digit methods for arbitrary number systems},
   journal = {IEEE Trans. on Inf. Theory},
   volume = {31},
   number = {1},
   pages = {102-105},
   month = jan,
   year = {1985},
   keywords = {parhash declustering}
}

@book{Hill86First,
   author = {R. Hill},
   title = {A First Course in Coding Theory},
   publisher = {Clarendon Press},
   address = {Oxford},
   year = {1986},
   keywords = {declustering parhash book}
}

@book{Hamming86Coding,
   author = {R.W. Hamming},
   title = {Coding and Information Theory},
   publisher = {Prentice-Hall},
   address = {Englewood Cliffs, N.J.},
   year = {1986},
   keywords = {parhash declustering},
   note = {2nd ed.}
}

@article{Wagner89Error,
   author = {N.R. Wagner and P.S. Putter},
   title = {Error Detecting Decimal Points},
   journal = {Comm. ACM (CACM)},
   volume = {32},
   number = {1},
   pages = {106-110},
   month = jan,
   year = {1989},
   keywords = {declustering parhash}
}

@article{Karlin88Parallel,
   author = {A.R. Karlin and E. Upfal},
   title = {Parallel Hashing: An Efficient Implementation of Shared Memory},
   journal = {Journal of ACM (JACM)},
   pages = {????},
   month = oct,
   year = {1988},
   keywords = {parhash declustering}
}

@article{Faloutsos91Disk,
   author = {C. Faloutsos and D. Metaxas},
   title = {Disk Allocation Methods Using Error Correcting Codes},
   journal = {IEEE Trans. on Computers},
   volume = {40},
   number = {8},
   pages = {907-914},
   month = aug,
   year = {1991},
   keywords = {parhash declustering},
   note = {early version available as UMIACS-TR-88-91 and CS-TR-2157}
}

@article{Rabin89Efficient,
   author = {M.O. Rabin},
   title = {Efficient Dispersal of Information for Security, Load Balancing and Fault Tolerance},
   journal = {JACM},
   month = apr,
   year = {1989},
   keywords = {RAID}
}

@article{Cooper89Inverted,
   author = {L.K.D. Cooper and A.L. Tharp},
   title = {Inverted Signature Trees and Text Searching on {CD}-ROMs},
   journal = {Information Processing and Management},
   volume = {25},
   number = {2},
   pages = {161-169},
   year = {1989}
}

@article{Christodoulaki89Analysis,
   author = {S. Christodoulakis and Y. Manolopoulos and P.-A. Larson},
   title = {Analysis of Overflow Handling for Variable Length Records},
   journal = {Information Systems},
   volume = {14},
   number = {2},
   pages = {151-162},
   year = {1989},
   keywords = {yannis yiannis}
}

@article{Lee89Partitioned,
   author = {D.L. Lee and C.-W. Leng},
   title = {Partitioned Signature File: Designs and Performance Evaluation},
   journal = {ACM Trans. on Information Systems (TOIS)},
   volume = {7},
   number = {2},
   pages = {158-180},
   month = apr,
   year = {1989},
   keywords = {toois}
}

@techreport{Rosenfeld82Application,
   author = {A. Rosenfeld and H. Samet and C. Shaffer and R.E. Webber},
   title = {Application of Hierarchical Data Structures to Geographical Information Systems},
   institution = {Univ. of Maryland, College Park, MD},
   type = {Computer Science},
   number = {TR-1197},
   month = jun,
   year = {1982},
   abstract = {Empirical results that the number of quadrants
depend on the perimeter of the geometric object.}
}

@article{Dyer82Space,
   author = {C.R. Dyer},
   title = {The Space Efficiency of Quadtrees},
   journal = {Computer Graphics and Image Processing},
   volume = {19},
   number = {4},
   pages = {335-348},
   month = aug,
   year = {1982},
   abstract = {Calculates the number of quadrants for a square
of 2**m x 2**m - best, worst and average case.}
}

@article{Gargantini82Effective,
   author = {I. Gargantini},
   title = {An Effective Way to Represent Quadtrees},
   journal = {Comm. of ACM (CACM)},
   volume = {25},
   number = {12},
   pages = {905-910},
   month = dec,
   year = {1982}
}

@article{Hunter79Operations,
   author = {G.M. Hunter and K. Steiglitz},
   title = {Operations on Images Using Quad Trees},
   journal = {IEEE Trans. on PAMI},
   volume = {PAMI-1},
   number = {2},
   pages = {145-153},
   month = apr,
   year = {1979},
   abstract = {give an upper bound on the number of nodes
in a guadtree that a polygon requires:
O(p+q), where p is the perimeter
and q the granularity.}
}

@article{Shaffer88Formula,
   author = {C.A. Shaffer},
   title = {A Formula for Computing the Number of Quadtree Node Fragments Created by a Shift},
   journal = {Pattern Recognition Letters},
   volume = {7},
   number = {1},
   pages = {45-49},
   month = jan,
   year = {1988},
   abstract = {Calculates average number of quadrants for a square of size 2**m.}
}

@book{Samet90Design,
   author = {H. Samet},
   title = {The Design and Analysis of Spatial Data Structures},
   publisher = {Addison-Wesley},
   year = {1990},
   keywords = {book}
}

@book{Rong90Spatial,
   author = {Y. Rong},
   title = {Spatial Access Methods Based on Fractals: Performance Analysis},
   publisher = {Univ. of Maryland, Dept. of Computer Science},
   year = {1990},
   keywords = {quadtrees},
   note = {under preparation}
}

@article{Shaffer90,
   author = {Clifford A. Shaffer and Hanan Samet and R.C. Nelson},
   title = {{QUILT}: {A} Geographic Information System Based on Quadtrees},
   journal = {International Journal of Geographic Information Systems},
   volume = {4},
   pages = {103-131},
   year = {1990},
   note = {Also available as CS-TR-1885.1, Univ. of Maryland, Dept. of Computer Science}
}

@article{Lipman85Rapid,
   author = {D.J. Lipman and W.R. Pearson},
   title = {Rapid and Sensitive Protein Similarity Searches},
   journal = {Science},
   volume = {227},
   pages = {1435-1441},
   publisher = {American Association for the Advancement of Science},
   month = {March 22},
   year = {1985},
   keywords = {dna}
}

@mastersthesis{Lewis89Animated,
   author = {E.P. Lewis},
   title = {Animated Images for a Multimedia Database},
   school = {Dept. of Computer Science, Univ. of Maryland},
   type = {Master's Thesis},
   address = {College Park},
   month = may,
   year = {1989},
   keywords = {masie scripted animation}
}

@inproceedings{Chang89Signature,
   author = {W.W. Chang and H.J. Schek},
   title = {A Signature Access Method for the Starbust Database System},
   booktitle = {Proc. VLDB Conference},
   pages = {145-153},
   address = {Amsterdam, Netherlands},
   month = {Aug. 22-25},
   year = {1989},
   keywords = {lindsay}
}

@article{Goddyn88Gray,
   author = {L. Goddyn and G.M. Lawrence and E. Nemeth},
   title = {Gray Codes with Optimized Run Lengths},
   journal = {Utilitas Mathematica},
   month = jan,
   year = {1988}
}

@article{Wan88algorithm,
   author = {S.J. Wan and S.K.M. Wong and P. Prusinkiewicz},
   title = {An algorithm for multidimensional data clustering},
   journal = {ACM Trans. on Math. Software},
   volume = {14},
   number = {2},
   pages = {153-162},
   month = jun,
   year = {1988}
}

@inproceedings{Orenstein89Redundancy,
   author = {J.A. Orenstein},
   title = {Redundancy in Spatial Databases},
   booktitle = {Proc. of ACM SIGMOD conf.},
   address = {Portland, Oregon},
   month = may,
   year = {1989}
}

@inproceedings{Orenstein90Comparison,
   author = {J.A. Orenstein},
   title = {A Comparison of Spatial Query Processing Techniques for Native and Parameter Spaces},
   booktitle = {Proc. of ACM SIGMOD conf.},
   pages = {343-352},
   address = {Atlantic City, New Jersey},
   year = {1990}
}

@inproceedings{Hutflesz88Twin,
   author = {A. Hutflesz and H.-W. Six and P. Widmayer},
   title = {Twin Grid Files: Space Optimizing Access Schemes},
   booktitle = {Proc. of ACM SIGMOD},
   pages = {183-190},
   address = {Chicago, Illinois},
   month = {June 1-3},
   year = {1988}
}

@article{Dyer80Computing,
   author = {C.R. Dyer},
   title = {Computing the Euler Number of an Image from Its Quadtree},
   journal = {Computer Graphics and Image Processing},
   volume = {13},
   pages = {270-276},
   year = {1980}
}

@article{Ooi87Spatial,
   author = {B.C Ooi and K.J. McDonell and R. Sacks-Davis},
   title = {Spatial kd-tree: An Indexing Mechanism for Spatial Database},
   journal = {COMPSAC conf.},
   pages = {433-438},
   address = {Tokyo},
   year = {1987}
}

@article{Faloutsos92Analytical,
   author = {C. Faloutsos},
   title = {Analytical Results on the Quadtree Decomposition of Arbitrary Rectangles},
   journal = {Pattern Recognition Letters},
   volume = {13},
   number = {1},
   pages = {31-40},
   month = jan,
   year = {1992},
   keywords = {prl},
   note = {Also available as UMIACS-TR-89-115, CS-TR-2354 Tech. Report, Univ. of Maryland at College Park},
   contents = {(not listed)}
}

@techreport{Lewis89,
   author = {E. Lewis and C. Faloutsos},
   title = {{MMS} - An Integrated Animation and Text Multimedia Database System},
   institution = {Dept. of Computer Science and UMIACS},
   type = {UMIACS-TR-89-125,},
   number = {CS-TR-2367},
   address = {University of Maryland, College Park},
   month = dec,
   year = {1989}
}

@article{Beckmann90R,
   author = {N. Beckmann and H.-P. Kriegel and R. Schneider and B. Seeger},
   title = {The  {R}*-tree: An Efficient  and Robust Access Method for Points and Rectangles},
   journal = {ACM SIGMOD},
   pages = {322-331},
   address = {Atlantic  City, NJ},
   month = {May 23-25},
   year = {1990},
   keywords = {spatial}
}

@article{Zanden87General,
   author = {B.T. Vander Zanden and H.M. Taylor and D. Bitton},
   title = {A General Framework for Computing Block Accesses},
   journal = {Information Systems},
   volume = {12},
   number = {2},
   pages = {177-190},
   year = {1987},
   keywords = {selectivities},
   abstract = {use approximations (taylor expansion, majorization, generating functions)
for skewed distributions.
Has some good references for text data.}
}

@article{Mullin90Optimal,
   author = {J.K. Mullin},
   title = {Optimal Semijoins for Distributed Database Systems},
   journal = {IEEE Trans. on Software Engineering (TSE)},
   volume = {16},
   number = {5},
   pages = {558-560},
   month = may,
   year = {1990},
   keywords = {bloom filter superimposed coding},
   abstract = {use bloom filters for more efficient semijoins},
   note = {correspondence}
}

@article{Faloutsos90Signature,
   author = {C. Faloutsos},
   title = {Signature-Based Text Retrieval Methods: {A} Survey},
   journal = {IEEE Data Engineering},
   volume = {13},
   number = {1},
   pages = {25-32},
   month = mar,
   year = {1990}
}

@article{Sunday90Very,
   author = {D.M. Sunday},
   title = {A Very Fast Substring Search Algorithm},
   journal = {Comm. of ACM (CACM)},
   volume = {33},
   number = {8},
   pages = {132-142},
   month = aug,
   year = {1990},
   keywords = {boyer moore text},
   abstract = {"Optimal Mismatch" algorithm consistently outperforms
BM - the speedup seems INDEPENDENT of the text,
depending only on the length of the search pattern.
Several ideas behind OM - one of the main ones:
start comparisons from the pattern character that
is most likely to create a mismatch.}
}

@incollection{Faloutsos92Signature,
   author = {Christos Faloutsos},
   editor = {William Bruce Frakes and Ricardo Baeza-Yates},
   title = {Signature Files},
   booktitle = {Information Retrieval: Data Structures and Algorithms},
   publisher = {Prentice Hall},
   address = {Englewood Cliffs, NJ},
   year = {1992},
   keywords = {book}
}

@article{Lin89Performance,
   author = {C-C. Lin and L. Mark and T. Sellis and C. Faloutsos},
   title = {Performance Issues in the Binary Relationship Model},
   journal = {Intern. Journal on Data and Knowledge Engineering},
   volume = {4},
   pages = {195-221},
   year = {1989}
}

@article{Roussopoulos91Architecture,
   author = {N. Roussopoulos and L. Mark and T. Sellis and C. Faloutsos},
   title = {An Architecture for High Performance Engineering Information Systems},
   journal = {IEEE Trans. on Software Engineering (TSE)},
   volume = {17},
   number = {1},
   pages = {22-33},
   month = jan,
   year = {1991},
   note = {early version available as UMIACS-TR-87-25, CS-TR-1870.}
}

@article{Nievergelt73Binary,
   author = {J. Nievergelt and E.M. Reingold},
   title = {Binary search trees of bounded balance},
   journal = {SIAM J. Comput.},
   volume = {2},
   pages = {33-43},
   year = {1973},
   keywords = {primary key}
}

@article{Brain90Perfect,
   author = {M.D. Brain and A.L. Tharp},
   title = {Perfect Hashing Using Sparse Matrix Packing},
   journal = {Information Systems},
   volume = {15},
   number = {3},
   pages = {281-290},
   year = {1990},
   abstract = {good survey of previous perfect hashing methods.
Compresses sparse 2-d arrays into minimal 1-D arrays in O(r**2) time.
Applicable up to sets of 5000 words (without segmenting them).}
}

@article{Rosenberg81Time,
   author = {A.L. Rosenberg and L. Snyder},
   title = {Time- and Space-Optimality in {B}-Trees},
   journal = {ACM TODS},
   volume = {6},
   number = {1},
   pages = {174-183},
   month = mar,
   year = {1981},
   abstract = {Compact B-trees maximize the space utilization (space optimal);
"visit optimal" B-trees minimize access time.
They show that compact trees can save up to 2.5 times more space,
incurring a cost of at most 1 extra disk access.
Compact trees have the leaves as full as possible;
"visit-optimal" ones have the higher level nodes as full as possible.
They provide an algorithm for compacting a 2-3 tree.}
}

@article{Mehlhorn90Bounded,
   author = {K. Mehlhorn and S. Naeher},
   title = {Bounded Ordered Dictionaries in {O}(loglogN) Time and {O}(n) Space},
   journal = {Information Processing Letters (IPL)},
   volume = {35},
   number = {4},
   pages = {183-189},
   month = aug,
   year = {1990},
   keywords = {priority queues naher}
}

@article{Jagadish90Linear,
   author = {H.V. Jagadish},
   title = {Linear Clustering of Objects with Multiple Attributes},
   journal = {ACM SIGMOD Conf.},
   pages = {332-342},
   address = {Atlantic City, NJ},
   month = {May 23-25},
   year = {1990},
   keywords = {fractals hilbert curve}
}

@article{Larson88Dynamic,
   author = {P.-A. Larson},
   title = {Dynamic Hash Tables},
   journal = {Comm. of ACM (CACM)},
   volume = {31},
   number = {4},
   pages = {446-457},
   month = apr,
   year = {1988},
   keywords = {primkey spiral hashing linear},
   abstract = {Compares linear hashing to spiral hashing, analytically \& experimentally.}
}

@article{Sheil78Median,
   author = {B.A. Sheil},
   title = {Median Split Trees: {A} Fast Lookup Technique for Frequently Occuring Keys},
   journal = {Comm. of ACM (CACM)},
   volume = {21},
   number = {11},
   month = nov,
   year = {1978},
   keywords = {primkey}
}

@article{Felician88Image,
   author = {L. Felician},
   title = {Image Base Management System: {A} Promising Tool in the Large Office System Environment},
   journal = {DATABASE ACM-SIGBDP},
   volume = {19},
   number = {1},
   pages = {29-36},
   year = {1988}
}

@article{Altschul90Basic,
   author = {S.F. Altschul and W. Gish and W. Miller and E.W. Myers and D.J. Lipman},
   title = {A Basic Local Alignment Search Tool},
   journal = {Journal of Molecular Biology},
   volume = {215},
   number = {3},
   pages = {403-410},
   year = {1990},
   keywords = {dna databases}
}

@techreport{Gibson88Coding,
   author = {G. Gibson and L. Hellerstein and R.M. Karp and R.H. Katz and D.A. Patterson},
   title = {Coding Techniques for Handling Failures in Large Disk Arrays},
   institution = {Comp. Science Div. (EECS), Univ. of California},
   type = {Rep. No. UCB/CSD},
   number = {88/477},
   address = {Berkeley, CA},
   month = dec,
   year = {1988},
   keywords = {error correcting codes declustering raid}
}

@article{Schuegraf76Compression,
   author = {E.J. Schuegraf},
   title = {Compression of Large Inverted Files with Hyperbolic Term Distribution},
   journal = {Information Processing and Management},
   volume = {12},
   pages = {377-384},
   year = {1976}
}

@article{Lee77Worst,
   author = {D.T. Lee and C.K. Wong},
   title = {Worst Case Analysis for Region and Partial Region Searches in Multidimensional Binary Search Trees and Balanced Quadtrees},
   journal = {Acta Informatica},
   volume = {9},
   number = {1},
   pages = {339-353},
   year = {1977}
}

@article{Manolopoulos90Analysis,
   author = {Y. Manolopoulos and C. Faloutsos},
   title = {Analysis for the End of Block Wasted Space},
   journal = {BIT},
   volume = {30},
   pages = {620-630},
   year = {1990},
   keywords = {umiacs}
}

@article{King74Binary,
   author = {D. R. King},
   title = {The Binary Vector as the Basis of an Inverted Index File},
   journal = {J. Lib. Autom.},
   volume = {7},
   number = {4},
   pages = {307},
   year = {1974}
}

@article{Abel90Comparative,
   author = {David J. Abel and David M. Mark},
   title = {A Comparative Analysis of some Two-Dimensional Orderings},
   journal = {Int. J. Geographical Information Systems},
   volume = {4},
   number = {1},
   pages = {21-31},
   year = {1990},
   keywords = {fractals},
   abstract = {compare hilbert to peano to row-wise ordering,
using several metrics (\# of clusters, distance of rook's-case neighbors etc)}
}

@article{Huang91Prioritized,
   author = {Yunming Huang and Howard M. Dreizen and Nikolas P. Galatsanos},
   title = {Prioritized Block {DCT} Image Coding},
   journal = {IEEE Trans. on Signal Processing},
   year = {1991},
   note = {working paper - to be submitted}
}

@inproceedings{Faloutsos91Predictive,
   author = {C. Faloutsos and R. Ng and T. Sellis},
   title = {Predictive Load Control for Flexible Buffer Allocation},
   booktitle = {VLDB conf. proceedings},
   pages = {265-274},
   address = {Barcelona, Spain},
   month = {Sept. 3-6},
   year = {1991},
   note = {Also available as UMIACS-TR-91-34 and CS-TR-2622}
}

@inproceedings{Ng91Flexible,
   author = {R. Ng and C. Faloutsos and T. Sellis},
   title = {Flexible Buffer Allocation Based on Marginal Gains},
   booktitle = {ACM-SIGMOD conf. proceedings},
   pages = {387-396},
   address = {Denver, Colorado},
   month = {May 29-31},
   year = {1991},
   note = {Also available as UMIACS-TR-91-35, CS-TR-2624.}
}

@article{Faloutsos92Hybrid,
   author = {C. Faloutsos and H.V. Jagadish},
   title = {Hybrid Index Organizations for Text Databases},
   journal = {EDBT '92},
   pages = {310-327},
   address = {Vienna, Austria},
   month = {March 23-27},
   year = {1992},
   keywords = {zipf signatures text retrieval},
   note = {Also available as UMIACS-TR-91-33 and CS-TR-2621}
}

@article{Wallace91JPEG,
   author = {Gregory K. Wallace},
   title = {The {JPEG} Still Picture Compression Standard},
   journal = {CACM},
   volume = {34},
   number = {4},
   pages = {31-44},
   month = apr,
   year = {1991}
}

@article{Jagadish89Object,
   author = {H.V. Jagadish and Lawrence O'Gorman},
   title = {An Object Model of Image Recognition},
   journal = {IEEE Computer},
   volume = {22},
   number = {12},
   pages = {33-41},
   month = dec,
   year = {1989}
}

@inproceedings{Jones91Integrating,
   author = {Mark A. Jones and Guy A. Story and Bruce W. Ballard},
   title = {Integrating Multiple Knowledge Sources in a Bayesian {OCR} Post-Processor},
   booktitle = {First International Conference on Document Analysis and Recognition},
   address = {Saint-Malo, France},
   month = sep,
   year = {1991},
   keywords = {jsb},
   note = {to appear}
}

@article{Story92RightPages,
   author = {G.A. Story and L. O'Gorman and D.S. Fox and L.L. Schaper and H.V. Jagadish},
   title = {The RightPages: An Electronic Library for Alerting and Browsing},
   journal = {IEEE Computer},
   volume = {25},
   number = {9},
   pages = {17-26},
   month = sep,
   year = {1992},
   keywords = {electronic publishing, document distribution}
}

@article{Millard83Developments,
   author = {K. Millard},
   title = {Developments on Automatic Fingerprints Recognition},
   journal = {Int. Carnahan Conf. on Security Technology},
   pages = {173-178},
   address = {Zurich, Switzerland},
   year = {1983}
}

@book{Becker88New,
   author = {Richard A. Becker and John M. Chambers and Allan R. Wilks},
   title = {The New {S} Language},
   publisher = {Wadsworth \& Brooks/Cole Advanced Books \& Software},
   address = {Pacific Grove, CA},
   year = {1988}
}

@book{Storer88Data,
   author = {James A. Storer},
   title = {Data Compression: Methods and Theory},
   publisher = {Computer Science Press, Inc.},
   year = {1988},
   keywords = {infotheory}
}

@article{Lomet90hB,
   author = {David B. Lomet and Betty Salzberg},
   title = {The hB-tree: {A} Multiattribute Indexing Method with Good Guaranteed Performance},
   journal = {ACM TODS},
   volume = {15},
   number = {4},
   pages = {625-658},
   month = dec,
   year = {1990},
   keywords = {spatial access methods},
   abstract = {Improvement over k-d-B-trees. Main ideas:
(1) use k-d trees to organize internal nodes
(2) allow nodes to have "holes"
My objections: no comparison with R-trees;
weak arguments against z-ordering;
deletion is "future work";
no implementation; no simulation results.}
}

@article{Witten87Arithmetic,
   author = {I.H. Witten and R.M. Neal and J.G. Cleary},
   title = {Arithmetic Coding for Data Compression},
   journal = {CACM},
   volume = {30},
   number = {6},
   pages = {520-540},
   month = jun,
   year = {1987},
   keywords = {infotheory getit}
}

@article{Can90Concepts,
   author = {Fazli Can and Esen A. Ozkarahan},
   title = {Concepts and Effectiveness of the Cover-Coefficient-Based Clustering Methodology for Text Databases},
   journal = {ACM TODS},
   volume = {15},
   number = {4},
   pages = {483-517},
   month = dec,
   year = {1990},
   abstract = {one of the ideas: try to guess the number of clusters,
using the average distance of document vectors}
}

@article{Omiecinski90Parallel,
   author = {Edward Omiecinski and Peter Scheuermann},
   title = {A Parallel Algorithm for Record Clustering},
   journal = {ACM TODS},
   volume = {15},
   number = {4},
   pages = {599-624},
   month = dec,
   year = {1990}
}

@inproceedings{Stanfill89Parallel,
   author = {Craig Stanfill and Robert Thau and David Waltz},
   title = {A Parallel Indexed Algorithm for Information Retrieval},
   booktitle = {Proc. of SIGIR},
   address = {Cambridge, MA},
   month = {June 26-28},
   year = {1989},
   keywords = {connection machine inverted indices text},
   abstract = {Uses inverted indices, and weighting schemes;
the postings are stored on the data-Vault (batch version)
Interesting ideas:
(1) Defines the "cost effectiveness" of a system
(2) on retrieval, reads a "strip" of postings in parallel,
also, it uses parallelism to compute relevance and to sort
BUT: Insertions are not mentioned; the management
of the postings list is also ignored}
}

@article{Shasha90New,
   author = {Dennis Shasha and Tsong-Li Wang},
   title = {New Techniques for Best-Match Retrieval},
   journal = {ACM TOIS},
   volume = {8},
   number = {2},
   pages = {140-158},
   month = apr,
   year = {1990},
   keywords = {approximate matching nearest neighbor dna text spatial},
   abstract = {Uses the triangle inequality and a set of precomputed distances
(Approximate Distance Map - ADM) to eliminate distant candidates.
Upon each new computation, the ADM is updated.
ADM contains the upper and lower bound for the distance
between every two data objects, as well as for the
query object and each data object.
Very useful bibliography.}
}

@book{Smith88Analysis,
   author = {Stephen Smith and Craig Stanfill},
   title = {An Analysis of the Effects of Data Corruption on Text Retrieval Performance},
   publisher = {Thinking Machines Corporation},
   address = {Cambridge, MA},
   month = dec,
   year = {1988},
   keywords = {ocr errors approximate matching connection machine},
   abstract = {They artificially corrupted ASCII english text,
by substituting at random the "@" character.
The major conclusion is that for 1-3\% rejection
rate on the character level,
the search performance (precision/recall) remains almost un-affected.
Among single word, boolean and relevance-feedback queries,
the latter shows the best resilience to errors.
Future research: try to explain the performance degradation
mathematically},
   note = {unpublished(?) manuscript}
}

@article{Kahan87Recognition,
   author = {Simon Kahan and Theo Pavlidis and Henry Baird},
   title = {On the Recognition of Printed Characters of Any Font and Size},
   journal = {IEEE Trans. on PAMI},
   volume = {PAMI-9},
   number = {2},
   pages = {274-287},
   month = mar,
   year = {1987},
   keywords = {getit}
}

@techreport{Stanfill88Parallel,
   author = {Craig Stanfill},
   title = {Parallel Computing for Information Retrieval: Recent Developments},
   institution = {Thinking Machines Corporation},
   type = {DR88-1},
   address = {Cambridge, MA},
   year = {1988},
   keywords = {text connection machine},
   abstract = {Nice overview of CM architectures;
Good arithmetic examples for several sizes of text databases}
}

@article{Ito83Hierarchical,
   author = {Tetsuro Ito and Makoto Kizawa},
   title = {Hierarchical File Organization and Its Applications to Similar-String Matching},
   journal = {ACM TODS},
   volume = {8},
   number = {3},
   pages = {410-433},
   month = sep,
   year = {1983},
   keywords = {approximate text nearest neighbor},
   abstract = {They propose the TSO=two dimensional sorting operation -
given a distance matrix, it is transformed to
a new distance matrix, where the distance between
two elements is their distance on the MST
On this new distance matrix, they sort,
effectively finding nearest neighbors;
This linear ordering is used to put records in pages
and to create higher level, B-tree type nodes
They provide an insertion algorithm - however,
it is not clear how they re-calculate the MST efficiently
They operate on words (spelling correction being the main motivation);
each letter is represented by a "1" in the appropriate}
}

@article{Morrissey90Imprecise,
   author = {J.M. Morrissey},
   title = {Imprecise Information and Uncertainty in Information Systems},
   journal = {ACM TOIS},
   volume = {8},
   number = {2},
   pages = {159-180},
   month = apr,
   year = {1990},
   keywords = {infotheory entropy},
   abstract = {use entropy to answer queries
on disjunctive facts}
}

@inproceedings{Aalbersberg91High,
   author = {Ijsbrand Jan Aalbersberg and Frans Sijstermans},
   title = {High-Quality and High-Performance Full-Text Document Retrieval: the Parallel InfoGuide System},
   booktitle = {Proceeding of the First International Conference on Parallel and Distributed Information Processing},
   pages = {142-150},
   address = {Miami Beach, Florida},
   month = {Dec. 4-6},
   year = {1991},
   keywords = {pooma philips},
   abstract = {use a transputer and the weighted vector model
to achieve fast similarity ranking.}
}

@inproceedings{Matsliach91Efficient,
   author = {G. Matsliach and O. Shmueli},
   title = {An Efficient Method for Distributing Search Structures},
   booktitle = {Proceeding of the First International Conference on Parallel and Distributed Information Processing},
   pages = {159-166},
   address = {Miami Beach, Florida},
   month = {Dec. 4-6},
   year = {1991},
   keywords = {getit}
}

@inproceedings{DeWitt91Parallel,
   author = {David J. DeWitt and Jeffrey F. Naughton and Donovan A. Schneider},
   title = {Parallel Sorting on a Shared-Nothing Architecture using Probabilistic Splitting},
   booktitle = {Proceeding of the First International Conference on Parallel and Distributed Information Processing},
   pages = {280-291},
   address = {Miami Beach, Florida},
   month = {Dec. 4-6},
   year = {1991}
}

@article{Friedman75Algorithm,
   author = {Jerome H. Friedman and Forest Baskett and Leonard H. Shustek},
   title = {An Algorithm for Finding Nearest Neighbors},
   journal = {IEEE Trans. on Computers (TOC)},
   volume = {C-24},
   pages = {1000-1006},
   month = oct,
   year = {1975},
   keywords = {approximate matching},
   abstract = {project points on 1-d; find nn there; do
a range search.
For high dimensionality (d$>$8), it behaves like linear search}
}

@article{Bentley80Optimal,
   author = {Jon Louis Bentley and Bruce W. Weide and Andrew C. Yao},
   title = {Optimal Expected-Time Algorithms for Closest Point Problems},
   journal = {ACM Trans. on Mathematical Software (TOMS)},
   volume = {6},
   number = {4},
   pages = {563-580},
   month = dec,
   year = {1980},
   keywords = {nearest neighbor approximate matching},
   abstract = {Main ideas: use cells and apply spiral searching.
The method assumes uniform distribution of points,
and low dimensionality (searching explodes exponentially
with dimensionality)}
}

@article{Fukunaga75Branch,
   author = {Keinosuke Fukunaga and Patrenahalli M. Narendra},
   title = {A Branch and Bound Algorithm for Computing k-Nearest Neighbors},
   journal = {IEEE Trans. on Computers (TOC)},
   volume = {C-24},
   number = {7},
   pages = {750-753},
   month = jul,
   year = {1975},
   keywords = {approximate matching},
   abstract = {Cluster objects; then build a tree structure
with "centroid-radius"; then use branch\&bound
to search.
VERY SIMILAR to our cluster tree;
it does not mention insertions and deletions, though.}
}

@article{Murtagh83Survey,
   author = {F. Murtagh},
   title = {A Survey of Recent Advances in Hierarchical Clustering Algorithms},
   journal = {The Computer Journal},
   volume = {26},
   number = {4},
   pages = {354-359},
   year = {1983},
   keywords = {approximate matching minimum spanning trees}
}

@article{Bentley78Fast,
   author = {Jon Louis Bentley and Jerome H. Friedman},
   title = {Fast Algorithms for Constructing Minimal Spanning Trees in Coordinate Spaces},
   journal = {IEEE Trans. on Computers (TOC)},
   volume = {C-27},
   number = {2},
   pages = {97-105},
   month = feb,
   year = {1978},
   keywords = {k-d trees clustering},
   abstract = {Main idea: use k-d trees, to find nearest neighbor
of each point; build chains of nearest neighbors
and combine them to MST.}
}

@article{Eastman82Partially,
   author = {C.M. Eastman and Maria Zemankova},
   title = {Partially Specified Nearest Neighbor Searches Using k-d Trees},
   journal = {Information Processing Letters (IPL)},
   volume = {15},
   number = {2},
   pages = {53-56},
   month = sep,
   year = {1982},
   keywords = {approximate matching},
   abstract = {extends a formula by Bentley, for the case
that some attributes are not specified.}
}

@article{Friedman77Algorithm,
   author = {Jerome H. Friedman and Jon Louis Bentley and R.A. Finkel},
   title = {An Algorithm for Finding Best Matches in Logarithmic Expected Time},
   journal = {ACM Trans. on Math. Software (TOMS)},
   volume = {3},
   number = {3},
   pages = {209-226},
   month = sep,
   year = {1977},
   keywords = {approximate getit}
}

@inproceedings{Eggers81Compression,
   author = {Susan J. Eggers and Frank Olken and Arie Shoshani},
   title = {A Compression Technique for Large Statistical Databases},
   booktitle = {Proc. VLDB},
   address = {Cannes, France},
   month = sep,
   year = {1981}
}

@article{Robinson81Another,
   author = {P. Robinson and D. Singer},
   title = {Another Spelling Correction Program},
   journal = {CACM},
   volume = {24},
   number = {5},
   pages = {296-297},
   month = may,
   year = {1981},
   keywords = {getit approximate matching}
}

@misc{Matias91Video,
   author = {Yossi Matias and Adi Shamir},
   title = {A Video Scrambling Technique Based on Space Filling Curves},
   year = {1991},
   keywords = {getmoreinfo},
   abstract = {use "randomized" versions of Hilbert curves
to scramble AND compress digitized video.}
}

@article{Ang91Video,
   author = {Peng H. Ang and Peter A. Ruetz and David Auld},
   title = {Video Compression Makes Big Gains},
   journal = {IEEE Spectrum},
   volume = {28},
   number = {10},
   pages = {16-19},
   month = oct,
   year = {1991},
   keywords = {mpeg jpeg}
}

@article{Morrison68,
   author = {Donald R. Morrison},
   title = {{PATRICIA}?????},
   journal = {Journal of ACM (JACM)},
   volume = {15},
   pages = {514-534},
   year = {1968},
   keywords = {fillitin}
}

@article{Eastman82Tree,
   author = {C.M Eastman and S.F. Weiss},
   title = {Tree Structures for High Dimensionality Nearest Neighbor Searching},
   journal = {Information Systems},
   volume = {7},
   number = {2},
   pages = {115-122},
   year = {1982},
   keywords = {kdtrees}
}

@book{Ripley81Spatial,
   author = {Brian P. Ripley},
   title = {Spatial Statistics},
   publisher = {John Wiley \& Sons},
   year = {1981},
   keywords = {hilbert voronoi diagram}
}

@article{Aurenhammer91Voronoi,
   author = {Franz Aurenhammer},
   title = {Voronoi Diagrams - {A} Survey of a Fundamental Geometric Data Structure},
   journal = {ACM Computing Surveys},
   volume = {23},
   number = {3},
   pages = {345-405},
   month = sep,
   year = {1991},
   keywords = {nearest neighbor approximate matching}
}

@inproceedings{Lin91,
   author = {Zheng Lin},
   title = {{CAT}: an Execution Model for Concurrent Full Text Search},
   booktitle = {Proceeding of the First International Conference on Parallel and Distributed Information Processing},
   pages = {151-158},
   address = {Miami Beach, Florida},
   month = {Dec. 4-6},
   year = {1991}
}

@article{Lander91Mapping,
   author = {Eric S. Lander and Robert Langridge and Damian M. Saccocio},
   title = {Mapping and Interpreting Biological Information},
   journal = {Comm. of ACM (CACM)},
   volume = {34},
   number = {11},
   pages = {33-39},
   month = nov,
   year = {1991}
}

@article{Durham83Spelling,
   author = {Ivor Durham and David A. Lamb and James B. Saxe},
   title = {Spelling Correction in User Interfaces},
   journal = {CACM},
   month = oct,
   year = {1983},
   keywords = {getit}
}

@book{Polya71How,
   author = {George Polya},
   title = {How to Solve It: {A} New Aspect of Mathematical Method},
   publisher = {Princeton University Press},
   year = {1971},
   keywords = {getit ord. number 99059412},
   note = {2nd ed.}
}

@techreport{Dar91,
   author = {S. Dar and N.H. Gehani and H.V. Jagadish},
   title = {{ODE}: Object Database and Environment},
   institution = {AT\&T Bell Laboratories},
   type = {Computer Technology and Research Laboratory Technical Reports},
   address = {Murray Hill, New Jersey},
   year = {1991}
}

@article{Deux91O2,
   author = {O. Deux},
   title = {The O2 System},
   journal = {Comm. of ACM (CACM)},
   volume = {34},
   number = {10},
   pages = {35-48},
   month = oct,
   year = {1991}
}

@inproceedings{Aref91Optimization,
   author = {Walid G. Aref and Hanan Samet},
   title = {Optimization Strategies for Spatial Query Processing},
   booktitle = {Proc. of VLDB (Very Large Data Bases)},
   pages = {81-90},
   address = {Barcelona, Spain},
   month = {Sept. 3-6},
   year = {1991}
}

@techreport{Story91,
   author = {Guy A. Story and Lawrence O'Gorman and H.V. Jagadish and Christos Faloutsos},
   title = {{DSA}: Approximate Searching with {OCR} errors},
   institution = {AT\&T Bell Laboratories},
   type = {Technical Memorandum},
   address = {Murry Hill, NJ},
   year = {1991}
}

@inproceedings{Jagadish91Retrieval,
   author = {H.V. Jagadish},
   title = {A Retrieval Technique for Similar Shapes},
   booktitle = {Proc. ACM SIGMOD Conf.},
   pages = {208-217},
   address = {Denver, Colorado},
   month = {May 29-31},
   year = {1991}
}

@inproceedings{Weikum91Dynamic,
   author = {Gerhard Weikum and Peter Zabback and Peter Scheuermann},
   title = {Dynamic File Allocation in Disk Arrays},
   booktitle = {Proc. ACM SIGMOD},
   pages = {406-415},
   address = {Denver, Colorado},
   month = {May 29-31},
   year = {1991}
}

@inproceedings{Chou85Evaluation,
   author = {H. Chou and D. DeWitt},
   title = {An Evaluation of Buffer Management Strategies for Relational Database Systems},
   booktitle = {Proc. of the 11th Intern. Conference Very Large Data Bases (VLDB)},
   pages = {127-141},
   address = {Stockholm, Sweden.},
   month = aug,
   year = {1985}
}

@inproceedings{Wolf91Effective,
   author = {Joel L. Wolf and Daniel M. Dias and Philip S. Yu and John Turek},
   title = {An Effective Algorithm for Parallelizing Hash Joins in the Presence of Data Skew},
   booktitle = {Proc. IEEE Conf. on Data Engineering},
   pages = {200-209},
   address = {Kobe, Japan},
   month = {April 8-12},
   year = {1991},
   keywords = {zipf}
}

@inproceedings{Ioannidis91Propagation,
   author = {Yannis E. Ioannidis and Stavros Christodoulakis},
   title = {On the Propagation of Errors in the  Size of Join Results},
   booktitle = {Proc. of ACM SIGMOD},
   pages = {268-277},
   address = {Denver, Colorado},
   month = {May 29-31},
   year = {1991},
   keywords = {zipf},
   abstract = {keep statistics of the most common values;
assume uniform for the rest}
}

@article{Silberschatz91Database,
   author = {Avi Silberschatz and Michael Stonebraker and Jeff Ullman},
   title = {Database Systems: Achievements and Opportunities},
   journal = {Comm. of ACM (CACM)},
   volume = {34},
   number = {10},
   pages = {110-120},
   month = oct,
   year = {1991}
}

@inproceedings{Kolovson91Segment,
   author = {Curtis P. Kolovson and Michael Stonebraker},
   title = {Segment Indexes: Dynamic Indexing Techniques for Multi-Dimensional Interval Data},
   booktitle = {Proc. ACM SIGMOD},
   pages = {138-147},
   address = {Denver, Colorado},
   month = {May 29-31},
   year = {1991},
   keywords = {temporal databases}
}

@inproceedings{Seeger91Multi,
   author = {Bernhard Seeger and Per-Ake Larson},
   title = {Multi-Disk {B}-trees},
   booktitle = {Proc. ACM SIGMOD},
   pages = {138-147},
   address = {Denver, Colorado},
   month = {May 29-31},
   year = {1991},
   keywords = {declustering}
}

@techreport{Rong91Analysis,
   author = {Yi Rong and Christos Faloutsos},
   title = {Analysis of the Clustering Property of Peano Curves},
   institution = {Univ. of Maryland},
   type = {Techn. Report CS-TR-2792,},
   number = {UMIACS-TR-91-151},
   month = dec,
   year = {1991}
}

@inproceedings{Kamel92Parallel,
   author = {Ibrahim Kamel and Christos Faloutsos},
   title = {Parallel {R}-trees},
   booktitle = {Proc. of ACM SIGMOD Conf.},
   pages = {195-204},
   address = {San Diego, CA},
   month = jun,
   year = {1992},
   note = {Also available as Tech. Report UMIACS TR 92-1, CS-TR-2820}
}

@misc{Faloutsos92Diamond,
   author = {Christos Faloutsos and H.V. Jagadish},
   title = {Diamond Tree: {A} High-Dimensionality Index Structure for Approximate Searching},
   year = {1992},
   note = {Available as SRC-TR-92-97}
}

@article{Meghini91Conceptual,
   author = {Carlo Meghini and Fausto Rabitti and Constantino Thanos},
   title = {Conceptual Modeling of Multimedia Documents},
   journal = {IEEE Computer},
   volume = {24},
   number = {10},
   pages = {23-30},
   month = oct,
   year = {1991}
}

@article{OGorman89approach,
   author = {Lawrence O'Gorman and Jeffrey V. Nickerson},
   title = {An approach to fingerprint filter design},
   journal = {Pattern Recognition},
   pages = {29-38},
   month = jan,
   year = {1989}
}

@article{Deerwester90Indexing,
   author = {S. Deerwester and S. T. Dumais and G. W. Furnas and T. K. Landauer and R. Harshman},
   title = {Indexing by Latent Semantic Analysis},
   journal = {Journal of the American Society for Information Science},
   volume = {41},
   number = {6},
   pages = {391-407},
   month = sep,
   year = {1990}
}

@misc{Jain92,
   author = {R. Jain and W. Niblack},
   title = {{NSF} Workshop on Visual Information Management},
   address = {Redwood City, CA},
   month = feb,
   year = {1992}
}

@inproceedings{Jagadish90Spatial,
   author = {H.V. Jagadish},
   title = {Spatial Search with Polyhedra},
   booktitle = {Proc. Sixth IEEE Int'l Conf. on Data Engineering},
   address = {Los Angeles, CA},
   month = feb,
   year = {1990}
}

@inproceedings{Agrawal89Ode,
   author = {R. Agrawal and N.H. Gehani},
   title = {Ode (Object Database and Environment): The Language and the Data Model},
   booktitle = {Proc. ACM-SIGMOD 1989 Int'l Conf. Management of Data},
   pages = {36-45},
   address = {Portland, Oregon},
   month = {May-June},
   year = {1989}
}

@inproceedings{Haas89Extensible,
   author = {L. M. Haas and J. C. Freytag and G. M. Lohman and H. Pirahesh},
   title = {Extensible Query Processing in Starburst},
   booktitle = {Proc. ACM-SIGMOD 1989 Int'l Conf. Management of Data},
   pages = {377-388},
   address = {Portland, Oregon},
   month = {May-June},
   year = {1989}
}

@incollection{Carey90EXODUS,
   author = {M.J. Carey and D.J. DeWitt and G. Graefe and D.M. Haight and J.E. Richardson and D.H. Schuh and E.J. Shekita and S.L. Vandenberg},
   editor = {S. Zdonik and D. Maier},
   title = {The {EXODUS} Extensible {DBMS} Project: An Overview},
   booktitle = {Readings in Object-Oriented Database Systems},
   publisher = {Morgan Kaufmann},
   year = {1990}
}

@book{Clocksin81Programming,
   author = {W.F. Clocksin and C.C. Mellish},
   title = {Programming in Prolog},
   publisher = {Springer Verlag},
   address = {New York},
   year = {1981}
}

@article{Fox91Order,
   author = {Edward A. Fox and Qi Fan Chen and Amjad M. Daoud and Lenwood S. Heath},
   title = {Order-Preserving Minimal Perfect Hash Functions and Information Retrieval},
   journal = {ACM Trans. on Information Systems (TOIS)},
   volume = {9},
   number = {3},
   pages = {281-308},
   month = jul,
   year = {1991},
   keywords = {hashing}
}

@article{Green92Evolution,
   author = {James L. Green},
   title = {The Evolution of {DVI} System Software},
   journal = {Comm. of ACM (CACM)},
   volume = {35},
   number = {1},
   pages = {52-67},
   month = jan,
   year = {1992},
   keywords = {mms multimedia}
}

@article{Bunton92Practical,
   author = {Suzanne Bunton and Gaetano Borriello},
   title = {Practical Dictionary Management for Hardware Data Compression},
   journal = {Comm. of ACM (CACM)},
   volume = {35},
   number = {1},
   pages = {95-104},
   month = jan,
   year = {1992},
   keywords = {infotheory},
   abstract = {FREEZE: stop adapting, once dict. is full.
FLUSH: double; drop the dictionary if it's full and restart ("compress" of UNIX)
LRU; SWAP: on overflow, use a secondary dict.; drop the primary
when the secondary overflows
SWAP*: do the same, but when the primary is half full.
TAGS: clever scheme to implement LRU: patterns are kept in a TRIE,
labelled ("tagged") according to the last time touched;
the root is labeled with the address of the leaf node that matched
N steps ago.
The method achieves almost identical performance with LRU,
superior to all other schemes.}
}

@article{Pennebaker88Optimal,
   author = {W. Pennebaker and J. Mitchell},
   title = {Optimal Hardware and Software Arithmetic Coding Procedures for the {Q}-coder},
   journal = {IBM Journal of Research and Development},
   volume = {32},
   number = {6},
   pages = {727-736},
   month = nov,
   year = {1988},
   keywords = {compression infotheory}
}

@article{Fox92Practical,
   author = {Edward A. Fox and Lenwood S. Heath and Qi Fan Chen and Amjad M. Daoud},
   title = {Practical Minimal Perfect Hash Functions for Large Databases},
   journal = {Comm. of ACM (CACM)},
   volume = {35},
   number = {1},
   pages = {105-121},
   month = jan,
   year = {1992},
   keywords = {hashing},
   abstract = {h(k) = $\lbrace$h0(k) + g(h1(k)) + g(h2(k))$\rbrace$ mod m, -
Three stages:
Mapping: the [h0(), h1() and h2()] triple is unique for each word
Ordering: look for a suitable g() function - divide the input
into "levels"
Searching: assign hash codes to the members of each level.
9h15min CPU, to find a mphf for 3.9M words, on a NeXT machine;
\~{}linear performance.}
}

@article{Bertino88Query,
   author = {Elisa Bertino and Fausto Rabitti and Simon Gibbs},
   title = {Query Processing in a Multimedia Document System},
   journal = {ACM Trans. on Office Information Systems (TOOIS)},
   volume = {6},
   number = {1},
   pages = {1-41},
   month = jan,
   year = {1988},
   keywords = {tois mms}
}

@article{Postel88Experimental,
   author = {Jonathan B. Postel and Gregory G. Finn and Alan R. Katz and Joyce K. Reynolds},
   title = {An Experimental Multimedia Mail System},
   journal = {ACM Trans. on Office Information Systems (TOOIS)},
   volume = {6},
   number = {1},
   pages = {63-81},
   month = jan,
   year = {1988},
   keywords = {tois mms}
}

@article{Baskett92Expert,
   author = {Forest Baskett},
   title = {Expert Opinion: PCs and workstations blend new kinds of applications},
   journal = {IEEE Spectrum},
   volume = {29},
   number = {1},
   pages = {28},
   month = jan,
   year = {1992},
   keywords = {virtual realities mms}
}

@article{Rosenblatt92PCs,
   author = {Alfred Rosenblatt},
   title = {PCs and Workstations},
   journal = {IEEE Spectrum},
   volume = {29},
   number = {1},
   pages = {27-29},
   month = jan,
   year = {1992},
   keywords = {mms multimedia}
}

@article{Jurgen92Consumer,
   author = {Ronald K. Jurgen},
   title = {Consumer Electronics},
   journal = {IEEE Spectrum},
   volume = {29},
   number = {1},
   pages = {52-54},
   month = jan,
   year = {1992},
   keywords = {mms multimedia compression}
}

@article{Narasimhalu91Multimedia,
   author = {A. Desai Narasimhalu and Stavros Christodoulakis},
   title = {Multimedia Information Systems: The Unfolding of a Reality},
   journal = {IEEE Computer},
   volume = {24},
   number = {10},
   pages = {6-8},
   month = oct,
   year = {1991},
   keywords = {mms}
}

@article{Fox91Advances,
   author = {Edward A. Fox},
   title = {Advances in Interactive Digital Multimedia Systems},
   journal = {IEEE Computer},
   volume = {24},
   number = {10},
   pages = {9-21},
   month = oct,
   year = {1991},
   keywords = {mms}
}

@article{Little91Spatio,
   author = {Thomas D.C. Little and Arif Ghafoor},
   title = {Spatio-Temporal Composition of Distributed Multimedia Objects for Value-Added Networks},
   journal = {IEEE Computer},
   volume = {24},
   number = {10},
   pages = {42-50},
   month = oct,
   year = {1991},
   keywords = {mms}
}

@article{Anderson91Continuous,
   author = {David P. Anderson and George Homsy},
   title = {A Continuous Media {I}/{O} Server and its Synchronization Mechanism},
   journal = {IEEE Computer},
   volume = {24},
   number = {10},
   pages = {51-57},
   month = oct,
   year = {1991},
   keywords = {mms multimedia}
}

@article{Palaniappan91InternetExpress,
   author = {Murugappan Palaniappan and George Fitzmaurice},
   title = {InternetExpress: An Inter-Desktop Multimedia Data-Transfer Service},
   journal = {IEEE Computer},
   volume = {24},
   number = {10},
   pages = {58-67},
   month = oct,
   year = {1991},
   keywords = {mms mail systems}
}

@article{Vin91Multimedia,
   author = {Harrick M. Vin and Polle T. Zellweger and Daniel C. Swinehart and P. Venkat Rangan},
   title = {Multimedia Conferencing in the Etherphone Environment},
   journal = {IEEE Computer},
   volume = {24},
   number = {10},
   pages = {69-79},
   month = oct,
   year = {1991},
   keywords = {mms}
}

@article{Lamb91ObjectStore,
   author = {Charles Lamb and Gordon Landis and Jack Orenstein and Dan Weinreb},
   title = {The ObjectStore Database System},
   journal = {Comm. of ACM (CACM)},
   volume = {34},
   number = {10},
   pages = {50-63},
   month = oct,
   year = {1991},
   keywords = {object oriented}
}

@article{Butterworth91GemStone,
   author = {Paul Butterworth and Allen Otis and Jacob Stein},
   title = {The GemStone Object Database System},
   journal = {Comm. of ACM (CACM)},
   volume = {34},
   number = {10},
   pages = {64-77},
   month = oct,
   year = {1991},
   keywords = {object oriented}
}

@article{Stonebraker91POSTGRES,
   author = {Michael Stonebraker and Greg Kemnitz},
   title = {The {POSTGRES} Next Generation Database Management System},
   journal = {Comm. of ACM (CACM)},
   volume = {34},
   number = {10},
   pages = {78-92},
   month = oct,
   year = {1991},
   keywords = {object oriented}
}

@article{Lohman91Extensions,
   author = {Guy M. Lohman and Bruce Lindsay and Hamid Pirahesh and K. Berhnard Schiefer},
   title = {Extensions to {STARBURST}: Objects, Types, Functions and Rules},
   journal = {Comm. of ACM (CACM)},
   volume = {34},
   number = {10},
   pages = {94-109},
   month = oct,
   year = {1991},
   keywords = {object oriented}
}

@article{Gall91,
   author = {D. Le Gall},
   title = {{MPEG}: {A} Video Compression Standard for Multimedia Applications},
   journal = {Comm. of ACM (CACM)},
   volume = {34},
   number = {4},
   pages = {46-58},
   month = apr,
   year = {1991}
}

@article{Newcomb91HyTime,
   author = {S. Newcomb and N. Kipp and V. Newcomb},
   title = {HyTime: The Hypermedia/Time-Based Document Structuring Language},
   journal = {Comm. of ACM (CACM)},
   volume = {34},
   number = {11},
   month = nov,
   year = {1991},
   keywords = {multimedia mms}
}

@article{Ghandeharizade91Object,
   author = {S. Ghandeharizadeh and L. Ramos and Z. Asad and W. Qureshi},
   title = {Object Placement in Parallel Hypermedia Systems},
   journal = {VLDB 1991},
   pages = {243-254},
   address = {Barcelona, Spain},
   month = sep,
   year = {1991},
   keywords = {multimedia parhash},
   abstract = {They distribute hypermedia database on m disks.
They do not use all the available processors.
They  prefetch data while the display is busy.}
}

@article{Jurgen92Digital,
   author = {Ronald K. Jurgen},
   title = {Digital Video},
   journal = {IEEE Spectrum},
   volume = {29},
   number = {3},
   pages = {24-30},
   month = mar,
   year = {1992},
   keywords = {multimedia mpeg jpeg}
}

@inproceedings{Ford91Optimizing,
   author = {Daniel Ford and Stavros Christodoulakis},
   title = {Optimizing Random Retrievals from {CLV} Format Optical Disks},
   booktitle = {Proc. of the 17th Int. Conf. on Very Large Data Bases (VLDB)},
   pages = {413-422},
   address = {Barcelona, Spain},
   month = {Sept. 3-6},
   year = {1991},
   keywords = {data placement}
}

@inproceedings{Palmer91Fido,
   author = {Mark Palmer and Stanley B. Zdonik},
   title = {Fido: {A} Cache that Learns to Fetch},
   booktitle = {Proc. of the 17th Int. Conf. on Very Large Data Bases (VLDB)},
   pages = {255-264},
   address = {Barcelona, Spain},
   month = {Sept. 3-6},
   year = {1991},
   keywords = {object oriented caching}
}

@book{Sankoff83Time,
   author = {David Sankoff and Joseph B. Kruskal},
   title = {Time warps, string edits and macromolecules: The theory and practice of sequence comparisons},
   publisher = {Addison-Wesley Publishing Company, Inc.},
   address = {Reading, MA},
   year = {1983},
   keywords = {QA292.T55, ISBN 0-201-07809-0}
}

@inproceedings{Freeston87BANG,
   author = {Michael Freeston},
   title = {The {BANG} File: {A} New Kind of Grid File},
   booktitle = {Proc. of ACM SIGMOD},
   pages = {260-269},
   address = {San Francisco, CA},
   month = {May 27-29},
   year = {1987},
   keywords = {seckey}
}

@book{Samet90Applications,
   author = {H. Samet},
   title = {Applications of Spatial Data Structures Computer Graphics, Image Processing and {GIS}},
   publisher = {Addison-Wesley},
   year = {1990},
   keywords = {second book}
}

@article{Burkhard73Some,
   author = {W.A. Burkhard and R.M. Keller},
   title = {Some Approaches to Best-Match File Searching},
   journal = {Comm. of the ACM (CACM)},
   volume = {16},
   number = {4},
   pages = {230-236},
   month = apr,
   year = {1973},
   keywords = {approximate matching nearest neighbor nn}
}

@article{Shapiro77Choice,
   author = {M. Shapiro},
   title = {The Choice of Reference Points in Best-Match File Searching},
   journal = {Comm. of the ACM (CACM)},
   volume = {20},
   number = {5},
   pages = {339-343},
   month = may,
   year = {1977},
   keywords = {approximate matching nearest neighbor nn}
}

@techreport{Mahanti90,
   author = {Ambuj Mahanti and Raghu Karinthi and Subrata Ghosh and Asim Pal},
   title = {{AI} Search for Minimum-Cost Set Cover and Multiple-Goal Plan Optimization Problems: Applications to Process Planning.},
   institution = {Univ. of Maryland},
   type = {UMIACS-TR-90-125,},
   number = {CS-TR2540},
   month = sep,
   year = {1990},
   keywords = {Astar search heuristics},
   abstract = {they propose the Nm heuristic for the cover-set problem:
Given a set of features (to drill),
and set-ups with costs for each feature,
find the best combination of set-ups to achieve all the features.
Their heuristic has better performance
than Timos's "average cost" heuristic
Could be useful for "approximate matching" applications},
   note = {to appear in AI Journal}
}

@inproceedings{DeWitt90Hybrid,
   author = {David J. DeWitt and Shahram Ghandeharizadeh},
   title = {Hybrid-Range Partitioning Strategy: {A} new Declustering Strategy for Multiprocessor Database Machine},
   booktitle = {Proc. 16th International Conference on VLDB},
   pages = {481-492},
   address = {Brisbane, Australia},
   month = aug,
   year = {1990}
}

@techreport{Srinivasan92Line,
   author = {Venkatachary Srinivasan},
   title = {On-Line Processing In Large-Scale Transaction Systems},
   institution = {Univ. of Wisconsin, Madison},
   type = {Computer Sciences Technical Report},
   number = {\#1071},
   month = jan,
   year = {1992},
   note = {Ph.D. Dissertation}
}

@article{Zhu89Technique,
   author = {Rui Feng Zhu and Tadao Takaoka},
   title = {A Technique for Two-Dimensional Pattern Matching},
   journal = {Comm. of ACM (CACM)},
   volume = {32},
   number = {9},
   pages = {1110-1120},
   month = sep,
   year = {1989},
   keywords = {boyer moore}
}

@misc{Wu92Approximate,
   author = {Sun Wu},
   title = {Approximate Pattern Matching and Its Applications},
   year = {1992},
   keywords = {boyer moore},
   note = {transparencies of presentation}
}

@inproceedings{Faloutsos92B,
   author = {Christos Faloutsos and H.V. Jagadish},
   title = {On {B}-tree Indices for Skewed Distributions},
   booktitle = {18th VLDB Conference},
   pages = {363-374},
   address = {Vancouver, British Columbia},
   month = {Aug. 23-27},
   year = {1992},
   keywords = {zipf fibonacci},
   abstract = {Studies schemes for maintaining postings lists
under zipf distributions.}
}

@article{Suetens92Computational,
   author = {Paul Suetens and Pascal Fua and Andrew J. Hanson},
   title = {Computational Strategies for Object Recognition},
   journal = {ACM Computing Surveys},
   volume = {24},
   number = {1},
   pages = {5-62},
   month = mar,
   year = {1992},
   keywords = {high dimensional indexing feature extraction}
}

@article{Mishra92Joing,
   author = {Priti Mishra and Margaret H. Eich},
   title = {Joing Processing in Relational Databases},
   journal = {ACM Computing Surveys},
   volume = {24},
   number = {1},
   pages = {63-113},
   month = mar,
   year = {1992}
}

@inproceedings{Saalfeld91New,
   author = {Alan Saalfeld},
   title = {New Proximity-Preserving Orderings for Spatial Data},
   booktitle = {Tenth Int. Symposium on Computer Assisted Cartography},
   pages = {59-76},
   address = {Baltimore, MD},
   month = mar,
   year = {1991},
   abstract = {use  eulerian tour on mst - then, sample systematically
.}
}

@inproceedings{Saalfeld91Construction,
   author = {Alan Saalfeld},
   title = {Construction of Spatially Articulated List Frames for Household Surveys},
   booktitle = {Symposium 91: Spatial Issues in Statistics},
   address = {Ottawa, Ontario, Canada},
   year = {1991}
}

@misc{Mavrovouniotis92Biochemical,
   author = {Michael Mavrovouniotis and Christos Faloutsos},
   title = {Biochemical Similarity in Databases Searches},
   year = {1992},
   note = {Working paper}
}

@techreport{Wu91Fast,
   author = {Sun Wu and Udi Manber},
   title = {Fast Text Searching With Errors},
   institution = {Department of Computer Science, University of Arizona},
   type = {Technical report},
   number = {\#91-11},
   month = jun,
   year = {1991}
}

@inproceedings{Wu92Agrep,
   author = {Sun Wu and Udi Manber},
   title = {Agrep - {A} Fast Approximate Pattern Searching Tool},
   booktitle = {USENIX Conference},
   month = jan,
   year = {1992}
}

@article{Tomasic93Performance,
   author = {Anthony Tomasic and Hector Garcia-Molina},
   title = {Performance of Inverted Indices in Distributed Text Document Retrieval Systems},
   journal = {PDIS},
   year = {1993},
   abstract = {they examine partitioning the postings file over
a multiprocessor system. Due to the skew,
they claim it is better to partition documents over the host machines,
and stripe postings lists across the disks of each machine.
They have very interesting results on the zipf distribution of words
(they do a quadratic approximation of the log-log curve,
which seems to work better)
and they describe the simulation very well (typical times
for communication delay, for seek time etc.)},
   note = {to appear}
}

@article{Ghandeharizade92Performance,
   author = {Shahram Ghandeharizadeh and David J. DeWitt and W. Qureshi},
   title = {A Performance Analysis of Alternative Multi-Attribute Declustering Strategies},
   journal = {SIGMOD Conf.},
   address = {San Diego, California},
   month = jun,
   year = {1992}
}

@inproceedings{Li92,
   author = {Jianzhong Li and Jaideep Srivastava and Doron Rotem},
   title = {{CMD}: {A} Multidimensional Declustering Method for Parallel Database Systems},
   booktitle = {Proc. of the 18th VLDB Conference},
   pages = {3-14},
   address = {Vancouver, British Columbia, Canada},
   month = {August 23-27},
   year = {1992}
}

@book{David81Order,
   author = {Herbert A. David},
   title = {Order Statistics},
   edition = {2nd},
   publisher = {Wiley and Sons, Inc.},
   year = {1981}
}

@book{Physical92Grand,
   author = {on Physical, Mathematical, Committee and Sciences, NSF, Engineering},
   title = {Grand Challenges: High Performance Computing and Communications},
   publisher = {National Science Foundation},
   year = {1992},
   note = {The FY 1992 U.S. Research and Development Program}
}

@article{Zobel92Efficient,
   author = {Justin Zobel and Alistair Moffat and Ron Sacks-Davis},
   title = {An Efficient Indexing Technique for Full-Text Database Systems},
   journal = {VLDB},
   pages = {352-362},
   address = {Vancouver, B.C., Canada},
   month = {Aug. 23-27},
   year = {1992},
   keywords = {compression inversion},
   abstract = {compress the postings lists, using run-length encoding.
Reports down to 6\% overhead, for document-level granularity.}
}

@book{Edwards66Technical,
   author = {Robert D. Edwards and John Magee},
   title = {Technical Analysis of Stock Trends},
   edition = {5th},
   publisher = {John Magee},
   address = {Springfield, Massachusetts},
   year = {1966},
   abstract = {Major points in the Dow theory:
1. Averages show no surprises;
2. Industrial avgs and Rails go together
3. 3 phases in the Bull Market; 3 phases in the Bear market
4. Primary trends: $>$20\% change; last $>$1 year
5. Secondary trends: 1/3-2/3 relative change over the primary trend;
last 3 weeks to months
6. Minor trends: last \~{}6 days
7. "Lines": $<$5\% fluctuations; last 2-3 weeks
My observation: Local minima/maxima seem to fall on same day, for Industrial
and for Rail averages},
   note = {second printing}
}

@book{Brooks84Investing,
   author = {Herb Brooks},
   title = {Investing with a Computer},
   publisher = {Petrocelli Books, Inc.},
   address = {New York, Princeton},
   year = {1984},
   keywords = {fourier fft stocks},
   abstract = {Suggests using filters (FIR, IIR) and FFTs to detect regularities in data.
Gives also Fortran programs to process time series.
Elementary introduction to digital filter design theory.
Useful points: 1. de-trend the data (ie., make x0 == xN)
2. apply windowing (eg., low pass filtering (?) )
Also, he uses Logs of the Dow index, almost always.}
}

@book{Chatfield84Analysis,
   author = {Christopher Chatfield},
   title = {The Analysis of Time Series: An Introduction},
   edition = {Third},
   publisher = {Chapman and Hall},
   address = {London \& New York},
   year = {1984}
}

@book{Schroeder91Fractals,
   author = {Manfred Schroeder},
   title = {Fractals, Chaos, Power Laws: Minutes from an Infinite Paradise},
   publisher = {W.H. Freeman and Company},
   address = {New York},
   year = {1991}
}

@book{Press88Numerical,
   author = {William H. Press and Brian P. Flannery and Saul A. Teukolsky and William T. Vetterling},
   title = {Numerical Recipes in {C}},
   publisher = {Cambridge University Press},
   year = {1988}
}

@book{Hamming77Digital,
   author = {Richard Wesley Hamming},
   title = {Digital Filters},
   publisher = {Prentice-Hall Signal Processing Series},
   address = {Englewood Cliffs, N.J.},
   year = {1977}
}

@book{Oppenheim75Digital,
   author = {Alan Victor Oppenheim and Ronald W. Schafer},
   title = {Digital Signal Processing},
   publisher = {Prentice-Hall},
   address = {Englewood Cliffs, N.J.},
   year = {1975}
}

@book{Rabiner75Theory,
   author = {Lawrence Richard Rabiner and Bernard Gold},
   title = {Theory and Application of Digital Signal Processing},
   publisher = {Prentice-Hall},
   address = {Englewood Cliffs, N.J.},
   year = {1975}
}

@book{Oppenheim83Signals,
   author = {Alan V. Oppenheim and Alan S. Willsky},
   title = {Signals and Systems},
   publisher = {Prentice-Hall Inc.},
   address = {Englewood Cliffs, N.J.},
   year = {1983}
}

@article{Baeza-Yates92New,
   author = {Ricardo Baeza-Yates and Gaston H. Gonnet},
   title = {A New Approach to Text Searching},
   journal = {Comm. of ACM (CACM)},
   volume = {35},
   number = {10},
   pages = {74-82},
   month = oct,
   year = {1992}
}

@article{Wu92Text,
   author = {Sun Wu and Udi Manber},
   title = {Text Searching Allowing Errors},
   journal = {Comm. of ACM (CACM)},
   volume = {35},
   number = {10},
   pages = {83-91},
   month = oct,
   year = {1992}
}

@techreport{Otterman92Approximate,
   author = {Michael Otterman},
   title = {Approximate Matching with High Dimensionality {R}-trees},
   institution = {Dept. of Computer Science, Univ. of Maryland},
   type = {M.Sc. scholarly paper},
   address = {College Park, MD},
   year = {1992},
   note = {supervised by C. Faloutsos}
}

@book{Wolfram91Mathematica,
   author = {Stephen Wolfram},
   title = {Mathematica},
   edition = {Second},
   publisher = {Addison Wesley},
   year = {1991}
}

@article{Motro88,
   author = {Amihai Motro},
   title = {{VAGUE}: {A} User Interface to Relational Databases that Permits Vague Queries},
   journal = {ACM Trans. on Information Systems (TOIS)},
   volume = {6},
   number = {3},
   pages = {187-214},
   month = jul,
   year = {1988}
}

@article{Stam88Bibliography,
   author = {R. Stam and Richard Snodgrass},
   title = {A Bibliography on Temporal Databases},
   journal = {IEEE Bulletin on Data Engineering},
   volume = {11},
   number = {4},
   month = dec,
   year = {1988}
}

@article{Collura92Epilog,
   author = {Thomas F. Collura and Ernest C. Jacobs and Richard C. Burgess and John P. Turnbull},
   title = {The Epilog System - Automated Long-Term {EEG} Monitoring for Epilepsy},
   journal = {IEEE Computer},
   volume = {25},
   number = {9},
   pages = {5-14},
   month = sep,
   year = {1992}
}

@article{Rioul91Wavelets,
   author = {Oliver Rioul and Martin Vetterli},
   title = {Wavelets and Signal Processing},
   journal = {IEEE SP Magazine},
   pages = {14-38},
   month = oct,
   year = {1991},
   abstract = {Introduction to wavelets}
}

@book{David90New,
   author = {{David Shepard Associates, Inc.}},
   title = {The New Direct Marketing},
   institution = {David Shepard Associates, Inc.},
   publisher = {Business One Irwin},
   address = {Homewood, IL 60430},
   year = {1990},
   keywords = {direct marketing data processing statistical methods},
   abstract = {Idea (among others): Cluster customers according to demographic etc
attributes
Other ideas: To compute correlations, use logarithms
(makes data distr.  closer to normal (?!));
multiple regression methods are explained and discussed;
principal component analysis (K-L (?)) is also discussed.}
}

@article{Elias75Universal,
   author = {P. Elias},
   title = {Universal Codeword Sets and Representations of Integers},
   journal = {IEEE Trans. on Information Theory},
   volume = {IT-21},
   pages = {194-203},
   year = {1975}
}

@inproceedings{Faloutsos93Declustering,
   author = {Christos Faloutsos and Pravin Bhagwat},
   title = {Declustering Using Fractals},
   booktitle = {2nd Int. Conference on Parallel and Distributed Information Systems (PDIS)},
   pages = {18-25},
   address = {San Diego, CA},
   month = {Jan. 20-22},
   year = {1993}
}

@inproceedings{Rabitti92Information,
   author = {Fausto Rabitti and Pascuale Savino},
   title = {An Information Retrieval Approach for Image Databases},
   booktitle = {VLDB Conf. Proceedings},
   pages = {574-584},
   address = {Vancouver, BC, Canada},
   month = {August 23-27},
   year = {1992}
}

@article{Wong92Digital,
   author = {Albert W.K. Wong and Ricky K. Taira and H.K. Huang},
   title = {Digital Archive Center: Implementation for a Radiology Department},
   journal = {AJR},
   volume = {159},
   pages = {1101-1105},
   month = nov,
   year = {1992}
}

@article{Helman89Representation,
   author = {James Helman and Lambertus Hesselink},
   title = {Representation and Display of Vector Field Topology in Fluid Flow Data Sets},
   journal = {IEEE Computer},
   volume = {22},
   number = {8},
   pages = {27-36},
   month = aug,
   year = {1989}
}

@article{Fuchs89Interactive,
   author = {Henry Fuchs and Marc Levoy and Stephen M. Pizer},
   title = {Interactive Visualization of 3D Medical Data},
   journal = {IEEE Computer},
   volume = {22},
   number = {8},
   pages = {46-51},
   month = aug,
   year = {1989}
}

@article{Reilly89Data,
   author = {Paul Reilly},
   title = {Data Visualization in Archeology},
   journal = {IBM Systems Journal},
   volume = {28},
   number = {4},
   pages = {569-579},
   year = {1989}
}

@article{DeFanti89Visualization,
   author = {Thomas A. DeFanti and Maxine D. Brown and Bruce H. McCormick},
   title = {Visualization: Expanding Scientific and Engineering Research Opportunities},
   journal = {IEEE Computer},
   volume = {22},
   number = {8},
   pages = {12-25},
   month = aug,
   year = {1989}
}

@article{Pelizzari89Accurate,
   author = {C.A. Pelizzari and G.T.Y. Chen and D.R. Spelbring and R.R. Weichselbaum and C.T. Chen},
   title = {Accurate three-dimensional registration of {CT}, {PET} and/or {MR} images of the brain},
   journal = {J. Comput. Assisted Tomogr.},
   volume = {13},
   pages = {20-26},
   year = {1989}
}

@article{Toga91Brain,
   author = {A.W. Toga and P. Banerjee and B.A. Payne},
   title = {Brain warping and averaging},
   journal = {Int. Symp. on Cereb. Blood Flow and Metab.},
   address = {Miami, FL},
   year = {1991}
}

@article{Toga90Warping,
   author = {A.W. Toga and P.K. Banerjee and E.M. Santori},
   title = {Warping 3D models for interbrain comparisons},
   journal = {Neurosc. Abs.},
   volume = {16},
   pages = {247},
   year = {1990}
}

@article{Agrawal92Interval,
   author = {Rakesh Agrawal and Sakti Ghosh and Tomasz Imielinski and Bala Iyer and Arun Swami},
   title = {An Interval Classifier for Database Mining Applications},
   journal = {VLDB Conf. Proc.},
   pages = {560-573},
   address = {Vancouver, BC, CANADA},
   month = {Aug. 23-27},
   year = {1992}
}

@inproceedings{Niblack93QBIC,
   author = {Wayne Niblack and Ron Barber and Will Equitz and Myron Flickner and Eduardo Glasman and Dragutin Petkovic and Peter Yanker and Christos Faloutsos and Gabriel Taubin},
   title = {The {QBIC} Project: Querying Images By Content Using Color, Texture and Shape},
   booktitle = {SPIE 1993 Intl. Symposium on Electronic Imaging: Science and Technology, Conf. 1908, Storage and Retrieval for Image and Video Databases},
   month = feb,
   year = {1993},
   note = {Also available as IBM Research Report RJ 9203 (81511), Feb. 1, 1993, Computer Science}
}

@article{Arya93,
   author = {Manish Arya and William Cody and Christos Faloutsos and Joel Richardson and Arthur Toga},
   title = {{QBISM}: {A} Prototype 3-{D} Medical Image Database System},
   journal = {IEEE Data Engineering Bulletin},
   volume = {16},
   number = {1},
   pages = {38-42},
   month = mar,
   year = {1993}
}

@article{Brown92Survey,
   author = {Lisa Gottesfeld Brown},
   title = {A Survey of Image Registration Techniques},
   journal = {ACM Computing Surveys},
   volume = {24},
   number = {4},
   pages = {325-376},
   month = dec,
   year = {1992}
}

@article{Kukich92Techniques,
   author = {Karen Kukich},
   title = {Techniques for Automatically Correcting Words in Text},
   journal = {ACM Computing Surveys},
   volume = {24},
   number = {4},
   pages = {377-440},
   month = dec,
   year = {1992}
}

@book{Fukunaga90Introduction,
   author = {Keinosuke Fukunaga},
   title = {Introduction to Statistical Pattern Recognition},
   edition = {2nd},
   publisher = {Academic Press},
   year = {1990}
}

@article{Hou92Content,
   author = {Hou and Hsu and Liu and Chiu},
   title = {A Content-based indexing technique using relative geometry features},
   journal = {SPIE 92},
   volume = {1662},
   pages = {59-68},
   year = {1992},
   abstract = {Their method is very domain-specific (medical images)
and it seems to allow for false misses:
given a picture, they (manually?) break it down to objects;
they keep the 4 'most important' objects, and use the center and area
of these objects as features.}
}

@article{Gary92Shape,
   author = {Gary and Mehrotra},
   title = {Shape Similarity-Based Retrieval in Image Database Systems},
   journal = {SPIE 92},
   volume = {1662},
   pages = {2-8},
   year = {1992},
   abstract = {They propose representing a polygonal shape as a high-dimensional point;
where the dimensionality is proportional to the number of sides of the polygon.
They make no attempt to reduce the dimensionality;
they mention PAMs, but they have not yet used one}
}

@misc{Cohen??Biomedical,
   author = {Arnon Cohen},
   title = {Biomedical Signal Processing},
   volume = {I \& II},
   ISBN = {0-8493-5933-3},
   note = {(v.1), 0-8493-5934-1 (v.2)}
}

@book{Ruskai92Wavelets,
   author = {Mary Beth Ruskai and Gregory Beylkin and Ronald Coifman and Ingrid Daubechies and Stephane Mallat and Yves Meyer and Louise Raphael},
   title = {Wavelets and their Applications},
   publisher = {Jones and Bartlett Publishers},
   address = {Boston, MA},
   year = {1992}
}

@book{Carroll71Word,
   author = {John B. Carroll and Peter Davies and Barry Richman},
   title = {Word Frequency Book},
   publisher = {American Heritage Publishing Co., Inc.},
   address = {New York},
   year = {1971},
   abstract = {Has frequencies of words in English text,
according to Grade of school, gender, subject etc.}
}

@misc{National78Theory,
   author = {{National Academy of Sciences, The National Research Council}},
   title = {On the Theory and Practice of Voice Identification},
   institution = {National Academy of Sciences, The National Research Council},
   address = {Washington, D.C. 20418},
   year = {1978},
   ISBN = {0-309-02873-6},
   note = {LOCCCN 79-63355}
}

@article{Jean90New,
   author = {Jack S.N. Jean},
   title = {New Distance Measure for Binary Images},
   journal = {Int. Conf. on Acoustics, Speech and Signal Processing (ICASSP)},
   volume = {4},
   address = {Albuquerque, NM},
   month = {April 3-6},
   year = {1990},
   abstract = {proposes a distance that takes into account
the neighborhood - most importantly,
it shows that this new distance function
corresponds to a Euclidean distance, after an eigen-decomposition
of the matrix of the generalized euclidean distance function.
It also shows that, after the transformation,
a neural network achieves consistently better performance
than the same neural network *without* the transformation
($>$95\% recognition rate, vs. 91-93\%)},
   note = {paper\#: M5.19}
}

@article{Schonfeld90Robust,
   author = {Dan Schonfeld and John Goutsias},
   title = {Robust Morphological Representation of Binary Images},
   journal = {Int. Conf. on Acoustics, Speech and Signal Processing (ICASSP)},
   volume = {4},
   address = {Albuquerque, NM},
   month = {April 3-6},
   year = {1990},
   note = {paper\# M5.20}
}

@book{Hartigan75Clustering,
   author = {John A. Hartigan},
   title = {Clustering Algorithms},
   publisher = {John Wiley \& Sons},
   year = {1975},
   keywords = {cluster analysis}
}

@techreport{Faloutsos93Packed,
   author = {Christos Faloutsos and Ibrahim Kamel},
   title = {Packed {R}-trees Using Fractals},
   institution = {Univ. of Maryland},
   type = {Systems Research Center (SRC)},
   number = {TR-93-1},
   address = {College Park},
   year = {1993}
}

@techreport{Kamel93Hilbert,
   author = {Ibrahim Kamel and Christos Faloutsos},
   title = {Hilbert {R}-tree: An Improved {R}-tree Using Fractals},
   institution = {Univ. of Maryland},
   type = {Systems Research Center (SRC)},
   number = {TR-93-19},
   address = {College Park},
   year = {1993}
}

@article{Abdel-Ghaffar93Optimal,
   author = {Khaled A.S. Abdel-Ghaffar and Amr El Abbadi},
   title = {Optimal Disk Allocation for Partial Match Queries},
   journal = {ACM TODS},
   volume = {18},
   number = {1},
   pages = {132-156},
   month = mar,
   year = {1993},
   keywords = {declustering error correcting codes},
   abstract = {They consider p-ary codes (-$>$ the domains have to be pXpX..p).
Also, they use the HAMMING distance between digits, as opposed to the Manhattan distance -
this won't work well for range queries!}
}

@article{Schank93Learning,
   author = {Roger C. Schank},
   title = {Learning via Multimedia Computers},
   journal = {Comm. ACM (CACM)},
   volume = {36},
   number = {5},
   pages = {54-56},
   month = may,
   year = {1993},
   keywords = {indexing},
   abstract = {highlights the need for indexing in multimedia databases}
}

@inproceedings{Chen93Declustering,
   author = {Ling Tony Chen and Doron Rotem},
   title = {Declustering Objects for Visualization},
   booktitle = {Proc. VLDB Conf.},
   address = {Dublin, Ireland},
   month = aug,
   year = {1993},
   keywords = {lattice},
   abstract = {uses a lattice model to do declustering - sounds great!},
   note = {to appear}
}

@inproceedings{Rao92Identifying,
   author = {A. Ravishankar Rao and Jerry Lohse},
   title = {Identifying high level features of texture perception},
   booktitle = {SPIE Conference},
   address = {San Jose},
   month = feb,
   year = {1992},
   abstract = {they turn a distance matrix into points in n-d space
The term is 'multidimensional scaling' - Kruskal, Psychometrica 29, 1964.}
}

@book{Gruenbaum87Tilings,
   author = {Branko Gruenbaum and Geoffrey Colin Shephard},
   title = {Tilings and Patterns},
   publisher = {W.H. Freeman and Company},
   address = {New York},
   year = {1987},
   keywords = {sphere packing error correcting codes}
}

@book{Strang80Linear,
   author = {Gilbert Strang},
   title = {Linear Algebra and Its Applications},
   edition = {2nd},
   publisher = {Academic Press},
   year = {1980},
   keywords = {svd singular value decomposition},
   abstract = {Excellent book on linear algebra}
}

@book{Grimson90Object,
   author = {W. Eric L. Grimson},
   title = {Object Recognition by Computer: The Role of Geometric Constraints},
   publisher = {MIT Press},
   address = {Cambridge, MA},
   year = {1990},
   abstract = {suggested by Myron Flickner}
}

@book{Conway88Sphere,
   author = {John Horton Conway and Neil James Alexander Sloane},
   title = {Sphere Packings, Lattices and Groups},
   publisher = {Springer Verlag},
   year = {1988},
   keywords = {QA166.7.C66, ISBN 0-387-96617-X},
   abstract = {useful for the vector quantization problem: how to place points in space
so that the average second moment of their Voronoi cells is as small as possible.
Seems extremely useful for declustering, using lattices and tilings.
(p. 4 dfn of fundamental parallelotope,
p. 41-42 difficulties in packings and coverings; reduction algorithms;
p. 75 minimal distance of a code)}
}

@book{MacWilliams77Theory,
   author = {Florence Jessie MacWilliams and Neil James Alexander Sloane},
   title = {The Theory of Error-Correcting Codes},
   publisher = {North Holland Publishing Company},
   year = {1977},
   keywords = {sphere packing lattice declustering}
}

@article{Preparata74Difference,
   author = {F.P. Preparata and Jurg Nievergelt},
   title = {Difference-Preserving Codes},
   journal = {IEEE Trans. on Information Theory},
   volume = {20},
   pages = {643-649},
   year = {1974}
}

@article{Cherkassky91Linear,
   author = {Vladimir Cherkassky and Karen Fassett and Nikolaos Vassilas},
   title = {Linear Algebra Approach to Neural Associative Memories and Noise Performance of Neural Classifiers},
   journal = {IEEE Trans. on Computers},
   volume = {40},
   number = {12},
   pages = {1429-1435},
   month = dec,
   year = {1991},
   keywords = {singular value decomposition svd}
}

@article{Agrawal93Efficient,
   author = {Rakesh Agrawal and Christos Faloutsos and Arun Swami},
   title = {Efficient Similarity Search In Sequence Databases},
   journal = {Fourth Int. Conf.  on Foundations of Data Organization and Algorithms (FODO)},
   pages = {69-84},
   address = {Evanston, Illinois},
   month = {Oct. 13-15},
   year = {1993},
   note = {also available through anonymous ftp, from olympos.cs.umd.edu: ftp/pub/TechReports/fodo.ps}
}

@inproceedings{Brinkhoff93Efficient,
   author = {Thomas Brinkhoff and Hans-Peter Kriegel and Bernhard Seeger},
   title = {Efficient Processing of Spatial Joins Using {R}-trees},
   booktitle = {Proc. of ACM SIGMOD},
   pages = {237-246},
   address = {Washington, D.C.},
   month = {May 26-28},
   year = {1993}
}

@article{Ubois93Location,
   author = {Jeff Ubois},
   title = {Location, Location, Location},
   journal = {SunWorld},
   volume = {6},
   number = {8},
   pages = {72-76},
   month = aug,
   year = {1993},
   abstract = {GIS market: \$2-3.5 Billion, growing 30 percent annually.
Also has several pointers to ftp-able cites with data.}
}

@article{Faloutsos94Efficient,
   author = {Christos Faloutsos and Ron Barber and Myron Flickner and J. Hafner and Wayne Niblack and Dragutin Petkovic and William Equitz},
   title = {Efficient and Effective Querying by Image Content},
   journal = {J. of Intelligent Information Systems},
   volume = {3},
   number = {3/4},
   pages = {231-262},
   month = jul,
   year = {1994}
}

@article{Kamel93Packing,
   author = {Ibrahim Kamel and Christos Faloutsos},
   title = {On Packing {R}-trees},
   journal = {Second Int. Conf. on Information and Knowledge Management (CIKM)},
   address = {Washington, D.C.},
   month = {Nov. 1-5},
   year = {1993}
}

@inproceedings{Kamel94Hilbert,
   author = {Ibrahim Kamel and Christos Faloutsos},
   title = {Hilbert {R}-tree: An improved {R}-tree using fractals},
   booktitle = {Proc. of VLDB Conference,},
   pages = {500-509},
   address = {Santiago, Chile},
   month = {Sept. 12-15},
   year = {1994}
}

@article{Faloutsos93High,
   author = {Christos Faloutsos and Ibrahim Kamel},
   title = {High Performance {R}-trees},
   journal = {IEEE Data Engineering Bulletin},
   volume = {16},
   number = {3},
   month = sep,
   year = {1993}
}

@book{Lekkerkerker69Geometry,
   author = {C. Gerrit Lekkerkerker},
   title = {Geometry of Numbers},
   publisher = {John Wiley and Sons, Inc.},
   year = {1969},
   keywords = {declustering minkowski lattice},
   abstract = {p 23, 51, 175}
}

@book{de-Weger89Algorithms,
   author = {Benne M. M. de-Weger},
   title = {Algorithms for Diophantine Equations},
   publisher = {Center for Mathematics and Computer Science (Centrum voor Wiskunde en Informatica - CWI)},
   year = {1989},
   keywords = {shortest lattice point},
   abstract = {p 45}
}

@article{Lin94TV,
   author = {King-Ip Lin and H.V. Jagadish and Christos Faloutsos},
   title = {The {TV}-tree - An Index Structure for High-dimensional Data},
   journal = {VLDB Journal},
   volume = {3},
   pages = {517-542},
   month = oct,
   year = {1994}
}

@article{Arya94,
   author = {Manish Arya and William Cody and Christos Faloutsos and Joel Richardson and Arthur Toga},
   title = {{QBISM}: Extending a {DBMS} to Support 3D  Medical Images},
   journal = {Tenth Int.  Conf. on Data Engineering (ICDE)},
   pages = {314-325},
   address = {Houston, TX},
   month = {Feb. 14-18},
   year = {1994}
}

@inproceedings{Panagopoulos94Bit,
   author = {George Panagopoulos and Christos Faloutsos},
   title = {Bit-Sliced Signature Files for Very Large Databases on a Parallel Machine Architecture},
   booktitle = {EDBT Conference},
   pages = {379-392},
   address = {Cambridge, UK},
   month = {March 28-31},
   year = {1994}
}

@book{Blum82Discovery,
   author = {Robert L. Blum},
   title = {Discovery and Representation of Causal Relationships from a Lart Time-Oriented Clinical Database: The {RX} project},
   publisher = {Springer-Verlag},
   year = {1982},
   keywords = {medical data mining},
   abstract = {describes a system that couples statistical and AI methods for data mining.}
}

@inproceedings{Vassiliadis93Input,
   author = {Dimitris Vassiliadis},
   title = {The Input-State Space Approach to the Prediction of Auroral Geomagnetic Activity from Solar Wind Variables},
   booktitle = {Int. Workshop on Applications of Artificial Intelligence in Solar Terrestrial Physics},
   address = {Lund, Sweden},
   month = {Sept. 22-24},
   year = {1993},
   keywords = {time series}
}

@article{Ioannidis93Optimal,
   author = {Yannis E. Ioannidis and Stavros Christodoulakis},
   title = {Optimal Histograms for Limiting Worst-Case Error Propagation in the Size of Join Results},
   journal = {ACM TODS},
   volume = {18},
   number = {4},
   pages = {709-748},
   year = {1993}
}

@inproceedings{Muralikrishna88Equi,
   author = {M. Muralikrishna and David J. DeWitt},
   title = {Equi-Depth Histograms For Estimating Selectivity Factors for Multi-Dimensional Queries},
   booktitle = {Proc. ACM SIGMOD},
   pages = {28-36},
   address = {Chicago, IL},
   month = {June 1-3},
   year = {1988}
}

@article{Faloutsos95Flexible,
   author = {Christos Faloutsos and Raymond Ng and Timos Sellis},
   title = {Flexible and Adaptable Buffer Management Techniques for Database Management Systems},
   journal = {IEEE Trans. on Computers},
   volume = {44},
   number = {4},
   pages = {546-560},
   month = apr,
   year = {1995},
   keywords = {queueing theory}
}

@incollection{DeFazio93Full,
   author = {Samuel DeFazio},
   editor = {Jim Gray},
   title = {Full-Text Document Retrieval Benchmark},
   booktitle = {The Benchmark Handbook for Database and Transaction Processing Systems},
   edition = {2nd},
   publisher = {Morgan Kaufmann},
   year = {1993},
   note = {chapter 8}
}

@inproceedings{Cutting90Optimizations,
   author = {Doug Cutting and Jan Pedersen},
   title = {Optimizations for Dynamic Inverted Index Maintenance},
   booktitle = {Proc. SIGIR},
   pages = {405-411},
   year = {1990}
}

@article{Harman90Retrieving,
   author = {Donna Harman and Gerald Candela},
   title = {Retrieving Records from a Gigabyte of Text on a Minicomputer Using Statistical Ranking},
   journal = {Journal of the American Society for Information Science (JASIS)},
   volume = {41},
   number = {8},
   pages = {581-589},
   year = {1990}
}

@inproceedings{Faloutsos94Fast,
   author = {Christos Faloutsos and M. Ranganathan and Yannis Manolopoulos},
   title = {Fast Subsequence Matching in Time-Series Databases},
   booktitle = {Proc. ACM SIGMOD},
   pages = {419-429},
   address = {Minneapolis, MN},
   month = {May 25-27},
   year = {1994},
   keywords = {fourier transform},
   note = {`Best Paper' award; also available as CS-TR-3190, UMIACS-TR-93-131, ISR TR-93-86.}
}

@inproceedings{Faloutsos94Beyond,
   author = {Christos Faloutsos and Ibrahim Kamel},
   title = {Beyond Uniformity and Independence: Analysis of {R}-trees Using the Concept of Fractal Dimension},
   booktitle = {Proc. ACM SIGACT-SIGMOD-SIGART PODS},
   pages = {4-13},
   address = {Minneapolis, MN},
   month = {May 24-26},
   year = {1994},
   note = {Also available as CS-TR-3198, UMIACS-TR-93-130}
}

@inproceedings{Agrawal93Mining,
   author = {Rakesh Agrawal and Tomasz Imielinski and Arun Swami},
   title = {Mining Association Rules between Sets of Items in Large Databases},
   booktitle = {Proc. ACM SIGMOD},
   pages = {207-216},
   address = {Washington, DC},
   month = {May 26-28},
   year = {1993}
}

@article{DeWitt92Parallel,
   author = {David DeWitt and Jim Gray},
   title = {Parallel Database Systems: The Future of High Performance Database Systems},
   journal = {CACM (Comm. of ACM)},
   volume = {35},
   number = {6},
   pages = {85-98},
   month = jun,
   year = {1992},
   keywords = {declustering}
}

@book{Kernighan84UNIX,
   author = {Brian W. Kernighan and Rob Pike},
   title = {The {UNIX} Programming Environment},
   publisher = {Prentice Hall},
   year = {1984}
}

@inproceedings{Brown94Supporting,
   author = {Eric W. Brown and James P. Callan and W. Bruce Croft and J. Elliot B. Moss},
   editor = {M. Jarke and Janis Bubenko and Keith Jeffery},
   title = {Supporting Full-Text Information Retrieval with a Persistent Object Store},
   booktitle = {Proc. of EDBT conference},
   pages = {365-378},
   publisher = {Springer Verlag},
   address = {Cambridge, U.K.},
   month = mar,
   year = {1994}
}

@article{Salton94Automatic,
   author = {Gerard Salton and James Allan and Chris Buckley},
   title = {Automatic Structuring and Retrieval of Large Text Files},
   journal = {Comm. of ACM (CACM)},
   volume = {37},
   number = {2},
   pages = {97-108},
   month = feb,
   year = {1994},
   keywords = {vector model syntactic},
   abstract = {It proposes to use document vectors for a first filtering,
followed by a comparison of section, paragraph and sentence vectors.
Thus, it is expected to capture the context better.
Strong points: \~{}10\% improvement in precision/recall.
Weak point: it needs the inverse document frequency -
not obvious how to apply it to a changing collection of documents.}
}

@article{Manolopoulos96Experimenting,
   author = {Yannis Manolopoulos and Christos Faloutsos},
   title = {Experimenting with Pattern Matching Algorithms},
   journal = {Information Science},
   volume = {90},
   pages = {75-89},
   year = {1996},
   keywords = {boyer moore text retrieval}
}

@techreport{Sarawagi93Efficient,
   author = {Sunita Sarawagi and Michael Stonebraker},
   title = {Efficient Organization of Large Multidimensional Arrays},
   institution = {Univ. of California, Berkeley},
   type = {UCB CS-TR},
   number = {S2K-93-32},
   address = {Berkeley, CA},
   year = {1993}
}

@inproceedings{Mackert86R,
   author = {Lothar M. Mackert and Guy M. Lohman},
   title = {R* Optimizer Validation and Performance Evaluation for Distributed Queries},
   booktitle = {Proc. of 12th Int. Conf. on Very Large Data Bases (VLDB)},
   address = {Kyoto, Japan},
   month = aug,
   year = {1986}
}

@book{Stewart93Fearful,
   author = {Ian Stewart and Martin Golubitsky},
   title = {Fearful Symmetry: Is God a Geometer?},
   publisher = {Penguin Books},
   year = {1993}
}

@book{Dougherty92Introduction,
   author = {Edward R. Dougherty},
   title = {An Introduction to Morphological Image Processing},
   volume = {TT9},
   publisher = {A publication of SPIE - the Int. Society for Optical Engineering (SPIE Press)},
   year = {1992},
   keywords = {morphology erosion dilation opening closing morphology}
}

@article{Delis93Performance,
   author = {Alexis Delis and Nick Roussopoulos},
   title = {Performance Comparison of Three Modern {DBMS} Architectures},
   journal = {IEEE Trans. on Software Engineering},
   volume = {19},
   number = {2},
   pages = {120-138},
   month = feb,
   year = {1993}
}

@book{Hardy54Introduction,
   author = {G.H. Hardy and E.M. Wright},
   title = {An Introduction to the Theory of Numbers},
   edition = {3rd},
   publisher = {Oxford University Press},
   year = {1954},
   keywords = {declustering},
   abstract = {farey numbers, p. 26 (integral lattice), p. 393.}
}

@book{Hancock39Development,
   author = {Harris Hancock},
   title = {Development of the Minkowski Geometry of Numbers},
   publisher = {Dover Publications Inc.},
   address = {New York},
   year = {1939},
   keywords = {declustering},
   abstract = {vol. I and II - call \# QA 241.H25, p. 22, 184, 671}
}

@book{Frakes92Information,
   author = {W. Frakes and R. Baeza-Yates},
   title = {Information Retrieval: Data Structures and Algorithms},
   publisher = {Prentice-Hall},
   year = {1992},
   keywords = {book}
}

@article{Gravano94Effectiveness,
   author = {Louis Gravano and Hector Garcia-Molina and Anthony Tomasic},
   title = {The Effectiveness of GlOSS for the Text Database Discovery Problem},
   journal = {ACM SIGMOD},
   pages = {126-137},
   address = {Minneapolis, MN},
   month = {May 24-27},
   year = {1994}
}

@article{Tomasic94Incremental,
   author = {Anthony Tomasic and Hector Garcia-Molina and Kurt Shoens},
   title = {Incremental Updates of Inverted Lists for Text Document Retrieval},
   journal = {ACM SIGMOD},
   pages = {289-300},
   address = {Minneapolis, MN},
   month = {May 24-27},
   year = {1994}
}

@article{Brinkhoff94Multi,
   author = {Thomas Brinkhoff and Hans-Peter Kriegel and Ralf Schneider and Bernhard Seeger},
   title = {Multi-Step Processing of Spatial Joins},
   journal = {ACM SIGMOD},
   pages = {197-208},
   address = {Minneapolis, MN},
   month = {May 24-27},
   year = {1994}
}

@article{Lo94Spatial,
   author = {Ming-Ling Lo and Chinya V. Ravishankar},
   title = {Spatial Joins Using Seeded Trees},
   journal = {ACM SIGMOD},
   pages = {209-220},
   address = {Minneapolis, MN},
   month = {May 24-27},
   year = {1994}
}

@inproceedings{Croft91Use,
   author = {W. Bruce Croft and Howard R. Turtle and David D. Lewis},
   title = {The Use of Phrases and Structured Queries in Information Retrieval},
   booktitle = {Proc. of ACM SIGIR},
   pages = {32-45},
   address = {Chicago, IL},
   month = {Oct. 13-16},
   year = {1991},
   keywords = {text syntactic}
}

@inproceedings{Lin91Self,
   author = {Xia Lin and Dagobert Soergel and Gary Marchionini},
   title = {A Self-Organizing Semantic Map for Information Retrieval},
   booktitle = {Proc. of ACM SIGIR},
   pages = {262-269},
   address = {Chicago, IL},
   month = {Oct. 13-16},
   year = {1991},
   keywords = {text neural nets kohonen}
}

@inproceedings{Rau91Creating,
   author = {Lisa F. Rau and Paul S. Jacobs},
   title = {Creating Segmented Databases from Free Text for Text Retrieval},
   booktitle = {Proc. of ACM SIGIR},
   pages = {337-346},
   address = {Chicago, IL},
   month = {Oct. 13-16},
   year = {1991},
   keywords = {natural language processing nlp}
}

@inproceedings{Mauldin91Performance,
   author = {Michael L. Mauldin},
   title = {Performance in {FERRET}: {A} Conceptual Information Retrieval System},
   booktitle = {Proc. of ACM SIGIR},
   pages = {347-355},
   address = {Chicago, IL},
   month = {Oct. 13-16},
   year = {1991},
   keywords = {natural language processing nlp}
}

@book{Salton89Automatic,
   author = {Gerard Salton},
   title = {Automatic Text Processing - The Transformation, Analysis and Retrieval of Information by Computer},
   publisher = {Addison-Wesley},
   address = {Readings, MA},
   year = {1989}
}

@article{Yan94Index,
   author = {Tak W. Yan and Hector Garcia-Molina},
   title = {Index Structures for Selective Dissemination of Information Under the Boolean Model},
   journal = {ACM TODS},
   volume = {19},
   number = {2},
   pages = {332-364},
   month = jun,
   year = {1994},
   keywords = {text}
}

@inproceedings{Manber94,
   author = {Udi Manber and Sun Wu},
   title = {{GLIMPSE}: {A} Tool to Search Through Entire File Systems},
   booktitle = {Proc. of USENIX Techn. Conf.},
   year = {1994},
   keywords = {text indexing information},
   note = {Also available as TR 93-94, Dept. of Comp. Sc., Univ. of Arizona, Tucson, or through anonymous ftp  (ftp://cs.arizona.edu/glimpse/glimpse.ps.Z)}
}

@article{Obraczka93,
   author = {Katia Obraczka and Peter B. Danzig and Shih-Hao Li},
   title = {{INTERNET} Resourse Discovery Services},
   journal = {IEEE Computer},
   month = sep,
   year = {1993}
}

@article{Berners-Lee92World,
   author = {Tim Berners-Lee and Robert Cailliau and Jean-F. Groff and Bernd Pollermann},
   title = {World-Wide Web: The Information Universe},
   journal = {Electronic Networking: Research, Applications and Policy},
   volume = {1},
   number = {2},
   year = {1992},
   keywords = {www},
   abstract = {from gravano sigmod 1994}
}

@article{Foltz92Personalized,
   author = {Peter W. Foltz and Susan T. Dumais},
   title = {Personalized Information Delivery: An Analysis of Information Filtering Methods},
   journal = {Comm. of ACM (CACM)},
   volume = {35},
   number = {12},
   pages = {51-60},
   month = dec,
   year = {1992},
   keywords = {latent semantic indexing LSI}
}

@incollection{Belkin87Retrieval,
   author = {N.J. Belkin and W.B. Croft},
   editor = {M. Williams},
   title = {Retrieval Techniques},
   booktitle = {Annual Review of Information Science and Technology},
   pages = {109-145},
   publisher = {Elsevier Science Publishers},
   address = {New York, NY},
   year = {1987},
   keywords = {text retrieval},
   abstract = {from D. Harman}
}

@techreport{Harman93First,
   author = {D. Harman},
   title = {The First Text REtrieval Conference ({TREC}-1)},
   institution = {National Institute of Standards and Technology},
   type = {Special Publication},
   number = {500-207},
   address = {Gaithersburg, MD.},
   year = {1993}
}

@techreport{Harman94Second,
   author = {D. Harman},
   title = {The Second Text REtrieval Conference ({TREC}-2)},
   institution = {National Institute of Standards and Technology},
   type = {Special Publication},
   number = {500-215},
   address = {Gaithersburg, MD.},
   year = {1994}
}

@techreport{Koudas94Declustering,
   author = {Nick Koudas and Christos Faloutsos and Ibrahim Kamel},
   title = {Declustering {R}-trees on multi-computer architectures},
   institution = {Dept. of Computer Science, UMIACS and ISR, Univ. of Maryland},
   type = {CS-TR- 3276, UMIACS-TR-94-60, ISR TR},
   number = {94-38},
   address = {College Park, MD},
   month = may,
   year = {1994}
}

@article{Bartell92Latent,
   author = {Brian T. Bartell and Garrison W. Cottrell and Richard K. Belew},
   title = {Latent Semantic Indexing is an Optimal Special Case of Multidimensional Scaling},
   journal = {SIGIR},
   pages = {161-167},
   address = {Denmark},
   year = {1992}
}

@article{Leland94Self,
   author = {W.E. Leland and M.S. Taqqu and W. Willinger and D.V. Wilson},
   title = {On the Self-Similar Nature of Ethernet Traffic},
   journal = {IEEE Transactions on Networking},
   volume = {2},
   number = {1},
   pages = {1-15},
   month = feb,
   year = {1994},
   keywords = {fractals power laws},
   note = {(earlier version in SIGCOMM '93, pp 183-193)}
}

@article{Oard94Automatic,
   author = {Douglas W. Oard and Nicholas DeClaris and Bonnie J. Dorr and Christos Faloutsos},
   title = {On Automatic Filtering of Multilingual Texts},
   journal = {IEEE Conf. on Systems, Man and Cybernetics},
   address = {San Antonio, TX},
   month = {Oct. 2-5},
   year = {1994}
}

@inproceedings{DeClaris94Information,
   author = {Nicholas DeClaris and Donna Harman and Christos Faloutsos and Susan Dumais and Douglas Oard},
   title = {Information Filtering and Retrieval: Overview, Issues and Directions},
   booktitle = {Annual Conference of the IEEE Engineering in Medicine and Biology Society},
   address = {Baltimore, MD},
   month = {Nov. 3-6},
   year = {1994},
   note = {to appear}
}

@article{Wang93Spectral,
   author = {Kuansan Wang and Shihab Shamma},
   title = {Spectral Shape Analysis in the Central Auditory System},
   journal = {NNSP},
   address = {Baltimore, MD},
   month = sep,
   year = {1993},
   keywords = {wavelets}
}

@article{Patrick68F,
   author = {Edward A. Patrick and Douglas R. Anderson},
   title = {F.{K}. Bechtel},
   journal = {IEEE Trans. on Computers},
   volume = {C-17},
   number = {10},
   pages = {949-953},
   month = oct,
   year = {1968},
   keywords = {fractals hilbert curve}
}

@article{Butz71Alternative,
   author = {Arthur R. Butz},
   title = {Alternative Algorithm for Hilbert's Space-Filling Curve},
   journal = {IEEE Trans. on Computers},
   volume = {C-20},
   number = {4},
   pages = {424-426},
   month = apr,
   year = {1971}
}

@article{Stevens83Manipulation,
   author = {J. Stevens and A.F. Lehar and F.H. Preston},
   title = {Manipulation and Presentation of Multidimensional Image Data Using the Peano Scan},
   journal = {IEEE Trans. on Pattern Analysis and Machine Intelligence (PAMI)},
   volume = {PAMI-5},
   number = {5},
   pages = {520-526},
   month = sep,
   year = {1983},
   keywords = {hilbert curve}
}

@article{Chang92Smart,
   author = {S.-K. Chang and T.Y. Hou and A. Hsu},
   title = {Smart Image Design for Large Image Databases},
   journal = {Journal of Visual Languages and Computing},
   volume = {3},
   pages = {323-342},
   publisher = {Academic Press Limited},
   year = {1992}
}

@book{Benson92Spatial,
   author = {Dan Benson},
   title = {Spatial and Symbolic Queries for 3-{D} Data},
   pages = {1-15},
   publisher = {Siemens Corporate Research, Inc.},
   address = {Princeton, New Jersey},
   year = {1992},
   note = {This paper is an adaptation of a previous paper presented at the 1992 SPIE Conference in San Jose.}
}

@inproceedings{Wilcox92Training,
   author = {Lynn D. Wilcox and Marcia A. Bush},
   title = {Training and Search Algorithms For An Interactive Wordspotting System},
   booktitle = {Proc. ICASSP},
   pages = {1-4},
   publisher = {IEEE},
   address = {Palo Alto, California},
   year = {1992},
   note = {This is not the proceedings version.  It says "to appear"}
}

@misc{Stein??Structural,
   author = {Fridtjof Stein and Gerard Medioni},
   title = {Structural Hashing:  Efficient Three Dimensional Object Recognition},
   volume = {PAMI Special Issue},
   pages = {1-58},
   address = {Los Angeles, California},
   year = {l992?}
}

@article{Vassilakopoulo93Efficiency,
   author = {Michael Vassilakopoulos and Yannis Manolopoulos},
   title = {Efficiency Analysis of Overlapped Quadtrees},
   journal = {Submitted to Image Understanding},
   pages = {1-21},
   address = {Thessaloniki, Greece},
   year = {1993}
}

@article{Dervos92Ranking,
   author = {D. Dervos and P. Linardis and Y. Manolopoulos},
   title = {A Ranking Technique for the Signature File Method},
   journal = {Submitted to Information \& Software Technology},
   pages = {1-20},
   address = {Thessaloniki, Greece},
   month = {Sept. 2},
   year = {1992}
}

@inproceedings{Schonfeld90Robusta,
   author = {Dan Schonfeld and John Goutsias},
   title = {Robust Morphological Representation of Binary Images},
   booktitle = {Proceedings ICASSP},
   pages = {4 pages},
   publisher = {IEEE},
   address = {Albuquerque, New Mexico},
   month = {April 3-6},
   year = {1990},
   note = {Published as Proceedings ICASSP, volume 4, number M}
}

@incollection{Carroll71Worda,
   author = {John B. Carroll},
   editor = {Peter Davies and Barry Richman},
   title = {Word Frequency Book},
   booktitle = {Word Frequency Book},
   pages = {various},
   publisher = {American Heritage Publishing Co.},
   address = {New York},
   year = {1971},
   abstract = {Statistical Analysis of the Corpus pp. xxi - xxxix
New Views of Lexicon by Peter Davies pp. xli - liv
Guide to Alphabetical List pp. 1 - 4
Alphabetical List ... one page
Rank List p. 565
Guide to the Frequency Distribution Graphs pp. 811 - 825}
}

@inproceedings{Aktug93Analysis,
   author = {Deniz Aktug and Fazli Can},
   title = {Analysis of Multiterm Queries in Dynamic Signature File Organization},
   booktitle = {Proceedings 16th ACM SIGIR Conference},
   pages = {1-11},
   address = {Oxford, OH},
   year = {1993}
}

@article{Cohen86Biomedical,
   author = {Arnon Cohen},
   title = {Biomedical Signal Processing},
   journal = {Compression and Automatic Recognition},
   volume = {II},
   pages = {48-86; 113-137},
   publisher = {CRC Press, Inc.},
   address = {Beer Sheva, Israel},
   year = {1986}
}

@misc{Jaffe91,
   author = {Laura Jaffe},
   title = {{DNA} Databases},
   address = {UMCP},
   month = {May 2},
   year = {1991},
   note = {CMSC 624 Term Project}
}

@inproceedings{Ng94Efficient,
   author = {Raymond T. Ng and Jiawei Han},
   title = {Efficient and Effective Clustering Methods for Spatial Data Mining},
   booktitle = {Proc. of VLDB Conf.},
   pages = {144-155},
   address = {Santiago, Chile},
   month = {Sept. 12-15},
   year = {1994}
}

@inproceedings{Brinkhoff94Impact,
   author = {Thomas Brinkhoff and Hans-Peter Kriegel},
   title = {The Impact of Global Clustering on Spatial Database Systems},
   booktitle = {Proc. of VLDB Conf.},
   pages = {168-179},
   address = {Santiago, Chile},
   month = {Sept. 12-15},
   year = {1994}
}

@inproceedings{Brown94Fast,
   author = {Eric W. Brown and James P. Callan and W. Bruce Croft},
   title = {Fast Incremental Indexing for Full-Text Information Retrieval},
   booktitle = {Proc. of VLDB Conf.},
   pages = {192-202},
   address = {Santiago, Chile},
   month = {Sept. 12-15},
   year = {1994}
}

@inproceedings{Johnson942Q,
   author = {Theodore Johnson and Dennis Shasha},
   title = {2Q: {A} Low Overhead High Performance Buffer Management Replacement Algorithm},
   booktitle = {Proc. of VLDB Conf.},
   pages = {439-450},
   address = {Santiago, Chile},
   month = {Sept. 12-15},
   year = {1994}
}

@inproceedings{Ng94Maximizing,
   author = {Raymond T. Ng and Jinhai Yang},
   title = {Maximizing Buffer and Disk Utilization for News On-Demand},
   booktitle = {Proc. of VLDB Conf.},
   pages = {451-462},
   address = {Santiago, Chile},
   month = {Sept. 12-15},
   year = {1994}
}

@inproceedings{Agrawal94Fast,
   author = {Rakesh Agrawal and Ramakrishnan Srikant},
   title = {Fast Algorithms for Mining Association Rules in Large Databases},
   booktitle = {Proc. of VLDB Conf.},
   pages = {487-499},
   address = {Santiago, Chile},
   month = {Sept. 12-15},
   year = {1994}
}

@inproceedings{Church88Stochastic,
   author = {K. Church},
   title = {A Stochastic Parts Program and Noun Phrase Parser for Unrestricted Text},
   booktitle = {Proc. of the Second conf. on Applied Natural Language Processing},
   pages = {136-143},
   year = {1988},
   keywords = {nlp}
}

@book{Quinlan93C4,
   author = {John Ross Quinlan},
   title = {C4.5: Programs for Machine Learning},
   publisher = {Morgan Kaufmann Publishers Inc.},
   address = {San Mateo, CA},
   year = {1993},
   keywords = {data mining, tree classifiers},
   abstract = {Describes a tree-based classifier.
It handles un-ordered domains and null/missing values.
It uses entropy arguments to choose attributes on which
to discriminate.
Interesting points:
1. pointer to a db repository: ml-repository@ics.uci.edu (UC Irvine)
2. comparison with statistics-based systems (p. 15)
"As a general rule [...] statistical techniques tend to focus
on tasks in which all the attributes have continuous
or ordinal values"
[S.M. Weiss and C.A. Kulikowski "Computer Systems that Learn",
Morgan Kaufmann, San Mateo CA, 1991] provide a comparison
3. MDL principle, to decide how to prune the tree
[Jorman Rissanen, Anal of Statistics, 11,2, 416-431]
[Quinlan and Rivest, "Inferring decision trees using the
Minimum Description Length principle" Information and Computation
80,3, 227-248
4. comparison of tree-based classifiers with neural nets (NN)
- they are both more robust
- they are about equally accurate (with NN slightly ahead);
but NN require much more computation (an order of magnitude more)
(p. 102)
5. CART is a statistics-based program
L. Breiman, J.H. Friedman, R.A. Olshen and C.J. Stone
"Classification and Regression Trees" Belmont, CA: Wadsworth (1984)
6. citation: Hunt 75 "Artificial Intelligence" NY, Academic Press
(pioneered the tree-based classification methods)}
}

@incollection{Baeza-Yates94Proximity,
   author = {Ricardo A. Baeza-Yates and Walter Cunto and Udi Manber and Sun Wu},
   editor = {M. Crochemore and D. Gusfield},
   title = {Proximity Matching using Fixed Queries Trees},
   booktitle = {5th Combinatorial Pattern Matching, LNCS 807},
   pages = {198-212},
   publisher = {Springer-Verlag},
   address = {Asilomar, CA},
   month = jun,
   year = {1994},
   keywords = {approximate},
   abstract = {Suggest to choose \$k\$ fixed objects and
build a tree using the distances of each object from
these \$k\$ anchor objects}
}

@article{Rosenburg90Spilling,
   author = {Susan Rosenburg},
   title = {Spilling a Little Ink The Animated Reviewer Tweening the Learning Curve},
   journal = {BYTE},
   pages = {174 \& 177},
   month = {March 1},
   year = {9990},
   note = {Found in Review "Art in Motion"}
}

@article{Stonebraker90Distributed,
   author = {Michael Stonebraker and Gerhard A. Schloss},
   title = {Distributed Raid - {A} New Multiple Copy Algorithm},
   journal = {IEEE Data Engineering},
   pages = {430-437},
   address = {Berkeley},
   year = {1990}
}

@misc{Wu92Approximatea,
   author = {Sun Wu},
   title = {Approximate Pattern Matching and Its Applications},
   address = {University of Arizona},
   year = {1992}
}

@misc{Wu92,
   author = {Sun Wu and Udi Manber},
   title = {{AGREP}- {A} Fast Approximate Pattern-Matching Tool},
   address = {Tucson},
   year = {1992}
}

@misc{Lindsay??Notes,
   author = {B. Lindsay and P. Selinger and C.A. Galtieri and J.N. Gray and R.A. Lorie and T.G. Price and F. Putzolu and I.L. Traiger and B.W. Wade},
   title = {Notes On Distributed Databases},
   address = {San Jose},
   note = {Research Report IBM}
}

@misc{Haas91Exploiting,
   author = {Laura M. Haas and William F. cody},
   title = {Exploiting Extensible {DBMS} in Integrated Geographic Info. Systems},
   pages = {423-450},
   address = {San Jose},
   year = {1991}
}

@misc{DeFazio92Document,
   author = {Samuel DeFazio},
   title = {Document Retrieval Benchmark},
   number = {Version 1.0},
   pages = {34},
   address = {Beaverton},
   year = {1992}
}

@techreport{Yang92Regularized,
   author = {Yongyi Yang and Nikolas P. Galatsanos and Aggelos K. Katsaggelos},
   title = {Regularized Reconstruction to Reduce Blocking Artifacts of Block Discrete Cosine Transform Compressed Images},
   institution = {Illinois Institute of Technology},
   type = {IIT-ECE-TR-11-92},
   pages = {32},
   address = {Chicago and Evanston},
   year = {1992}
}

@article{Huang92Prioritized,
   author = {Yunming Huang and Howard M. Dreizen and Nikolas P. Galatsanos},
   title = {Prioritized {DCT} for Compression and Progressive Transmission of Images},
   journal = {IEEE Transactions on Image Processing},
   volume = {1},
   number = {4},
   pages = {477-487},
   month = oct,
   year = {1992}
}

@article{Manber88Using,
   author = {Udi Manber},
   title = {Using Induction to Design Algorithms},
   journal = {Communications of the ACM},
   volume = {31},
   number = {11},
   pages = {1300-1313},
   month = nov,
   year = {1988},
   abstract = {An analogy between proving mathematical theorems and designing computer
algorithms provides an elegant methodology for designing algorithms,
explaining their behavior, and understanding their key ideas.}
}

@article{Manber90Suffix,
   author = {Udi Manber and Gene Myers},
   title = {Suffix Arrays: {A} New Method for On-Line String Searches},
   journal = {1st Annual ACM-SIAM Symp on Discrete Algorithms},
   address = {Tucson},
   month = jan,
   year = {1990}
}

@misc{Manber92Approximate,
   author = {Udi Manber and Sun Wu},
   title = {Approximate String Matching with Arbitrary Costs for Text \& Hypertext},
   pages = {12},
   address = {Tucson},
   month = mar,
   year = {1992},
   abstract = {A new algorithm for finding the best approximate match of a pattern inside
a text, or all matches within a given error cost, is given......}
}

@article{Manber91Algorithm,
   author = {Udi Manber and Ricardo Baeza-Yates},
   editor = {G.R. Andrews},
   title = {An Algorithm for string matching with a sequence of don't cares},
   journal = {Information Processing Letters},
   number = {37},
   pages = {133-136},
   publisher = {North-Holland},
   address = {Tucson and Santiago},
   year = {1991}
}

@techreport{Wu91Improving,
   author = {Sun Wu and Udi Manber and Eugene Myers},
   title = {Improving the Running Times for Some String-Matching Problems},
   institution = {The University of Arizona},
   type = {TR},
   number = {91-20},
   pages = {17},
   address = {Tucson},
   month = aug,
   year = {1991}
}

@inproceedings{Lai91Flying,
   author = {Patrick Lai and Udi Manber},
   title = {Flying Through Hypertex},
   booktitle = {Hypertext '91 Proceedings},
   pages = {123-132},
   address = {Tucson},
   month = dec,
   year = {1991}
}

@inproceedings{Amir92Efficient,
   author = {Amihood Amir and Martin Farach and Yossi Matias},
   title = {Efficient Randomized Dictionary-Matching Algorithms},
   booktitle = {Symposium Combinated Pattern Matching},
   pages = {13},
   month = {Symp. April},
   year = {1992}
}

@article{Leung84Approximate,
   author = {Clement Leung},
   title = {Approximate Storage Utilization of {B}-trees: {A} Simple Derivation and Generalizations},
   journal = {Information Processing Letters (IPL)},
   volume = {19},
   number = {4},
   pages = {199-201},
   month = nov,
   year = {1984}
}

@article{Comer82Flat,
   author = {Douglas Comer},
   title = {The Flat File System {FFG}: {A} Database System Consisting of Primitives},
   journal = {Software-Practice and Experience},
   month = nov,
   year = {1982},
   keywords = {getit}
}

@article{Hunt77Fast,
   author = {J.W. Hunt and T.G. Szymanski},
   title = {A Fast Algorithm for Computing Longest Common Subsequences},
   journal = {CACM},
   month = may,
   year = {1977},
   abstract = {describes the "diff" algorithm of UNIX}
}

@book{Ritter92Neural,
   author = {Helge Ritter and Thomas Martinetz and Klaus Schulten},
   title = {Neural Computation and Self-Organizing Maps},
   publisher = {Addison Wesley},
   address = {Reading, MA},
   year = {1992},
   keywords = {neural networks kohonen learning}
}

@article{Cutting93Constant,
   author = {Douglass R. Cutting and David R. Karger and Jan O. Pedersen},
   title = {Constant Interaction-Time Scatter/Gather Browsing of Very Large Document Collections},
   journal = {ACM-SIGIR},
   pages = {126-134},
   address = {Pittsburgh, PA},
   month = {June 27 - July 1},
   year = {1993}
}

@article{Cutting92Scatter,
   author = {Douglass R. Cutting and David R. Karger and Jan O. Pedersen and John W. Tukey},
   title = {Scatter/Gather: a Cluser-Based Approach to Browsing Large Document Collections},
   journal = {ACM-SIGIR},
   pages = {318-329},
   address = {Copenhagen, Denmark},
   month = {June 21-24},
   year = {1992}
}

@book{Golub89Matrix,
   author = {G. H. Golub and C. F. Van-Loan},
   title = {Matrix Computations},
   edition = {2nd},
   publisher = {The Johns Hopkins University Press},
   address = {Baltimore},
   year = {1989},
   keywords = {svd},
   abstract = {For SVD, if you only need the singular values, the complexety is O(N\^{}2*M)
where N$<$=M.
If you also need the singular vectors, it becomes O(N*M\^{}2).}
}

@book{Press92Numerical,
   author = {William H. Press and Saul A. Teukolsky and William T. Vetterling and Brian P. Flannery},
   title = {Numerical Recipes in {C}},
   edition = {2nd},
   publisher = {Cambridge University Press},
   year = {1992}
}

@article{Faloutsos95FastMap,
   author = {Christos Faloutsos and King-Ip (David) Lin},
   title = {FastMap: {A} Fast Algorithm for Indexing, Data-Mining and Visualization of Traditional and Multimedia Datasets},
   journal = {ACM SIGMOD},
   pages = {163-174},
   address = {San Jose, CA},
   month = {May 23-25},
   year = {1995},
   keywords = {mds karhunen-loeve svd}
}

@article{Yang94Example,
   author = {Yiming Yang and Christopher G. Chute},
   title = {An Example-Based Mapping Method for Text Categorization and Retrieval},
   journal = {ACM Trans. On Inf. Systems (TOIS)},
   volume = {12},
   number = {3},
   pages = {252-277},
   month = jul,
   year = {1994},
   keywords = {lsi svd}
}

@inproceedings{Burdett93Nonlinear,
   author = {Christina J. Burdett and Harold G. Longbotham and Mita Desai and Walter B. Richardson and John F. Stoll},
   editor = {Raj S. Acharaya and Dmitry B. Goldgof},
   title = {Nonlinear Indicators of Malignancy},
   booktitle = {Proc. SPIE 1993 - Biomedical Image Processing and Biomedical Visualization},
   volume = {1905 (part two of two)},
   pages = {853-860},
   address = {San Jose, CA},
   month = {Feb. 1-4},
   year = {1993},
   keywords = {fractal dimension},
   abstract = {They compute the fractal dimension;
malignant tumors have higher fractal dimension
(2.56, vs 2.37 for the benign ones)}
}

@book{Schuster88Deterministic,
   author = {Heinz Georg Schuster},
   title = {Deterministic Chaos: An Introduction},
   edition = {revised},
   publisher = {VCH Verlagsgesellschaft mbH},
   address = {Weinheim, Germany},
   year = {1988},
   keywords = {fractal dimension},
   abstract = {Contains an interesting proof about the correlation dimension
and the selectivity of a spatial join (p. 122)},
   note = {Second}
}

@article{Tsangaris92Performance,
   author = {Manolis M. Tsangaris and Jeffrey F. Naughton},
   title = {On the Performance of Object Clustering Techniques},
   journal = {ACM SIGMOD},
   pages = {144-153},
   address = {San Diego, CA},
   month = {June 2-5},
   year = {1992},
   note = {also available as U. of Wisconsin-Madison TR-1090}
}

@article{Hume91Fast,
   author = {Andrew Hume and Daniel Sunday},
   title = {Fast String Searching},
   journal = {Software - Practice and Experience},
   volume = {21},
   number = {11},
   pages = {1221-1248},
   month = nov,
   year = {1991},
   keywords = {boyer moore},
   abstract = {They propose some smart 'C' tricks to accelerate
the Boyer-Moore algorithm,
making it 4.5 times faster}
}

@book{Barnsley88Fractals,
   author = {Michael Barnsley},
   title = {Fractals Everywhere},
   publisher = {Academic Press Inc.},
   address = {San Diego, CA},
   year = {1988},
   keywords = {iterated function systems topology contractive mappings image compression},
   abstract = {The authoritative book on the iterated function systems.
Rigorous mathematics with excellent drawings to explain ideas.
Major points:
1. the theorem that contractive mappings lead to a fixed point
(attractor!)
2. measurement of the Hausdorff fractal dimension
3. fractal interpolation
4. Definition of Hausdorff distance between two shapes (==
minimum radius of expansion that one shape needs, in order to cover
the other)}
}

@book{Reitgen92Fractals,
   author = {Heinz-Otto Reitgen and Hartmut Juergens and Dietmar Saupe},
   title = {Fractals for the Classroom: Part One - Introduction to Fractals and Chaos},
   publisher = {Springer Verlag},
   address = {New York},
   year = {1992},
   keywords = {fractal dimension brownian motion},
   abstract = {Interesting discussion of
1. algorithms for generating fractal motion and fractal landscapes
2. percolation
3. DLA (diffusion limited aggregation)
4. universality of Sierpinski gasket}
}

@book{Kaye93Chaos,
   author = {Brian Kaye},
   title = {Chaos and Complexity: Discovering the Surprising Patterns of Science and Technology},
   publisher = {VCH Verlagsgesellschaft mbH},
   address = {New York},
   year = {1993},
   keywords = {fractal dimension},
   abstract = {Very interesting and relatively easy-to-read book.
Major point: it gives examples of shapes that have two different
fractal dimensions, such as rocks after tumbling (p. 312)
Also, it gives the Macintosh-Pascal code for DLA (p. 292-297)}
}

@article{Faloutsos97Analysis,
   author = {Christos Faloutsos and H.V. Jagadish and Yannis Manolopoulos},
   title = {Analysis of the n-dimensional quadtree decomposition for arbitrary hyper-rectangles},
   journal = {IEEE-TKDE},
   volume = {9},
   number = {3},
   pages = {373-383},
   month = may,
   year = {1997},
   keywords = {magic interpolation},
   note = {Also available as CS-TR-3381, UMIACS-TR-94-130 from UMD-CS.}
}

@misc{Petrakis94Similarity,
   author = {Euripides G.M. Petrakis and Christos Faloutsos},
   title = {Similarity Searching in Large Image DataBases},
   year = {1994},
   keywords = {medical attribute graphs R-tree spatial access method},
   note = {submitted for publication. Also available as technical report at MUSIC with \# TR-01-94, UMIACS-TR-94-134, CS-TR-3388}
}

@article{Vassilakopoulo93Overlapping,
   author = {M. Vassilakopoulos and Y. Manolopoulos and K. Economou},
   title = {Overlapping Quadtrees for the Representation of Similar Images},
   journal = {Image and Vision Computing},
   volume = {11},
   number = {5},
   pages = {257-262},
   month = jun,
   year = {1993}
}

@article{Vassilakopoulo95Efficiency,
   author = {M. Vassilakopoulos and Y. Manolopoulos},
   title = {Efficiency Analysis of Overlapped Quadtrees},
   journal = {Nordic Journal on Computing},
   year = {1995},
   note = {in press}
}

@article{Vassilakopoulo95Dynamic,
   author = {M. Vassilakopoulos and Y. Manolopoulos},
   title = {Dynamic Inverted Quadtree: a Structure for Pictorial Databases},
   journal = {Information Systems},
   year = {1995},
   note = {in press}
}

@book{Rabiner93Fundamentals,
   author = {Lawrence Rabiner and Biing-Hwang Juang},
   title = {Fundamentals of Speach Recognition},
   publisher = {Prentice Hall},
   year = {1993},
   contents = {(not listed)}
}

@article{Kass88Snakes,
   author = {Michael Kass and Andrew Witkin and Demetri Terzopoulos},
   title = {Snakes: Active Contour Models},
   journal = {International Journal of Computer Vision},
   volume = {1},
   number = {4},
   pages = {321-331},
   year = {1988}
}

@article{Terzopoulos86Image,
   author = {Demetri Terzopoulos},
   title = {Image Analysis Using Multigrid Relaxation Methods},
   journal = {IEEE PAMI},
   volume = {8},
   number = {2},
   pages = {129-139},
   month = mar,
   year = {1986}
}

@article{Lin87Classification,
   author = {C. C. Lin and R. Chellappa},
   title = {Classification of Partial 2-{D} Shapes Using Fourier Descriptors},
   journal = {IEEE PAMI},
   volume = {PAMI-9},
   number = {5},
   pages = {686-690},
   month = sep,
   year = {1987}
}

@inproceedings{Witkin86Signal,
   author = {A. Witkin and D. Terzopoulos and M. Kaas},
   title = {Signal Matching through scale space},
   booktitle = {Proc. Am. Assoc. Artif. Intel.},
   pages = {714-719},
   address = {Philadelphia},
   year = {1986},
   contents = {(not listed)}
}

@article{Grassberger90Optimized,
   author = {Peter Grassberger},
   title = {An Optimized Box-Assisted Algorithm for Fractal Dimensions},
   journal = {Phys. Lett. A},
   volume = {148},
   pages = {63-68},
   year = {1990},
   contents = {(not listed)}
}

@book{Tong90Non,
   author = {Howell Tong},
   title = {Non-linear Time Series: {A} Dynamical System Approach},
   publisher = {Clarendon Press},
   address = {Oxford},
   year = {1990},
   contents = {(not listed)}
}

@incollection{Smith92Optimal,
   author = {Richard Smith},
   editor = {M. Castagli and S. Eubank},
   title = {Optimal Estimation of Fractal Dimension},
   booktitle = {Nonlinear Modeling and Forecasting},
   pages = {115-135},
   publisher = {Addison Wesley},
   year = {1992},
   note = {Proc. Vol. XII},
   contents = {(not listed)}
}

@article{Rabiner90Tutorial,
   author = {L.R. Rabiner},
   title = {A Tutorial on Hidden Markov Models and Selected Applications},
   journal = {IEEE Proc.},
   volume = {77},
   pages = {257},
   year = {1990},
   contents = {(not listed)}
}

@article{Friedman91Multrivariate,
   author = {Jerome H. Friedman},
   title = {Multrivariate Adaptive Regression Splines (with discussion)},
   journal = {Annals of Statistics},
   volume = {19},
   pages = {1-141},
   year = {1991},
   contents = {(not listed)}
}

@article{Murphy86Efficiency,
   author = {O.J. Murphy and S.M. Selkow},
   title = {The Efficiency of Using k-d Trees for Finding Nearest Neighbors in Discrete Space},
   journal = {Information Proc. Letters},
   volume = {23},
   pages = {215-218},
   year = {1986},
   contents = {(not listed)}
}

@article{Wolfram84Computation,
   author = {Stephen Wolfram},
   title = {Computation Theory of Cellular Automata},
   journal = {Comm. Math. Phys.},
   volume = {96},
   pages = {15},
   year = {1984},
   contents = {(not listed)}
}

@book{Castagli92Nonlinear,
   author = {M. Castagli and S. Eubank},
   title = {Nonlinear Modeling and Forecasting},
   publisher = {Addison Wesley},
   year = {1992},
   abstract = {Main ideas:
for prediction in time series, the 'standard' procedure is
- use embedding space with dimension dE
- try to approximate the function y = f( vec-x ),
using nearest neighbors, smoothing splines, Voronoi tesselations,
neural networks, genetic algorithms
For the stock market:
- "it seems stochastic, not low-order chaotic" [Granger and Terasvirta];
- [Brock+Potter] stock returns do not fit
* linear models
* non-stationary models
* deterministic chaos
- [LeBaron] "for the S\&P index, successive returns are uncorrelated;
volatilies (eg., abs. values of returns) seem to be correlated",
ie., large jumps are typically followed by large jumps.},
   note = {Proc. Vol. XII}
}

@incollection{LeBaron92Nonlinear,
   author = {Blake LeBaron},
   editor = {M. Castagli and S. Eubank},
   title = {Nonlinear Forecasts for the {SP} Stock Index},
   booktitle = {Nonlinear Modeling and Forecasting},
   pages = {381-393},
   publisher = {Addison Wesley},
   year = {1992},
   note = {Proc. Vol. XII}
}

@incollection{Mead92Prediction,
   author = {W.C. Mead and R.D. Jones and Y.C. Lee and C.W. Barnes and G.W. Flake and L.A. Lee and M.K. O'Rourke},
   editor = {M. Castagli and S. Eubank},
   title = {Prediction of Chaotic Time Series using {CNLS}-Net-Example: The Mackey-Glass Equation},
   booktitle = {Nonlinear Modeling and Forecasting},
   pages = {39-71 381-393},
   publisher = {Addison Wesley},
   year = {1992},
   note = {Proc. Vol. XII}
}

@inproceedings{Roussopoulos95Nearest,
   author = {Nick Roussopoulos and Steve Kelley and F. Vincent},
   title = {Nearest Neighbor Queries},
   booktitle = {Proc. of ACM-SIGMOD},
   pages = {71-79},
   address = {San Jose, CA},
   month = may,
   year = {1995}
}

@inproceedings{Faloutsos95FastMapa,
   author = {Christos Faloutsos and King-Ip (David) Lin},
   title = {FastMap: {A} Fast Algorithm for Indexing, Data-Mining and Visualization of Traditional and Multimedia Datasets},
   booktitle = {Proc. of ACM-SIGMOD},
   address = {San Jose, CA},
   month = may,
   year = {1995},
   note = {to appear}
}

@techreport{Faloutsos94FastMap,
   author = {Christos Faloutsos and King-Ip (David) Lin},
   title = {FastMap: {A} Fast Algorithm for Indexing, Data-Mining and Visualization of Traditional and Multimedia Datasets},
   institution = {Dept. of Computer Science, Univ. of Maryland},
   type = {CS-TR-3383 UMIACS-TR-94-132 ISR TR},
   number = {94-80},
   address = {College Park},
   year = {1994},
   note = {also available from mosaic (URL ftp: //olympos.cs.umd.edu /pub/TechReports /sigmod95.ps)}
}

@book{Buja91Computing,
   author = {Andreas Buja and Paul A. Tukey},
   title = {Computing and Graphics in Statistics},
   publisher = {Springer Verlag},
   address = {New York},
   year = {1991},
   abstract = {Main ideas:
1) parallel axis
2) data-flow style of interface
3) using icons or texture to encode a third variable
4) A.S.H. (= smoothing the distribution)
5) Grand/guided tours
6) Brushing/Painting (ie., selection of some points in one plot
highlights them in another, too)
Minor ideas: (a) Interface between 'S' and Mathematica
My ideas: (a) wavelets, instead of A.S.H. for smoothing
(b) contact Alberto for UNIX/dbms}
}

@article{Toga91digital,
   author = {Arthur W. Toga},
   title = {A digital three-dimensional atlas of structure/function relationships},
   journal = {J. Chem. Neuroanat.},
   volume = {4},
   number = {5},
   pages = {313-318},
   year = {1991}
}

@inproceedings{Schwarz86Extensibility,
   author = {P. Schwarz and W. Chang and J.C. Freytag and G. Lohman and J. McPherson and C. Mohan and H. Pirahesh},
   title = {Extensibility in the Starburst Database System},
   booktitle = {Proc. 1986 Int'l Workshop on Object-Oriented Database Systems},
   pages = {85-92},
   address = {Pacific Grove},
   month = sep,
   year = {1986}
}

@article{Lehman89Starburst,
   author = {T.J. Lehman and B. Lindsay},
   title = {The Starburst Long Field Manager},
   journal = {VLDB Conf. Proc.},
   pages = {375-383},
   address = {Amsterdam},
   month = aug,
   year = {1989}
}

@misc{IBM92,
   author = {{IBM}},
   title = {{IBM} {AIX} Visualization Data Explorer/6000 User's Guide,},
   institution = {IBM},
   edition = {Second},
   year = {1992},
   note = {Publication No. SC38-0496-1}
}

@misc{Talairach88Co,
   author = {J. Talairach and P. Tournoux},
   title = {Co-planar stereotactic atlas of the human brain},
   address = {Thieme, Stuttgart},
   year = {1988}
}

@incollection{Field93Scale,
   author = {D.J. Field},
   editor = {M. Farge and J.C.R. Hunt and J.C. Vassilicos},
   title = {Scale-invariance and Self-similar `Wavelet' Transforms: an Analysis fo Natural Scenes and Mammalian Visual Systems},
   booktitle = {Wavelets, Fractals, and Fourier Transforms},
   pages = {151-193},
   publisher = {Clarendon Press},
   address = {Oxford},
   year = {1993},
   abstract = {interesting idea: the mammalian optical neurons
are organized so that each neuron is tuned to a specific wavelet;
moreover, natural images excite only *few* of the neurons
(despite the fact that all neurons have the same probability
of becoming excited). This means that the wavelet transform
achieves excellent compression for natural images.
Also interesting - Kuansan Wang had mentioned that the auditory
system should behave similarly, since the visual and the auditory
cortex are interchangeable!}
}

@incollection{Muller93Morphology,
   author = {J. Muller},
   editor = {M. Farge and J.C.R. Hunt and J.C. Vassilicos},
   title = {Morphology of Disordered Materials Studied by Multifractal Analysis},
   booktitle = {Wavelets, Fractals, and Fourier Transforms},
   pages = {397-403},
   publisher = {Clarendon Press},
   address = {Oxford},
   year = {1993},
   abstract = {uses the multifractal spectrum, to separate malignant from benign tumors -
Looks very interesting for the work that Alberto Belussi is doing.
Maybe the width of the f(a) vs a curve depends on the \$p\$ parameter
of Schroeder.}
}

@book{Chambers92Statistical,
   author = {John M. Chambers and Trevor J. Hastie},
   title = {Statistical Models in {S}},
   publisher = {Wadsworth and Brooks/Cole},
   address = {Pacific Grove, CA},
   year = {1992},
   abstract = {uses k-d trees for 'local regression methods',
to do interpolation and blending.
We could substitute them with R-trees.}
}

@inproceedings{Pagel93Towards,
   author = {B. Pagel and H. Six and H. Toben and P. Widmayer},
   title = {Towards an Analysis of Range Query Performance},
   booktitle = {Proc. of ACM SIGACT-SIGMOD-SIGART Symposium on Principles of Database Systems (PODS)},
   pages = {214-221},
   address = {Washington, D.C.},
   month = may,
   year = {1993}
}

@inproceedings{Belussi95Estimating,
   author = {Alberto Belussi and Christos Faloutsos},
   title = {Estimating the Selectivity of Spatial Queries Using the `Correlation' Fractal Dimension},
   booktitle = {Proc. of VLDB},
   pages = {299-310},
   address = {Zurich, Switzerland},
   month = sep,
   year = {1995}
}

@book{Brockwell87Time,
   author = {Peter J. Brockwell and Richard A. Davis},
   title = {Time Series: Theory and Methods},
   publisher = {Springer Verlag},
   address = {New York},
   year = {1987},
   abstract = {Discusses ARMA, ARIMA models with a very strong math view point.
Gives the Yule-Walker equations for ARMA models;
the Wold decomposition;
the Akaike Information Criterion (AIC).
It presents Hilbert spaces with inner products,
fractional differencing models (FARMA) (\~{}pink noise),
random variables with infinite variance, and
Kalman filtering.}
}

@book{Box76Time,
   author = {George E.P. Box and Gwilym M. Jenkins},
   title = {Time Series Analysis: Forecasting and Control},
   edition = {Revised},
   publisher = {Holden-Day Inc.},
   address = {San Francisco},
   year = {1976}
}

@incollection{Rissanen90Coding,
   author = {J. Rissanen},
   editor = {Renato M. Capocelli},
   title = {Coding and Complexity},
   booktitle = {Sequences: Combinatorics, Compression, Security and Transmission},
   pages = {312-325},
   publisher = {Springer Verlag Inc.},
   address = {New York},
   year = {1990}
}

@book{Chaitin87Algorithmic,
   author = {G.J. Chaitin},
   title = {Algorithmic Information Theory},
   publisher = {Cambridge University Press},
   year = {1987}
}

@article{Chaitin66Length,
   author = {G.J. Chaitin},
   title = {On the Length of Programs for Computing Finite Binary Sequences},
   journal = {J. of ACM},
   volume = {13},
   pages = {547-569},
   year = {1966}
}

@article{Chaitin75Theory,
   author = {G.J. Chaitin},
   title = {A Theory of Program Size Formally Identical to Information Theory},
   journal = {J. of ACM},
   volume = {22},
   pages = {329-340},
   year = {1975}
}

@incollection{Rissanen85Minimum,
   author = {J. Rissanen},
   editor = {S. Kotz and N. L. Johnson},
   title = {Minimum Description Length Principle},
   booktitle = {Encyclopedia of Statistical Sciences},
   volume = {V},
   pages = {523-527},
   publisher = {John Wiley and Sons},
   address = {New York},
   year = {1985}
}

@article{Kolmogorov83Combinatorial,
   author = {A.N. Kolmogorov},
   title = {Combinatorial Foundation of Information Theory and the Calculus of Probabilities},
   journal = {Russian Math. Surveys},
   volume = {38},
   pages = {29-40},
   year = {1983}
}

@article{Flickner95Query,
   author = {Myron Flickner and Harpreet Sawhney and Wayne Niblack and Jon Ashley and Qian Huang and Byron Dom and Monika Gorkani and Jim Hafner and Denis Lee and Dragutin Petkovic and David Steele and Peter Yanker},
   title = {Query by Image and Video Content: The {QBIC} System},
   journal = {IEEE Computer},
   volume = {28},
   number = {9},
   pages = {23-32},
   month = sep,
   year = {1995},
   contents = {(not listed)}
}

@article{Arman93Model,
   author = {Farshid Arman and J.K. Aggarwal},
   title = {Model-Based Object Recognition in Dense-Range Images - {A} Review},
   journal = {ACM Computing Surveys},
   volume = {25},
   number = {1},
   pages = {5-43},
   month = mar,
   year = {1993}
}

@article{Grosky92pictorial,
   author = {William I. Grosky and Peter Neo and Rajiv Mehrotra},
   title = {A pictorial index mechanism for model-based matching},
   journal = {Data and Knowledge Engineering},
   volume = {8},
   pages = {309-327},
   publisher = {North Holland},
   year = {1992}
}

@article{Wu94Identifying,
   author = {Jian Kang Wu and Arcot Desai Narasimhalu},
   title = {Identifying Faces Using Multiple Retrievals},
   journal = {IEEE Multimedia},
   volume = {1},
   number = {2},
   pages = {27-38},
   year = {1994}
}

@book{Box94Time,
   author = {George E.P. Box and Gwilym M. Jenkins and Gregory C. Reinsel},
   title = {Time Series Analysis: Forecasting and Control},
   edition = {3rd},
   publisher = {Prentice Hall},
   address = {Englewood Cliffs, NJ},
   year = {1994}
}

@article{Wang94Combinatorial,
   author = {Jason Tsong-Li Wang and Gung-Wei Chirn and Thomas G. Marr and Bruce Shapiro and Dennis Shasha and Kaizhong Zhang},
   title = {Combinatorial Pattern Discovery for Scientific Data: Some Preliminary Results},
   journal = {ACM SIGMOD},
   address = {Minneapolis, MN},
   month = may,
   year = {1994}
}

@book{Tong93Dimension,
   author = {Howell Tong},
   title = {Dimension Estimation and Models},
   publisher = {World Scientific},
   year = {1993},
   abstract = {I think it mentioned FARMA models (ie, with fractional integration)}
}

@article{Pagel95Window,
   author = {Bernd-Uwe Pagel and Hans-Werner Six and Mario Winter},
   title = {Window Query-Optimal Clustering of Spatial Objects},
   journal = {ACM SIGACT-SIGMOD-SIGART Symp. on Principles of Database Systems (PODS)},
   pages = {86-94},
   address = {San Jose, CA},
   month = {May 22-25},
   year = {1995}
}

@inproceedings{Freeston95General,
   author = {Michael Freeston},
   title = {A General Solution of the n-dimensional {B}-tree Problem},
   booktitle = {Proc. of ACM-SIGMOD},
   pages = {80-91},
   address = {San Jose, CA},
   month = {May 23-25},
   year = {1995}
}

@inproceedings{Houtsma95Set,
   author = {M. Houtsma and A. Swami},
   title = {Set-Oriented Mining for Association Rules},
   booktitle = {Proceedings of IEEE Data Engineering Conference},
   month = mar,
   year = {1995},
   note = {Also appeared as IBM Research Report RJ 9567}
}

@article{Zhang95Video,
   author = {Hongjiang Zhang and Chien Yong Low and Stephen W. Smoliar},
   title = {Video Parsing and Browsing Using Compressed Data},
   journal = {Multimedia Tools and Applications},
   volume = {1},
   pages = {89-111},
   year = {1995}
}

@article{Hampapur95Production,
   author = {Arun Hampapur and Ramesh Jain and Terry E. Weymouth},
   title = {Production Model Based Digital Video Segmentation},
   journal = {Multimedia Tools and Applications},
   volume = {1},
   pages = {9-46},
   year = {1995}
}

@article{Bowman94Scalable,
   author = {C. Mic Bowman and Peter B. Danzig and Udi Manber and Michael F. Schwartz},
   title = {Scalable Internet Resource Discovery: Research Problems and Approaches},
   journal = {Comm. of ACM (CACM)},
   volume = {37},
   number = {8},
   pages = {98-114},
   month = aug,
   year = {1994}
}

@article{Bach93Visual,
   author = {Jeffrey R. Bach and Santanu Paul and Ramesh Jain},
   title = {A Visual Information Management System for the Interactive Retrieval of Faces},
   journal = {IEEE Trans. on Knowledge and Data Engineering (TKDE)},
   volume = {5},
   number = {4},
   pages = {619-628},
   month = aug,
   year = {1993}
}

@article{Hernandez95Merge,
   author = {Mauricio A. Hernandez and Salvatore Stolfo},
   title = {The Merge/Purge Problem for Large Databases},
   journal = {ACM-SIGMOD},
   pages = {127-138},
   month = may,
   year = {1995}
}

@book{Daubechies92Ten,
   author = {Ingrid Daubechies},
   title = {Ten Lectures on Wavelets},
   publisher = {Capital City Press},
   address = {Montpelier, Vermont},
   year = {1992},
   abstract = {chapter 1: Haar wavelet},
   note = {Society for Industrial and Applied Mathematics (SIAM), Philadelphia, PA}
}

@inproceedings{Yao85General,
   author = {Andrew C. Yao and F. Frances Yao},
   title = {A General Approach to d-Dimensional Geometric Queries},
   booktitle = {Proc. of the 17th Annual ACM Symposium on Theory of Computing (STOC)},
   pages = {163-168},
   address = {Providence, RI},
   month = {May 6-8},
   year = {1985},
   keywords = {dimensionality curse}
}

@book{Peitgen92Chaos,
   author = {Heinz-Otto Peitgen and Hartmut Juergens and Dietmar Saupe},
   title = {Chaos and Fractals: New Frontiers of Science},
   publisher = {Springer-Verlag New York Inc.},
   year = {1992},
   abstract = {excellent, comprehensive treatment of the topic.
Appendix B is a great introduction to multifractals.}
}

@book{Hastings93Fractals,
   author = {Harold M. Hastings and George Sugihara},
   title = {Fractals: {A} User's Guide for the Natural Sciences},
   publisher = {Oxford University Press},
   year = {1993},
   abstract = {Korcak's law of island area distribution;
Intersting applications for time series, biological data,
seismological data}
}

@article{Vassilakopoulo96Random,
   author = {Michael Vassilakopoulos and Yannis Manolopoulos},
   title = {A Random Model for Analyzing Region Quadtrees},
   journal = {Pattern Recognition Letters},
   year = {1996},
   note = {to appear}
}

@article{Sidiropoulos96Viterbi,
   author = {Nikolaos Sidiropoulos},
   title = {The Viterbi Optimal Runlength-Constrained Approximation Nonlinear Filter},
   journal = {IEEE Trans. on Signal Processing},
   year = {1996},
   abstract = {clever algorithm to find a piece-wise flat approximation of a function
under minimum-runlength restriction. Also, it has a wonderful
survey on older function-approximation algorithms,
typically using dynamic programming (aka Viterbi)},
   note = {to appearalso available as ISR TR}
}

@book{Weigend94Time,
   author = {Andreas S. Weigend and Neil A. Gerschenfeld},
   title = {Time Series Prediction: Forecasting the Future and Understanding the Past},
   publisher = {Addison Wesley},
   year = {1994},
   abstract = {has the winners of the SFI forecasting competition}
}

@article{Gudivada95Guest,
   author = {Venkat N. Gudivada and Vijay V. Raghavan},
   title = {Guest Editors' Introduction: Content-Based Image Retrieval Systems},
   journal = {IEEE Computer},
   volume = {28},
   number = {9},
   pages = {18-22},
   month = sep,
   year = {1995}
}

@article{Ogle95Chabot,
   author = {Virginia E. Ogle and Michael Stonebraker},
   title = {Chabot: Retrieval from a Relational Database of Images},
   journal = {IEEE Computer},
   volume = {28},
   number = {9},
   pages = {40-48},
   month = sep,
   year = {1995},
   keywords = {qbic query by image content}
}

@inproceedings{Yan95Duplicate,
   author = {Tak W. Yan and Hector Garcia-Molina},
   title = {Duplicate Removal in Information Dissemination},
   booktitle = {Proc. of VLDB},
   pages = {66-77},
   address = {Zurich, Switzerland},
   month = sep,
   year = {1995}
}

@techreport{Ester95Database,
   author = {Martin Ester and Hans-Peter Kriegel and Xiaowei Xu},
   title = {A Database Interface for Clustering in Large Spatial Databases},
   institution = {Univ. of Munich, Inst. for Informatics},
   type = {TR},
   number = {95-10},
   year = {1995}
}

@techreport{Seidl953D,
   author = {Thomas Seidl and Hans-Peter Kriegel},
   title = {A 3D Molecular Surface Representation Supporting Neighborhood Queries},
   institution = {Univ. of Munich, Inst. for Informatics},
   type = {TR},
   number = {95-09},
   year = {1995}
}

@inproceedings{Brin95Near,
   author = {Sergei Brin},
   title = {Near Neighbor Search in Large Metric Spaces},
   booktitle = {Proc. of VLDB Conf.},
   pages = {574-584},
   address = {Zurich, Switzerland},
   month = {Sept. 11-15},
   year = {1995}
}

@inproceedings{Hellerstein95Generalized,
   author = {Joseph M. Hellerstein and Jeffrey F. Naughton and Avi Pfeffer},
   title = {Generalized Search Trees for Database Systems},
   booktitle = {Proc. of VLDB Conf.},
   pages = {562-573},
   address = {Zurich, Switzerland},
   month = {Sept. 11-15},
   year = {1995}
}

@article{Goldin95Similarity,
   author = {Dina Q. Goldin and Paris C. Kanellakis},
   title = {On Similarity Queries for Time-Series Data: Constraint Specification and Implementation},
   journal = {Int. Conf. on Principles and Practice of Constraint Programming (CP95)},
   address = {Cassis, France},
   month = {Sept. 19-22},
   year = {1995}
}

@inproceedings{DeWitt94Client,
   author = {David J. DeWitt and Navin Kabra and Jun Luo and Jignesh Patel and Jie-Bing Yu},
   title = {The Client/Server Paradise},
   booktitle = {Proceedings of the VLDB},
   address = {Santiago, Chile},
   month = sep,
   year = {1994}
}

@inproceedings{Berson95Fault,
   author = {Steven Berson and Leana Golubchik and Richard R. Muntz},
   title = {Fault Tolerant Design of Multimedia Servers},
   booktitle = {Proc. of ACM SIGMOD},
   pages = {364-375},
   address = {San Jose, CA},
   month = jun,
   year = {1995}
}

@incollection{Rigney94Multi,
   author = {David R. Rigney and Ary L. Goldberger and Wendell C. Ocasio and Yuhei Ichimaru and George B. Moody and Roger G. Mark},
   editor = {Andreas S. Weigend and Neil A. Gerschenfeld},
   title = {Multi-Channel Physiological Data: Description and Analysis},
   booktitle = {Time Series Prediction: Forecasting the Future and Understanding the Past},
   pages = {105-130},
   publisher = {Addison Wesley},
   year = {1994},
   abstract = {describes the respiratory data of set B of the competition,
related to patients with apnea.
Also, they mention that "over long time periods,
the heart rate has 1/f Fourier Spectrum.}
}

@incollection{Granger94Forecasting,
   author = {Clive W. J. Granger},
   editor = {Andreas S. Weigend and Neil A. Gerschenfeld},
   title = {Forecasting in Economics},
   booktitle = {Time Series Prediction: Forecasting the Future and Understanding the Past},
   pages = {529-538},
   publisher = {Addison Wesley},
   year = {1994}
}

@article{Evangelidis95hBP,
   author = {Georgios Evangelidis and David Lomet and Betty Salzberg},
   title = {The hBP-tree: {A} Modified hB-tree Supporting Concurrency, Recovery and Node Consolidation},
   journal = {VLDB},
   pages = {551-561},
   address = {Zurich, Switzerland},
   month = sep,
   year = {1995}
}

@inproceedings{Freedman95SPIFFI,
   author = {Craig S. Freedman and David J. DeWitt},
   title = {The {SPIFFI} Scalable Video-On-Demand System},
   booktitle = {Proc. of ACM SIGMOD},
   pages = {352-363},
   address = {San Jose, CA},
   month = jun,
   year = {1995}
}

@inproceedings{Aref95Handwritten,
   author = {Walid Aref and Daniel Barbara and Padmavathi Vallabhaneni},
   title = {The Handwritten Trie: Indexing Electronic Ink},
   booktitle = {Proc. of ACM SIGMOD},
   pages = {151-162},
   address = {San Jose, CA},
   month = jun,
   year = {1995}
}

@inproceedings{Jagadish95Similarity,
   author = {H.V. Jagadish and Alberto O. Mendelzon and Tova Milo},
   title = {Similarity-Based Queries},
   booktitle = {Proc. ACM SIGACT-SIGMOD-SIGART PODS},
   pages = {36-45},
   address = {San Jose, CA},
   month = {May 22-25},
   year = {1995}
}

@inproceedings{Kanellakis95Constraint,
   author = {Paris Kanellakis},
   title = {Constraint Programming and Database Languages: {A} Tutorial},
   booktitle = {Proc. ACM SIGACT-SIGMOD-SIGART PODS},
   pages = {46-53},
   address = {San Jose, CA},
   month = {May 22-25},
   year = {1995}
}

@inproceedings{Brodsky95Separability,
   author = {Alexander Brodsky and Catherine Lassez and Jean-Louis Lassez and Michael Maher},
   title = {Separability of Polyhedra for Optimal Filtering of Spatial and Constraint Data},
   booktitle = {Proc. ACM SIGACT-SIGMOD-SIGART PODS},
   pages = {54-65},
   address = {San Jose, CA},
   month = {May 22-25},
   year = {1995}
}

@article{Samet85Model,
   author = {Hanan Samet and Clifford A. Shaffer},
   title = {A Model for the Analysis of Neighbor Finding in Pointer-Based Quadtrees},
   journal = {IEEE Trans. on Pattern Analysis and Machine Intelligence (PAMI)},
   volume = {7},
   number = {6},
   pages = {717-720},
   year = {1985},
   abstract = {"in a quadtree, each leaf node is equally likely
to appear at any position and level in the tree"
(from Vassilakopoulos and  Manolopoulos, PRL 96)}
}

@article{Mathieu87Average,
   author = {C. Mathieu and C. Puech and H. Yahia},
   title = {Average Efficiency of data structures for Binary Image Processing},
   journal = {Information Processing Letters},
   volume = {26},
   number = {2},
   pages = {89-93},
   year = {1987}
}

@article{Manolopoulos96Creation,
   author = {Yannis Manolopoulos and Enrico Nardelli and Guido Proietti and Michael Vassilakopoulos},
   title = {On the Creation of Quadtrees by Using a Branching Process},
   journal = {Image and Vision Computing},
   year = {1996},
   note = {to appear}
}

@inproceedings{Agrawal95Fast,
   author = {Rakesh Agrawal and King-Ip Lin and Harpreet S. Sawney and Kyuseok Shim},
   title = {Fast Similarity Search in the Presence of Noise, Scaling and Translation in Time-Series Databases},
   booktitle = {Proc. of VLDB},
   pages = {490-501},
   address = {Zurich, Switzerland},
   month = sep,
   year = {1995}
}

@inproceedings{Agrawal95Querying,
   author = {Rakesh Agrawal and Giuseppe Psaila and Edward L. Wimmers and Mohamed Zait},
   title = {Querying Shapes of Histories},
   booktitle = {Proc. of VLDB},
   pages = {502-514},
   address = {Zurich, Switzerland},
   month = sep,
   year = {1995}
}

@article{Faloutsos96Signature,
   author = {Christos Faloutsos and H.V. Jagadish and Alberto Mendelzon and Tova Milo},
   title = {A Signature Technique for Similarity-Based Queries},
   journal = {ACM PODS},
   year = {1996},
   note = {submitted; also available as ATT Bell Labs Tech. report}
}

@book{Walrand88Introduction,
   author = {Jean Walrand},
   title = {An Introduction to Queueing Networks},
   publisher = {Prentice-Hall},
   year = {1988}
}

@techreport{Faloutsos95Analysis,
   author = {Christos Faloutsos and Volker Gaede},
   title = {Analysis of the z-ordering Method Using the Hausdorff Fractal Dimension},
   institution = {ATT Bell Labs},
   type = {112530-951108-15TM},
   month = nov,
   year = {1995}
}

@article{Morrison68a,
   author = {Donald R. Morrison},
   title = {{PATRICIA} - Practical Algorithm to Retrieve Information Coded in Alphanumeric},
   journal = {Journal of ACM (JACM)},
   volume = {15},
   number = {4},
   pages = {514-534},
   month = oct,
   year = {1968}
}

@book{Sedgewick83Algorithms,
   author = {Robert Sedgewick},
   title = {Algorithms},
   publisher = {Addison Wesley},
   year = {1983},
   abstract = {has about hashing, on chapter 16}
}

@book{Gonnet91Handbook,
   author = {Gaston H. Gonnet and Ricardo Baeza-Yates},
   title = {Handbook of Algorithms and Data Structures In Pascal and {C}},
   edition = {2nd},
   publisher = {Addison-Wesley},
   year = {1991},
   abstract = {hashing, on chapter 3.3}
}

@book{Korth91Database,
   author = {Henry F. Korth and Abraham Silberschatz},
   title = {Database System Concepts},
   edition = {2nd},
   publisher = {McGraw Hill},
   year = {1991},
   abstract = {ISBN 0-07-044754-3}
}

@book{Date86Introduction,
   author = {Chris J. Date},
   title = {An Introduction to Database Systems},
   publisher = {Addison-Wesley},
   year = {1986},
   note = {Vol. I; 4th ed.}
}

@book{Bertino93Object,
   author = {Elisa Bertino and Lorenzo Martino},
   title = {Object-Oriented Database Systems: Concepts and Architectures},
   publisher = {Addison-Wesley},
   year = {1993}
}

@book{Kemper94Object,
   author = {Alfons Kemper and Guido Moerkotte},
   title = {Object Oriented Database Management: Applications in Engineering and Computer Science},
   publisher = {Prentice Hall},
   year = {1994}
}

@inproceedings{Shum88Information,
   author = {Chung-Dak Shum and Richard Muntz},
   title = {An Information-Theoretic Study on Aggregate Responses},
   booktitle = {Proc. of VLDB},
   pages = {479-490},
   address = {Los Angeles, CA},
   month = aug,
   year = {1988},
   abstract = {Seems like it creates groups of records (in form of a tree)
so that the entropy is maximized, for a given number of nodes.
Reminds of CART etc machine learning algorithms.}
}

@article{Hou91Statistical,
   author = {Wen-Chi Hou and Gultekin Ozsoyoglu},
   title = {Statistical Estimators for Aggregate Relational Algebra Queries},
   journal = {ACM TODS},
   volume = {16},
   number = {4},
   pages = {600-654},
   month = dec,
   year = {1991},
   abstract = {formula for the estimation of the number of distinct values
from a sample. Also mentioned in [Haas et al, VLDB 95]}
}

@article{Malvestuto93Universal,
   author = {Francesco M. Malvestuto},
   title = {A Universal-Scheme Approach to Statistical Databases Containing Homgeneous Summary Tables},
   journal = {ACM TODS},
   volume = {18},
   number = {4},
   pages = {678-708},
   month = dec,
   year = {1993},
   abstract = {Tries to infer the exact answers (or ranges of answers)
to queries, given summary tables.}
}

@techreport{Taylor94Analyzing,
   author = {Stephen Taylor and Nabil Hachem and Stanley Selkow},
   title = {Analyzing {NIBGF} and {BANG} Files Using the Capitalist Model},
   institution = {Worcester Polytechnic Institute},
   type = {WPI-CS-TR-94-4},
   month = nov,
   year = {1994}
}

@article{Ioannidis95Balancing,
   author = {Yannis E. Ioannidis and Viswanath Poosala},
   title = {Balancing Histogram Optimality and Practicality for Query Result Size Estimation},
   journal = {ACM SIGMOD},
   pages = {233-244},
   address = {San Jose, CA},
   month = jun,
   year = {1995},
   abstract = {"our overall conclusion is that the moste effective approach is to focus
on the class of histograms that accurately maintain
the frequencies of a few attributed
values and assume the uniform distribution for the rest,
and choose for each relation the histogram in that class that is optimal
for the self-join query."}
}

@unpublished{Gibbons95Practical,
   author = {P. Gibbons and Yossi Matias and A. Witkowski},
   title = {Practical Maintenance Algorithms for High-Biased Histograms Using Probabilistic Filtering},
   type = {unpublished manuscript},
   month = oct,
   year = {1995},
   note = {ATT Bell Laboratories}
}

@unpublished{Alon95space,
   author = {N. Alon and Yossi Matias and M. Szegedi},
   title = {The space complexity of approximating the frequency moments},
   type = {unpublished manuscript},
   month = jul,
   year = {1995},
   note = {ATT Bell Laboratories}
}

@article{Abiteboul95power,
   author = {S. Abiteboul and C. Beeri},
   title = {On the power of languages for the manipulation of complex objects},
   journal = {Journal on Very Large Databases - VLDB.},
   year = {1995},
   note = {To appear - INRIA research report n 846.}
}

@inproceedings{Beeri92Functional,
   author = {C. Beeri and T. Milo},
   title = {Functional and predicative programming in {OODB}'s.},
   booktitle = {Proc. 11th Symp. on Principles of Database Systems},
   address = {San-Diego},
   year = {1992}
}

@inproceedings{Colby94Query,
   author = {L.S. Colby and E.L. Robertson and L.V. Saxton and D. Van Gucht},
   title = {A Query Language for List Based Complex-Objects},
   booktitle = {Proc. 13th ACM Symp. on Principles of Database Systems},
   pages = {179-189},
   address = {Minneapolis, MN},
   month = may,
   year = {1994}
}

@article{Gonnet92Exhaustive,
   author = {G.H Gonnet and M.A. Cohen and S.A. Benner},
   title = {Exhaustive Matching of the Entire Protein Sequence Database},
   journal = {Science},
   volume = {256},
   number = {5},
   month = jun,
   year = {1992}
}

@inproceedings{Grumbach95algebra,
   author = {S. Grumbach and T. Milo},
   title = {An algebra for pomsets},
   booktitle = {Proc. of ICDT 95},
   year = {1995}
}

@inproceedings{Grahne94Reasoning,
   author = {G. Grahne and M. Nykanen and and E. Ukkonen.},
   title = {Reasoning about strings in databases},
   booktitle = {Proc. 13th ACM Symp. on Principles of Database Systems},
   pages = {303-312},
   address = {Minneapolis, MN},
   month = may,
   year = {1994}
}

@inproceedings{Ginsburg92Towards,
   author = {S. Ginsburg and X. Wang},
   title = {Towards a unified approach to querying sequenced data},
   booktitle = {Proc. 11th ACM Symp. on Principles of Database Systems},
   pages = {293-300},
   address = {San Diego, CA},
   year = {1992}
}

@inproceedings{Richardson92Supporting,
   author = {J. Richardson},
   title = {Supporting lists in a data model (a timely approach)},
   booktitle = {Proc. 18th Intl. Conf. on Very Large Databases},
   address = {Vancouver},
   month = aug,
   year = {1992}
}

@inproceedings{Seshadri94Sequence,
   author = {Praveen Seshadri and Miron Livny and Raghu Ramakrishnan},
   title = {Sequence query processing},
   booktitle = {Proc. Int. Conf. on Management of Data SIGMOD},
   pages = {430-441},
   address = {Minneapolis, MN},
   month = may,
   year = {1994}
}

@book{Ullman88Database,
   author = {J.D. Ullman},
   title = {Database and Knowledge Base Systems},
   publisher = {Computer Science Press},
   year = {1988}
}

@inproceedings{Haas95Sampling,
   author = {Peter J. Haas and Jeffrey F. Naughton and S. Seshadri and Lynne Stokes},
   title = {Sampling-Based Estimation of the Number of Distinct Values of an Attribute},
   booktitle = {Proc. of VLDB},
   pages = {311-322},
   address = {Zurich, Switzerland},
   month = sep,
   year = {1995}
}

@inproceedings{Gray94Quickly,
   author = {Jim Gray and Prakash Sundaresan and Susan Englert and Ken Baclawski and Peter Weinberger},
   title = {Quickly Generating Billion-Record Synthetic Databases},
   booktitle = {Proc. ACM-SIGMOD},
   pages = {243-252},
   address = {Minneapolis, Minnesota},
   month = may,
   year = {1994}
}

@article{Mandelbrot63Stable,
   author = {B.B. Mandelbrot},
   title = {The Stable Paretian Income Distribution when the Apparent Exponent is Near Zero},
   journal = {Int. Econ. Rev.},
   volume = {4},
   pages = {111-115},
   year = {1963},
   abstract = {from schroeder91}
}

@book{Pareto96Oeuvres,
   author = {V. Pareto},
   title = {Oeuvres Completes},
   publisher = {Droz},
   address = {Geneva},
   year = {1896},
   abstract = {from schroeder91}
}

@inproceedings{Selinger79Access,
   author = {P.G. Selinger and D.D. Astrahan and R.A. Chamberlain and R.A. Lorie and T.G. Price},
   title = {Access Path Selection in a Relational Database Management System},
   booktitle = {Proc. ACM-SIGMOD},
   pages = {23-34},
   year = {1979}
}

@inproceedings{Dumais94Latent,
   author = {Susan T. Dumais},
   editor = {D. K. Harman},
   title = {Latent Semantic Indexing ({LSI}) and {TREC}-2},
   booktitle = {The Second Text Retrieval Conference (TREC-2)},
   pages = {105-115},
   publisher = {NIST},
   address = {Gaithersburg, MD},
   month = mar,
   year = {1994},
   note = {Special publication 500-215}
}

@incollection{Arya96Design,
   author = {Manish Arya and William Cody and Christos Faloutsos and Joel Richardson and Arthur Toga},
   editor = {V.S. Subrahmanian and Sushil Jajodia},
   title = {Design and Implementation of {QBISM}, a 3D Medical Image Database System},
   booktitle = {Multimedia Database Systems: Issues and Research Directions},
   pages = {79-100},
   publisher = {Springer Verlag},
   year = {1996},
   ISBN = {3-540-58710-1}
}

@book{Peters91Chaos,
   author = {Edgar E. Peters},
   title = {Chaos and Order in Capital Markets: {A} New View of Cycles, Prices and Market Volatility},
   publisher = {John Wiley and Sons Inc.},
   year = {1991},
   abstract = {Shows that the Hurst exponent of financial time series
is not 1.5 (ie, not a random walk), but typically higher.
Thus, there is a long-term memory in such systems.
It is only after a long time that the R/S slope becomes 1.5,
which means that, after a long time,
the system behaves like a random walk.
The break-point is at a time-length which corresponds to 'statistical cycle'
(the notion is not well defined, but seems interesting)}
}

@article{Callan74Theory,
   author = {E. Callan and D. Shapiro},
   title = {A Theory of Social Imitation},
   journal = {Physics Today},
   volume = {27},
   year = {1974},
   abstract = {according to Edgar Peters ("chaos and order in capital markets"),
they claim that humans behavior can be modeled as Ising spins}
}

@inproceedings{Ganguly96Sampling,
   author = {Sumit Ganguly and Phil Gibbons and Yossi Matias and Avi Silberschatz},
   title = {A Sampling Algorithm for Estimating Join Size},
   booktitle = {Proc. of ACM SIGMOD},
   address = {Montreal, Canada},
   year = {1996},
   note = {to appear}
}

@article{Widom95Research,
   author = {Jennifer Widom},
   title = {Research Problems in Data Warehousing},
   journal = {CIKM},
   month = nov,
   year = {1995},
   note = {Invited paper}
}

@inproceedings{Theodoridis96Model,
   author = {Yannis Theodoridis and Timos Sellis},
   title = {A Model for the Prediction of {R}-tree Performance},
   booktitle = {Proc. of ACM PODS},
   address = {Montreal, Canada},
   year = {1996}
}

@inproceedings{Chen94Adaptive,
   author = {Chungmin M. Chen and Nick Roussopoulos},
   title = {Adaptive Selectivity Estimation Using Query Feedback},
   booktitle = {Proc. of the ACM-SIGMOD},
   pages = {161-172},
   address = {Minneapolis, MN},
   month = may,
   year = {1994}
}

@inproceedings{Ng95Information,
   author = {Wee-Keong Ng and Chinya V. Ravishankar},
   title = {Information Synthesis in Statistical Databases},
   booktitle = {Proc. CIKM},
   pages = {355-361},
   address = {Baltimore, MD},
   month = nov,
   year = {1995}
}

@inproceedings{Harinarayan96Implementing,
   author = {Venky Harinarayan and Anand Rajaraman and Jeffrey D. Ullman},
   title = {Implementing Data Cubes Efficiently},
   booktitle = {Proc. ACM SIGMOD},
   address = {Montreal, Canada},
   year = {1996},
   note = {to appear}
}

@techreport{Gray95Data,
   author = {J. Gray and A. Bosworth and A. Layman and H. Pirahesh},
   title = {Data Cube: {A} Relational Aggregation Operator Generalizing Group-By, Cross-Tab, and Sub-Totals},
   institution = {Microsoft},
   type = {Technical Report No.},
   number = {MSR-TR-95-22},
   year = {1995}
}

@article{Koudas96Declustering,
   author = {Nick Koudas and Christos Faloutsos and Ibrahim Kamel},
   title = {Declustering Spatial Databases on a Multi-Computer Architecture},
   journal = {EDBT Conf. Proc.},
   pages = {592-614},
   address = {Avignon, France},
   month = mar,
   year = {1996}
}

@techreport{Colton94Illustra,
   author = {Malcolm Colton},
   title = {Illustra, Relational Databases and Spatial Data},
   institution = {Illustra Information Technologies, Inc.},
   type = {An Illustra Technical White Paper},
   month = nov,
   year = {1994}
}

@article{Faloutsos96Relaxing,
   author = {Christos Faloutsos and Ibrahim Kamel},
   title = {Relaxing the Uniformity and Independence Assumptions, Using the Concept of Fractal Dimension},
   journal = {Journal of Computer and System Sciences (JCSS)},
   year = {1996},
   note = {to appear}
}

@techreport{Gaede95Survey,
   author = {Volker Gaede and Oliver Guenther},
   title = {Survey on Multidimensional Access Methods},
   institution = {Institut fuer Wirtschaftsinformatik, Humboldt-Universitaet zu Berlin},
   type = {ISS-16},
   month = aug,
   year = {1995}
}

@book{Crandall94Projects,
   author = {Richard E. Crandall},
   title = {Projects in Scientific Computation},
   publisher = {Springer-Verlag New York, Inc.},
   year = {1994},
   keywords = {wavelets dwt}
}

@book{Vidakovic91Wavelets,
   author = {Brani Vidakovic and Peter Mueller},
   title = {Wavelets for Kids},
   publisher = {Duke University},
   address = {Durham, NC},
   year = {1991},
   keywords = {wavelets Haar},
   note = {ftp://ftp.isds.duke.edu/pub/Users/brani/papers/}
}

@inproceedings{Williams96Indexing,
   author = {Hugh Williams and Justin Zobel},
   title = {Indexing Nucleotide Databases for Fast Query Evaluation},
   booktitle = {Proc. of 5-th Intl. Conf. on Extending Database Technology (EDBT)},
   pages = {275-288},
   address = {Avignon, France},
   month = mar,
   year = {1996},
   note = {Eds. P. Apers, M. Bouzeghoub, G. Gardarin}
}

@article{Codd70Relational,
   author = {E. F. Codd},
   title = {A Relational Model of Data for Large Shared Data Banks},
   journal = {Comm. of ACM},
   volume = {13},
   number = {6},
   pages = {377-387},
   year = {1970}
}

@inproceedings{Kumar95Access,
   author = {Anil Kumar and Vassilis J. Tsotras and Christos Faloutsos},
   title = {Access Methods for Bi-Temporal Databases},
   booktitle = {Int. Workshop on Temporal Databases},
   address = {Zurich, Switzerland},
   month = sep,
   year = {1995}
}

@techreport{Moon96Analysis,
   author = {Bongki Moon and H.V. Jagadish and Christos Faloutsos and Joel H. Saltz},
   title = {Analysis of the Clustering Properties of Hilbert Space-filling Curve},
   institution = {Dept. of Computer Science, Univ. of Maryland},
   type = {Technical Report},
   number = {CS-TR-3611},
   address = {College Park, MD},
   year = {1996}
}

@article{Korn96Fast,
   author = {Flip Korn and Nikolaos Sidiropoulos and Christos Faloutsos and Eliot Siegel and Zenon Protopapas},
   title = {Fast Nearest-Neighbor Search in Medical Image Databases},
   journal = {Conf. on Very Large Data Bases (VLDB)},
   address = {Bombay, India},
   month = sep,
   year = {1996},
   note = {Also available as Univ. of Maryland tech. report: CS-TR-3613, ISR-TR-96-13}
}

@book{Horn86Robot,
   author = {Berthold Horn},
   title = {Robot Vision},
   publisher = {MIT Press},
   address = {Cambridge, Mass.},
   year = {1986}
}

@article{Faloutsos96Modeling,
   author = {Christos Faloutsos and Yossi Matias and Avi Silberschatz},
   title = {Modeling skewed distributions using multifractals and the `80-20 law'},
   journal = {VLDB},
   address = {Bombay, India},
   month = sep,
   year = {1996}
}

@article{Faloutsos96Analysis,
   author = {Christos Faloutsos and Volker Gaede},
   title = {Analysis of the z-ordering Method Using the Hausdorff Fractal Dimension},
   journal = {VLDB},
   address = {Bombay, India},
   month = sep,
   year = {1996}
}

@incollection{Rasmussen92Clustering,
   author = {Edie Rasmussen},
   editor = {William B. Frakes and Ricardo Baeza-Yates},
   title = {Clustering Algorithms},
   booktitle = {Information Retrieval: Data Structures and Algorithms},
   pages = {419-442},
   publisher = {Prentice Hall},
   year = {1992}
}

@book{Sigmund93Games,
   author = {Karl Sigmund},
   title = {Games of Life: Explorations in Ecology, Evolution and Behaviour},
   publisher = {Oxford University Press},
   address = {Oxford, U.K.},
   year = {1993},
   abstract = {mentions about the lynx dataset, on p. 45}
}

@article{Schatz96Building,
   author = {Bruce Schatz and Hsinchun Chen},
   title = {Building Large-Scale Digital Libraries},
   journal = {IEEE Computer},
   volume = {29},
   number = {5},
   pages = {22-26},
   month = may,
   year = {1996}
}

@article{Wactlar96Intelligent,
   author = {Howard D. Wactlar and Takeo Kanade and Michael A. Smith and Scott M. Stevens},
   title = {Intelligent Access to Digital Video: Informedia Project},
   journal = {IEEE Computer},
   volume = {29},
   number = {5},
   pages = {46-52},
   month = may,
   year = {1996}
}

@article{Petrakis97Similarity,
   author = {Euripides G.M. Petrakis and Christos Faloutsos},
   title = {Similarity Searching in Medical Image DataBases},
   journal = {IEEE Trans. on Knowledge and Data Engineering (TDKE)},
   volume = {9},
   number = {3},
   pages = {435-447},
   month = may,
   year = {1997},
   keywords = {medical attribute graphs R-tree spatial access method},
   note = {Also available as technical report at MUSIC with TR-01-94, UMIACS-TR-94-134, CS-TR-3388}
}

@incollection{Gonnet92New,
   author = {Gaston H. Gonnet and Ricardo A. Baeza-Yates and Tim Snider},
   editor = {William B. Frakes and Ricardo Baeza-Yates},
   title = {New Indices for Text: {PAT} Trees and {PAT} Arrays},
   booktitle = {Information Retrieval},
   publisher = {Prentice Hall},
   year = {1992}
}

@article{Agrawal93Database,
   author = {Rakesh Agrawal and Tomasz Imielinski and Arun Swami},
   title = {Database mining: a performance perspective},
   journal = {IEEE Trans. on Knowledge and Data Engineering},
   volume = {5},
   number = {6},
   pages = {914-925},
   year = {1993}
}

@book{Aho88AWK,
   author = {Alfred V. Aho and Brian W. Kernighan and Peter J. Weinberger},
   title = {The {AWK} Programming Language},
   publisher = {Addison Wesley},
   year = {1988}
}

@article{Zezula91Dynamic,
   author = {P. Zezula and F. Rabitti and P. Tiberio},
   title = {Dynamic Partitioning of Signature Files},
   journal = {ACM TOIS},
   volume = {9},
   number = {4},
   pages = {336-369},
   month = oct,
   year = {1991}
}

@article{Bentley79Multidimensional,
   author = {Jon L. Bentley},
   title = {Multidimensional Binary Search Trees in Database Applications},
   journal = {IEEE Trans. on Software Engineering (TSE)},
   volume = {SE-5},
   number = {4},
   pages = {333-340},
   month = jul,
   year = {1979}
}

@techreport{Zolnowsky78Topics,
   author = {J. E. Zolnowsky},
   title = {Topics in Computational Geometry},
   institution = {Stanford University (Ph.D. dissertation)},
   type = {Stanford Comp. Sc. Rep. STAN-CS-78-659,},
   number = {SLAC-206},
   year = {1978},
   note = {proved that best-match search is on the average at most O( (logN)**k), for the best match search for every existing point in a given set.}
}

@article{Langenhop89Model,
   author = {Carl E. Langenhop and William E. Wright},
   title = {A Model of the Dynamic Behavior of {B}-trees},
   journal = {Acta Informatica},
   volume = {27},
   pages = {41-59},
   year = {1989},
   abstract = {B-tree utilization: approx 60-67 percent}
}

@article{Yao78Random,
   author = {A. C. Yao},
   title = {On Random 2,3 Trees},
   journal = {Acta Informatica},
   volume = {9},
   pages = {159-170},
   year = {1978},
   abstract = {2-3 tree average utilization, under pure insertions: 69 per cent}
}

@inproceedings{Johnson89Utilization,
   author = {Theodore Johnson and Dennis Shasha},
   title = {Utilization of {B}-trees with Inserts, Deletes and Modifies},
   booktitle = {Proc. of ACM SIGACT-SIGMOD-SIGART PODS},
   pages = {235-246},
   address = {Philadelphia, PA},
   month = {March 29-31},
   year = {1989}
}

@inproceedings{Seeger90Buddy,
   author = {Bernhard Seeger and Hans-Peter Kriegel},
   title = {The Buddy-Tree: An Efficient and Robust Access Method for Spatial Database Systems},
   booktitle = {Proc. of VLDB},
   pages = {590-601},
   address = {Brisbane, Australia},
   month = {Aug. 13-16},
   year = {1990}
}

@article{Gudivada95Content,
   author = {Venkat N. Gudivada and Vijay V. Raghavan},
   title = {Content-Based Image Retrieval Systems},
   journal = {IEEE Computer},
   volume = {28},
   number = {9},
   pages = {18-22},
   month = sep,
   year = {1995}
}

@article{Mehrotra95Similar,
   author = {Rajiv Mehrotra and James E. Gary},
   title = {Similar-Shape Retrieval in Shape Data Management},
   journal = {IEEE Computer},
   volume = {28},
   number = {9},
   pages = {57-62},
   month = sep,
   year = {1995}
}

@book{Jolliffe86Principal,
   author = {I.T. Jolliffe},
   title = {Principal Component  Analysis},
   publisher = {Springer Verlag},
   year = {1986},
   abstract = {seems like a great book on PCA -
it shows the connection between PCA and SVD;
talks about how to choose the number of eigenvectors to keep;
discusses outlier detection;
uses PCA for stock prices (Dow Jones)},
   contents = {(not listed)}
}

@article{Garcia-Ruiz93Growth,
   author = {Juan Manuel Garcia-Ruiz and Enrique Louis and Paul Meakin and Leonard M. Sander},
   title = {Growth Patterns in Physical Science and Biology},
   journal = {NATO ASI series, Series B, Physics; vol. 304},
   publisher = {Plenum Press},
   address = {New York},
   year = {1993},
   note = {Proceedings of a NATO Advanced Research Workshop, Oct. 7-11, 1991, Granada, Spain}
}

@book{Faloutsos96Searching,
   author = {Christos Faloutsos},
   title = {Searching Multimedia Databases by Content},
   publisher = {Kluwer Academic Inc.},
   year = {1996},
   ISBN = {0-7923-9777-0}
}

@inproceedings{Ester95Knowledge,
   author = {Martin Ester and Hans-Peter Kriegel and Xiaowei Xu},
   title = {Knowledge Discovery in Large Spatial Databases: Focusing Techniques for Efficient Class Identification},
   booktitle = {Proc. of 4th International Symposium on Large Spatial Databases},
   address = {Portland, Maine},
   year = {1995}
}

@misc{Salzberg94Comparison,
   author = {Betty Salzberg and Vassilis J. Tsotras},
   title = {A Comparison of Access Methods for Time-Evolving Data},
   year = {1994},
   note = {available as Tech. report (CATT-TR-94-81) from Brooklyn Polytechnic University, or (NU-CCS-94-21) from Northeastern University.}
}

@inproceedings{Kobla??Archiving,
   author = {Vikrant Kobla and David Doermann and King-Ip (David) Lin},
   title = {Archiving, indexing and retrieval of video in the compressed domain},
   booktitle = {SPIE conference},
   address = {Boston, MA},
   year = {Nov. 18-22},
   note = {Published as SPIE conference, volume 2916, number 9}
}

@techreport{Faloutsos97Information,
   author = {Christos Faloutsos and H.V. Jagadish and Nikolaos Sidiropoulos},
   title = {Information Recovery from Partial Data},
   institution = {Inst. for Systems Research, Univ. of Maryland},
   type = {Tech. Report},
   number = {ISR-TR-97-7},
   address = {College Park, MD},
   year = {1997}
}

@inproceedings{Korn96Fasta,
   author = {Flip Korn and Nicholaos Sidiropoulos and Christos Faloutsos and Eliot Siegel and Zenon Protopapas},
   title = {Fast and Effective Similarity Search in Medical Tumor Databases using Morphology},
   booktitle = {SPIE Proceedings Vol. 2916},
   address = {Boston, MA},
   month = nov,
   year = {1996},
   keywords = {nearest neighbor query, morphology}
}

@inproceedings{Kobla97Compressed,
   author = {Vikrant Kobla and David S. Doermann and King-Ip (David) Lin and Christos Faloutsos},
   title = {Compressed domain video indexing techniques using {DCT} and motion vector information in {MPEG} video},
   booktitle = {Proceedings of the SPIE conference on Storage and Retrieval for Image and Video Databases V},
   volume = {3022},
   month = feb,
   year = {1997}
}

@article{Sidiropoulos96Theorem,
   author = {Nikolaos Sidiropoulos and H.V. Jagadish and Christos Faloutsos},
   title = {A Theorem on Signal Reconstruction from Partial Sums},
   journal = {IEEE Signal Processing Letters},
   month = nov,
   year = {1996},
   note = {Submitted}
}

@inproceedings{Fagin??Combining,
   author = {Ronald Fagin},
   title = {Combining Fuzzy Information from Multiple Systems},
   booktitle = {Proc. of ACM PODS},
   pages = {216-226},
   address = {Montreal, Canada},
   year = {June 3-5}
}

@article{Papadopoulos97Performance,
   author = {Apostolos Papadopoulos and Yannis Manolopoulos},
   title = {Performance of nearest neighbor queries in {R}-trees},
   journal = {6th Int. Conf. on Database Theory (ICDT '97)},
   address = {Delphi, Greece},
   month = {Jan. 8-10},
   year = {1997}
}

@book{Faloutsos95Signature,
   author = {Christos N. Faloutsos and H.V. Jagadish and Alberto O. Mendelzon and Tova Milo},
   title = {Signature Technique for Similarity-Based Queries},
   number = {112530-951110-16TM},
   publisher = {AT\&T},
   address = {Murray Hill, NJ},
   month = nov,
   year = {1995}
}

@article{Arya963D,
   author = {Manish Arya and William Cody and Christos Faloutsos and Joel Richardson and Arthur Toga},
   title = {A 3D Medical Image Database Management System},
   journal = {Computerized Medical Imaging and Graphics},
   volume = {20},
   number = {4},
   pages = {269-284},
   month = apr,
   year = {1996}
}

@article{Manolopoulos96creationa,
   author = {Y. Manolopoulos and E. Nardelli and G. Proietti and M. Vassilakopoulos},
   title = {On the creation of quadtrees by using a branching process},
   journal = {Image and Vision Computing},
   volume = {14},
   number = {2},
   pages = {159-164},
   month = mar,
   year = {1996},
   abstract = {should be related to multifractals}
}

@article{Crovella96Self,
   author = {M. Crovella and A. Bestavros},
   title = {Self-Similarity in World Wide Web Traffic, Evidence and Possible Causes},
   journal = {Sigmetrics},
   pages = {160-169},
   year = {1996}
}

@article{Faloutsos97Recovering,
   author = {Christos Faloutsos and H.V. Jagadish and Nikolaos D. Sidiropoulos},
   title = {Recovering Information from Summary Data},
   journal = {VLDB},
   address = {Athens, Greece},
   month = aug,
   year = {1997}
}

@techreport{Sclaroff97ImageRover,
   author = {S. Sclaroff and L. Taycher and M. La Cascia},
   title = {ImageRover: {A} Content-Based Image Browser for the World Wide Web},
   institution = {Boston University Computer Science Dept.},
   type = {TR97-005},
   month = apr,
   year = {1997},
   note = {to appear in Proc. IEEE Workshop on Content-based Access of Image and Video Libraries, June 1997.}
}

@article{Smith97Searching,
   author = {J.R. Smith and S.-F. Chang},
   title = {Searching for Images and Videos on the World-Wide Web},
   journal = {IEEE Multimedia Magazine},
   year = {1997},
   note = {(to appear - also Columbia U. CU/CTR Technical Report \#459-96-25)}
}

@inproceedings{Agnew97Web,
   author = {Brent Agnew and Christos Faloutsos and Zhengyu Wang and Don Welch and Xiaogang Xue},
   title = {Web-based Multi-media Retrieval by Content},
   booktitle = {Proceedings of the SPIE conference on Storage and Retrieval for Image and Video Databases V},
   address = {San Jose, CA},
   month = {Feb. 8-14},
   year = {1997}
}

@article{Korn97Efficiently,
   author = {Flip Korn and H.V. Jagadish and Christos Faloutsos},
   title = {Efficiently Supporting Ad Hoc Queries in Large Datasets of Time Sequences},
   journal = {ACM SIGMOD},
   pages = {289-300},
   address = {Tucson, AZ},
   month = {May 13-15},
   year = {1997}
}

@article{Faloutsos97Signature,
   author = {Christos Faloutsos and H.V. Jagadish and Alberto O. Mendelzon and Tova Milo},
   title = {A Signature Technique for Similarity-Based Queries},
   journal = {SEQUENCES 97},
   pages = {2-20},
   publisher = {IEEE Computer Society},
   address = {Positano-Salerno, Italy},
   month = {June 11-13},
   year = {1997}
}

@article{Zhang96,
   author = {Tian Zhang and Raghu Ramakrishnan and Miron Livny},
   title = {{BIRCH}: An Efficient Data Clustering Method for Very Large Databases},
   journal = {ACM SIGMOD},
   pages = {103-114},
   address = {Montreal, Canada},
   month = may,
   year = {1996}
}

@article{Faloutsos97Analysisa,
   author = {Christos Faloutsos and H.V. Jagadish and Yannis Manolopoulos},
   title = {Analysis of the n-dimensional quadtree decomposition for arbitrary hyper-rectangles},
   journal = {IEEE-TKDE},
   volume = {9},
   number = {3},
   pages = {373-383},
   month = may,
   year = {1997},
   keywords = {magic interpolation},
   note = {Also available as CS-TR-3381, UMIACS-TR-94-130 from UMD-CS.}
}

@article{Petrakis97Similaritya,
   author = {Euripides G.M. Petrakis and Christos Faloutsos},
   title = {Similarity Searching in Medical Image DataBases},
   journal = {IEEE Trans. on Knowledge and Data Engineering (TDKE)},
   volume = {9},
   number = {3},
   pages = {435-447},
   month = may,
   year = {1997},
   keywords = {medical attribute graphs R-tree spatial access method},
   note = {Also available as technical report at MUSIC with TR-01-94, UMIACS-TR-94-134, CS-TR-3388}
}

@techreport{Kobla96Feature,
   author = {Vikrant Kobla and David S. Doermann and King-Ip (David) Lin and Christos Faloutsos},
   title = {Feature Normalization for Video Indexing and Retrieval},
   institution = {Univ. of Maryland, College Park},
   type = {CAR-TR-847,},
   number = {CS-TR-3732},
   month = nov,
   year = {1996}
}

@article{Kumar98Designing,
   author = {Anil Kumar and Vassilis J. Tsotras and Christos Faloutsos},
   title = {Designing Access Methods for Bitemporal Databases},
   journal = {IEEE Trans. on Knowledge and Data Engineering (IEEE-TKDE),},
   volume = {10},
   number = {1},
   pages = {1-20},
   month = jan,
   year = {1998}
}

@article{Hou92Contenta,
   author = {T.Y. Hou and A. Hsu and P. Liu and M.Y. Chiu},
   title = {A Content-based indexing technique using relative geometry features},
   journal = {SPIE 92},
   volume = {1662},
   pages = {59-68},
   year = {1992}
}

@article{Willinger95Self,
   author = {Walter Willinger and Murad Taqqu and Robert Sherman and Daniel V. Wilson},
   title = {Self-similarity through high variability: statistical analysis of Ethernet {LAN} traffic at the source level},
   journal = {ACM SIGCOMM'95. Computer Communication Review},
   volume = {25},
   pages = {100-113},
   year = {1995}
}

@article{Ishikawa98MindReader,
   author = {Yoshiharu Ishikawa and Ravishankar Subramanya and Christos Faloutsos},
   title = {MindReader: Querying databases through multiple examples},
   journal = {VLDB},
   address = {New York, NY.},
   month = aug,
   year = {1998},
   note = {Also available as CMU-CS-98-119, Carnegie Mellon University}
}

@article{Yi98Efficient,
   author = {Byoung-Kee Yi and H.V. Jagadish and Christos Faloutsos},
   title = {Efficient Retrieval of Similar Time Sequences Under Time Warping},
   journal = {ICDE 98},
   address = {Orlando, Florida},
   month = {Feb. 23 - 27},
   year = {1998}
}

@article{Kobla97VideoTrails,
   author = {Vikrant Kobla and David S. Doermann and Christos Faloutsos},
   title = {VideoTrails: Representing and Visualizing Structure in Video Sequences},
   journal = {ACM Multimedia 97},
   address = {Seattle, WA},
   month = nov,
   year = {1997}
}

@inproceedings{Hellerstein97Online,
   author = {Joseph M. Hellerstein and Peter J. Haas and Helen Wang},
   title = {Online Aggregation},
   booktitle = {SIGMOD Conference},
   pages = {171-182},
   year = {1997}
}

@techreport{Sato96Name,
   author = {S. Sato and T. Kanade},
   title = {Name-It: Association of Face and Name in Video},
   institution = {CMU School of CS,},
   type = {TR},
   number = {CMU-CS-96-205},
   month = dec,
   year = {1996}
}

@inproceedings{Berchtold98Pyramid,
   author = {Stefan Berchtold and Christian Boehm and Hans-Peter Kriegel},
   title = {The Pyramid-Tree: Breaking the Curse of Dimensionality},
   booktitle = {SIGMOD Conference},
   pages = {142-153},
   year = {1998}
}

@inproceedings{Berchtold97Fast,
   author = {Stefan Berchtold and Christian Boehm and Bernhard Braunmueller and Daniel A. Keim and Hans-Peter Kriegel},
   title = {Fast Similarity Search in Multimedia Databases},
   booktitle = {SIGMOD Conference},
   pages = {1-12},
   year = {1997}
}

@article{Berchtold97Cost,
   author = {Stefan Berchtold and Christian Boehm and Daniel A. Keim and Hans-Peter Kriegel},
   title = {A Cost Model For Nearest Neighbor Search in High-Dimensional Data Space},
   journal = {PODS},
   pages = {78-86},
   year = {1997}
}

@article{Yi98Efficienta,
   author = {Byoung-Kee Yi and H. V. Jagadish and Christos Faloutsos},
   title = {Efficient Retrieval of Similar Time Sequences Under Time Warping},
   journal = {ICDE},
   pages = {201-208},
   year = {1998}
}

@inproceedings{Bozkaya97Distance,
   author = {Tolga Bozkaya and Z. Meral Ozsoyoglu},
   title = {Distance-Based Indexing for High-Dimensional Metric Spaces},
   booktitle = {SIGMOD Conference},
   pages = {357-368},
   year = {1997}
}

@article{Katayama97SR,
   author = {Norio Katayama and Shin'ichi Satoh},
   title = {The {SR}-tree: An Index Structure for High-Dimensional Nearest Neighbor Queries},
   journal = {SIGMOD},
   pages = {369-380},
   year = {1997}
}

@article{Ciaccia98Cost,
   author = {Paolo Ciaccia and Marco Patella and Pavel Zezula},
   title = {A Cost Model for Similarity Queries in Metric Spaces.},
   journal = {PODS},
   pages = {59-68},
   year = {1998}
}

@article{Ciaccia97M,
   author = {Paolo Ciaccia and Marco Patella and Pavel Zezula},
   title = {M-tree: An Efficient Access Method for Similarity Search in Metric Spaces.},
   journal = {VLDB},
   pages = {426-435},
   year = {1997}
}

@book{Kruskal78Multidimensional,
   author = {Joseph B. Kruskal and Myron Wish},
   title = {Multidimensional scaling},
   publisher = {SAGE publications},
   address = {Beverly Hills},
   year = {1978}
}

@article{Hellerstein95Generalizeda,
   author = {Joseph M. Hellerstein and Jeffrey F. Naughton and Avi Pfeffer},
   title = {Generalized Search Trees for Database Systems},
   journal = {VLDB},
   pages = {562-573},
   year = {1995}
}

@inproceedings{Koudas97Size,
   author = {Nick Koudas and Kenneth C. Sevcik},
   title = {Size Separation Spatial Join},
   booktitle = {SIGMOD Conference},
   pages = {324-335},
   year = {1997}
}

@article{Lo96Spatial,
   author = {Ming-Ling Lo and Chinya V. Ravishankar},
   title = {Spatial Hash-Joins},
   journal = {SIGMOD Conf.},
   pages = {247-258},
   year = {1996}
}

@article{Berchtold96X,
   author = {Stefan Berchtold and Daniel A. Keim and Hans-Peter Kriegel},
   title = {The {X}-tree: An Index Structure for High-Dimensional Data},
   journal = {VLDB},
   pages = {28-39},
   year = {1996}
}

@inproceedings{Chang97VideoQ,
   author = {S.-F. Chang and W. Chen and H.J. Meng and H. Sundaram and D. Zhong},
   title = {VideoQ- An Automatic Content-Based Video Search System Using Visual Cues},
   booktitle = {ACM Multimedia Conference},
   year = {Nov. 1997 Seattle, WA},
   note = {also Columbia University/CTR Technical Report, CTR-TR \# 478-97-12.}
}

@inproceedings{Beigi98MetaSEEk,
   author = {Mandis Beigi and Ana Benitez and S.-F. Chang},
   title = {MetaSEEk: {A} Content-Based Meta Search Engine for Images},
   booktitle = {SPIE Conference on Storage and Retrieval for Image and Video Database},
   address = {San Jose},
   month = feb,
   year = {1998},
   note = {also Columbia University/CTR Technical Report, CTR-TR \# 480-97-14}
}

@book{Mitchell97Machine,
   author = {Tom Mitchell},
   title = {Machine Learning},
   publisher = {McGraw Hill},
   year = {1997}
}

@inproceedings{Hellerstein94Practical,
   author = {Joseph M. Hellerstein},
   title = {Practical Predicate Placement},
   booktitle = {SIGMOD Conference},
   pages = {325-335},
   year = {1994}
}

@article{Belussi98Self,
   author = {Alberto Belussi and Christos Faloutsos},
   title = {Self-Spacial Join Selectivity Estimation Using Fractal Concepts},
   journal = {ACM TOIS},
   volume = {16},
   number = {2},
   pages = {161-201},
   year = {1998}
}

@article{Riedel98Active,
   author = {Erik Riedel and Garth A. Gibson and Christos Faloutsos},
   title = {Active Storage for Large-Scale Data Mining and Multimedia},
   journal = {VLDB},
   pages = {62-73},
   year = {1998}
}

@article{Yi00Online,
   author = {Byoung-Kee Yi and N.D. Sidiropoulos and Theodore Johnson and H.V. Jagadish and Christos Faloutsos and Alexandros Biliris},
   title = {Online Data Mining for Co-Evolving Time Sequences},
   journal = {ICDE},
   pages = {13-22},
   address = {San Diego, CA},
   year = {2000}
}

@misc{Johnson99Accurate,
   author = {Theodore Johnson and Christos Faloutsos},
   title = {Accurate block selectivities using the `Recursive Biased Distribution'},
   year = {1999},
   note = {submitted}
}

@article{Korn98Fast,
   author = {Philip (Flip) Korn and Nicholas Sidiropoulos and Christos Faloutsos and Eliot Siegel and Zenon Protopapas},
   title = {Fast and Effective Retrieval of Medical Tumor Shapes},
   journal = {IEEE TKDE},
   volume = {10},
   number = {6},
   pages = {889-904},
   month = nov,
   year = {1998}
}

@article{Korn98Ratio,
   author = {Flip Korn and Alexandros Labrinidis and Yannis Kotidis and Christos Faloutsos},
   title = {Ratio Rules: {A} New Paradigm for Fast, Quantifiable Data Mining},
   journal = {VLDB},
   address = {New York, NY},
   year = {1998}
}

@inproceedings{Proietti98Selectivity,
   author = {Guido Proietti and Christos Faloutsos},
   title = {Selectivity Estimation of Window Queries for Line Segment Datasets},
   booktitle = {Seventh Int. Conference on Information and Knowledge Management (CIKM),},
   address = {Washington, DC},
   month = {Nov. 3-7},
   year = {1998}
}

@inproceedings{Proietti99I,
   author = {Guido Proietti and Christos Faloutsos},
   title = {I/{O} complexity for range queries on region data stored using an {R}-tree},
   booktitle = {International Conference on Data Engineering (ICDE)},
   address = {Sydney, Australia},
   month = {March 23-26},
   year = {1999}
}

@article{Bradley98Scaling,
   author = {Paul S. Bradley and Usama M. Fayyad and Cory Reina},
   title = {Scaling Clustering Algorithms to Large Databases},
   journal = {KDD},
   pages = {9-15},
   address = {New York, NY},
   year = {1998}
}

@inproceedings{Guha98,
   author = {Sudipto Guha and Rajeev Rastogi and Kyuseok Shim},
   title = {{CURE}: An Efficient Clustering Algorithm for Large Databases},
   booktitle = {SIGMOD Conference},
   pages = {73-84},
   address = {Seattle, Washington},
   year = {1998}
}

@article{Sander98Density,
   author = {Joerg Sander and Martin Ester and Hans-Peter Kriegel and Xiaowei Xu},
   title = {Density-Based Clustering in Spatial Databases: The Algorithm {GDBSCAN} and Its Applications.},
   journal = {Data Mining and Knowledge Discovery},
   volume = {2},
   number = {2},
   pages = {169-194},
   year = {1998}
}

@article{Wactlar99Lessons,
   author = {Howard D. Wactlar and Michael G. Christel and Yihong Gong and Alexander G. Hauptmann},
   title = {Lessons Learned from Building a Terabyte Digital Video Library},
   journal = {IEEE Computer},
   volume = {32},
   number = {2},
   pages = {66-73},
   year = {1999}
}

@article{Natsev99,
   author = {Apostol Natsev and Rajeev Rastogi and Kyuseok Shim},
   title = {{WALRUS}: {A} Similarity Retrieval Algorithm for Image Databases},
   journal = {ACM SIGMOD},
   address = {Phildelphia, PA},
   year = {1999},
   note = {to appear}
}

@inproceedings{Kleinberg98Authoritative,
   author = {Jon Kleinberg},
   title = {Authoritative sources in a hyperlinked environment},
   booktitle = {Proc. 9th ACM-SIAM Symposium on Discrete Algorithms},
   year = {1998},
   note = {Also appears as IBM Research Report RJ 10076, May 1997.}
}

@inproceedings{Henrich94distance,
   author = {A. Henrich},
   title = {A distance-scan algorithm for spatial access structures},
   booktitle = {Proceedings of the 2nd ACM Workshop on Advances in Geographic Information Systems, ACM Press,},
   pages = {136-143},
   address = {Gaithersburg, Maryland},
   month = dec,
   year = {1994}
}

@inproceedings{Rafiei97Similarity,
   author = {Davood Rafiei and Alberto O. Mendelzon},
   title = {Similarity-Based Queries for Time Series Data},
   booktitle = {SIGMOD Conference},
   pages = {13-25},
   address = {Tucson, AZ},
   year = {1997}
}

@inproceedings{Feldmann98Data,
   author = {A. Feldmann and A. C. Gilbert and W. Willinger},
   title = {Data networks as cascades: Investigating the multifractal nature of Internet {WAN} traffic},
   booktitle = {Proc. of the ACM/SIGCOMM},
   pages = {25-38},
   address = {Vancouver, B.C.},
   year = {1998}
}

@misc{Bak96How,
   author = {Per Bak},
   title = {How Nature Works : The Science of Self-Organized Criticality},
   pages = {Copernicus Books},
   month = sep,
   year = {1996},
   ISBN = {0387947914}
}

@article{Gaede98Multidimensional,
   author = {Volker Gaede and Oliver Guenther},
   title = {Multidimensional Access Methods},
   journal = {Computing Surveys},
   volume = {30},
   number = {2},
   pages = {170-231},
   year = {1998}
}

@article{Sheikholeslami98WaveCluster,
   author = {Gholamhosein Sheikholeslami and Surojit Chatterjee and Aidong Zhang},
   title = {WaveCluster: {A} Multi-Resolution Clustering Approach for Very Large Spatial Databases},
   journal = {VLDB},
   pages = {428-439},
   address = {New York, NY},
   year = {1998}
}

@article{Traina00Slim,
   author = {Caetano Traina and Agma J. M. Traina and Bernhard Seeger and Christos Faloutsos},
   title = {Slim-Trees: High Performance Metric Trees Minimizing Overlap Between Nodes},
   journal = {EDBT},
   pages = {51-65},
   address = {Konstanz, Germany},
   year = {2000}
}

@article{Traina00Distance,
   author = {Caetano Traina and Agma J. M. Traina and Christos Faloutsos},
   title = {Distance Exponent: {A} New Concept for Selectivity Estimation in Metric Trees},
   journal = {ICDE},
   pages = {195},
   address = {San Diego, CA},
   year = {2000},
   note = {(poster presentation)}
}

@article{Pagel00Deflating,
   author = {Bernd-Uwe Pagel and Flip Korn and Christos Faloutsos},
   title = {Deflating the Dimensionality Curse Using Multiple Fractal Dimensions},
   journal = {ICDE},
   pages = {589-598},
   address = {San Diego, CA},
   year = {2000}
}

@inproceedings{Riedel00Data,
   author = {Erik Riedel and Christos Faloutsos and Gregory R. Ganger and David Nagle},
   title = {Data Mining on an {OLTP} System (Nearly) for Free},
   booktitle = {SIGMOD Conference},
   pages = {13-21},
   address = {Dallas, TX},
   year = {2000}
}

@inproceedings{Faloutsos00Spatial,
   author = {Christos Faloutsos and Bernhard Seeger and Agma J. M. Traina and Caetano Traina},
   title = {Spatial Join Selectivity Using Power Laws.},
   booktitle = {SIGMOD Conference},
   pages = {177-188},
   address = {Dallas, TX},
   year = {2000}
}

@inproceedings{Palmer00Density,
   author = {Christopher R. Palmer and Christos Faloutsos},
   title = {Density Biased Sampling: An Improved Method for Data Mining and Clustering.},
   booktitle = {SIGMOD Conference},
   pages = {82-92},
   address = {Dallas, TX},
   year = {2000}
}

@article{Wu00,
   author = {Leejay Wu and Christos Faloutsos and Katia P. Sycara and Terry R. Payne},
   title = {{FALCON}: Feedback Adaptive Loop for Content-Based Retrieval.},
   journal = {VLDB},
   address = {Kairo, Egypt},
   year = {2000:},
   note = {to appear}
}

@article{Yi00Fast,
   author = {Byoung-Kee Yi and Christos Faloutsos},
   title = {Fast Time Sequence Indexing for Arbitrary Lp Norms.},
   journal = {VLDB},
   address = {Kairo, Egypt},
   year = {2000},
   note = {to appear}
}

@article{Korn00Quantifiable,
   author = {Flip Korn and Alexandros Labrinidis and Yannis Kotidis and Christos Faloutsos},
   title = {Quantifiable Data Mining Using Ratio Rules.},
   journal = {VLDB Journal},
   volume = {8},
   number = {3-4},
   pages = {254-266},
   year = {2000}
}

@article{Filho01Similarity,
   author = {Roberto F. Santos Filho and Agma Traina and Caetano Traina and Christos Faloutsos},
   title = {Similarity search without tears: the {OMNI} family of all-purpose access methods},
   journal = {ICDE},
   address = {Heidelberg, Germany},
   year = {2001},
   note = {to appear}
}

@article{Traina00Slima,
   author = {Caetano Traina and Agma J. M. Traina and Bernhard Seeger and Christos Faloutsos},
   title = {Slim-Trees: High Performance Metric Trees Minimizing Overlap Between Nodes},
   journal = {EDBT},
   pages = {51-65},
   address = {Constanz, Germany},
   year = {2000}
}

@inproceedings{Natsev99a,
   author = {Apostol Natsev and Rajeev Rastogi and Kyuseok Shim},
   title = {{WALRUS}: {A} Similarity Retrieval Algorithm for Image Databases.},
   booktitle = {SIGMOD Conference},
   pages = {395-406},
   address = {Philadelphia, PA},
   year = {1999}
}

@article{Rastogi98,
   author = {Rajeev Rastogi and Kyuseok Shim},
   title = {{PUBLIC}: {A} Decision Tree Classifier that Integrates Building and Pruning},
   journal = {VLDB},
   pages = {404-415},
   address = {New York, NY},
   year = {1998}
}

@book{Han00Data,
   author = {Jiawei Han and Micheline Kamber},
   title = {Data Mining: Concepts and Techniques},
   publisher = {Morgan Kaufmann},
   month = apr,
   year = {2000},
   ISBN = {1-55860-489-8}
}

@article{Das98Rule,
   author = {Gautam Das and King-Ip Lin and Heikki Mannila and Gopal Renganathan and Padhraic Smyth},
   title = {Rule Discovery from Time Series},
   journal = {KDD},
   pages = {16-22},
   address = {New York, NY},
   year = {1998}
}

@article{Hatonen96Knowledge,
   author = {Kimmo Hatonen and Mika Klemettinen and Heikki Mannila and Pirjo Ronkainen and Hannu Toivonen},
   title = {Knowledge Discovery from Telecommunication Network Alarm Databases},
   journal = {ICDE},
   pages = {115-122},
   address = {New Orleans, Louisiana},
   year = {1996}
}

@article{Agrawal95Mining,
   author = {Rakesh Agrawal and Ramakrishnan Srikant},
   title = {Mining Sequential Patterns},
   journal = {ICDE},
   pages = {3-14},
   address = {Taipei, Taiwan},
   year = {1995}
}

@article{Ester98Algorithms,
   author = {Martin Ester and Alexander Frommelt and Hans-Peter Kriegel and Joerg Sander},
   title = {Algorithms for Characterization and Trend Detection in Spatial Databases},
   journal = {KDD},
   pages = {44-50},
   address = {New York, NY},
   year = {1998}
}

@article{Albert99Diameter,
   author = {Reka Albert and Hawoong Jeong and Albert-Laszlo Barabasi},
   title = {Diameter of the World Wide Web},
   journal = {Nature},
   number = {401},
   pages = {130-131},
   year = {1999}
}

@article{Kumar99Extracting,
   author = {S. Ravi Kumar and Prabhakar Raghavan and Sridhar Rajagopalan and Andrew Tomkins},
   title = {Extracting Large-Scale Knowledge Bases from the Web},
   journal = {VLDB},
   pages = {639-650},
   address = {Edinburgh, Scotland},
   year = {1999}
}

@article{Kumar99Trawling,
   author = {S. Ravi Kumar and Prabhakar Raghavan and Sridhar Rajagopalan and Andrew Tomkins},
   title = {Trawling the Web for Emerging Cyber-Communities},
   journal = {Computer Networks},
   volume = {31},
   number = {11-16},
   pages = {1481-1493},
   year = {1999}
}

@article{Ailamaki99DBMSs,
   author = {A. Ailamaki and D. J. DeWitt and M. D. Hill and D. A. Wood},
   title = {DBMSs On {A} Modern Processor: Where Does Time Go?},
   journal = {VLDB},
   pages = {266-277},
   month = sep,
   year = {1999}
}

@article{Faloutsos99Power,
   author = {Michalis Faloutsos and Petros Faloutsos and Christos Faloutsos},
   title = {On Power-law Relationships of the Internet Topology},
   journal = {SIGCOMM},
   pages = {251-262},
   address = {Cambridge, MA},
   month = {Aug-Sept.},
   year = {1999}
}

@techreport{Riedel99Active,
   author = {Erik Riedel},
   title = {Active Disks - Remote Execution for Network-Attached Storage},
   institution = {CMU},
   type = {CMU-CS-99-177},
   address = {Pittsburgh, PA},
   month = nov,
   year = {1999},
   note = {Doctoral Dissertation}
}

@article{Boehm00Cost,
   author = {Christian Boehm},
   title = {A Cost Model for Query Processing in High Dimensional Data Spaces},
   journal = {ACM TODS},
   volume = {25},
   number = {2},
   pages = {129-178},
   month = jun,
   year = {2000}
}

@article{Korn01Dimensionality,
   author = {Flip Korn and Bernd-Uwe Pagel and Christos Faloutsos},
   title = {On the "Dimensionality Curse" and the "Self-Similarity Blessing"},
   journal = {IEEE Trans. on Knowledge and Data Engineering (TKDE)},
   volume = {13},
   number = {1},
   pages = {96-11},
   month = jan,
   year = {2001},
   note = {("special section: best of ICDE 2000")}
}

@article{Moon01Analysis,
   author = {Bongki Moon and H.V. Jagadish and Christos Faloutsos and Joel H. Saltz},
   title = {Analysis of the Clustering Properties of the Hilbert Space-Filling Curve},
   journal = {IEEE Trans. on Knowledge and Data Engineering (TKDE)},
   volume = {13},
   number = {1},
   pages = {124-141},
   month = jan,
   year = {2001}
}

@article{Breslau99Web,
   author = {L. Breslau and P. Cao and L. Fan and G. Phillips and S. Shenker},
   title = {Web Caching and Zipf-like Distributions: Evidence and Implications},
   journal = {IEEE Infocom'99},
   address = {New York, NY},
   month = mar,
   year = {1999}
}

@article{Pan01VideoGraph,
   author = {Jia-Yu Pan and Christos Faloutsos},
   title = {VideoGraph: a new tool for video mining and classification},
   journal = {JCDL},
   pages = {116-117},
   address = {Roanoke, VA},
   year = {2001}
}

@inproceedings{Traina00Fast,
   author = {Caetano Traina and Agma Traina and Leejay Wu and Christos Faloutsos},
   title = {Fast feature selection using the fractal dimension,},
   booktitle = {XV Brazilian Symposium on Databases (SBBD)},
   address = {Paraiba, Brazil},
   month = oct,
   year = {2000}
}

@article{Yi00Fasta,
   author = {Byoung-Kee Yi and Christos Faloutsos},
   title = {Fast Time Sequence Indexing for Arbitrary Lp Norms},
   journal = {VLDB},
   address = {Cairo, Egypt},
   month = {Sept. 10-14},
   year = {2000}
}

@article{Wu00a,
   author = {Leejay Wu and Christos Faloutsos and Katia Sycara and Terry Payne},
   title = {{FALCON}: Feedback Adaptive Loop for Content-Based Retrieval},
   journal = {VLDB},
   address = {Kairo, Egypt},
   month = {Sept. 10-14},
   year = {2000}
}

@inproceedings{Palmer01connectivity,
   author = {Chris Palmer and Georgios Siganos and Michalis Faloutsos and Christos Faloutsos and Phil Gibbons},
   title = {The connectivity and fault-tolerance of the Internet topology},
   booktitle = {Workshop on Network Related Data Management (NRDM 2001)},
   address = {Santa Barbara, CA},
   month = {May 25},
   year = {2001}
}

@article{Traina01Tri,
   author = {Agma Traina and Caetano Traina and Spiros Papadimitriou and Christos Faloutsos},
   title = {Tri-Plots: Scalable Tools for Multidimensional Data Mining},
   journal = {KDD},
   address = {San Francisco, CA},
   month = aug,
   year = {2001}
}

@article{Bi01DGX,
   author = {Zhiqiang Bi and Christos Faloutsos and Flip Korn},
   title = {The "{DGX}" Distribution for Mining Massive, Skewed Data},
   journal = {KDD},
   address = {San Francisco, CA},
   month = aug,
   year = {2001},
   note = {Runner up for Best Paper Award}
}

@article{Jin01Meta,
   author = {Rong Jin and Christos Faloutsos and Alex G. Hauptmann},
   title = {Meta-scoring: Automatically Evaluating Term Weighting Schemas in {IR} without Precision-Recall},
   journal = {SIGIR},
   address = {New Orleans LA},
   month = {Sept. 9-13},
   year = {2001}
}

@article{Margaritis01NetCube,
   author = {Dimitris Margaritis and Christos Faloutsos and Sebastian Thrun},
   title = {NetCube: {A} Scalable Tool for Fast Data Mining and Compression},
   journal = {VLDB},
   address = {Rome Italy},
   month = sep,
   year = {2001}
}

@article{Wang02Data,
   author = {Mengzhi Wang and Tara Madhyastha and Ngai Hang Chang and Spiros Papadimitriou and Christos Faloutsos},
   title = {Data Mining Meets Performance Evaluation: Fast Algorithms for Modeling Bursty Traffic},
   journal = {ICDE},
   address = {San Jose, CA},
   month = feb,
   year = {2002}
}

@book{Zaniolo97Advanced,
   author = {Carlo Zaniolo},
   title = {Advanced Database Systems},
   publisher = {Morgan Kaufmann},
   address = {San Francisco, CA},
   month = may,
   year = {1997}
}

@incollection{Faloutsos99Multimedia,
   author = {Christos Faloutsos},
   editor = {Ricardo Baeza-Yates and Berthier Ribeiro-Neto.},
   title = {Multimedia {IR}: Indexing and Searching,},
   booktitle = {Modern Information Retrieval},
   publisher = {Addison Wesley Longman},
   year = {1999}
}

@inproceedings{Ankerst99,
   author = {Mihael Ankerst and Markus M. Breunig and Hans-Peter Kriegel and Joerg Sander},
   title = {{OPTICS}: Ordering Points To Identify the Clustering Structure},
   booktitle = {SIGMOD Conference},
   pages = {49-60},
   year = {1999}
}

@article{Zhang00K,
   author = {Bin Zhang and Meichun Hsu and Umeshwar Dayal},
   title = {K-Harmonic Means - {A} Spatial Clustering Algorithm with Boosting},
   journal = {TSDM},
   pages = {31-45},
   year = {2000}
}

@inproceedings{Gunopulos01Time,
   author = {Dimitrios Gunopulos and Gautam Das},
   title = {Time Series Similarity Measures and Time Series Indexing},
   booktitle = {SIGMOD Conference},
   address = {Santa Barbara, CA},
   year = {2001},
   note = {Tutorial}
}

@inproceedings{Keogh01Locally,
   author = {Eamonn J. Keogh and Kaushik Chakrabarti and Sharad Mehrotra and Michael J. Pazzani},
   title = {Locally Adaptive Dimensionality Reduction for Indexing Large Time Series Databases},
   booktitle = {SIGMOD Conference},
   address = {Santa Barbara, CA},
   year = {2001}
}

@article{Proietti01Accurate,
   author = {Guido Proietti and Christos Faloutsos},
   title = {Accurate Modeling of Region Data},
   journal = {IEEE TKDE},
   volume = {13},
   number = {6},
   pages = {874-883},
   month = nov,
   year = {2001}
}

@inproceedings{Fayyad96Knowledge,
   author = {U. Fayyad and G. Piatetsky-Shapiro and Padhraic Smyth},
   title = {Knowledge discovery and data mining: Towards a unifying framework.},
   booktitle = {In Proceedings, Second International Conference on Knowledge Discovery and Data Mining.},
   pages = {AAAI Press},
   address = {Portland, Oregon},
   year = {1996}
}

@inproceedings{Bonnet01Towards,
   author = {Philippe Bonnet and J. E. Gehrke and Praveen Seshadri},
   title = {Towards Sensor Database Systems},
   booktitle = {In Proceedings of the Second International Conference on Mobile Data Management},
   address = {Hong Kong},
   month = jan,
   year = {2001}
}

@article{Babu01Continuous,
   author = {Shivnath Babu and Jennifer Widom},
   title = {Continuous Queries over Data Streams},
   journal = {SIGMOD Record 109-120},
   volume = {30},
   number = {3},
   year = {2001}
}

@inproceedings{Knorr01Robust,
   author = {Edwin Knorr and Raymond Ng and Ruben Zamar},
   title = {Robust Space Transformations for Distance-based Operations},
   booktitle = {KDD conference},
   address = {San Francicso},
   month = aug,
   year = {2001}
}

@inproceedings{Agrawal01Vinci,
   author = {R. Agrawal and R. J. Bayardo and D. Gruhl and S. Papadimitriou},
   title = {Vinci: {A} Service-Oriented Architecture for Rapid Development of Web Applications.},
   booktitle = {In Proc.  of the 10th Int'l World Wide Web Conference},
   address = {Hong Kong},
   year = {2001}
}

@article{May76Simple,
   author = {Robert May},
   title = {Simple Mathematical Models with Very Complicated Dynamics},
   journal = {Nature},
   volume = {261 459},
   year = {1976},
   note = {Lessons of chaos in population biology.}
}

@article{Gehrke01Computing,
   author = {J. E. Gehrke and Flip Korn and Divesh Srivastava},
   title = {On Computing Correlated Aggregates Over Continual Data Streams},
   journal = {ACM Sigmod},
   address = {Santa Barbara, California},
   month = may,
   year = {2001}
}

@article{Bonnet01Towardsa,
   author = {Philippe Bonnet and J. E. Gehrke and Praveen Seshadri},
   title = {Towards Sensor Database Systems},
   journal = {Intern. Conf. on Mobile Data Management},
   address = {Hong Kong},
   month = jan,
   year = {2001}
}

@article{Babu01Continuousa,
   author = {S. Babu and J. Widom},
   title = {Continuous Queries over Data Streams},
   journal = {In Sigmod Record},
   month = sep,
   year = {2001}
}

@inproceedings{Breunig00,
   author = {Markus M. Breunig and Hans-Peter Kriegel and Raymond T. Ng and Joerg Sander},
   title = {{LOF}: Identifying Density-Based Local Outliers},
   booktitle = {SIGMOD Conference},
   pages = {93-104},
   address = {Dallas, TX},
   year = {2000}
}

@incollection{Takens81Detecting,
   author = {F. Takens},
   title = {Detecting strange attractors in fluid turbulence},
   booktitle = {Dynamical Systems and Turbulence},
   publisher = {Springer-Verlag.},
   address = {Berlin},
   year = {1981}
}

@inproceedings{Oppenheim02MEMS,
   author = {I.J. Oppenheim and A. Jain and D. W. Greve},
   title = {A {MEMS} Ultrasonic Transducer for Resident Monitoring of Steel Structures},
   booktitle = {SPIE Smart Structures Conference SS05},
   address = {San Diego},
   month = mar,
   year = {2002}
}

@incollection{Sauer94Time,
   author = {T. Sauer},
   editor = {A. S. Weigend and N. A. Gershenfeld},
   title = {Time series prediction using delay coordinate embedding},
   booktitle = {Time Series Prediction: Forecasting the Future and Understanding the Past},
   publisher = {Addison-Wesley},
   year = {1994}
}

@article{Papadimitriou03,
   author = {Spiros Papadimitriou and Hiroyuki Kitagawa and Phil Gibbons and Christos Faloutsos},
   title = {{LOCI}: Fast Outlier Detection Using the Local Correlation Integral},
   journal = {ICDE},
   address = {Bangalore, India},
   month = {March 2003 5 - March 8},
   year = {2003}
}

@article{Wang02Capturing,
   author = {Mengzhi Wang and Anastassia Ailamaki and Christos Faloutsos},
   title = {Capturing the spatio-temporal behavior of real traffic data},
   journal = {Performance 2002 (IFIP Int. Symp. on Computer Performance Modeling, Measurement and Evaluation),},
   address = {Rome, Italy},
   month = sep,
   year = {2002},
   note = {Best Student paper award}
}

@article{Pan02GeoPlot,
   author = {Jia-Yu (Tim) Pan and Christos Faloutsos},
   title = {GeoPlot: Spatial Data Mining on Video Libraries},
   journal = {CIKM 2002},
   address = {Washington DC},
   month = nov,
   year = {2002}
}

@article{Chakrabarti02F4,
   author = {Deepay Chakrabarti and Christos Faloutsos},
   title = {F4: Large-Scale Automated Forecasting using Fractals},
   journal = {CIKM 2002},
   address = {Washington DC},
   month = nov,
   year = {2002}
}

@article{Pan02VideoCubes,
   author = {Jia-Yu (Tim) Pan and Christos Faloutsos},
   title = {VideoCubes: {A} Novel Tool for Video Mining and Classification},
   journal = {ICADL 2002},
   address = {Singapore},
   month = {Dec. 11-14},
   year = {2002}
}

@article{Pan02FastCARS,
   author = {Jia-Yu (Tim) Pan and Srinivasan Seshan and Christos Faloutsos},
   title = {FastCARS: Fast, Correlation-Aware Sampling for Network Data Mining},
   journal = {GlobeCom 2002},
   address = {Taipei, Taiwan},
   month = {Nov. 17-21},
   year = {2002}
}

@article{Papadimitriou03a,
   author = {Spiros Papadimitriou and Hiroyuki Kitagawa and Phil Gibbons and Christos Faloutsos},
   title = {{LOCI}: Fast Outlier Detection Using the Local Correlation Integral},
   journal = {ICDE 2003},
   address = {Bangalore, India},
   month = {March 5 - March 8},
   year = {2003}
}

@article{Palmer03Electricity,
   author = {Christopher R. Palmer and Christos Faloutsos},
   title = {Electricity Based External Similarity of Categorical Attributes},
   journal = {PAKDD 2003},
   address = {Seoul, South Korea},
   month = {April-May},
   year = {2003}
}

@article{Jiang02Accelerating,
   author = {Changhao Jiang and Minglong Shao and Yiheng Li and Peng Jia},
   title = {Accelerating Clustering Methods through Fractal Based Analysis},
   journal = {Fractal KDD, (in conjunction with SIGKDD)},
   address = {Edmonton, Canada},
   year = {2002}
}

@article{Siganos03Power,
   author = {Georgos Siganos and Michalis Faloutsos and Petros Faloutsos and Christos Faloutsos},
   title = {Power-Laws and the {AS}-level Internet Topology},
   journal = {IEEE/ACM Transactions on Networking},
   volume = {11},
   number = {4},
   pages = {514-524},
   year = {2003}
}

@article{Sidiropoulos00Blind,
   author = {N.D. Sidiropoulos and G.B. Giannakis and R. Bro},
   title = {Blind {PARAFAC} Receivers for {DS}-{CDMA} Systems},
   journal = {IEEE Trans. on Signal Processing},
   volume = {48},
   number = {3},
   pages = {810-823},
   month = mar,
   year = {2000}
}

@article{Olston03Adaptive,
   author = {C. Olston and J. Jiang and J. Widom},
   title = {Adaptive Filters for Continuous Queries over Distributed Data Streams},
   journal = {ACM SIGMOD},
   year = {2003}
}

@article{Dobra02Processing,
   author = {Alin Dobra and Minos Garofalakis and Johannes Gehrke and Rajeev Rastogi},
   title = {Processing Complex Aggregate Queries over Data Streams},
   journal = {ACM SIGMOD},
   pages = {61-72},
   address = {Madison, Wisconsin},
   month = jun,
   year = {2002}
}

@article{Garofalakis02Querying,
   author = {Minos N. Garofalakis and Johannes Gehrke and Rajeev Rastogi},
   title = {Querying and Mining Data Streams: You Only Get One Look},
   journal = {ACM SIGMOD},
   pages = {635},
   address = {Madison, Wisconsin},
   month = jun,
   year = {2002},
   note = {(tutorial)}
}

@article{Papadimitriou03Adaptive,
   author = {Spiros Papadimitriou and Anthony Brockwell and Christos Faloutsos},
   title = {Adaptive, Hands-Off Stream Mining},
   journal = {VLDB},
   address = {Berlin, Germany},
   month = sep,
   year = {2003}
}

@article{Montgomery01Identifying,
   author = {Alan L. Montgomery and Christos Faloutsos},
   title = {Identifying Web Browsing Trends and Patterns},
   journal = {IEEE Computer},
   volume = {34},
   number = {7},
   pages = {94-95},
   month = jul,
   year = {2001}
}

@article{Wu02Making,
   author = {Leejay Wu and Christos Faloutsos},
   title = {Making every bit count: Fast nonlinear axis scaling},
   journal = {KDD},
   address = {Edmonton, Alberta, Canada},
   month = jul,
   year = {2002}
}

@book{Buchanan02Nexus,
   author = {Mark Buchanan},
   title = {Nexus: Small Worlds and the Groundbreaking Theory of Networks},
   publisher = {W.W. Norton and Co.},
   year = {2002},
   ISBN = {0-393-04153-0}
}

@inproceedings{Brandes03Experiments,
   author = {Ulrik Brandes and Marco Gaertler and Dorothea Wagner},
   title = {Experiments on Graph Clustering Algorithms},
   booktitle = {Proc. 11th Europ. Symp. Algorithms (ESA '03), LNCS 2832},
   pages = {568-579},
   publisher = {Springer-Verlag},
   year = {2003},
   abstract = {Compare MCL with ICC and GMC, and show that their GMC is better}
}

@article{Flake02Self,
   author = {Gary Flake and Steve Lawrence and C. Lee Giles and Frans Coetzee},
   title = {Self-Organization and Identification of Web Communities},
   journal = {IEEE Computer},
   volume = {35},
   number = {3},
   month = mar,
   year = {2002}
}

@article{Virtanen03Clustering,
   author = {Satu Virtanen},
   title = {Clustering the Chilean Web},
   journal = {LA-WEB 2003},
   address = {Santiago, Chile},
   month = nov,
   year = {2003}
}

@inproceedings{Virtanen03Stochastic,
   author = {Satu Virtanen},
   title = {Stochastic Online Clustering of Massive Graphs},
   booktitle = {NIPS 2003 workshop on Robust Communication Dynamics in Complex Networks},
   address = {Vancouver, Canada},
   month = dec,
   year = {2003}
}

@phdthesis{Dongen00Graph,
   author = {Stijn van Dongen},
   title = {Graph Clustering by Flow Simulation},
   school = {University of Utrecht},
   type = {Ph.{D}. Thesis},
   month = may,
   year = {2000},
   abstract = {proposes the Markov Cluster Algorithm}
}

@article{Chakrabarti04R,
   author = {Deepayan Chakrabarti and Yiping Zhan and Christos Faloutsos},
   title = {R-{MAT}: {A} Recursive Model for Graph Mining},
   journal = {SIAM Int. Conf. on Data Mining},
   address = {Lake Buena Vista, FL},
   month = apr,
   year = {2004}
}

@inproceedings{Dhillon03Information,
   author = {Inderjit S. Dhillon and Subramanyam Mallela and Dharmendra S. Modha},
   title = {Information-theoretic Co-Clustering},
   booktitle = {The Ninth ACM SIGKDD International Conference on Knowledge Discovery and Data Mining (KDD 03)},
   address = {Washington, DC},
   month = {August 24-27},
   year = {2003}
}

@article{Karypis99Parallel,
   author = {G. Karypis and V. Kumar},
   title = {Parallel multilevel k-way partitioning for irregular graphs},
   journal = {SIAM Review},
   volume = {41},
   number = {2},
   pages = {278-300},
   year = {1999}
}

@inproceedings{Koyuturk03,
   author = {M. Koyuturk and A. Grama},
   title = {{PROXIMUS}: {A} Framework for Analyzing Very High Dimensional Discrete-Attributed Datasets},
   booktitle = {Proc. Ninth ACM SIGKDD Intl. Conf. on Knowledge Discovery and Data Mining (KDD 2003)},
   pages = {147-156},
   year = {2003}
}

@inproceedings{Neville03Learning,
   author = {J. Neville and D. Jensen and L. Friedland and M. Hay},
   title = {Learning Relational Probability Trees},
   booktitle = {Proceedings of the 9th ACM SIGKDD International Conference on Knowledge Discovery and Data Mining},
   address = {Washington, DC},
   year = {2003}
}

@book{Dzeroski01Relational,
   author = {Saso Dzeroski and Nada Lavrac},
   title = {Relational Data Mining},
   publisher = {Springer, Berlin},
   year = {2001}
}

@book{Wasserman94Social,
   author = {Stanley Wasserman and Katherine Faust and Dawn Iacobucci},
   title = {Social Network Analysis : Methods and Applications},
   publisher = {Cambridge University Press},
   month = nov,
   year = {1994}
}

@article{Jeh03Scaling,
   author = {Glen Jeh and Jennifer Widom},
   title = {Scaling personalized web search},
   journal = {WWW},
   pages = {271-279},
   year = {2003}
}

@article{Chandra89Electrical,
   author = {Ashok K. Chandra and Prabhakar Raghavan and Walter L. Ruzzo and Roman Smolensky and Prasoon Tiwari},
   title = {The Electrical Resistance of a Graph Captures its Commute and Cover Times},
   journal = {STOC},
   pages = {574-586},
   year = {1989}
}

@article{Haveliwala02Topic,
   author = {Taher H. Haveliwala},
   title = {Topic-sensitive PageRank},
   journal = {WWW},
   pages = {517-526},
   year = {2002}
}

@article{Kempe03Maximizing,
   author = {D. Kempe and J. Kleinberg and E. Tardos},
   title = {Maximizing the Spread of Influence through a Social Network},
   journal = {KDD},
   address = {Washington, DC},
   year = {2003}
}

@article{Domingos01Mining,
   author = {Pedro Domingos and Matt Richardson},
   title = {Mining the network value of customers},
   journal = {KDD},
   pages = {57-66},
   address = {San Francisco},
   year = {2001}
}

@article{Newman03structure,
   author = {M. E. J. Newman},
   title = {The structure and function of complex networks},
   journal = {SIAM Review},
   volume = {45},
   pages = {167-256},
   year = {2003}
}

@article{Dorogovtsev02Evolution,
   author = {S.N. Dorogovtsev and J.F.F. Mendes},
   title = {Evolution of networks},
   journal = {Advances in Physics},
   volume = {51},
   pages = {1079-1187},
   year = {2002}
}

@inproceedings{Brin98anatomy,
   author = {S. Brin and L. Page},
   title = {The anatomy of a large-scale hypertextual (Web) search engine.},
   booktitle = {Proc. 7th International World Wide Web Conference (WWW7)/Computer Networks},
   pages = {107--117},
   year = {1998},
   note = {Published as Proc. 7th International World Wide Web Conference (WWW7)/Computer Networks, volume 30, number 1-7}
}

@article{Chen03,
   author = {H. Chen and J. Schroeder and R Hauck and L. Ridgeway and H. Atabakhsh and H. Gupta and C. Boarman and K. Rasmussen and A. Clements},
   title = {{COPLIN} COnnect: Information and knowledge management for law enforcement},
   journal = {CACM},
   volume = {46},
   number = {1},
   pages = {28-34},
   month = jan,
   year = {2003}
}

@book{Ramakrishnan02Database,
   author = {Raghu Ramakrishnan and Johannes Gehrke},
   title = {Database Management Systems},
   publisher = {McGraw-Hill},
   year = {2002},
   note = {3rd ed}
}

@book{Barabasi02Linked,
   author = {Albert-Laszlo Barabasi},
   title = {Linked: The New Science of Networks},
   edition = {1st},
   publisher = {Perseus Publishing},
   month = may,
   year = {2002}
}

@book{Mitchell97Machinea,
   author = {Tom Mitchell},
   title = {Machine Learning},
   publisher = {McGraw Hill},
   year = {1997}
}

@article{Vilalta02Predicting,
   author = {Ricardo Vilalta and Sheng Ma},
   title = {Predicting Rare Events In Temporal Domains},
   journal = {ICDM},
   pages = {474-481},
   address = {Maebashi City, Japan},
   year = {2002}
}

@article{Brodie03Active,
   author = {Mark Brodie and Irina Rish and Sheng Ma and Natalia Odintsova},
   title = {Active Probing Strategies for Problem Diagnosis in Distributed Systems},
   journal = {IJCAI},
   pages = {1337-1338},
   year = {2003}
}

@article{Milgram67Small,
   author = {Stanley Milgram},
   title = {The Small World Problem},
   journal = {Psychology Today},
   volume = {1},
   number = {1},
   pages = {60-67},
   year = {1967}
}

@article{Watts98Collective,
   author = {Duncan J. Watts and Steven H. Strogatz},
   title = {Collective Dynamics of 'Small-World' Networks.},
   journal = {Nature},
   number = {393},
   pages = {440-442},
   year = {1998}
}

@article{Vilalta02Predictinga,
   author = {Ricardo Vilalta and Sheng Ma},
   title = {Predicting Rare Events In Temporal Domains},
   journal = {ICDM},
   pages = {474-481},
   address = {Maebashi City, Japan},
   year = {2002}
}

@article{Brodie03Activea,
   author = {Mark Brodie and Irina Rish and Sheng Ma and Natalia Odintsova},
   title = {Active Probing Strategies for Problem Diagnosis in Distributed Systems},
   journal = {IJCAI},
   pages = {1337-1338},
   year = {2003}
}

@article{Milgram67Smalla,
   author = {Stanley Milgram},
   title = {The Small World Problem},
   journal = {Psychology Today},
   volume = {1},
   number = {1},
   pages = {60-67},
   year = {1967}
}

@article{Watts98Collectivea,
   author = {Duncan J. Watts and Steven H. Strogatz},
   title = {Collective Dynamics of 'Small-World' Networks.},
   journal = {Nature},
   number = {393},
   pages = {440-442},
   year = {1998}
}

@inproceedings{Pan04AutoSplit,
   author = {Jia-Yu Pan and Hiroyuki Kitagawa and Christos Faloutsos and Masafumi Hamamoto},
   title = {AutoSplit: Fast and Scalable Discovery of Hidden Variables in Stream and Multimedia Databases.},
   booktitle = {Proc. of the The Eighth Pacific-Asia Conference on Knowledge Discovery and Data Mining (PAKDD 2004),},
   address = {Sydney, Australia},
   month = {May 26-28},
   year = {2004},
   note = {(Best Student Paper Award)}
}

@inproceedings{Manjhi05Finding,
   author = {Amit Manjhi and Vladislav Shkapenyuk and Kedar Dhamdhere and Christopher Olston},
   title = {Finding (Recently) Frequent Items in Distributed Data Streams},
   booktitle = {Proc. of the IEEE International Conference on Data Engineering},
   address = {Tokyo, Japan},
   year = {2005}
}

@inproceedings{McWherter05Improving,
   author = {David T. McWherter and Bianca Schroeder and Anastassia Ailamaki and Mor Harchol-Balter},
   title = {Improving Preemptive Prioritization via Statistical Characterization of {OLTP} Locking},
   booktitle = {Proc. of the IEEE International Conference on Data Engineering},
   address = {Tokyo, Japan},
   year = {2005}
}

@article{Wang03Epidemic,
   author = {Yang Wang and Deepayan Chakrabarti and Chenxi Wang and Christos Faloutsos},
   title = {Epidemic Spreading in Real Networks: An Eigenvalue Viewpoint},
   journal = {SRDS},
   address = {Florence, Italy},
   year = {2003}
}

@inproceedings{Vasilescu02Multilinear,
   author = {M. Alex O. Vasilescu and Demetri Terzopoulos},
   title = {Multilinear Image Analysis for Facial Recognition},
   booktitle = {International Conference on Pattern Recognition (ICPR'02)},
   volume = {Volume 2},
   address = {Quebec City, QC, Canada},
   year = {2002}
}

@book{Sole00Signs,
   author = {Ricard Sole and Brian Goodwin},
   title = {Signs of Life: How Complexity Pervades Biology},
   publisher = {Perseus Books Group},
   address = {New York, NY},
   year = {2000}
}

@inproceedings{Rui98Human,
   author = {Yong Rui and Thomas S. Huang and Sharad Mehrotra},
   title = {Human Perception Subjectivity and Relevance Feedback in Multimedia Information Retrieval},
   booktitle = {Proceedings of IST and SPIE Storage and Retrieval of Image and Video Databases VI},
   address = {San Jose, CA},
   month = jan,
   year = {1998}
}

@inproceedings{Prokaew99Similarity,
   author = {Kriengkrai Prokaew and Sharad Mehrotra and Michael Ortega and Kaushik Chakrabarti},
   title = {Similarity Search Using Multiple Examples in {MARS}},
   booktitle = {1999 International Conference on Visual Information Systems},
   month = jun,
   year = {1999}
}

@article{Tong01Support,
   author = {Simon Tong and Edward Y. Chang},
   title = {Support vector machine active learning for image retrieval},
   journal = {ACM Multimedia},
   pages = {107-118},
   address = {Ottawa, ON},
   year = {2001}
}

@book{Castelli02Image,
   author = {V. Castelli and L.D. Bergman},
   title = {Image Databases: Search and Retrieval of Digital Imagery},
   publisher = {John Wiley and Sons},
   year = {2002}
}

@article{Pan04Automatic,
   author = {Jia-Yu Pan and Hyung-Jeong Yang and Christos Faloutsos and Pinar Duygulu},
   title = {Automatic Multimedia Cross-modal Correlation Discovery},
   journal = {ACM SIGKDD},
   address = {Seattle, WA},
   month = aug,
   year = {2004}
}

@article{Mehta96,
   author = {Manish Mehta and Rakesh Agrawal and Jorma Rissanen},
   title = {{SLIQ}: {A} Fast Scalable Classifier for Data Mining},
   journal = {EDBT},
   pages = {18-32},
   year = {1996}
}

@article{Gonnet81Expected,
   author = {Gaston H. Gonnet},
   title = {Expected Length of the Longest Probe Sequence in Hash Code Searching},
   journal = {JACM},
   volume = {28},
   pages = {289--304},
   year = {1981}
}

@article{Yang94Examplea,
   author = {Yiming Yang and Christopher G. Chute},
   title = {An Example-Based Mapping Method for Text Categorization and Retrieval},
   journal = {ACM TOIS},
   volume = {12},
   pages = {281--295},
   year = {1994}
}

@article{Shafer96,
   author = {John C. Shafer and Rakesh Agrawal and Manish Mehta},
   title = {{SPRINT}: {A} Scalable Parallel Classifier for Data Mining},
   journal = {VLDB'96},
   pages = {544-555},
   address = {Mumbai (Bombay), India},
   month = {September 3-6},
   year = {1996}
}

@article{Gehrke98RainForest,
   author = {Johannes Gehrke and and Raghu Ramakrishnan and and Venkatesh Ganti},
   title = {RainForest - {A} Framework for Fast Decision Tree Construction of Large Datasets},
   journal = {VLDB'98},
   pages = {416-427},
   address = {New York City, New York USA},
   month = {August 24-27},
   year = {1998}
}

@article{Huang04From,
   author = {K. Huang and R. F. Murphy},
   title = {From Quantitative Microscopy to Automated Image Understanding},
   journal = {J. Biomed. Optics},
   volume = {9},
   pages = {893-912},
   year = {2004}
}

@misc{Bhattacharya05ViVos,
   author = {Arnab Bhattacharya and Vebjorn Ljosa and Jia-Yu Pan and Mark R. Verardo and Hyungjeong Yang and Christos Faloutsos and Ambuj K. Singh},
   title = {ViVos: Visual Vocabulary Construction for Mining Biomedical Images},
   year = {2005},
   note = {under review}
}

@book{Baeza-Yates99Modern,
   author = {Ricardo Baeza-Yates and Berthier Ribeiro-Neto},
   title = {Modern Information Retrieval},
   publisher = {Addison Wesley Longman},
   year = {1999}
}

@inproceedings{Jain04Adaptive,
   author = {Ankur Jain and Edward Y. Chang and Yuan-Fang Wang},
   title = {Adaptive Stream Resource Management Using Kalman Filters},
   booktitle = {SIGMOD Conference},
   pages = {11-22},
   year = {2004}
}

@inproceedings{Tao04Prediction,
   author = {Yufei Tao and Christos Faloutsos and Dimitris Papadias and Bin Liu},
   title = {Prediction and Indexing of Moving Objects with Unknown Motion Patterns},
   booktitle = {SIGMOD Conference},
   pages = {611-622},
   address = {Paris, FR},
   year = {2004}
}

@inproceedings{Chakrabarti04Fully,
   author = {Deepayan Chakrabarti and Spiros Papadimitriou and Dharmendra S. Modha and Christos Faloutsos},
   title = {Fully automatic cross-associations},
   booktitle = {KDD Conference},
   pages = {79-88},
   address = {Seattle, WA},
   year = {2004}
}

@article{Kollios99Indexing,
   author = {George Kollios and Dimitrios Gunopulos and Vassilis J. Tsotras},
   title = {On Indexing Mobile Objects},
   journal = {PODS},
   pages = {261-272},
   year = {1999}
}

@inproceedings{Keogh04Towards,
   author = {Eamonn J. Keogh and Stefano Lonardi and Chotirat (Ann) Ratanamahatana},
   title = {Towards parameter-free data mining},
   booktitle = {KDD Conference},
   pages = {206-215},
   address = {Seattle, WA},
   year = {2004}
}

@article{Jacobs95Fast,
   author = {Charles E. Jacobs and Adam Finkelstein and David H. Salesin},
   title = {Fast Multiresolution Image Querying},
   journal = {ACM SIGGRAPH},
   pages = {277--286},
   address = {Los Angeles, CA},
   year = {1995}
}

@article{Ailamaki03environmental,
   author = {Anastassia Ailamaki and Christos Faloutsos and Paul S. Fischbeck and Mitchell J. Small and Jeanne VanBriesen},
   title = {An environmental sensor network to determine drinking water quality and security},
   journal = {SIGMOD Record},
   volume = {32},
   number = {4},
   pages = {47-52},
   year = {2003}
}

@book{Misslinger05Internet,
   author = {Stefan Misslinger},
   title = {Internet Worm Propagation},
   publisher = {Techn. Univ. of Munich},
   year = {2005}
}

@article{Papadimitriou05Streaming,
   author = {Spiros Papadimitriou and Jimeng Sun and Christos Faloutsos},
   title = {Streaming Pattern Discovery in Multiple Time-Series},
   journal = {VLDB},
   address = {Trondheim, Norway},
   year = {2005}
}

@article{Leskovec05Realistic,
   author = {Jure Leskovec and Deepayan Chakrabarti and Jon M. Kleinberg and Christos Faloutsos},
   title = {Realistic, Mathematically Tractable Graph Generation and Evolution, Using Kronecker Multiplication},
   journal = {PKDD},
   pages = {133-145},
   address = {Porto, Portugal},
   year = {2005}
}

@inproceedings{Leskovec05Graphs,
   author = {Jure Leskovec and Jon Kleinberg and Christos Faloutsos},
   title = {Graphs over time: densification laws, shrinking diameters and possible explanations},
   booktitle = {Proc. of ACM SIGKDD},
   pages = {177-187},
   publisher = {ACM Press},
   address = {Chicago, Illinois, USA},
   year = {2005}
}

@article{Faloutsos04Fast,
   author = {Christos Faloutsos and Andrew Tomkins and Kevin McCurley},
   title = {Fast Discovery of 'Connection Subgraphs'},
   journal = {KDD},
   address = {Seattle, WA},
   month = aug,
   year = {2004}
}

@article{Sun05Neighborhood,
   author = {Jimeng Sun and Huiming Qu and Deepayan Chakrabarti and Christos Faloutsos},
   title = {Neighborhood Formation and Anomaly Detection in Bipartite Graph},
   journal = {ICDM},
   address = {Houston, TX, U.S.A.},
   month = {November 27-30},
   year = {2005}
}

@article{Chung03Spectra,
   author = {Fan Chung and Linyuan Lu and Van Vu},
   title = {Spectra of random graphs with given expected degrees},
   journal = {PNA},
   volume = {100},
   number = {11},
   pages = {6313-6318},
   month = {May 27},
   year = {2003}
}

@inproceedings{Berger05Spread,
   author = {Noam Berger and Christian Borgs and Jennifer T. Hayes and Amin Saberi},
   title = {On the Spread of Viruses in the Internet},
   booktitle = {Proceedings of the 16th ACM-SIAM Symposium on Discrete Algorithm (SODA)},
   pages = {301-310},
   address = {Vancouver, BC, Canada},
   year = {2005}
}

@article{Briesemeister03Epidemic,
   author = {Linda Briesemeister and Patric Lincoln and Philip Porras},
   title = {Epidemic Profiles and Defense of Scale-Free Networks},
   journal = {WORM 2003},
   address = {Washington, DC},
   month = {Oct. 27},
   year = {2003}
}

@misc{Chung??Eigenvalues,
   author = {Fan Chung and Linyuan Lu and Van Vu},
   title = {Eigenvalues of Random Power Law Graphs},
   note = {unpublished manuscript}
}

@article{Equiluz02Epidemic,
   author = {Victor M. Equiluz and Konstantin Klemm},
   title = {Epidemic threshold in structured scale-free networks},
   journal = {arXiv:cond-mat/02055439},
   month = {May 21},
   year = {2002}
}

@article{Cohen03Efficient,
   author = {Reuven Cohen and Shlomo Havlin and Daniel ben-Avraham},
   title = {Efficient Immunization Strategies for Computer Networks and Populations},
   journal = {Physical Review Letters},
   volume = {91},
   number = {24},
   month = dec,
   year = {2003}
}

@article{Hayashi03Recoverable,
   author = {Yukio Hayashi and Masato Minoura and Jun Matsukubo},
   title = {Recoverable prevalence in growing scale-free networks and the effective immunization},
   journal = {arXiv:cond-mat/0305549 v2},
   month = {Aug. 6},
   year = {2003}
}

@article{Madar04Immunization,
   author = {Nilly Madar and Tomer Kalisky and Reuven Cohen and Daniel ben-Avraham and Shlomo Havlin},
   title = {Immunization and Epidemic Dynamics in Complex Networks},
   journal = {Eur. Phys. J. B},
   volume = {38},
   number = {2},
   pages = {269-276},
   year = {2004}
}

@article{Hethcote00mathematics,
   author = {Herbert W. Hethcote},
   title = {The mathematics of infectious diseases},
   journal = {SIAM Review},
   volume = {42},
   pages = {599-653},
   year = {2000}
}

@book{Hethcote84Gonorrhea,
   author = {Herbert W. Hethcote and James A. Yorke},
   title = {Gonorrhea Transmission Dynamics and Control},
   volume = {56},
   publisher = {Springer},
   month = dec,
   year = {1984},
   annote = {http://www.math.uiowa.edu/ftp/hethcote/lnb56.pdf},
   note = {Lecture Notes in Biomathematics}
}

@article{Chung03Eigenvalues,
   author = {Fan Chung and Lincoln Lu and Van Vu},
   title = {Eigenvalues of random power law graphs},
   journal = {Annals of Combinatorics},
   volume = {7},
   pages = {21-33},
   year = {2003}
}

@article{Sun06Beyond,
   author = {Jimeng Sun and Dacheng Tao and Christos Faloutsos},
   title = {Beyond streams and graphs: dynamic tensor analysis},
   journal = {KDD},
   pages = {374-383},
   address = {Philadelphia, PA, USA},
   year = {2006}
}

@article{Tong06Center,
   author = {Hanghang Tong and Christos Faloutsos},
   title = {Center-piece subgraphs: problem definition and fast solutions},
   journal = {KDD},
   pages = {404-413},
   address = {Philadelphia, PA, USA},
   year = {2006}
}

@article{Leskovec06Sampling,
   author = {Jure Leskovec and Christos Faloutsos},
   title = {Sampling from large graphs},
   journal = {KDD},
   pages = {631-636},
   address = {Philadelphia, PA, USA},
   year = {2006}
}

@article{Boehm06Robust,
   author = {Christian Boehm and Christos Faloutsos and Jia-Yu Pan and Claudia Plant},
   title = {Robust information-theoretic clustering},
   journal = {KDD},
   pages = {65-75},
   address = {Philadelphia, PA, USA},
   year = {2006}
}

@article{Pan06Automatic,
   author = {Jia-Yu Pan and Andre Guilherme Ribeiro Balan and Eric P. Xing and Agma J. M. Traina and Christos Faloutsos},
   title = {Automatic mining of fruit fly embryo images},
   journal = {KDD},
   pages = {693-698},
   address = {Philadelphia, PA, USA},
   year = {2006}
}

@article{Sun06Distributed,
   author = {Jimeng Sun and Spiros Papadimitriou and Christos Faloutsos},
   title = {Distributed Pattern Discovery in Multiple Streams},
   journal = {PAKDD},
   pages = {713-718},
   address = {Singapore},
   year = {2006}
}

@article{Chau06Detecting,
   author = {Duen Horng Chau and Shashank Pandit and Christos Faloutsos},
   title = {Detecting Fraudulent Personalities in Networks of Online Auctioneers},
   journal = {PKDD},
   pages = {103-114},
   address = {Berlin, Germany},
   year = {2006}
}

@inproceedings{Keim96Databases,
   author = {Daniel A. Keim},
   title = {Databases and Visualization},
   booktitle = {SIGMOD Conference},
   pages = {543},
   address = {Montreal, Canada},
   year = {1996},
   note = {tutorial}
}

@book{Tan05Introduction,
   author = {Pang-Ning Tan and Michael Steinbach and Vipin Kumar},
   title = {Introduction to Data Mining},
   publisher = {Addison-Wesley},
   year = {2005}
}

@article{Ellson01Graphviz,
   author = {John Ellson and Emden R. Gansner and Eleftherios Koutsofios and Stephen C. North and Gordon Woodhull},
   title = {Graphviz - Open Source Graph Drawing Tools},
   journal = {Graph Drawing},
   pages = {483-484},
   address = {Vienna, Austria},
   year = {2001}
}

@article{Rodrigues06GMine,
   author = {Rodrigues, Jr., Jose F. and Hanghang Tong and Agma J. M. Traina and Christos Faloutsos and Jure Leskovec},
   title = {GMine: a system for scalable, interactive graph visualization and mining},
   journal = {VLDB},
   pages = {1195 - 1198},
   address = {Seoul, Korea},
   year = {2006}
}

@inproceedings{Sakurai05,
   author = {Yasushi Sakurai and Spiros Papadimitriou and Christos Faloutsos},
   title = {{BRAID}: Stream Mining through Group Lag Correlations},
   booktitle = {SIGMOD Conference},
   pages = {599-610},
   address = {Baltimore, MD, USA},
   year = {2005}
}

@article{Sun07Less,
   author = {Jimeng Sun and Yinglian Xie and Hui Zhang and Christos Faloutsos},
   title = {Less is More: Compact Matrix Decomposition for Large Sparse Graphs},
   journal = {SDM 2007},
   address = {Minneapolis, MN, USA},
   month = apr,
   year = {2007}
}

@article{Karagiannis03User,
   author = {Thomas Karagiannis and Michalis Faloutsos and Mart Molle},
   title = {A User-Friendly Self-Similarity Analysis Tool.},
   journal = {Special Section on Tools and Technologies for Networking Research and Education, ACM SIGCOMM Computer Communication Review,},
   year = {2003}
}

@inproceedings{McGlohon07Finding,
   author = {M. McGlohon and J. Leskovec and C. Faloutsos and N. Glance and M. Hurst},
   title = {Finding patterns in blog shapes and blog evolution},
   booktitle = {International Conference on Weblogs and Social Media.},
   address = {Boulder, Colo.},
   month = mar,
   year = {2007}
}

@article{Sun07GraphScope,
   author = {Jimeng Sun and Spiros Papadimitriou and Philip S. Yu and Christos Faloutsos},
   title = {GraphScope: Parameter-Free Mining of Large Time-Evolving Graphs},
   journal = {KDD},
   address = {San Jose, CA, USA},
   month = aug,
   year = {2007}
}

@article{Chakrabarti07Epidemic,
   author = {Deepayan Chakrabarti and Yang Wang and Chenxi Wang and Jure Leskovec and Christos Faloutsos},
   title = {Epidemic Thresholds in Real Networks},
   journal = {ACM Transactions on Information and System Security (ACM TISSEC)},
   volume = {10},
   number = {4},
   year = {2007}
}

@book{May07Theoretical,
   author = {Robert May and Angela McLean},
   title = {Theoretical Ecology: Principles and Applications},
   edition = {3rd},
   publisher = {Oxford University Press},
   year = {2007},
   abstract = {ISBN-10: 0199209995 ISBN-13: 978-0199209996}
}

@book{Jackson91Perspectives,
   author = {E. Atlee Jackson},
   title = {Perspectives of nonlinear dynamics},
   publisher = {Cambridge University Press},
   year = {1991}
}

@article{Chakrabarti06Graph,
   author = {Deepayan Chakrabarti and Christos Faloutsos},
   title = {Graph mining: Laws, generators, and algorithms},
   journal = {ACM Comput. Surv.},
   volume = {38},
   number = {1},
   year = {2006}
}

@article{Akoglu08,
   author = {Leman Akoglu and Mary McGlohon and Christos Faloutsos},
   title = {{RTM}: Laws and a Recursive Generator for Weighted Time-Evolving Graphs},
   journal = {ICDM},
   address = {Pisa, Italy},
   year = {2008}
}

@article{Tsourakakis08Counting,
   author = {Charalampos Tsourakakis},
   title = {Counting of Triangles in Large Real Networks, without counting: Algorithms and Laws},
   journal = {ICDM},
   address = {Pisa, Italy},
   year = {2008}
}

@inproceedings{Gallagher08Using,
   author = {Brian Gallagher and Hanghang Tong and Tina Eliassi-Rad and Christos Faloutsos},
   title = {Using ghost edges for classification in sparsely labeled networks},
   booktitle = {Proceeding of the 14th ACM SIGKDD international conference on Knowledge discovery and data mining},
   pages = {256-264},
   publisher = {ACM},
   address = {Las Vegas, Nevada, USA},
   year = {2008},
   note = {http://doi.acm.org/10.1145/1401890.1401925}
}

@inproceedings{McGlohon08Weighted,
   author = {Mary McGlohon and Leman Akoglu and Christos Faloutsos},
   title = {Weighted graphs and disconnected components: patterns and a generator},
   booktitle = {Proceeding of the 14th ACM SIGKDD international conference on Knowledge discovery and data mining},
   pages = {524-532},
   publisher = {ACM},
   address = {Las Vegas, Nevada, USA},
   year = {2008},
   note = {http://doi.acm.org/10.1145/1401890.1401955}
}

@inproceedings{Seshadri08Mobile,
   author = {Mukund Seshadri and Sridhar Machiraju and Ashwin Sridharan and Jean Bolot and Christos Faloutsos and Jure Leskove},
   title = {Mobile call graphs: beyond power-law and lognormal distributions},
   booktitle = {Proceeding of the 14th ACM SIGKDD international conference on Knowledge discovery and data mining},
   pages = {596-604},
   publisher = {ACM},
   address = {Las Vegas, Nevada, USA},
   year = {2008},
   note = {http://doi.acm.org/10.1145/1401890.1401963}
}

@inproceedings{Tong08Colibri,
   author = {Hanghang Tong and Spiros Papadimitriou and Jimeng Sun and Philip S. Yu and Christos Faloutsos},
   title = {Colibri: fast mining of large static and dynamic graphs},
   booktitle = {Proceeding of the 14th ACM SIGKDD international conference on Knowledge discovery and data mining},
   pages = {686-694},
   publisher = {ACM},
   address = {Las Vegas, Nevada, USA},
   year = {2008},
   note = {http://doi.acm.org/10.1145/1401890.1401973}
}

@techreport{Ganger03Self,
   author = {Gregory R. Ganger and John D. Strunk and Andrew J. Klosterman},
   title = {Self-* Storage: Brick-based Storage with Automated Administration},
   institution = {Carnegie Mellon University},
   type = {Technical Report,},
   number = {CMU-CS-03-178,},
   month = aug,
   year = {2003}
}

@inproceedings{Leskovec07Cost,
   author = {Jure Leskovec and Andreas Krause and Carlos Guestrin and Christos Faloutsos and Jeanne VanBriesen and Natalie Glance},
   title = {Cost-effective outbreak detection in networks},
   booktitle = {Proceedings of the 13th ACM SIGKDD international conference on Knowledge discovery and data mining},
   pages = {420-429},
   publisher = {ACM},
   address = {San Jose, California, USA},
   year = {2007},
   note = {http://doi.acm.org/10.1145/1281192.1281239}
}

@article{Dean04MapReduce,
   author = {Jeffrey Dean and Sanjay Ghemawat},
   title = {MapReduce: Simplified Data Processing on Large Clusters},
   journal = {OSDI'04},
   address = {San Francisco, CA},
   month = dec,
   year = {2004}
}

@incollection{Das07Detecting,
   author = {Kaustav Das and Jeff Schneider},
   title = {Detecting anomalous records in categorical datasets},
   booktitle = {ACM SIGKDD},
   pages = {220-229},
   publisher = {ACM},
   address = {San Jose, California, USA},
   year = {2007}
}

@article{Ferlez08Monitoring,
   author = {Jure Ferlez and Christos Faloutsos and Jure Leskovec and Dunja Mladenic and Marko Grobelnik},
   title = {Monitoring Network Evolution using {MDL}},
   journal = {ICDE},
   year = {2008}
}

@article{Ferlez08Monitoringa,
   author = {Jure Ferlez and Christos Faloutsos and Jure Leskovec and Dunja Mladenic and Marko Grobelnik},
   title = {Monitoring Network Evolution using {MDL}},
   journal = {ICDE},
   address = {Cancun, Mexico},
   year = {2008}
}

@article{Prakash09,
   author = {B. Aditya Prakash and Nicholas Valler and David Andersen and Michalis Faloutsos and Christos Faloutsos},
   title = {{BGP}-lens: patterns and anomalies in internet routing updates},
   journal = {KDD},
   pages = {1315-1324},
   address = {Paris, France},
   year = {2009}
}

@article{Li09DynaMMo,
   author = {Lei Li and James McCann and Nancy S. Pollard and Christos Faloutsos},
   title = {DynaMMo: mining and summarization of coevolving sequences with missing values},
   journal = {KDD},
   pages = {507-516},
   address = {Paris, France},
   year = {2009}
}

@article{Du09Large,
   author = {Nan Du and Christos Faloutsos and Bai Wang and Leman Akoglu},
   title = {Large Human Communication Networks: Patterns and a Utility-Driven Generator.},
   journal = {KDD},
   address = {Paris, France},
   year = {2009}
}

@book{Garofalakis09Data,
   author = {Minos Garofalakis and Johannes Gehrke and Rajeev Rastogi},
   title = {Data Stream Management: Processing High-Speed Data Streams},
   publisher = {Springer},
   year = {2009}
}


\begin{thebibliography}{1}

\bibitem{juan2014beyond}
D.-C. Juan, L.~Li, H.-K. Peng, D.~Marculescu, and C.~Faloutsos.
\newblock Beyond poisson: Modeling inter-arrival time of requests in a
  datacenter.
\newblock In {\em Advances in Knowledge Discovery and Data Mining}, pages
  198--209. Springer, 2014.

\end{thebibliography}


\begin{thebibliography}{20}

\bibitem{2016costofdatadowntime}
2016 cost of data center outages report. 
\newblock https://goo.gl/OeNM4U.

\bibitem{GoogleClusterData}
Google cluster data - discussions, 2011.
\newblock https://groups.google.com/forum/\#!forum/googleclusterdata-discuss.

\bibitem{barroso2013datacenter}
L. A. Barroso, J. Clidaras, and U. H ¨ olzle. 
\newblock The datacenter as a computer: An introduction to the design of warehouse-scale machines. 
\newblock \em{Synthesis lectures on computer architecture}, 8(3):1–154, 2013.

\bibitem{bishop2006pattern}
C. Bishop. 
\newblock \em{Pattern Recognition and Machine Learning.}
\newblock Information Science and Statistics.
\newblock Springer, 2006.

\bibitem{box2015time}
G. E. Box, G. M. Jenkins, G. C. Reinsel, and G. M. Ljung. 
\newblock \em{Time series analysis: forecasting and control.}
\newblock John Wiley \& Sons, 2015.

\bibitem{breiman2001random}
L. Breiman. 
\newblock Random forests. 
\newblock \em{Machine learning}, 45(1):5–32, 2001.

\bibitem{chen2014failure}
X. Chen, C.-D. Lu, and K. Pattabiraman. 
\newblock Failure analysis of jobs in compute clouds: A google cluster case study. 
\newblock In \em{2014 IEEE 25th International Symposium on Software Reliability Engineering}, pages 167–177. IEEE, 2014.

\bibitem{guan2013adaptive} 
Q. Guan and S. Fu. 
\newblock Adaptive anomaly identification by exploring metric subspace in cloud computing infrastructures. 
\newblock In \em{Reliable Distributed Systems (SRDS), 2013 IEEE 32nd International Symposium on}, pages 205–214. IEEE, 2013.

\bibitem{juan2014beyond}
D.-C. Juan, L.~Li, H.-K. Peng, D.~Marculescu, and C.~Faloutsos.
\newblock Beyond poisson: Modeling inter-arrival time of requests in a
  datacenter.
\newblock In {\em Advances in Knowledge Discovery and Data Mining}, pages 198--209. Springer, 2014.

\bibitem{liu2012characterizing}
Z. Liu and S. Cho. 
\newblock Characterizing machines and workloads on a google cluster. 
\newblock In \em{2012 41st International Conference on Parallel Processing Workshops}, pages 397–403. IEEE, 2012.

\bibitem{miller2010terminating}
T. D. Miller and I. L. Crawford Jr. 
\newblock Terminating a non-clustered workload in response to a failure of a system with a clustered workload, Jan. 26 2010. 
\newblock US Patent 7,653,833.

\bibitem{powers2011evaluation}
D. M. Powers. 
\newblock Evaluation: from precision, recall and f-measure to roc, informedness, markedness and correlation. 2011.

\bibitem{reiss2012heterogeneity}
C. Reiss, A. Tumanov, G. R. Ganger, R. H. Katz, and M. A. Kozuch. 
\newblock Heterogeneity and dynamicity of clouds at scale: Google trace analysis. 
\newblock In \em{SOCC}, page 7. ACM, 2012.

\bibitem{clusterdata:Reiss2011}
C. Reiss, J. Wilkes, and J. L. Hellerstein. 
\newblock Google cluster-usage traces: format + schema.
\newblock Technical report, Google Inc., Mountain View, CA, USA, Nov. 2011. 
\newblock Revised 2014-11-17 for version 2.1. 
\newblock Posted at https://github.com/google/cluster-data.

\bibitem{scholkopf1997comparing}
B. Scholkopf, K.-K. Sung, C. J. Burges, F. Girosi, P. Niyogi, T. Poggio, and V. Vapnik.
\newblock Comparing support vector machines with gaussian kernels to radial basis function classifiers.
\newblock \em{IEEE transactions on Signal Processing}, 45(11):2758–2765, 1997.

\bibitem{Van-Rijsbergen79Information}
C. Van-Rijsbergen. 
\newblock \em{Information Retrieval.} 
\newblock Butterworths, London, England, 2nd edition, 1979.

\bibitem{verma2015large}
A. Verma, L. Pedrosa, M. Korupolu, D. Oppenheimer, E. Tune, and J. Wilkes. 
\newblock Large-scale cluster management at google with borg. 
\newblock In \em{Proceedings of the Tenth European Conference on Computer Systems}, page 18. ACM, 2015.

\end{thebibliography}


\end{document}